\documentclass[aps,prc,superscriptaddress,longbibliography, showpacs,floatfix,letterpaper,preprintnumbers,nofootinbib,notitlepage]{revtex4-1}
\usepackage[letterpaper, total={7in,9.5in}]{geometry}
\usepackage{subcaption, caption}
\usepackage[colorlinks=true, linkcolor=blue, citecolor=blue, urlcolor=blue]{hyperref}
\usepackage{amssymb}
\usepackage[title]{appendix}
\usepackage{color}
\usepackage{graphicx}
\usepackage{dcolumn}
\usepackage{bm}
\usepackage{amsmath,mathrsfs}
\usepackage{braket}
\usepackage{xfrac}
\usepackage{soul}
\usepackage{lineno}
\usepackage{slashed}
\usepackage{appendix}
\usepackage{tabularx,enumitem,paralist}
\newcolumntype{R}{>{\raggedleft\arraybackslash}X}%
\newcolumntype{I}[1]{>{\begin{compactitem}}p{#1}<{\end{compactitem}}}

\begin{document}



\title{Bremsstrahlung photon contributions to parton energy loss at high virtuality $Q^2$: a perturbative calculation at $\mathcal{O}(\alpha_s \alpha_{em})$}
\author{Amit Kumar}
\email[Corresponding author: ]{amit.kumar@uregina.ca}
\affiliation{Department of Physics, University of Regina, Regina, Saskatchewan S4S 0A2, Canada}

\author{Gojko Vujanovic}
\email[Corresponding author: ]{gojko.vujanovic@uregina.ca}
\affiliation{Department of Physics, University of Regina, Regina, Saskatchewan S4S 0A2, Canada}

\date{\today}

\begin{abstract}
In this work, real photon production scattering kernels from jet-medium interactions in the QCD medium are perturbatively calculated using the higher-twist (HT) formalism. Focus is given towards real photon production from a highly virtual (and highly energetic) quark, taking into account heavy-quark mass scales [\href{https://journals.aps.org/prc/abstract/10.1103/PhysRevC.94.054902}{Phys. Rev. C 94, 054902 (2016)}], fermion-boson conversion processes [\href{http://dx.doi.org/10.1016/j.nuclphysa.2007.06.009}{Nucl. Phys. A 793, 128 (2007)}], as well as coherence effects [\href{https://journals.aps.org/prc/abstract/10.1103/PhysRevC.105.024908}{Phys. Rev. C 105, 024908 (2022)}]. A generalized factorization procedure, such as that used in $e$-$A$ deep-inelastic scattering, is employed to derive an improved single-scattering medium-induced photon emission kernels that go beyond the traditional in-medium gluon exchange approximation. Diagrams with real-photon emission from the hard quark are classified based on the final-state particles, and include two types of scattering kernels at $\mathcal{O}(\alpha_{em}\alpha_{s})$ giving the following final states: (i) real photon and real quark, (ii) real photon and real gluon. The collisional kernels, thus derived, include full phase factors from all nonvanishing diagrams and complete second-order derivative terms in the transverse momentum gradient expansion. Moreover, the calculation includes heavy-quark mass effects, thus exploring heavy-quark energy loss. The in-medium parton distribution functions and the related jet transport coefficients have a hard transverse momentum dependence (of the emitted gluon or photon) present within the phase factor. It is observed that the jet transport coefficients resemble the transverse-momentum-dependent parton distribution functions. 
\end{abstract}

\maketitle

\section{Introduction}
Ultrarelativistic heavy-ions collisions performed at the Relativistic Heavy-Ion Collider (RHIC) and the Large Hadron Collider (LHC) produce a deconfined state of quarks and gluons, called quark-gluon plasma (QGP). One of the primary goals of these collisions is to constrain properties of QGP, through, e.g., the modifications it imparts on high-energy jets and photons. In past decades, many observables have been proposed to constrain the patron energy loss, including high-$p_{\rm T}$ hadrons \cite{PHENIX:2001hpc,PHENIX:2003djd,PHENIX:2003qdj,STAR:2002ggv,STAR:2003fka}, single-inclusive jets \cite{ATLAS:2018gwx,ALICE:2019qyj, CMS:2016uxf, CMS:2021vui}, $\gamma$-triggered jets \cite{ATLAS:2018dgb,ATLAS:2023iad,CMS:2017ehl}, $\gamma$-hadron correlation \cite{STAR:2016jdz,PHENIX:2020alr}, flow observables \cite{CMS:2017xgk,CMS:2012tqw,ATLAS:2018ezv} and so on. The challenge is to describe multiple observables simultaneously. The JETSCAPE framework \cite{Putschke:2019yrg,JETSCAPE:2019udz} has emerged as a unified and modular framework comprehensively and simultaneously studying multiple observables, allowing to obtain novel constraints on parton energy loss. Parton energy-loss models, such as MATTER \cite{JETSCAPE:2022jer}, LBT \cite{JETSCAPE:2022jer}, and MARTINI \cite{Park:2019sdn} are based on medium-induced gluon bremsstrahlung kernel and have been implemented in the JETSCAPE framework. However, a comprehensive implementation of the multi-scale dynamics responsible for medium-induced photon bremsstrahlung computed in parton energy loss simulations is still lacking. This paper focuses on providing a comprehensive calculation of photon production from highly virtual quarks of all flavors.

As the electromagnetic coupling is much smaller than the strong coupling, electromagnetic radiation can leave the QGP as soon as it is produced and with negligible rescattering, thus carrying detailed information on the QGP state at production time. So far, two approaches have been considered for the real-photon production. Perturbative calculations at low virtualities require extensive resummations, to account for infrared behavior composed of hard thermal loops \cite{Braaten:1991gm} as well as the Landau-Pomeranchuck-Migdal effect \cite{Arnold:2002zm,Arnold:2001ms}. Electromagnetic radiation from the QGP has shown a remarkable convergence when going from leading \cite{Arnold:2001ms} to next-to-leading order in perturbative QCD (pQCD) corrections \cite{Ghiglieri:2013gia}, an observation that is not universal across all QGP-related observables. Indeed, perturbative calculations of the transport coefficient $\hat{q}$ \cite{Caron-Huot:2008zna}, encapsulating transverse momentum broadening of the high-energy partons traversing the QGP, show significant corrections when comparing leading order and next-to-leading order pQCD calculations, at typical QGP temperature scales reached in heavy-ion collisions. A recent comparison of the electromagnetic spectral function in perturbative, i.e., next-to-leading order (NLO) pQCD calculations \cite{Jackson:2019yao}, and nonperturbative \cite{Ali:2024xae} approaches shows remarkable reliability of thermal photon perturbative calculations, especially when higher temperatures are considered. 

Given this behavior of perturbative calculations of electromagnetic radiation, this work focuses on extending photon production rates by including additional sources of jet-medium real photon production. To date, phenomenological calculations of jet-medium photons \cite{Schenke:2009gb,Yazdi:2022cuk} only include electromagnetic production from nearly on-shell light mass partons. Of course, other photon sources, such as prompt photons \cite{Gale:2021emg}, photons from the hydrodynamical evolution \cite{Paquet:2015lta}, and hadronic transport emissions \cite{Gotz:2021dco} have been considered. In this contribution, focus is given towards calculating real photon production rates from highly virtual partons, specifically obtaining a photon production rate from highly virtual quarks of light and heavy flavors, through the higher-twist formalism, thus giving an in-medium correction to prompt photon production.~\footnote{Note that the Arnold-Moore-Yaffe (AMY) formalism \cite{Arnold:2002zm,Arnold:2000dr,Arnold:2003zc}, upon which jet-medium photon simulations \cite{Schenke:2009gb,Yazdi:2022cuk} were devised, has yet to be extended to include the mass scales of heavy quark flavors. A similar statement holds true for the next-to-leading order extension of the AMY formalism \cite{Ghiglieri:2013gia}. In the higher-twist formalism, heavy-quark mass scales have been taken into account only in the context of gluon radiation from a massive quark \cite{Abir:2015hta}.} 
\begin{figure}[!h]
    \centering 
    \begin{subfigure}[t]{0.42\textwidth}
        \centering        \includegraphics[height=0.9in]{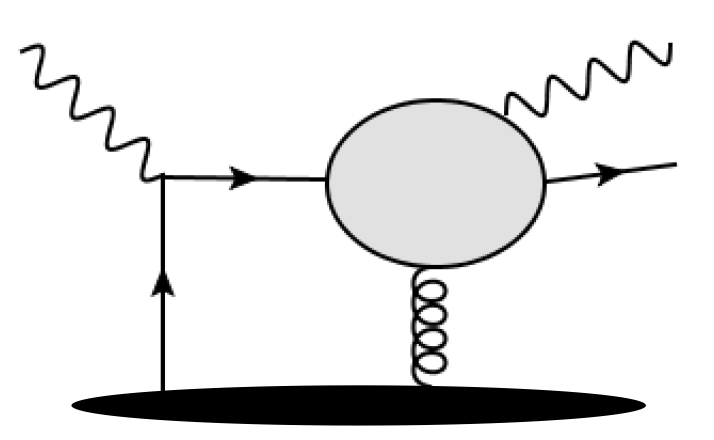}
        \caption{A photon and quark in the final state.}
    \end{subfigure}%
    \begin{subfigure}[t]{0.42\textwidth}
        \centering        \includegraphics[height=0.9in]{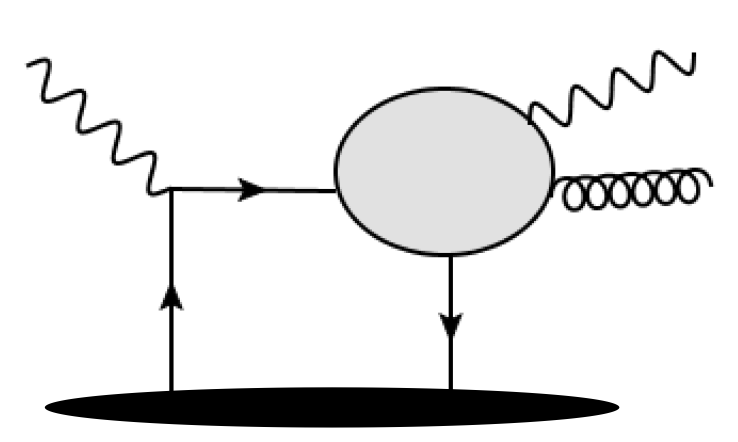}
        \caption{A photon and gluon in the final state.}
    \end{subfigure}
\caption{Single-scattering-induced photon emission processes in a DIS between a highly virtual photon and the nucleus at next-to-leading order (NLO).}\label{fig:HT_photons}
\end{figure}

Schematically, photon production investigated herein stems from processes illustrated in Fig.~\ref{fig:HT_photons}. These will be studied using the higher-twist (HT) formalism in the single-scattering-induced radiation limit. The processes depicted in the blob diagram Fig.~\ref{fig:HT_photons} (a) are referred to as kernel-1 and involve Glauber gluon exchange with the nuclear medium, whereas processes in Fig.~\ref{fig:HT_photons} (b) involve quark-to-gluon conversion processes~\cite{Ghiglieri:2015ala,Qin:2008rd}. In addition to including photon emissions from light and heavy quarks, attention will also be given towards phase coherence effects \cite{Sirimanna:2021sqx} that were recently included in the context of gluon emissions from highly virtual quarks. The calculation herein will extend the results in Refs.~\cite{Sirimanna:2021sqx,Schafer:2007xh} by including additional interactions with in-medium partons depicted in Fig.~\ref{fig:HT_photons} (b). These will be referred to as Kumar-Vujanovic (KV) kernels.

\section{Hadronic Tensor in Deep Inelastic Scattering}
\begin{figure}[!h]
\includegraphics[width=0.32\textwidth]{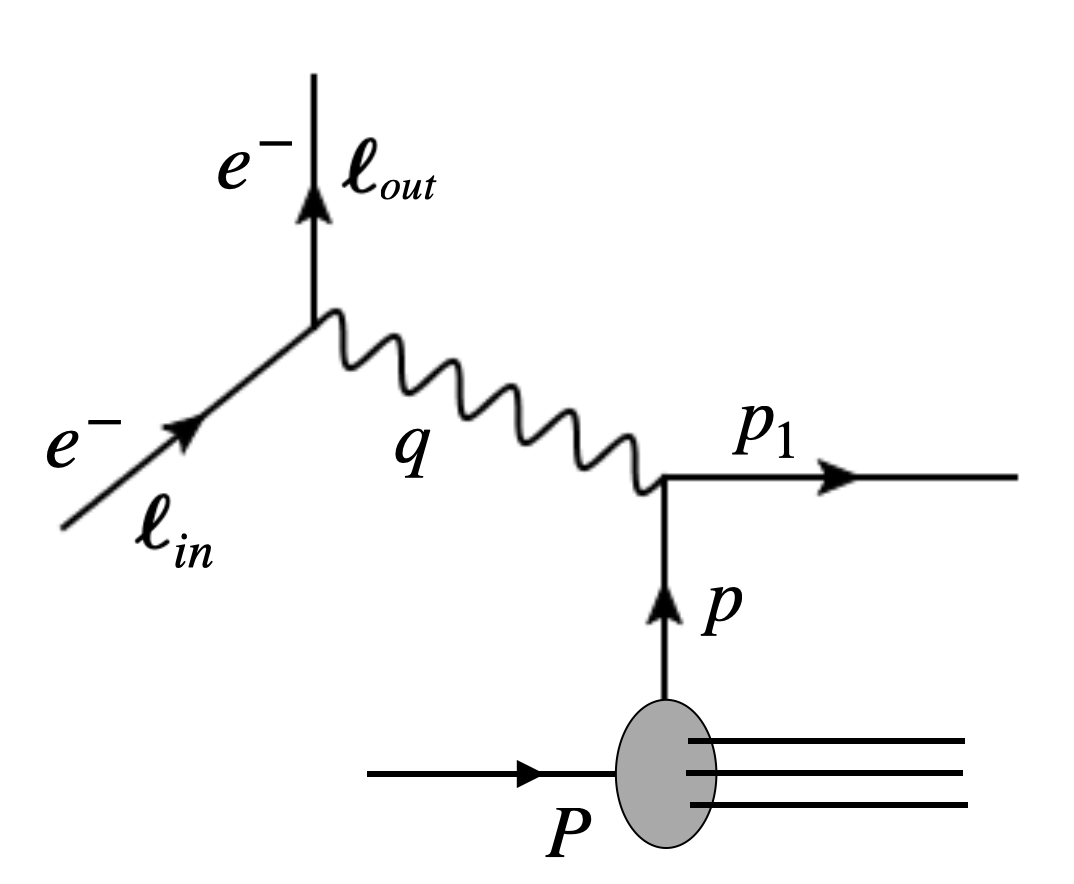}
\caption{A schematic diagram of deep-inelastic scattering between an electron and a nucleon inside the nucleus. The virtual photon carries momentum $q$, whereas the struck quark carries momentum $p$. The nucleus momentum is $\mathcal{P}=AP$, where $P$ is the momentum of the nucleon.}
\label{fig:DIS_eA_collision}
\end{figure}

The goal of this paper is to study the propagation of a hard quark through QCD nuclear matter and derive a medium-modified scattering kernel for single photon emission from the jet. It is studied in terms of the hadronic tensor ($W^{\mu\nu}$) in the context of deep-inelastic scattering (DIS) between the energetic electron and the nucleus carrying mass number $A$.  In general, the DIS between the electron and a large nucleus results in the production of a scattered electron and a remnant $X$ containing hadronic states. Its equation is given below:
\begin{equation}
e^-(\ell_{\rm in})+A(\mathcal{P}) \to e^-(\ell_{\rm out})+X. 
\label{eq:DIS_reaction}
\end{equation}
The difference of the outgoing $(\ell_{\rm out})$ and incoming $(\ell_{\rm in})$ electron momenta allows to define the virtual photon momentum $q^\mu$, as illustrated in Fig.~\ref{fig:DIS_eA_collision}, which is both highly energetic (i.e., a hard photon) and highly virtual. The study presented herein is carried out in the Breit frame, where the momentum of the virtual photon has the following form:
\begin{eqnarray}
q^\mu &\equiv& \ell^\mu_{\rm out}-\ell^\mu_{\rm in}\nonumber\\
      &=&\left[q^+,q^-,{\bf q}_\perp={\bf 0}_\perp\right] \nonumber\\
      &=&\left[\frac{-Q^2}{2q^-},q^-,0,0\right],
\end{eqnarray}
where $-Q^2=q^2=q^\mu q_\mu$ defines the virtuality of the photon. The connection between the light-cone coordinates and Cartesian coordinates is
\begin{eqnarray}
\left[q^+,q^-,{\bf q}_\perp\right]&=&\left[\frac{q^0+q^z}{\sqrt{2}},\frac{q^0-q^z}{\sqrt{2}},{\bf q}_\perp\right],
\end{eqnarray}
thus giving the expected $q^2=2q^{+}q^{-} - {\bf q}_\perp \cdot {\bf q}_\perp$. 

In Fig.~\ref{fig:DIS_eA_collision}, the incoming virtual photon can strike light- and heavy-quark flavors alike, both of which are considered herein. The nucleus momentum is labeled by $\mathcal{P}^\mu$, the average momentum of the nucleon momentum is $P^\mu$ such that $\mathcal P^\mu =AP^\mu$, while the quark momentum is $p^\mu$. The cross section for the process in Eq.(\ref{eq:DIS_reaction}) can be separated into a QED portion of the scattering and a QCD portion as
\begin{eqnarray}
\ell^0_{\rm out}\frac{d^3 \sigma}{d^3\ell_{\rm out} } = \frac{\alpha^2_{\rm em}}{2\pi s}\frac{L_{\mu\nu} W^{\mu\nu}_{\rm had} }{Q^4},
\label{eq:crossection_eX}
\end{eqnarray}
where the fine-structure constant is $\alpha_{\rm em}=e^2/(4\pi)$. In Eq.~\ref{eq:crossection_eX}, the interesting QCD portion is encoded in the hadronic tensor $W^{\mu\nu}_{\rm had}$. The label ``had" indicates that the final state consists of hadronic states. The rest of the expression consists of QED interaction and kinematics. More specifically, $\ell^0_{\rm out}$ is the energy of the outgoing electron, $s=\left(\mathcal{P}+\ell_{\rm in}\right)^2$ is the usual Mandelstam variable, the QED leptonic tensor is given by $L^{\mu\nu}=\frac{1}{2}{\rm Tr}\left[\slashed{\ell}_{\rm in}\gamma^\mu\slashed{\ell}_{\rm out}\gamma^\nu \right]$, while $\slashed{\ell}=\ell_\mu\gamma^\mu$ with $\gamma^\mu$ are the usual Dirac matrices. The hadronic tensor in Eq.~\ref{eq:crossection_eX} is defined as a complete matrix element given as \cite{Collins:2011zzd} 
\begin{eqnarray}
W^{\mu\nu}_{\rm had} = \sum_{X} \delta^{(4)}\left(p_{X}-\mathcal{P}-q\right) \left\langle AP, S\left| j^{\mu}(0) \right|X \right\rangle \left\langle X\left| j^{\nu}(0)\right| AP, S\right\rangle
\end{eqnarray}
where the sum over $X$ denotes the usual Lorentz-invariant sum and integral over all hadronic states. The QCD electromagnetic current is defined as $j^{\mu}(x) = \sum_{f} e_{f} \bar{\psi}_{f}(x) \gamma^{\mu} \psi_{f}(x)$ and $S$ is the spin state of the target. The fractional charge $e_f=2/3$ for up, charm, and top quarks, while it is $e_f=-1/3$ for down, strange, and bottom quarks. The factors of the electric unit charge $e$ have been collected, both from $L_{\mu\nu}$ and $W^{\mu\nu}_{\rm had}$, giving $\alpha^2_{\rm em}$ in Eq.~\ref{eq:crossection_eX}. The hard scale is given by $Q=\sqrt{-q^2} \gg \Lambda_{QCD}$ such that the $q^+$ and $q^-$ components of the incoming virtual momentum are large, i.e. $\mathcal{O}(1)$,  giving $q^\mu\sim[\mathcal{O}(1),\mathcal{O}(1),{\bf 0}_\perp]Q$. In this setup, the struck nucleon is traveling in positive $z$-direction and hence the struck quark has a very small $p^-\sim \lambda^2 Q$ momentum,  where the dimensionless parameter $\lambda$ is a small quantity $\lambda^2 \llless 1$, while the large component is $p^+\sim Q$, thus $p^\mu\sim[\mathcal{O}(1),\mathcal{O}(\lambda^2),{\bf 0}_\perp]Q$. The momentum components of the quark after the scattering are organized as $p^\mu_1\sim[\mathcal{O}(\lambda^2),\mathcal{O}(1),{\bf 0}_\perp]Q$.~\footnote{Note that, $p^\mu_1=\left[\frac{M^2-Q^2+2\left(p^{+}q^{-}-\frac{M^2}{2p^+}\frac{Q^2}{2q^-}\right)}{2p^{-}_1},p^-_1,{\bf 0}_\perp\right]$, where $M$ is the mass of the quark which is not neglected herein as $\frac{M}{Q}\sim\mathcal{O}(\lambda)$.} Thus, $\lambda$ is used to establish a perturbation series expansion.  

In Fig.~\ref{fig:DIS_eA_collision}, we depict the DIS process where the virtual photon strikes a quark inside a nucleon and flips the parton's direction.  We consider the next-to-leading-order (NLO) process consisting of a photon emission from the struck quark. We also consider processes where the struck quark undergoes single rescattering with the nuclear medium in addition to photon emission from the quark. To incorporate the real photon emission processes from the hard quark, we decompose the hadronic tensor ($W^{\mu\nu}$) as follows:  
\begin{eqnarray}
\frac{dW^{\mu\nu}}{dy}
&=& \frac{dW^{\mu\nu}_0}{dy}+\sum_{i=1,2} \frac{dW^{\mu\nu}_i}{dy}\nonumber\\
&=& \sum_{f}\int dx F^{A}_{_f}(x) \mathcal{H}^{\mu\nu}_0 \mathcal{K}_0 \nonumber\\
&+& \sum_{i=1,2} \sum_{f} \int dx F^{A}_{_f}(x) \mathcal{H}^{\mu\nu}_0 \mathcal{K}_i,\nonumber\\
\label{eq:W_total}
\end{eqnarray}
where $\frac{dW^{\mu\nu}_0}{dy}$ is the vacuum contribution to photon radiation from the jet, while the in-medium correction is encapsulated by $\sum_i \frac{dW^{\mu\nu}_i}{dy}$ and is depicted in Figs.~\ref{fig:HT_photons}.~\footnote{Virtual corrections at $\mathcal{O}(\alpha_{\rm em}\alpha_s)$ are included in Appendixes ~\ref{append:kernel3} and \ref{append:kernel4}.} The radiated photon ($\ell_2$) is traveling in the negative $z$ direction and is collinear to the direction of the final state quark; its momentum fraction is $y=\ell^{-}_{2}/q^-$. In Eq.~\ref{eq:W_total}, the first common factor for both vacuum and in-medium contributions is the parton distribution function (PDF)
\begin{eqnarray}
F^{A}_{f}(x)=A\int \frac{dy^-}{2\pi} \frac{e^{-ixP^+y^-}}{2} \langle P \vert \bar{\psi}_{_f} \left(y^-\right)\gamma^+\psi_{_f}(0) \vert P \rangle,
\end{eqnarray}
giving the probability of finding a quark of given flavor ($f$) with momentum fraction $x$ in the nucleus $A$ with which the virtual photon can collide. The momentum fraction $x$ carried by the struck quark is $x=p^+/P^+$, where $p^+$ is the first component of its momentum in light-cone coordinates, with $P^+$ being the corresponding momentum of the nucleon in the nucleus. $\langle P \vert \bar{\psi}_{f}\left(y^-\right)\gamma^+\psi_{f}(0) \vert P \rangle$ is a two-point fermionic correlator with a light-cone separation $y^-$ along negative $z$ direction.
The partonic hard part with no emission and no in-medium scattering is encoded in the function $\mathcal{H}^{\mu\nu}_0$ given by
\begin{eqnarray}
\mathcal{H}^{\mu\nu}_0=\frac{e^2_f}{2}(2\pi)\delta\left[\left(q+xP\right)^2\right]{\rm Tr}\left[\slashed{P}\gamma^\mu\left(\slashed{q}+x\slashed{P}\right)\gamma^\nu\right],
\end{eqnarray}
The goal of the subsequent sections is to explore the functions $\mathcal{K}_i$, where $\mathcal{K}_0$ describes vacuum contributions, while $\mathcal{K}_{i=1,2}$ correspond to the in-medium interactions described in Fig.~\ref{fig:HT_photons} (a) and (b), respectively. 
Many possible diagrams can affect photon radiation from the final state quark in a strongly interacting nuclear medium, each explored in a dedicated section. Therefore, our discussion is separated into two categories: hadronic tensor for the vacuum photon radiation and medium-modified photon emission.

\subsection{Single photon emission from the hard quark without in-medium scattering: The vacuum contribution}
\begin{figure}[!h]
    \centering
    \includegraphics[width=0.85\textwidth]{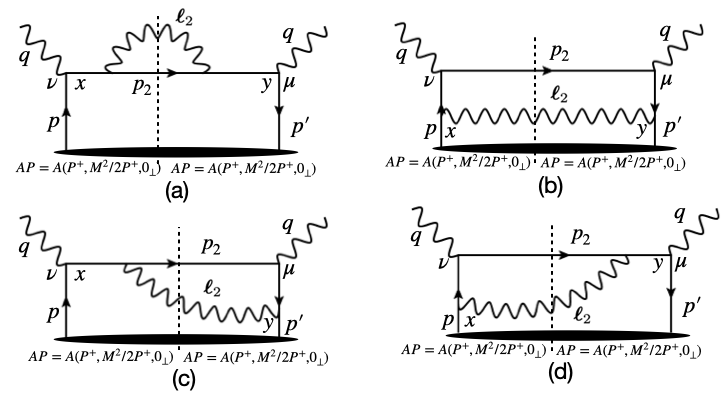}
    \caption{Forward scattering diagrams of leading order photon production from the quark. The cut-line (i.e., dashed line) represents the final state.}
    \label{fig:W_0_vacuum}
\end{figure}

For the case of single photon emission without in-medium scattering, there are a total of four diagrams. However, in the light-cone gauge $(A^-=0)$, the dominant contribution comes from the diagram in Figs.~\ref{fig:W_0_vacuum}(a), i.e., the emission of the photon from the final-state quark. The hadronic tensor for the diagrams shown in Fig.~\ref{fig:W_0_vacuum}(b), (c), and (d) does not give a leading log contribution; it is a subleading correction shown in Appendix \ref{append:wmunu_vacuum}. The hadronic tensor for the diagram shown in Fig.~\ref{fig:W_0_vacuum}(a) is given as
\begin{eqnarray}
\begin{split}
W^{\mu\nu}_{0,a} & =\sum_f 2 \left[-g^{\mu\nu}_{\perp\perp}\right] e^2 e^4_f  \int d (\Delta X^{-}) e^{iq^{+}\Delta X^{-}} \left\langle AP \left| \bar{\psi}_{_f}\left(\Delta X^-\right) \frac{\gamma^{+}}{4} \psi_{_f}(0)\right| AP\right\rangle\\
  & \times  \int  \frac{dy}{2\pi}\frac{d^2 \ell_{2\perp}}{(2\pi)^2} e^{ -i\left\{ \frac{\pmb{\ell}^2_{2\perp} + yM^2}{2y(1-y)q^-}\right\} \Delta X^{-} } \left[\frac{1+ \left(1-y\right)^2}{y}\right]   \frac{[\pmb{\ell}^2_{2\perp} +M^2 y^4 \kappa ]}{ \left[  \pmb{\ell}^2_{2\perp}  + y^2  M^2 \right]^2} ,
\end{split}\label{eq:W_0_vacuum_a}
\end{eqnarray}
where $M$ is the mass of the struck quark, while
\begin{eqnarray}
\kappa=\frac{1}{1+(1-y)^2}.
\label{eq:kappa}
\end{eqnarray}
The emitted photon ($\ell_2$) is traveling in the negative $z$ direction, having momentum fraction $y=\ell^{-}_2/q^-$. Note, in Eq.~\ref{eq:crossection_eX} and \ref{eq:W_0_vacuum_a}, one factor of $e^2$ has already been taken out from the hadronic tensor. 

Any calculation of $W^{\mu\nu}$ proceeds by first obtaining the full $T$-matrix amplitude $T^{\mu\nu}$ of a given process before extracting the forward scattering limit using
\begin{eqnarray}
W^{\mu\nu}&=&\frac{1}{2\pi}{\rm Disc}\left[T^{\mu\nu}\right],\nonumber\\
\label{eq:T_disc}
\end{eqnarray}
The various components of the momentum $\ell^\mu_2=\left[\frac{\ell^{2}_{2\perp}}{2yq^-},yq^-,\pmb{\ell}_{2\perp}\right]$ have different powers of the small scale $\lambda$, specifically $\ell^\mu_2\sim\left[\mathcal{O}(\lambda^2),\mathcal{O}(1),\mathcal{O}(\lambda),\mathcal{O}(\lambda) \right]Q$.~\footnote{The $\ell_2$ photon is on-shell $\ell^2_2=0$ after the imaginary part of the $T$-matrix amplitude has been taken via Eq.~\ref{eq:T_disc}.} The outgoing quark $p^\mu_2=\left[\frac{\ell^2_{2\perp}+M^2}{2\left(1-y\right)q^-},\left(1-y\right)q^-,-\pmb{\ell}_{2\perp}\right]$ scales as $p^\mu_2\sim [\mathcal{O}(\lambda^2),\mathcal{O}(1),\mathcal{O}(\lambda),\mathcal{O}(\lambda)]Q$ and takes into account that $p^2_2=M^2$. Having established the result in the vacuum, along with providing details about the size of different contributions relative to the scale $\lambda$, the manner in which the nuclear environment affects the vacuum result is considered next. 
\subsection{Classification of single-scattering-induced photon emission diagrams}
In this section, we consider DIS between the virtual photon and the nucleus in which the struck quark, after hard scattering, undergoes Bremsstrahlung photon radiation and in-medium scattering with the remainder of the nucleus. The in-medium scattering kernels for real photon emission from the quark that contribute at $\mathcal{O}$ ($\alpha_s \alpha_{em}$) are classified based on the identity of the particles in the final state. The first kind of kernel ($\mathcal{K}_{1}$) contains a real photon and a quark in the final state shown in Fig.~\ref{fig:HT_photons} (a). In this kernel, the hard quark undergoes real photon emission and in-medium Glauber gluon scattering. There are a total of eight possible diagrams for this process, and these are shown in Fig.~\ref{fig:SESS_diagram_photon_quark}. Their calculation is discussed in Sec. \ref{sec:photon_quark}. 

The second kind of kernel $(\mathcal{K}_{2})$ of interest consists of a real photon and real gluon emission with an in-medium Glauber quark exchange with the medium, as depicted in Fig.~\ref{fig:HT_photons} (b). There are a total of six possible central cut diagrams, shown in Fig.~\ref{fig:SESS_diagram_photon_gluon} contributing to this kernel, which are discussed in Sec. \ref{sec:photon_gluon}. 
\section{Single-scattering-induced emission: the single-photon single-quark final state}\label{sec:photon_quark}
\begin{figure}[!h]
    \centering
    \includegraphics[width=0.5\textwidth]{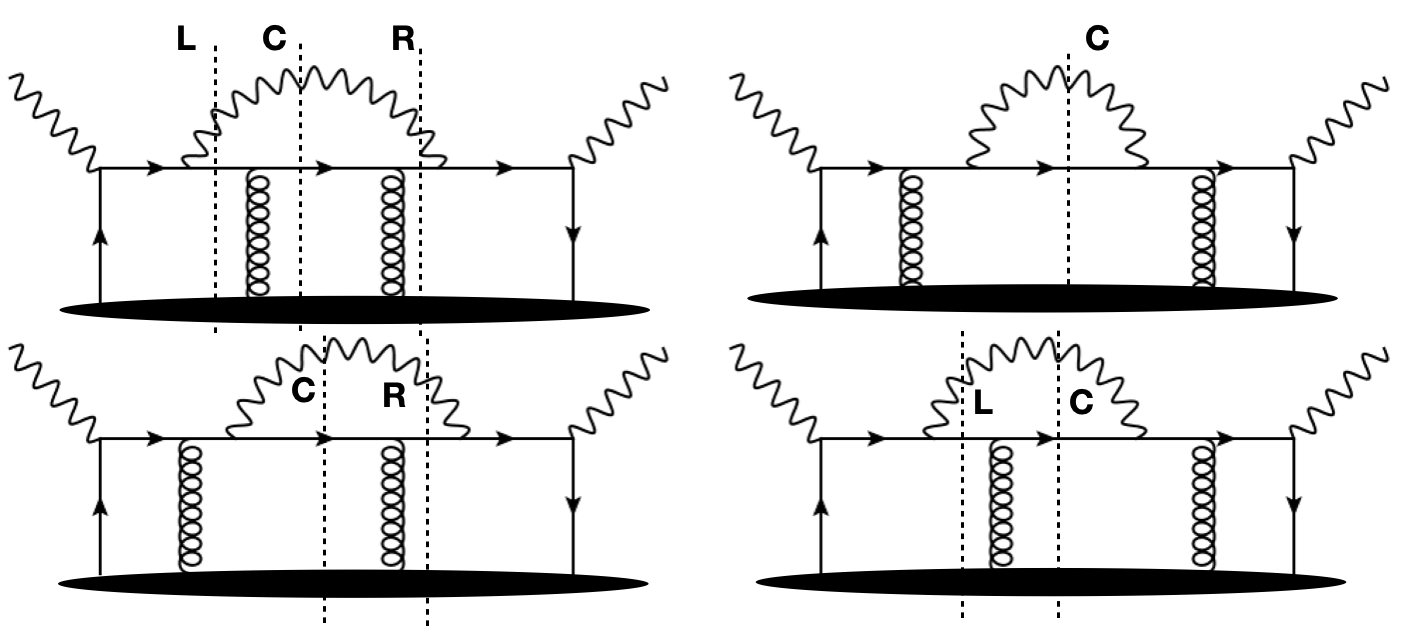}
    \caption{Forward scattering diagrams for single photon emission with a single Glauber gluon scattering, giving a final state consisting of a real photon and a quark. These diagrams contribute to kernel-1. The cut-lines L, C, R represent the left-cut, the center-cut, and the right-cut, respectively.  }
    \label{fig:SESS_diagram_photon_quark}
\end{figure}

Considered below are solely cases where the hard quark produced in the primary hard scattering undergoes a single photon emission and single Glauber gluon scattering with the nuclear medium, as depicted in Fig.~\ref{fig:SESS_diagram_photon_quark}. The (dashed) cut-line represents the final state and gives rise to a total of eight diagrams. The full analytic calculation considering all possible diagrams is now presented, including complete phase factors and quark-mass effects. We performed the calculation in light-cone gauge $n \cdot A =A^{-}=0$, where light-cone vector $n=[1,0,{\bf 0}_{\perp}]$, and the photon with four-momentum $X$ has a polarization tensor
\begin{eqnarray}
d^{(X)}_{\mu\nu}=-g_{\mu\nu}+\frac{X_{\mu}n_{\nu}+n_{\mu}X_{\nu}}{n\cdot X}.
\label{eq:d_X_sigma}
\end{eqnarray}
To illustrate the manner in which our results are obtained, we will present one calculation in great detail corresponding to the top left diagram in Fig.~\ref{fig:SESS_diagram_photon_quark}. All other diagrams are calculated in Appendix~\ref{append:kernel1}.

\begin{figure}[!h]
    \centering
    \includegraphics[width=0.45\textwidth]{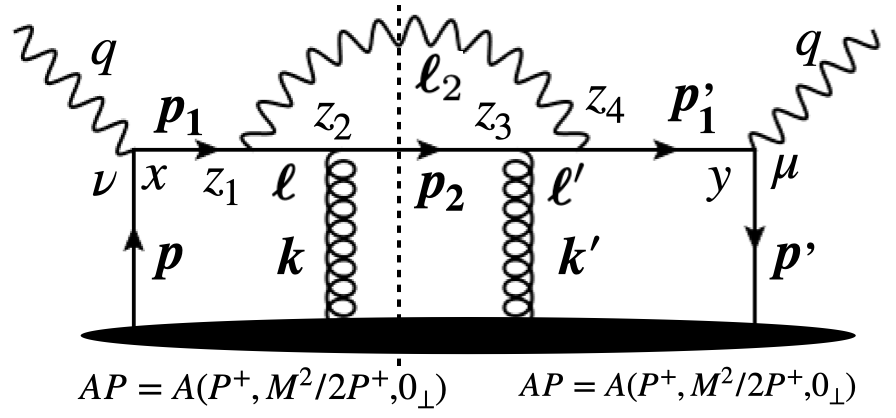}
    \caption{A single-scattering-induced photon emission process at next-to-leading order for kernel-1. }    
\label{fig:kernel1_ph_qqgm_qqgm_ph_central}
\end{figure}

The process shown in Fig.~\ref{fig:kernel1_ph_qqgm_qqgm_ph_central}, where the hard quark undergoes bremsstrahlung photon radiation followed by in-medium Glauber gluon scattering, is now considered.
The amplitude ($T^{\mu\nu}_{1,c}$) of the central-cut forward scattering diagram in Fig.~\ref{fig:kernel1_ph_qqgm_qqgm_ph_central} is
\begin{equation}
\begin{split}
T^{\mu\nu}_{1,c} & = \sum_{f} e^2 e^4_f g^2_s  \int d^{4}y d^{4}x d^{4} z_{1} d^{4} z_{2} d^{4}z_{3} d^{4}z_{4} e^{iq(y-x)} \langle AP| \bar{\psi}_{_f}(y) \gamma^\mu \int \frac{d^4p'_1}{(2\pi)^4} \frac{i(\slashed{p}'_1+M)}{\left[\left(p'_1\right)^2-M^2+i\epsilon\right]} e^{-ip'_1(y-z_4)} i\gamma^{\sigma_4} \\
& \times \int \frac{d^4\ell'}{(2\pi)^4} \frac{i\left(\slashed{\ell}'+M\right)}{\left[\left(\ell'\right)^2-M^2+i\epsilon\right]} e^{-i\ell'(z_4-z_3)} i\gamma^{\sigma_3} A_{\sigma_3}(z_3) \int \frac{d^4p_2}{(2\pi)^4} \frac{i(\slashed{p}_2+M)}{(p^2_2-M^2+i\epsilon)} e^{-ip_2\left(z_3-z_2\right)} i\gamma^{\sigma_2} A_{\sigma_2} (z_2)\\
& \times \int \frac{d^4\ell}{(2\pi)^4} \frac{i\left(\slashed{\ell}+M\right)}{\left(\ell^2-M^2+i\epsilon\right)} e^{-i\ell\left(z_2-z_1\right)} i\gamma^{\sigma_1} \int \frac{d^4p_1}{(2\pi)^4} \frac{i(\slashed{p}_1+M)}{\left(p^2_1-M^2+i\epsilon\right)} e^{-ip_1\left(z_1-x\right)} \gamma^\nu \psi_{_f}(x) | AP\rangle  \\
& \times \int \frac{d^4\ell_2}{(2\pi)^4} \frac{id^{(\ell_2)}_{\sigma_1\sigma_4}}{\left(\ell^2_2+i\epsilon\right)} e^{-i\ell_2\left(z_4 - z_1\right)},
\end{split}
\end{equation}
where $A_{\sigma}=t^{a}A^{a}_{\sigma}$ represents the gluonic gauge field with $t^a$ being the Gell-Mann SU(3) matrix. 
The above equation is rearranged below to better highlight the operator structure. That is,   
\begin{eqnarray}
\begin{split}
T^{\mu \nu}_{1,c} & = \sum_{f} e^2 e^4_f g^2_s \int d^4x\,d^4y\,d^4z_1\,  d^{4} z_{2}  d^{4} z_{3}  d^{4} z_{4}  \int \frac{d^{4} p_{1}}{(2 \pi)^{4}}  \frac{d^{4} p'_{1}}{(2 \pi)^{4}}  \frac{d^{4} \ell}{(2 \pi)^{4}}  \frac{d^{4} \ell'}{(2 \pi)^{4}}  \frac{d^{4} \ell_{2}}{(2 \pi)^{4}}  \frac{d^{4} p_{2}}{(2 \pi)^{4}} \\
& \times e^{iy\left(q-p'_{1}\right) }  e^{-ix\left(q-p_{1}\right) }  e^{iz_{1}(\ell - p_{1} + \ell_{2})} e^{iz_{2}(-\ell+p_{2})}  e^{iz_{3}(\ell' -p_{2})} e^{iz_{4}(p'_{1} -\ell' -\ell_{2})} \\
& \times \left\langle AP\left| \mathrm{Tr} \left[  \psi_{_f}(x) \bar{\psi}_{_f}(y) \gamma^{\mu} \frac{ i(\slashed{p}'_{1} +M) }{\left[\left(p'_{1}\right)^{2} -M^2 + i \epsilon\right]} i\gamma^{\sigma_{4}} \frac{i(\slashed{\ell}' +M)}{\left[\left(\ell'\right)^{2}-M^2+i\epsilon\right]} i\gamma^{\sigma_{3}} A_{\sigma_{3}}(z_{3}) \right.\right.\right. \\
& \left. \left.  \left. \times \frac{i(\slashed{p}_{2} +M)}{( p^{2}_{2} -M^2 + i \epsilon )} i\gamma^{\sigma_{2}}  A_{\sigma_{2}}(z_{2}) \frac{i(\slashed{\ell}+M) }{( \ell^{2} -M^2 + i \epsilon )} i\gamma^{\sigma_{1}} \frac{i(\slashed{p}_{1}+M)}{(p^{2}_{1} -M^2 + i \epsilon) } \gamma^{\nu} \right] \right| AP\right\rangle \times  \frac{id^{(\ell_2)}_{\sigma_4 \sigma_1}}{\left(\ell^2_2+i\epsilon\right)}.
\end{split}
\label{eq:T_quark_photon}
\end{eqnarray}
The trace ${\rm Tr}[...]$ in Eq.~\ref{eq:T_quark_photon} sums over the Dirac matrices and the Gell-Mann SU(3) color matrices. The notation $\sum_f$ represents the sum over quark flavors. To obtain the hadronic tensor, we apply Cutkosky's~\cite{Cutkosky:1960sp,Peskin:1995ev} procedure and evaluate the discontinuity along the cut-line, i.e., in the photon and quark propagators, namely
\begin{eqnarray}
\mathrm{Disc} \left[ \frac{1}{\ell^2_2 +i\epsilon}\right] &=& -2\pi i \delta\left(\ell^2_2\right),\nonumber\\
\mathrm{Disc}\left[\frac{1}{p^2_2-M^2+i\epsilon}\right] &=& -2\pi i\delta(p^2_2 -M^2).
\end{eqnarray}
The resulting hadronic tensor has the following form
\begin{eqnarray}
\begin{split}
W^{\mu \nu}_{1,c} & = \sum_{f} e^2 e^4_f g^2_s \int d^4x\,d^4y\,d^4z_1\,  d^{4} z_{2}  d^{4} z_{3}  d^{4} z_{4}  \int \frac{d^{4} p_{1}}{(2 \pi)^{4}}  \frac{d^{4} p'_{1}}{(2 \pi)^{4}}  \frac{d^{4} \ell}{(2 \pi)^{4}}  \frac{d^{4} \ell'}{(2 \pi)^{4}}  \frac{d^{4} \ell_{2}}{(2 \pi)^{4}}  \frac{d^{4} p_{2}}{(2 \pi)^{4}} \\
& \times e^{iy\left(q-p'_{1}\right) }  e^{-ix\left(q-p_{1}\right) }  e^{iz_{1}(\ell - p_{1} + \ell_{2})} e^{iz_{2}(-\ell+p_{2})}  e^{iz_{3}(\ell' -p_{2})} e^{iz_{4}(p'_{1} -\ell' -\ell_{2})} \\
& \times \left\langle AP\left| \mathrm{Tr} \left[  \psi_{_f}(x) \bar{\psi}_{_f}(y) \gamma^{\mu} \frac{ (-i)(\slashed{p}'_{1} +M) }{\left[\left(p'_{1}\right)^{2}-M^2-i\epsilon\right]} (-i)\gamma^{\sigma_{4}} \frac{(-i)(\slashed{\ell}' +M)}{\left[\left(\ell'\right)^{2} -M^2 - i\epsilon \right]} (-i)\gamma^{\sigma_{3}} A_{\sigma_{3}}(z_{3}) \right.\right.\right. \\
& \left. \left.  \left. \times (\slashed{p}_{2} +M) i\gamma^{\sigma_{2}}  A_{\sigma_{2}}(z_{2}) \frac{i(\slashed{\ell}+M) }{( \ell^{2} -M^2 + i \epsilon )} i\gamma^{\sigma_{1}} \frac{i(\slashed{p}_{1}+M)}{(p^{2}_{1} -M^2 + i \epsilon) } \gamma^{\nu} \right] \right| AP\right\rangle \\
& \times  d^{(\ell_2)}_{\sigma_4 \sigma_1} (2\pi)\delta(p^{2}_{2}-M^2) (2\pi)\delta(\ell^{2}_{2}) .
\end{split}
\label{eq:Wmunu_quark_photon_1}
\end{eqnarray}

In order to separate out the perturbative and non-perturbative portions of this calculation, a power-counting scheme is established. The incoming quark before primary scattering is moving in positive direction, i.e., $p=\left[p^+, M^2/2p^{+},\pmb{0}_{\perp}\right]$, and thus $p^\mu\sim\left[\mathcal{O}(1),\mathcal{O}(\lambda),\pmb{0}_\perp\right]Q$. The same $\lambda$-scales also hold for $p'$ since $p'^+/Q\sim 1$ and $p'=\left[p'^{+}, M^{2}/2p'^{+}, \pmb{0}_{\perp}\right]$. Isolating the leading nonperturbative component, which is $\psi(x) \bar{\psi}(y)$ for the first scattering correlator herein, allows to write $\psi(x)\bar{\psi}(y)$ in terms of a scalar function $T(x,y)$:
\begin{equation}
\psi(x) \bar{\psi}(y)=\slashed{p}T(x,y)=p^+\gamma^- T(x,y)\Longrightarrow {\rm Tr}[\gamma^+\psi(x)\bar{\psi}(y)]=p^+{\rm Tr}[\gamma^+\gamma^-] T(x,y) \Longrightarrow \psi(x)\bar{\psi}(y)=\gamma^-{\rm Tr}\left[\bar{\psi}(y)\frac{\gamma^+}{4}\psi(x)\right].
\label{eq:psi_psi_bar}
\end{equation}
In the light-cone gauge $A^{-}=0$, the Glauber gluon emanating from the medium has $A^+ \ggg A_\perp$, and thus $\gamma^{\sigma_3}A_{\sigma_3}(z_3)\approx\gamma^- A^+(z_3)$ and $\gamma^{\sigma_2}A_{\sigma_2}(z_2)\approx\gamma^- A^+(z_2)$. In addition, we assume that the hard quark produced from the primary hard scattering with the nucleon struck by the virtual photon, will undergo further re-scatterings while traversing the remaining $A-1$ nucleons. As $Q\ggg 1$, any scatterings following the first one are assumed independent, and thus the correlators are factorized:
\begin{equation}
\left\langle AP \left| {\rm Tr}\left[\bar{\psi}_{_f}(y)\frac{\gamma^+}{4}\psi_{_f}(x)\right]A^+(z_3)A^+(z_2)\right| AP \right\rangle \approx
\left\langle P \left| {\rm Tr}\left[\bar{\psi}_{_f}(y)\frac{\gamma^+}{4}\psi_{_f}(x)\right]\right| P \right\rangle \left\langle P_{A-1} \left| A^+(z_3)A^+(z_2)\right|P_{A-1}\right\rangle,
\label{eq:transport_coeff}
\end{equation}
where the first term will be absorbed in the definition of the nuclear parton distribution function, while the second term will be included in the scattering kernel. The resulting $W^{\mu\nu}$ is
\begin{eqnarray}
W^{\mu \nu}_{1,c} &=& \sum_{f} e^2 e^4_f g^2_s \int d^4x\,d^4y\,d^4z_1\,  d^{4} z_{2}  d^{4} z_{3}  d^{4} z_{4}  \int \frac{d^{4} p_{1}}{(2 \pi)^{4}}  \frac{d^{4} p'_{1}}{(2 \pi)^{4}}  \frac{d^{4} \ell}{(2 \pi)^{4}}  \frac{d^{4} \ell'}{(2 \pi)^{4}}  \frac{d^{4} \ell_{2}}{(2 \pi)^{4}}  \frac{d^{4} p_{2}}{(2 \pi)^{4}} e^{iy\left(q-p'_{1}\right) }  e^{-ix\left(q-p_{1}\right) }\nonumber\\
&\times&  \left\langle P \left| \bar{\psi}_{_f}(y) \frac{\gamma^{+}}{4} \psi_{_f}(x)\right| P\right\rangle e^{iz_{1}(\ell - p_{1} + \ell_{2})} e^{iz_{2}(-\ell+p_{2})}  e^{iz_{3}(\ell' -p_{2})} \langle P_{A-1} | {\rm Tr}[A^{+}(z_3)A^+(z_2)] | P_{A-1} \rangle e^{iz_{4}(p'_{1} -\ell' -\ell_{2})} \nonumber\\
&\times&  \mathrm{Tr} \left[\gamma^{-} \gamma^{\mu} \frac{ (\slashed{p}'_{1} +M) }{\left[\left(p'_{1}\right)^{2}-M^2-i\epsilon\right]} \gamma^{\sigma_{4}} \frac{(\slashed{\ell}' +M)}{\left[\left(\ell'\right)^{2} -M^2 - i\epsilon \right]} \gamma^{-}   (\slashed{p}_{2} +M) \gamma^{-}   \frac{(\slashed{\ell}+M) }{( \ell^{2} -M^2 + i \epsilon )} \gamma^{\sigma_{1}} \frac{(\slashed{p}_{1}+M)}{(p^{2}_{1} -M^2 + i \epsilon) } \gamma^{\nu} \right] \nonumber \\
&\times&   d^{(\ell_2)}_{\sigma_4 \sigma_1} (2\pi)\delta(p^{2}_{2}-M^2) (2\pi)\delta(\ell^{2}_{2}).
\label{eq:Wmu_quark_photon_12}
\end{eqnarray}
In Eq.~\ref{eq:Wmu_quark_photon_12}, the first trace ${\rm Tr}[A^{+}(z_3)A^{+}(z_2)]$ is over the Gell-Mann matrices, whereas the second trace (c.f. third line) is over the Dirac matrices. After performing the change of variable $p'_1=q+p'$ and $p_1=q+p$ in Eq.~\ref{eq:Wmu_quark_photon_12} (see also Fig.~\ref{fig:kernel1_ph_qqgm_qqgm_ph_central}), which stems from energy and momentum conservation, the integration measure $d^4 p'_1$ transforms as $d^4 p'_1\to d^4 p'$, while $d^4 p_1\to d^4 p$. Combining these results yields
\begin{eqnarray}
\begin{split}
W^{\mu \nu}_{1,c} & = \sum_{f} e^2 e^4_f g^2_s \int d^4x\,d^4y\,d^4z_1\,d^4z_2\,d^4z_3\,d^4z_4\,\int \frac{d^4p}{(2\pi)^4} \frac{d^4p'}{(2\pi)^4}\frac{d^4\ell}{(2\pi)^4}\frac{d^4\ell'}{(2\pi)^4}\frac{d^4\ell_2}{(2\pi)^4}\frac{d^4p_2}{(2\pi)^4} e^{-iyp'} e^{ixp} \\
& \times  \left\langle P \left| \bar{\psi}_{_f}(y) \frac{\gamma^{+}}{4} \psi_{_f}(x)\right| P\right\rangle e^{iz_1(\ell-q-p+\ell_2)} e^{iz_2(-\ell+p_2)} e^{iz_3(\ell'-p_2)} \langle P_{A-1} | {\rm Tr}[ A^{+}(z_3)A^+(z_2)] | P_{A-1} \rangle  e^{iz_4(p'+q-\ell'-\ell_2)} \\
& \times \mathrm{Tr} \left[\frac{\gamma^-\gamma^\mu(\slashed{q}+\slashed{p}'+M)}{\left[\left(q+p'\right)^2-M^2-i\epsilon\right]} \gamma^{\sigma_4} \frac{(\slashed{\ell}'+M)}{\left[\left(\ell'\right)^2-M^2-i\epsilon\right]} \gamma^-(\slashed{p}_2+M) \frac{\gamma^- (\slashed{\ell}+M) }{(\ell^2-M^2+i\epsilon)} \gamma^{\sigma_1} \frac{(\slashed{q}+\slashed{p}+M)}{\left[\left(q+p\right)^2-M^2+i\epsilon\right]} \gamma^{\nu} \right]\\
& \times  d^{(\ell_2)}_{\sigma_4 \sigma_1} (2\pi)\delta(p^{2}_{2}-M^2) (2\pi)\delta(\ell^{2}_{2}).
\end{split}
\label{eq:Wmunu_quark_photon_3}
\end{eqnarray}

Performing the integrals over $d^4z_1$ and $d^4z_4$ gives  
$(2 \pi)^{4} \delta^{(4)}\left(-q-p+\ell+\ell_{2}\right) (2 \pi)^{4} \delta^{(4)}\left(q+p'-\ell'-\ell_{2}\right)$,
which allows for the $d^4 \ell$ and $d^4 \ell'$ integration in Eq.~\ref{eq:Wmunu_quark_photon_3} to be performed, generating 
\begin{eqnarray}
\begin{split}
W^{\mu\nu}_{1,c} & = \sum_f e^2 e^4_f g^2_{s} \int d^4 x d^4 y d^4 z_{2} d^4 z_{3} \int \frac{d^4 p}{(2\pi)^4} \frac{d^4 p'}{(2\pi)^4}  \frac{d^4 \ell_{2}}{(2\pi)^4} \frac{d^4 p_{2}}{(2\pi)^4} e^{ip'(z_3-y)} e^{ip(x-z_2)}\left\langle P \left| \bar{\psi}_{_f}(y) \frac{\gamma^{+}}{4} \psi_{_f}(x)\right| P\right\rangle  \\
& \times e^{iz_3\left(q-p_{2}-\ell_{2}\right) } e^{iz_2(\ell_2 + p_2 - q)} \langle P_{A-1} | {\rm Tr}[ A^{+}(z_3)A^+(z_2)] | P_{A-1} \rangle (2 \pi) \delta\left(\ell_{2}^{2}\right) (2 \pi) \delta\left(p_{2}^{2}-M^{2}\right) d^{(\ell_2)}_{\sigma_{1} \sigma_{4}} \\
& \times \frac{{\rm Tr} \left[\gamma^{-} \gamma^{\mu} \left(\slashed{q}+\slashed{p}'+M\right) \gamma^{\sigma_{4}}\left(\slashed{q} +\slashed{p}' - \slashed{\ell}_2 + M\right)\gamma^-  \left(\slashed{p}_2+M\right) \gamma^-\left(\slashed{q}+\slashed{p}-\slashed{\ell}_2+M\right)\gamma^{\sigma_1} \left(\slashed{q}+\slashed{p}+M\right)\gamma^{\nu} \right]}{\left[\left(q+p'\right)^{2}-M^2-i\epsilon\right] \left[\left(q+p'-\ell_2\right)^2 - M^2 - i\epsilon\right] \left[\left(q+p-\ell_2\right)^2 -M^2+i\epsilon\right]\left[\left(q+p\right)^2-M^2+i\epsilon\right]}.
\end{split}\label{eq:W_quark_photon_reduced}
\end{eqnarray}

The expression in Eq.~\ref{eq:W_quark_photon_reduced} becomes singular when the denominator of the quark propagator for $p_1$, $\ell$, $\ell'$ and $p'_1$ vanishes. Computing this integral is easiest in the complex plane of $p^{+}$ and $p'^{+}$, where both $p^{+}$ and $p'^{+}$ have two simple poles.~\footnote{One of the propagators takes the form 
\begin{eqnarray}
\left[\left(q+p\right)^{2} - M^{2} + i \epsilon\right]^{-1}&=&\left[2\left(q^{+}+p^+\right)\left(q^{-}+p^-\right)-\left\vert\pmb{q}_\perp+\pmb{p}_\perp\right\vert^2-M^2+i\epsilon\right]^{-1} 
\approx \left[2\left(q^{+}+p^+\right)q^{-}\left[1+\mathcal{O}\left(\lambda^2\right)\right]-M^2+i\epsilon\right]^{-1} \nonumber \\ 
\Rightarrow \left[\left(q+p\right)^{2} - M^{2} + i \epsilon\right]^{-1} &\approx& \frac{1}{2q^-}\left[q^{+}+p^{+}-\frac{M^2}{2q^-}+i\epsilon\right]^{-1} 
\end{eqnarray}
where the established power counting $p^-/q^-\sim \lambda^2$ together with $\pmb{p}_\perp=\pmb{0}_\perp$ was used to simplify the full propagator to the expression above. A similar procedure is used for $\left[\left(q+p-\ell_{2}\right)^2-M^2+i\epsilon\right]$.
} The contour integration for $p^{+}$ can be carried out as
\begin{equation}
\begin{split}
C_{1} & = \oint \frac{dp^{+}}{(2\pi)} \frac{e^{ip^{+}\left(x^{-}-z^{-}_{2}\right)}}{\left[\left(q+p\right)^{2}-M^{2}+i\epsilon\right]\left[\left(q+p-\ell_{2}\right)^{2} -M^2 + i\epsilon\right]} \\
      & = \oint \frac{dp^{+}}{(2\pi)} \frac{e^{ip^{+}\left(x^{-} - z^{-}_{2}\right)}}{2q^{-}\left[q^{+}+p^{+} -\frac{M^2}{2q^{-}}+i\epsilon\right]2\left(q^{-}-\ell^{-}_{2}\right) \left[q^{+} + p^{+} -\ell^{+}_{2}-\frac{\pmb{\ell}^{2}_{2\perp}+M^2}{2\left(q^{-}-\ell^{-}_{2}\right)} + i\epsilon\right]}   \\
      & = \frac{(2\pi i)}{2\pi} \frac{\theta\left(x^{-} - z^{-}_{2}\right)}{4q^{-}(q^{-}-\ell^{-}_{2})} \left[\frac{e^{i\left(-q^{+} + \frac{M^2}{2q^-}\right)\left(x^{-}-z^{-}_2\right)}}{\left\{\frac{M^2}{2q^-} -\ell^+_2 -\frac{\pmb{\ell}^2_{2\perp}+M^2}{2\left(q^{-}-\ell^{-}_{2}\right)}\right\}}+
      \frac{e^{i\left(-q^{+}+\ell^+_2 + \frac{\pmb{\ell}^2_{2\perp}+M^2}{2\left(q^{-}-\ell^-_2\right)}\right)\left(x^{-}-z^-_2\right)}}{\left\{\ell^+_2 +\frac{\pmb{\ell}^2_{2\perp}+M^2}{2\left(q^{-}-\ell^-_2\right)}-\frac{M^2}{2q^-}\right\}}\right]   \\
      & = \frac{(2\pi i)}{2\pi} \frac{\theta(x^{-} - z^{-}_{2})}{4q^{-}(q^{-}-\ell^{-}_{2})}  e^{i\left(-q^{+}+\frac{M^2}{2q^-}\right)\left(x^{-}-z^-_2\right)}\left[ \frac{ -1 + e^{i \mathcal{G}^{(\ell_2)}_M\left(x^{-}-z^{-}_2\right)} }{ \mathcal{G}^{(\ell_2)}_M}  \right],
\end{split}
\label{eq:C_1_in_W1}
\end{equation}
where
\begin{equation}
\mathcal{G}^{(\ell_2)}_M = \ell^{+}_{2} + \frac{\pmb{\ell}^{2}_{2\perp}+M^2}{2(q^{-}-\ell^{-}_{2})} -\frac{M^2}{2q^{-}}. 
\end{equation}
The contour integration for $p'^+$ proceeds analogously, giving
\begin{equation}
\begin{split}
C_{2} & = \oint \frac{dp'^{+}}{(2\pi)} \frac{e^{-ip'^{+}\left(y^{-}-z^{-}_{3}\right)}}{\left[\left(q+p'\right)^{2} - M^{2} - i \epsilon\right]\left[\left(q+p'-\ell_{2}\right)^{2} -M^2 - i\epsilon \right]} \\
      & = \oint \frac{dp'^{+}}{(2\pi)} \frac{e^{-ip'^{+}\left(y^{-} - z^{-}_{3}\right) }}{2q^{-}\left[q^{+} + p'^{+} - \frac{M^2}{2q^{-}} - i \epsilon\right] 2\left(q^{-} -\ell^{-}_{2} \right)\left[q^{+} + p'^{+} -\ell^{+}_{2} - \frac{\pmb{\ell}^{2}_{2\perp}+M^2}{2\left(q^{-}-\ell^{-}_{2}\right)} - i\epsilon\right]}  \\
      & = \frac{(-2\pi i)}{2\pi} \frac{\theta\left(y^{-} - z^{-}_{3}\right)}{4q^{-}(q^{-}-\ell^{-}_{2})} e^{-i\left(-q^{+} + \frac{M^2}{2q^-}\right)\left(y^{-}-z^-_3\right)}\left[\frac{ -1 + e^{-i\mathcal{G}^{(\ell_2)}_M\left(y^{-}-z^-_3\right)}}{\mathcal{G}^{(\ell_2)}_M}\right].
\end{split}
\label{eq:}
\end{equation}
As the final expression for $C_1$ and $C_2$ is independent of $p$ and $p'$, respectively, the dependence on these variables in Eq.~\ref{eq:W_quark_photon_reduced} remains within $e^{ip(x-z_2)}$ and $e^{ip'(z_3-y)}$ as well as the trace over $\gamma$-matrices. While our $\lambda$-power counting scheme constrains the size of momentum variables, the same cannot be said about position variables. Thus, the $e^{ip(x-z_2)}$ and $e^{ip'(z_3-y)}$ phase factors must remain intact. As the trace in Eq.~\ref{eq:W_quark_photon_reduced} only contributes at $\mathcal{O}(\lambda^2)$ in $p$ and $p'$, the only non-trivial contribution remaining to the $p$ and $p'$ integrals stems solely from $e^{ip(x-z_2)}$ and $e^{ip'(z_3-y)}$ phase factors. To perform the remaining integrals for $p$ and $p'$, the following substitutions are used $p=\left[p^+,p^{-},\pmb{0}_\perp\right]=\left[p^+,\frac{M^2}{2p^{+}}+ \delta p^-, \pmb{0}_{\perp}\right]$ and $p'=\left[p'^{+},p'^{-},\pmb{0}_\perp\right] = \left[p'^{+},\frac{M^2}{2p'^{+}} +\delta p'^{-},\pmb{0}_{\perp}\right]$, where $\delta p^-\sim \mathcal{O}(\lambda^2)$ and $\delta p'^-\sim \mathcal{O}(\lambda^2)$. Thus, the integrals over $dp^- d^2 p_{\perp} d p'^{-} d^2p'_{\perp}$ simply become integrals over $d(\delta p^-) d^2 p_{\perp} d(\delta p'^{-}) d^2p'_{\perp}$ yielding 
\begin{equation}
(2\pi)^3 \delta\left(x^{+}-z^+_2\right) \delta^2\left(\pmb{x}_\perp - \pmb{z}_{2\perp}\right) (2\pi)^3\delta\left(-y^{+} + z^+_3\right)\delta^2\left(-\pmb{y}_\perp+\pmb{z}_{3\perp}\right).
\label{eq:spacetime_delta}
\end{equation}
Performing the integral over spacetime variables $(x^{+}, \pmb{x}_{\perp})$ and $(y^+,\pmb{y}_{\perp})$ using $\delta$-functions in Eq.~\ref{eq:spacetime_delta} yields
\begin{equation}
\begin{split}
W^{\mu\nu}_{1,c} & = \sum_f e^2 e^4_f g^2_{s} \int dx^-\,dy^-\,d^4z_2\,d^4z_3 \int \frac{d^4 \ell_{2}}{(2\pi)^4} \frac{d^4 p_{2}}{(2\pi)^4} \left\langle P \left| \bar{\psi}_{_f}\left(z^+_3,y^-,\pmb{z}_{3\perp}\right) \frac{\gamma^{+}}{4} \psi_{_f}\left(z^+_2,x^-,\pmb{z}_{2\perp}\right)\right| P\right\rangle  \\
& \times \frac{\theta(x^{-} - z^{-}_{2}) \theta(y^{-} - z^{-}_{3})}{ \left(2q^{-}\right)^2 \left[2(q^{-}-\ell^-_2)\right]^2} \left[\mathcal{G}^{(\ell_2)}_M\right]^{-2} e^{i\left(q^{+}-\frac{M^2}{2q^-}\right)\left(y^{-}-x^{-}-z^{-}_3+z^{-}_2\right)} \left[ -1 + e^{i\mathcal{G}^{(\ell_2)}_{M}(x^{-}-z^{-}_2)} \right] \left[-1 + e^{-i\mathcal{G}^{(\ell_2)}_{M}(y^{-}-z^{-}_3)} \right]\\
& \times e^{iz_3\left(q-p_{2}-\ell_{2}\right) } e^{iz_2(\ell_2 + p_2 - q)} \langle P_{A-1} | {\rm Tr}[ A^{+}(z_3)A^+(z_2)] | P_{A-1} \rangle (2 \pi) \delta\left(\ell_{2}^{2}\right) (2 \pi)  \delta\left(p_{2}^{2}-M^{2}\right) d^{(\ell_2)}_{\sigma_{1} \sigma_{4}} \\
& \times {\rm Tr} \left[\gamma^{-} \gamma^{\mu} \left(\slashed{q}+\slashed{p}'+M\right) \gamma^{\sigma_{4}}\left(\slashed{q} +\slashed{p}' - \slashed{\ell}_2 + M\right)\gamma^-  \left(\slashed{p}_2+M\right) \gamma^-\left(\slashed{q}+\slashed{p}-\slashed{\ell}_2+M\right)\gamma^{\sigma_1} \left(\slashed{q}+\slashed{p}+M\right)\gamma^{\nu} \right],
\end{split}
\label{eq:W_quark_photon_2}
\end{equation}
where $x^{+}=z^{+}_{2}$, $y^{+}=z^{+}_{3}$, $\pmb{x}_{\perp}=\pmb{z}_{2\perp}$, and $\pmb{y}_{\perp}=\pmb{z}_{3\perp}$ were used to simplify the expression for $W^{\mu\nu}$. Note that the trace inside $W^{\mu\nu}$ has been left intact owing to power counting. The next step is to perform the $d\ell^+_2$ as well as the $dp^+_2 dp^-_2$ integrals aided by the presence of the $\delta$-functions $\delta(\ell^{2}_{2})$ and $\delta(p^{2}_{2} - M^2)$ and $\lambda$-power counting. Indeed,
\begin{eqnarray}
\delta\left(\ell^2_2\right)\nonumber&=&\delta\left(2\ell^+_2\ell^-_2 -\pmb{\ell}^2_{2\perp}\right)=\frac{1}{2\ell^-_2} \delta\left(\ell^+_2-\frac{\pmb{\ell}^2_{2\perp}}{2\ell^-_2}\right)\nonumber\\
\delta(p^2_2 -M^2)&=&\delta\left(2p^+_2 p^-_2 -\pmb{p}^2_{2\perp}-M^2\right)=\frac{1}{2p^-_2}\delta\left(p^+_2 -\frac{\pmb{p}^2_{2\perp} +M^2}{2p^-_2} \right),
\label{eq:kernel1_deltafunction_cut-line}
\end{eqnarray}
which, when inserted in $W^{\mu\nu}$, gives 
\begin{equation}
\begin{split}
W^{\mu\nu}_{1,c} & = \sum_f e^2 e^4_f g^2_{s} \int dx^-\,dy^-\,d^4z_2\,d^4z_3 \int \frac{d^4 \ell_{2}}{(2\pi)^3} \frac{d^4 p_{2}}{(2\pi)^3} \left\langle P \left| \bar{\psi}_{_f}\left(z^+_3,y^-,\pmb{z}_{3\perp}\right) \frac{\gamma^{+}}{4} \psi_{_f}\left(z^+_2,x^-,\pmb{z}_{2\perp}\right)\right| P\right\rangle  \\
& \times \frac{\theta(x^{-} - z^{-}_{2}) \theta(y^{-} - z^{-}_{3})}{ \left(2q^{-}\right)^2 \left[2(q^{-}-\ell^-_2)\right]^2} \left[\mathcal{G}^{(\ell_2)}_M\right]^{-2} e^{i\left(q^{+}-\frac{M^2}{2q^-}\right)\left(y^{-}-x^{-}-z^{-}_3+z^{-}_2\right)} \left[ -1 + e^{i\mathcal{G}^{(\ell_2)}_{M}(x^{-}-z^{-}_2)} \right] \left[-1 + e^{-i\mathcal{G}^{(\ell_2)}_{M}(y^{-}-z^{-}_3)} \right]\\
& \times e^{i\left(p_2+\ell_2\right)\left(z_2-z_3\right) } e^{iq\left(z_3-z_2\right)} \frac{1}{2\ell^-_2} \delta\left(\ell^+_2-\frac{\pmb{\ell}^2_{2\perp}}{2\ell^-_2}\right) \frac{1}{2p^-_2}\delta\left(p^+_2 -\frac{\pmb{p}^2_{2\perp} +M^2}{2p^-_2} \right) \langle P_{A-1} | {\rm Tr}[A^{+}(z_3)A^+(z_2)] | P_{A-1} \rangle d^{(\ell_2)}_{\sigma_{1} \sigma_{4}}\\
& \times {\rm Tr} \left[\gamma^{-} \gamma^{\mu} \left(\slashed{q}+\slashed{p}'+M\right) \gamma^{\sigma_{4}}\left(\slashed{q} +\slashed{p}' - \slashed{\ell}_2 + M\right)\gamma^-  \left(\slashed{p}_2+M\right) \gamma^-\left(\slashed{q}+\slashed{p}-\slashed{\ell}_2+M\right)\gamma^{\sigma_1} \left(\slashed{q}+\slashed{p}+M\right)\gamma^{\nu} \right].
\end{split}\label{eq:W_quark_photon_2}
\end{equation}
Defining the momentum fraction $y$ as $\ell^{-}_{2}=yq^{-}$ allows one to rewrite $d\ell^{-}_{2} = q^{-} dy$. Furthermore, energy and momentum conservation in Fig.~\ref{fig:kernel1_ph_qqgm_qqgm_ph_central} implies that 
\begin{equation}
q+p=p_1=\ell_2+\ell=\ell_2+(p_2-k) \Longleftrightarrow  q+p-\ell_2-p_2+k=0.
\label{eq:lpq_4_mom_cons}
\end{equation}
While the $\delta$-functions can be used to perform the $\ell^+_2$ and $p^+_2$ integrals, $p^-_2$ can also be performed using $\lambda$-power counting. Indeed, as $k^\mu\sim\left[\mathcal{O}(\lambda^2),\mathcal{O}(\lambda^2),\mathcal{O}(\lambda),\mathcal{O}(\lambda)\right]Q$, while $\ell^\mu_2\sim\left[\mathcal{O}(\lambda^2),\mathcal{O}(1),\mathcal{O}(\lambda),\mathcal{O}(\lambda) \right]Q$ and $p^\mu_2\sim [\mathcal{O}(\lambda^2),\mathcal{O}(1),\mathcal{O}(\lambda),\mathcal{O}(\lambda)]Q$, using energy and momentum conservation implies
\begin{eqnarray}
0&=&q^{-}+p^{-}-\ell^{-}_{2}-p^{-}_{2} + k^{-} \nonumber\\
0&=&q^-+\mathcal{O}(\lambda^2)-\ell^-_2-p^-_2+\mathcal{O}(\lambda),
\label{eq:kernel1-l2_p2_k}
\end{eqnarray}
and thus the following change of variable $p^-_2=q^{-} - \ell^{-}_{2} + k^{-} + \delta p^{-}_{2}$, where $\delta p^-_2\sim\mathcal{O}(\lambda^2)$ is a small quantity, induces a change in the integration measure $dp^{-}_{2} = d(\delta p^{-}_{2})$. Thus, the integration over $dp^{-}_{2}$ yields a $\delta(z^{+}_{2} -z^{+}_{3})$, as the only function in Eq.~\ref{eq:W_quark_photon_2} that is not small is $e^{ip^{-}_{2}(z^+_3-z^+_2)}$, since $(z^+_3-z^+_2)$ is not subjet to the power counting in $\lambda$. Any other dependence on $p^-_2$ seen in Eq.~\ref{eq:W_quark_photon_2} can simply be set to $q^{-}-\ell^-_2+k^-$. Thus,
\begin{equation}
\begin{split}
W^{\mu\nu}_{1,c} & =\sum_f e^2 e^4_f g^2_{s} \int dx^-\,dy^-\,d^4z_2\,d^4z_3 \int \frac{dy d^2\ell_{2\perp}}{(2\pi)^3} \frac{d^2 p_{2\perp}}{(2\pi)^2}\delta\left(z^+_2-z^+_3\right) \left\langle P \left| \bar{\psi}_{_f}\left(z^+_3,y^-,\pmb{z}_{3\perp}\right) \frac{\gamma^{+}}{4} \psi_{_f}\left(z^+_2,x^-,\pmb{z}_{2\perp}\right)\right| P\right\rangle  \\
& \times \frac{\theta(x^{-} - z^{-}_{2}) \theta(y^{-} - z^{-}_{3})}{\left(2q^{-}\right)^2 \left[2q^{-}\left(1-y\right)\right]^2} \left[\mathcal{G}^{(\ell_2)}_M\right]^{-2} e^{i\left(q^{+}-\frac{M^2}{2q^-}\right)\left(y^{-}-x^{-}\right)} \left[ -1 + e^{i\mathcal{G}^{(\ell_2)}_{M}(x^{-}-z^{-}_2)} \right] \left[-1 + e^{-i\mathcal{G}^{(\ell_2)}_{M}(y^{-}-z^{-}_3)} \right]\\
& \times e^{i\left(z^-_2-z^-_3\right)\mathcal{H}^{\left(\ell_2,p_2\right)}_M} e^{-i\left(\pmb{p}_{2\perp}+\pmb{\ell}_{2\perp}\right)\cdot\left(\pmb{z}_{2\perp}-\pmb{z}_{3\perp}\right)} e^{-ik^-\left(z^+_3-z^+_2\right)} \frac{1}{2y}\frac{1}{2q^-(1-y+\eta y)} \langle P_{A-1} | {\rm Tr}[A^{+}(z_3)A^+(z_2)] | P_{A-1} \rangle\\
& \times  d^{(\ell_2)}_{\sigma_{1} \sigma_{4}} {\rm Tr} \left[\gamma^{-} \gamma^{\mu} \left(\slashed{q}+\slashed{p}'+M\right) \gamma^{\sigma_{4}}\left(\slashed{q} +\slashed{p}' - \slashed{\ell}_2 + M\right)\gamma^-  \left(\slashed{p}_2+M\right) \gamma^-\left(\slashed{q}+\slashed{p}-\slashed{\ell}_2+M\right)\gamma^{\sigma_1} \left(\slashed{q}+\slashed{p}+M\right)\gamma^{\nu} \right],
\end{split}\label{eq:W_quark_photon_3}
\end{equation}
where 
\begin{eqnarray}
\eta = \frac{k^-}{\ell^-_2 }=\frac{k^-}{ y q^-},
\label{eq:eta_definition}
\end{eqnarray}
and 
\begin{eqnarray}
\mathcal{G}^{(\ell_2)}_M &=& \frac{\pmb{\ell}^2_{2\perp} + y^2M^2}{2y\left(1-y\right)q^-},\\
\mathcal{H}^{(\ell_2,p_2)}_M &=& \ell^+_2+p^+_2-\frac{M^2}{2q^-}=\frac{ \pmb{\ell}^{2}_{2\perp} }{2yq^{-}} + \frac{\pmb{p}^2_{2\perp} + M^2  }{2q^{-}(1-y + \eta y)} - \frac{M^2}{2q^-}.
\end{eqnarray}
Applying the following transformation  $\pmb{p}_{2\perp} + \pmb{\ell}_{2\perp} = \pmb{k}_{\perp}$ allows to express $d^{2}p_{2\perp} \to d^{2}k_{2\perp}$, for a fixed $\ell_2$. Furthermore, the following transformation of coordinates is useful
\begin{eqnarray}
z &=& \frac{z_3+z_2}{2}, \nonumber\\
\Delta z&=& z_3-z_2,
\label{eq:kernel1_z_deltaz_coordinate}
\end{eqnarray}
as the integration measure remains unchanged, i.e. $d^4 z_3 d^4 z_2 = d^4 z d^4 (\Delta z)$. The resulting hadronic tensor has the following form:
\begin{eqnarray}
\begin{split}
W^{\mu\nu}_{1,c} & =\sum_f e^2 e^4_f g^2_{s} \int d x^{-}\, dy^{-}\, d^4z\, d^4 (\Delta z) \int \frac{dy}{2\pi} \frac{d^2 \ell_{2\perp}}{(2\pi)^2} \frac{d^2 k_{\perp}}{(2\pi)^2} \delta(\Delta z^+) e^{-i\Delta z^+ k^-} \\
  & \times  \left\langle P_{A-1} \left| {\rm Tr}[A^{+}\left(z+\frac{\Delta z}{2}\right)A^+\left(z-\frac{\Delta z}{2}\right)] \right| P_{A-1} \right\rangle  \\
  & \times e^{i\left(q^{+} - \frac{M^2}{2q^-}\right)(y^{-}-x^{-})} \left\langle P \left| \bar{\psi}_{_f}\left(z^{+}+\frac{\Delta z^+}{2},y^-,\pmb{z}_{\perp}+\frac{\pmb{\Delta z}_\perp}{2}\right) \frac{\gamma^{+}}{4} \psi_{_f}\left(z^+-\frac{\Delta z^+}{2},x^-,\pmb{z}_{\perp}-\frac{\pmb{\Delta z}_\perp}{2}\right)\right| P\right\rangle\\
  & \times \left[ -1 + e^{i\mathcal{G}^{(\ell_2)}_{M}(x^{-}-z^{-}_{2})} \right] \left[ -1 + e^{-i\mathcal{G}^{(\ell_2)}_{M}(y^{-}-z^{-}_{3})} \right] e^{-i\Delta z^{-}\mathcal{H}^{(\ell_2,p_2)}_M} e^{i\pmb{k}_{\perp}\cdot \pmb{\Delta z}_{\perp}}\\ 
  & \times \frac{\theta(x^{-} - z^{-}_{2}) \theta(y^{-} - z^{-}_{3})}{ (2q^{-})^2 [2\left(1-y\right)q^{-} ]^2} \left[\mathcal{G}^{(\ell_2)}_M\right]^{-2}  \frac{1}{2y}\frac{1}{2(1-y+\eta y)q^-}\\
& \times d^{(\ell_2)}_{\sigma_{1} \sigma_{4}} {\rm Tr} \left[\gamma^{-} \gamma^{\mu} \left(\slashed{q}+\slashed{p}'+M\right)\gamma^{\sigma_{4}}\left(\slashed{q} +\slashed{p}' - \slashed{\ell}_2 + M\right)\gamma^- \left(\slashed{p}_2+M\right) \gamma^{-}\left(\slashed{q}+\slashed{p}-\slashed{\ell}_2+M\right)\gamma^{\sigma_{1}} \left(\slashed{q}+\slashed{p}+M\right)\gamma^{\nu} \right]. 
\end{split}
\label{eq:W_quark_photon_4}
\end{eqnarray}
Note that the two-point gauge field operator $ \langle P_{A-1}|{\rm Tr}[A^+(z+\Delta z/2) A^+(z-\Delta z/2)] |P_{A-1}\rangle$  is invariant under translation by four-vector $z$. This is primarily true owing to the fact that the incoming state $|P_{A-1}\rangle$ and the outgoing state $\langle P_{A-1}|$ are identical. Therefore, any dependence on $z$ seen in that operator expectation value is not physical.~\footnote{A similar statement can be made to hold true for the $z^+$ and $\pmb{z}_\perp$ dependence within the $\langle\bar{\psi}\gamma^+\psi\rangle$ operator in Eq.~\ref{eq:W_quark_photon_4}, by undoing the spacetime integrals associated with the $\delta$-functions in Eq.~\ref{eq:spacetime_delta}.} The phases that depend on the relative distances $\Delta X^-=y^{-}-x^{-}$  such as $e^{i\left(q^{+} - \frac{M^2}{2q^-}\right)(y^{-}-x^{-})}$ are absorbed in the definition of the quark PDF, and phases $e^{-i\Delta z^{-}\mathcal{H}^{(\ell_2,p_2)}_M} e^{i\pmb{k}_{\perp}\cdot \pmb{\Delta z}_{\perp}}$ are included within the nuclear medium's distribution function.
 
While translational invariance was helpful for dealing with expectation values of operator products, quantum coherence (or interference) effects are more sensitive to positional information, as seen in the phase factor $\left[ -1 + e^{i\mathcal{G}^{(\ell_2)}_{M}(x^{-}-z^{-}_{2})} \right]$ and $\left[ -1 + e^{-i\mathcal{G}^{(\ell_2)}_{M}(y^{-}-z^{-}_{3})} \right]$. Since the process (Fig.~\ref{fig:kernel1_ph_qqgm_qqgm_ph_central}) in the amplitude is identical to the process on the complex conjugate, the $W^{\mu\nu}$ is required to be a real number. Therefore, the remaining phase factors must be real-valued:
\begin{equation}
\begin{split}
  & \mathcal{R}  = \left[ -1 + e^{i\mathcal{G}^{(\ell_2)}_{M}\left(x^{-}-z^{-}_{2}\right)} \right] \left[ -1 + e^{-i\mathcal{G}^{(\ell_2)}_{M}\left(y^{-}-z^{-}_{3}\right)} \right] \\
  & \mathcal{R} = \left[ 1  - e^{i\mathcal{G}^{(\ell_2)}_{M}(x^{-}-z^{-}_{2})}  - e^{-i\mathcal{G}^{(\ell_2)}_{M}(y^{-}-z^{-}_{3})}  + e^{i\mathcal{G}^{(\ell_2)}_{M}\left(x^{-}-z^{-}_{2} - y^{-} + z^{-}_{3}\right)}   \right] \in \mathbb{R}\\
  & \Longrightarrow \mathcal{G}^{(\ell_2)}_{M}\left(x^{-}-z^{-}_{2} - y^{-} + z^{-}_{3}\right) = 2n\pi , \text{ where, } n \in \mathbb{Z} \\
  & \Longrightarrow \mathcal{G}^{(\ell_2)}_{M}\left(x^{-}-z^{-}_{2}\right) = \mathcal{G}^{(\ell_2)}_{M}\left(y^{-} - z^{-}_{3}\right) + 2n\pi , \\
  & \Longrightarrow \mathcal{R} = \left[ 2 -2 \cos\left\{\mathcal{G}^{(\ell_2)}_{M}\left(y^{-}-z^{-}_{3}\right)\right\}\right] = \left[ 2 -2 \cos\left\{\mathcal{G}^{(\ell_2)}_{M}\left(x^{-}-z^{-}_{2}\right)\right\}\right].
\label{eq:2_2cos_k1}
\end{split}    
\end{equation}
The above derivation entails that $y^{-} - z^{-}_{3} = x^{-} - z^{-}_{2}$, which is expected as $x^{-} - z^{-}_{2}$ represents the distance between first scattering and second scattering on the amplitude side, while $y^{-} - z^{-}_{3}$ is the same distance on the complex conjugate side. As $\theta(x^{-} - z^{-}_{2})$ suggests that $x^{-} - z^{-}_{2}>0$, while $\theta(y^{-} - z^{-}_{3})$ implies $y^{-} - z^{-}_{3}>0$, a new length integration variable $\zeta^-=y^{-} - z^{-}_{3} = x^{-} - z^{-}_{2}$ is defined to encapsulate that spacetime distance, and ensure that the scattering probability is real-valued. Introducing new variables for distance
\begin{eqnarray}
\Delta X^{-} &=& y^{-} - x^{-}, \nonumber\\
X^{-} &=& \frac{ y^{-} + x^{-} }{2}, \nonumber\\
\zeta^{-} &=& y^{-} - z^{-}_{3} = x^{-} - z^{-}_{2} = X^{-} - z^{-},
\label{eq:X_Delta_X_zeta}
\end{eqnarray}
and incorporating them in $W^{\mu\nu}$, allows one to perform the integrals over $d^4 z$,~\footnote{The integral over $d^4 z$ just gives an overall normalization factor, which is absorbed in the redefinition of the operator product expectation value.} $\Delta z^+$, and $X^-$,~\footnote{In Eq.~\ref{eq:W_final_quark_photon}, the expectation value $\langle\bar{\psi}\gamma^+\psi\rangle$ was translated by a different amount than $\langle A^+A^+ \rangle$.}
giving
\begin{eqnarray}
\begin{split}
W^{\mu\nu}_{1,c} & =\sum_f e^2 e^4_f g^2_{s} \int d (\Delta X^{-}) d \zeta^{-} d (\Delta z^{-})d^2 \Delta z_{\perp} \int \frac{dy}{2\pi}\frac{d^2 \ell_{2\perp}}{(2\pi)^2} \frac{d^2 k_{\perp}}{(2\pi)^2} e^{i\Delta X^{-}\left(q^{+}-\frac{M^2}{2q^-}\right)} \left\langle P \left| \bar{\psi}_{_f}(\Delta X^-) \frac{\gamma^{+}}{4} \psi_{_f}(0)\right| P\right\rangle  \\
  & \times  \left[2-2\cos\left\{\mathcal{G}^{(\ell_2)}_{M}\zeta^-\right\}\right] e^{-i\Delta z^{-} \mathcal{H}^{(\ell_2,p_2)}_M}  e^{i\pmb{k}_{\perp}\cdot \Delta\pmb{z}_{\perp}}\\ 
  & \times \frac{\theta(\zeta^-) }{ (2q^{-})^2 [2\left(1-y\right)q^{-} ]^2} \left[ \mathcal{G}^{(\ell_2)}_{M} \right]^{-2} \left[\frac{1}{2y}\frac{1}{2(1-y+\eta y)q^-}\right] \langle P_{A-1} | {\rm Tr}[A^{+}(\zeta^-, \Delta z^-, \Delta z_{\perp})A^+(\zeta^-, 0)] | P_{A-1}\rangle  \\
  & \times d^{(\ell_2)}_{\sigma_{1} \sigma_{4}} {\rm Tr} \left[\gamma^{-} \gamma^{\mu} \left(\slashed{q}+\slashed{p}'+M\right)\gamma^{\sigma_{4}}\left(\slashed{q} +\slashed{p}' - \slashed{\ell}_2 + M\right)\gamma^- \left(\slashed{p}_2+M\right) \gamma^{-}\left(\slashed{q}+\slashed{p}-\slashed{\ell}_2+M\right)\gamma^{\sigma_{1}} \left(\slashed{q}+\slashed{p}+M\right)\gamma^{\nu} \right].
\end{split}
\label{eq:W_quark_photon_5}   
\end{eqnarray}
The trace in the above equation can be simplified to get
\begin{equation}
\begin{split}
    & {\rm Tr} \left[\gamma^{-} \gamma^{\mu} \left(\slashed{q}+\slashed{p}'+M\right)\gamma^{\sigma_{4}}\left(\slashed{q} +\slashed{p}' - \slashed{\ell}_2 + M\right)\gamma^- \left(\slashed{p}_2+M\right) \gamma^{-}\left(\slashed{q}+\slashed{p}-\slashed{\ell}_2+M\right)\gamma^{\sigma_{1}} \left(\slashed{q}+\slashed{p}+M\right)\gamma^{\nu} \right] d^{(\ell_2)}_{\sigma_{1} \sigma_{4}} \\
  & = {\rm Tr} \left[\gamma^{-} \gamma^{\mu} \left(\slashed{q}+\slashed{p}'+M\right)\gamma^{\sigma_{4}}\left(\slashed{q} +\slashed{p}' - \slashed{\ell}_2 + M\right)\gamma^- \left(\slashed{p}_2+M\right) \gamma^{-}\left(\slashed{q}+\slashed{p}-\slashed{\ell}_2+M\right)\gamma^{\sigma_{1}} \left(\slashed{q}+\slashed{p}+M\right)\gamma^{\nu} \right] \\ 
  & \times 
  \left[  -g_{\sigma_{1} \sigma_{4}} + \frac{n_{\sigma_{1}} \ell_{2\sigma_{4}} + n_{\sigma_{4}} \ell_{2\sigma_{1}} } {n \cdot \ell_2} \right]   \\
  & = 32 (q^-)^3 \left[ -g^{\mu\nu}_{\perp\perp} \right] \left[\frac{1-y+\eta y}{y} \right] \left[ \frac{1+ \left(1-y\right)^2}{y}\right] \left[ \pmb{\ell}^2_{2\perp} + M^2 y^4 \kappa \right],
\end{split}\label{eq:trace_ph_qqgm_qqgm_ph}
\end{equation}
where $\kappa$ is defined in  
Eq.~\ref{eq:kappa}.
Using the expression in Eq.~\ref{eq:trace_ph_qqgm_qqgm_ph}, the hadronic tensor becomes
\begin{eqnarray}
\begin{split}
W^{\mu\nu}_{1,c} & =\sum_f 2 \left[-g^{\mu\nu}_{\perp\perp}\right] e^2 e^4_f g^2_{s} \int d (\Delta X^{-}) e^{i\Delta X^{-}\left(q^{+}-\frac{M^2}{2q^-}\right)} \left\langle P \left| \bar{\psi}_{_f}(\Delta X^-) \frac{\gamma^{+}}{4} \psi_{_f}(0)\right| P\right\rangle\\
  & \times  \int d (\Delta z^{-})d^2 \Delta z_{\perp} \frac{dy}{2\pi}\frac{d^2 \ell_{2\perp}}{(2\pi)^2} \frac{d^2 k_{\perp}}{(2\pi)^2} \left[\frac{1+ \left(1-y\right)^2}{y}\right] e^{-i\Delta z^{-}\mathcal{H}^{(\ell_2,p_2)}_M} e^{i\pmb{k}_{\perp}\cdot \Delta\pmb{z}_{\perp} } \\ 
  & \times \int d \zeta^- \theta(\zeta^-) \left[2 - 2\cos\left\{\mathcal{G}^{(\ell_2)}_{M}\zeta^-\right\}\right] \frac{[\pmb{\ell}^2_{2\perp} +M^2 y^4 \kappa ]}{ \left[  \pmb{\ell}^2_{2\perp}  + y^2  M^2 \right]^2} \langle P_{A-1} | {\rm Tr}[A^{+}(\zeta^-, \Delta z^-, \Delta z_{\perp})A^+(\zeta^-,0)] | P_{A-1} \rangle,
\end{split}\label{eq:W_final_quark_photon}
\end{eqnarray}
where $\kappa$ is defined in Eq.~\ref{eq:kappa}, while, for completeness, 
\begin{eqnarray}
\mathcal{G}^{(\ell_2)}_M &=& \ell^{+}_{2} + \frac{\pmb{\ell}^{2}_{2\perp}+M^2}{2(q^{-}-\ell^{-}_{2})} - \frac{M^2}{2q^{-}} = \frac{\pmb{\ell}^2_{2\perp} + y^2M^2}{2y(1-y)q^-}\nonumber,\\
\mathcal{H}^{(\ell_2,p_2)}_M &=& \ell^{+}_{2} + p^{+}_{2} - \frac{M^2}{2q^-} =  \frac{ \pmb{\ell}^{2}_{2\perp} -yM^2}{2yq^{-}} +\frac{\left(\pmb{\ell}_{2\perp} - \pmb{k}_{\perp}\right)^2+M^2}{2q^{-}\left(1-y+\eta y\right)}.
\label{eq:H_l2-p2-M}
\end{eqnarray} 
There are seven other diagrams, including non-central-cut diagrams, present in kernel-1, whose contributions to the hadronic tensor $W^{\mu\nu}_1$ are in Appendix~\ref{append:kernel1}. 

\section{Single-scattering-induced emission: one photon and one gluon in the final state}
\label{sec:photon_gluon}
\begin{figure}[!h]
    \centering
    \includegraphics[width=0.45\textwidth]{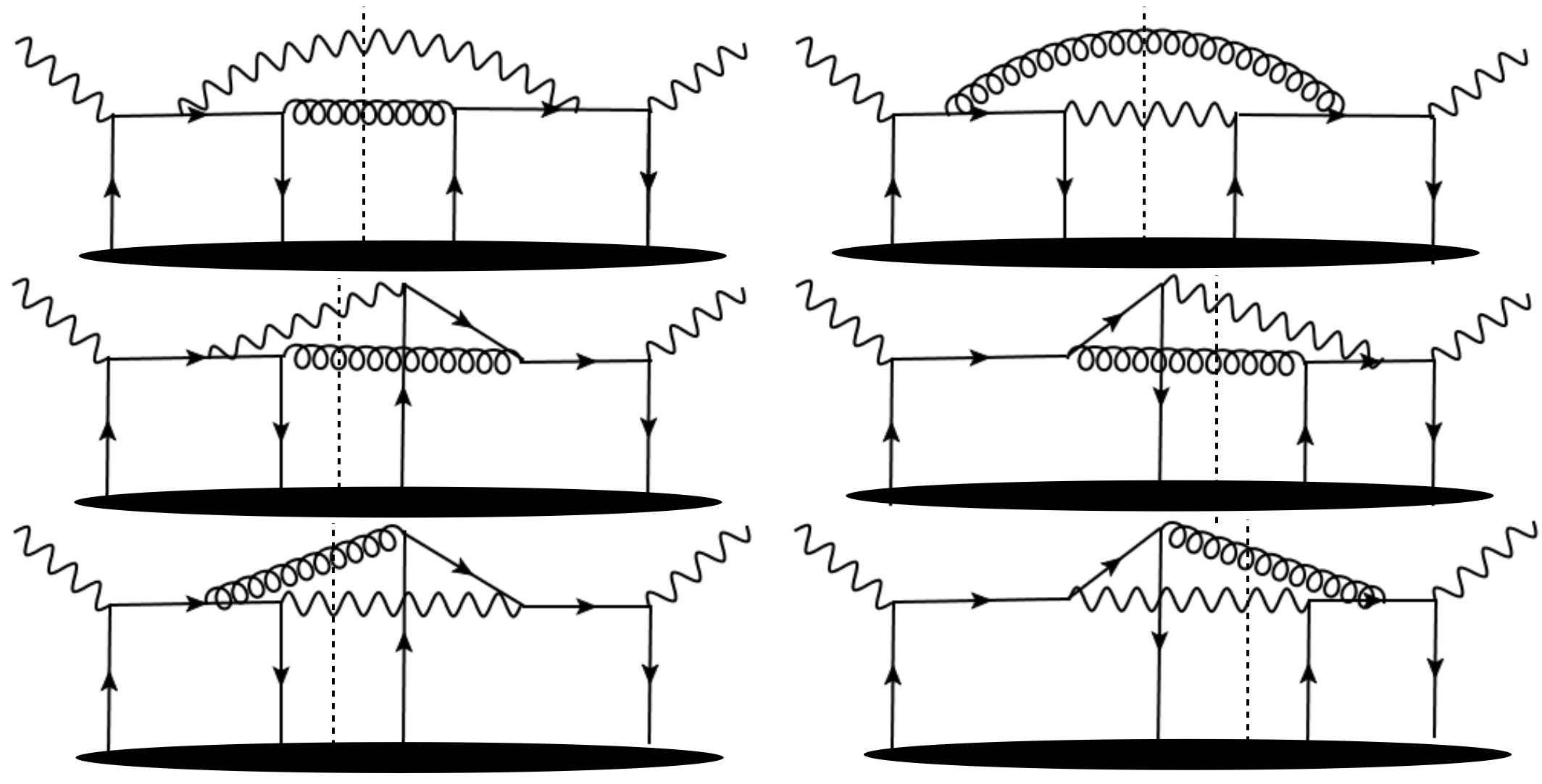}
    \caption{Diagrams in kernel-2 giving a real photon and a gluon final states. The second scattering with the nuclear medium is mediated by the exchange of a Glauber quark.}
    \label{fig:SESS_diagram_photon_gluon}
\end{figure}
The possible diagrams that give rise to the photon-gluon final state are illustrated in Fig.~\ref{fig:SESS_diagram_photon_gluon}. There are a total of six central-cut diagrams. These diagrams are referred to as kernel-2 in the remainder of the paper. As in the previous section, the calculation of the central cut with the most salient information is given before the final hadronic tensor  $W^{\mu\nu}_2$ is quoted.

\begin{figure}[!h]
    \centering
    \includegraphics[width=0.45\textwidth]{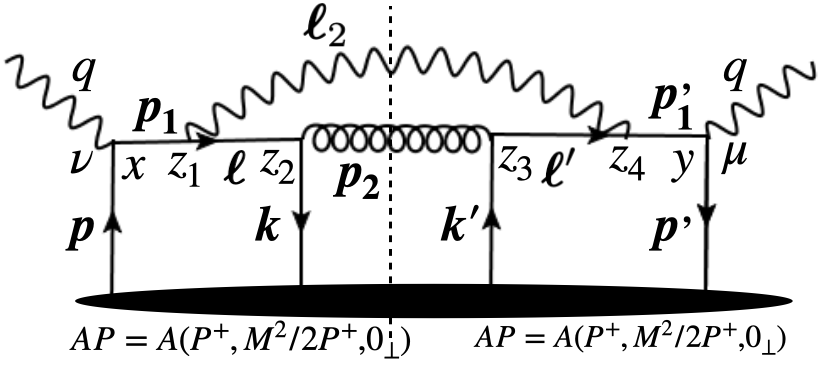}
    \caption{A forward scattering diagram in kernel-2. It consists of a fermion-to-boson conversion process, giving a real photon and a gluon final states.  }
    \label{fig:kernel2_ph_qgqm_gqqm_ph}
\end{figure}

The diagram under consideration is shown in Fig. \ref{fig:kernel2_ph_qgqm_gqqm_ph}. In that diagram, the hard quark produced from the initial-state hard scattering undergoes collinear emission followed by in-medium fermion-to-boson conversion. The forward scattering amplitude of this process is given as
\begin{equation}
\begin{split}
T^{\mu \nu}_{2,c} &= \sum_{f} e^2 e^4_f g^{2}_s\int\,d^4x\,d^4y\,d^4z_1\,d^4z_2\,d^4z_3\,d^4z_4\,\int\frac{d^4p_1}{(2\pi)^4}\frac{d^4p'_1}{(2\pi)^4}\frac{d^4\ell}{(2\pi)^4}\frac{d^4\ell'}{(2\pi)^4}\frac{d^4\ell_2}{(2\pi)^4}\frac{d^4p_2}{(2\pi)^4} \\
& \times e^{iy\left(q-p'_{1}\right)} e^{-ix\left(q-p_{1}\right)} e^{iz_{1}(\ell - p_{1} + \ell_{2})} e^{iz_{2}(-\ell+p_{2})} e^{iz_{3}(\ell'-p_{2})} e^{iz_{4}(p'_{1} -\ell' -\ell_{2})} \\
& \times \left\langle AP\left| {\rm Tr} \left[\psi_{_f}(x)\bar{\psi}_{_f}(y)\gamma^{\mu}\frac{i\slashed{p}'_1}{\left[\left(p'_1\right)^2+i\epsilon\right]}i\gamma^{\sigma_4}\frac{i\slashed{\ell}'}{\left[\left(\ell'\right)^2+i\epsilon\right]}i\gamma^{\sigma_3}\psi_{_f}(z_3)\bar{\psi}_{_f}(z_2)i\gamma^{\sigma_2}\frac{i\slashed{\ell}}{\left(\ell^2+i\epsilon\right)}i\gamma^{\sigma_1}\frac{i\slashed{p}_1}{\left(p^2_1+i\epsilon\right)}\gamma^\nu\right]\right| AP\right\rangle \\
& \times \frac{id^{(\ell_2)}_{\sigma_4 \sigma_1}}{\ell^2_2+i\epsilon} \frac{id^{(p_2)}_{\sigma_3 \sigma_2}}{p^2_2+i\epsilon} {\rm Tr}\left[t^a t^b\right]\delta^{ab},
\end{split}
\label{eq:T_gluon_photon}
\end{equation}
where $d^{(\ell_2)}_{\sigma_4 \sigma_1}$ and $d^{(p_2)}_{\sigma_3 \sigma_2}$ are defined as in Eq.~\ref{eq:d_X_sigma}, while the color algebra can be simplified to
\begin{equation}
\mathrm{Tr} \left[t^a t^b \right] \delta^{ab} = \frac{\delta^{ab}}{2}\delta^{ab} = \frac{1}{2}\sum^8_{i=1} \delta^{ii} = \frac{N^2_c-1}{2} = C_F N_c.\label{eq:color_factor}
\end{equation}
The integral over $d^4z_1$ and $d^4z_4$ can be carried out to yield $(2\pi)^{(4)}\delta^4(\ell-p_1+\ell_2) (2\pi)^4 \delta^{(4)}\left(p'_1-\ell'-\ell_2\right)$, allowing the integration over $d^4\ell$ and $d^4\ell'$ to be performed. Moreover, instituting the change of variable $p'_1=q+p'$ and $p_1=q+p$ in Eq.~\ref{eq:T_gluon_photon} modifies the integration $d^4 p'_1\to d^4 p'$, while $d^4 p_1\to d^4 p$. Applying Cutkosky's rule \cite{Cutkosky:1960sp,Peskin:1995ev}, the hadronic tensor for the central-cut diagram is
\begin{equation}
\begin{split}
W^{\mu\nu}_{2,c} &= \sum_{f} e^2e^4_f g^2_s C_F N_c\int d^4x\,d^4y\,d^4z_2\,d^4z_3\,\int\frac{d^4 p}{(2\pi)^4} \frac{d^4p'}{(2\pi)^{4}}\frac{d^4\ell_2}{(2\pi)^4}\frac{d^4p_2}{(2\pi)^4}e^{-iyp'}e^{ipx}\left\langle P\left| \bar{\psi}_{_f}(y)\frac{\gamma^+}{4}\psi_{_f}(x)\right|P\right\rangle\\
             & \times (2\pi)\delta\left(\ell^2_2\right)(2\pi)\delta\left(p^2_2\right)\times e^{iz_2\left(p_2+\ell_2-q-p\right)} e^{iz_3\left(q+p'-\ell_2-p_2\right)} \left\langle P_{A-1} \middle|\bar{\psi}_{_f}\left(z_2\right)\frac{\gamma^+}{4}\psi_{_f}\left(z_3\right)\middle| P_{A-1} \right\rangle\\
             & \times \frac{\mathrm{Tr}\left[\gamma^{-} \gamma^{\mu} (\slashed{q}+\slashed{p}') \gamma^{\sigma_4} \left(\slashed{q}+\slashed{p}'-\slashed{\ell}_2\right) \gamma^{\sigma_3}\gamma^- \gamma^{\sigma_2} \left(\slashed{q} + \slashed{p} - \slashed{\ell}_2\right) \gamma^{\sigma_1} \left(\slashed{q} + \slashed{p}\right) \gamma^\nu\right]}{\left[\left(q+p'\right)^2-i\epsilon\right]\left[\left(q+p'-\ell_2\right)^2-i\epsilon\right]\left[\left(q+p-\ell_2\right)^2+i\epsilon\right]\left[\left(q+p\right)^2+i\epsilon\right]} d^{(\ell_2)}_{\sigma_4\sigma_1}d^{(p_2)}_{\sigma_3\sigma_{2}},
  \end{split}\label{eq:kernel2_W2_intermediatet-1}
\end{equation}
where the discontinuity in the final state photon and gluon propagators has been applied, namely
\begin{eqnarray}
\mathrm{Disc} \left[ \frac{1}{\ell^2_2 +i\epsilon}\right] &=& -2\pi i\delta\left(\ell^2_2\right),\nonumber\\
\mathrm{Disc}\left[\frac{1}{p^2_2+i\epsilon}\right] &=& -2\pi i \delta(p^2_2 ).
\end{eqnarray}
Equation~\ref{eq:kernel2_W2_intermediatet-1} exhibits a singularity when the quark propagator for $p_1$, $\ell$, $\ell'$ and $p'_1$ becomes on-shell. There are two simple poles for $p^+$ and $p'^+$, respectively. The contour integration for $p^+$ in the complex plane gives
\begin{equation}
\begin{split}
\tilde{C}_1&= \oint \frac{dp^+}{(2\pi)}\frac{e^{ip^+\left(x^{-}-z^{-}_2\right)}}{\left[\left(q+p\right)^2+i\epsilon\right]\left[\left(q+p-\ell_2\right)^2+i\epsilon\right]} \\
   &= \frac{(2\pi i)}{2\pi} \frac{\theta(x^{-}-z^{-}_{2})}{4q^-\left(q^{-}-\ell^-_2\right)}e^{-iq^+\left(x^{-}-z^-_2\right)}\left[\frac{-1 + e^{i\mathcal{G}^{(\ell_2)}_0\left(x^{-}-z^-_2\right)}}{\mathcal{G}^{(\ell_2)}_0}\right],
\end{split}
\label{eq:contour-c1-nlo2-central}
\end{equation}
where 
\begin{equation}
\mathcal{G}^{(\ell_2)}_0 = \ell^+_2 + \frac{\pmb{\ell}^2_{2\perp}}{2\left(q^{-}-\ell^-_2\right)} = \frac{\pmb{\ell}^2_{2\perp}}{2y(1-y)q^-} .
\label{eq:kernel-2_G0L2}
\end{equation}
On the other hand, the contour integration over $p'^+$ yields
\begin{equation}
\begin{split}
\tilde{C}_{2} & = \oint \frac{dp'^+}{(2\pi)} \frac{e^{-ip'^+\left(y^{-}-z^-_3\right)}}{\left[\left(q+p'\right)^2-i\epsilon\right]\left[\left(q+p'-\ell_2\right)^2-i\epsilon\right]}\\
      & = \frac{(-2\pi i)}{2\pi} \frac{\theta(y^{-}-z^-_3)}{4q^-\left(q^{-}-\ell^-_2\right)}e^{iq^{+}\left(y^{-}-z^-_3\right)}\left[\frac{-1+e^{-i \mathcal{G}^{(\ell_2)}_0\left(y^{-}-z^{-}_{3}\right)}}{\mathcal{G}^{(\ell_2)}_0}\right].
\end{split}
\label{eq:contour-c2-nlo2-central}
\end{equation}

To perform the remaining integrals for $p$ and $p'$, the same procedure as in Eq.~\ref{eq:spacetime_delta} was followed yielding $(2\pi)^3 \delta\left(x^{+}-z^+_2\right) \delta^2\left(\pmb{x}_\perp - \pmb{z}_{2\perp}\right) (2\pi)^3\delta\left(-y^{+} + z^+_3\right)\delta^2\left(-\pmb{y}_\perp+\pmb{z}_{3\perp}\right)$.
%
%
Using these $\delta$-functions allows to performing the integral over spacetime variables $(x^{+}, \pmb{x}_{\perp})$ and $(y^+,\pmb{y}_{\perp})$ to give 
\begin{equation}
\begin{split}
W^{\mu\nu}_{2,c} & = \sum_{f} e^2 e^4_f g^2_{s} [C_F N_c] \int dx^-\,dy^-\,d^4z_2\,d^4z_3 \int \frac{d^4 \ell_{2}}{(2\pi)^4} \frac{d^4 p_{2}}{(2\pi)^4} \left\langle P \left| \bar{\psi}_{_f}\left(z^+_3,y^-,\pmb{z}_{3\perp}\right) \frac{\gamma^{+}}{4} \psi_{_f}\left(z^+_2,x^-,\pmb{z}_{2\perp}\right)\right| P\right\rangle  \\
& \times \frac{\theta(x^{-} - z^{-}_{2}) \theta(y^{-} - z^{-}_{3})}{ \left(2q^{-}\right)^2 \left[2(q^{-}-\ell^-_2)\right]^2} \left[\mathcal{G}^{(\ell_2)}_0\right]^{-2} e^{iq^{+}\left(y^{-}-x^{-}-z^{-}_3+z^{-}_2\right)} \left[ -1 + e^{i\mathcal{G}^{(\ell_2)}_{0}(x^{-}-z^{-}_2)} \right] \left[-1 + e^{-i\mathcal{G}^{(\ell_2)}_{0}(y^{-}-z^{-}_3)} \right]\\
& \times e^{iz_3\left(q-p_{2}-\ell_{2}\right) } e^{iz_2(\ell_2 + p_2 - q)} \left\langle P_{A-1} \left| \bar{\psi}_{_f}\left(z_2\right)\frac{\gamma^+}{4}\psi_{_f}\left(z_3\right)\right| P_{A-1} \right\rangle (2 \pi) \delta\left(\ell_{2}^{2}\right) (2 \pi)  \delta\left(p_{2}^{2}\right) d^{(\ell_2)}_{\sigma_{1} \sigma_{4}} d^{(p_2)}_{\sigma_{3} \sigma_{2}} \\
& \times \mathrm{Tr}\left[\gamma^{-} \gamma^{\mu} (\slashed{q}+\slashed{p}') \gamma^{\sigma_4} \left(\slashed{q}+\slashed{p}'-\slashed{\ell}_2\right) \gamma^{\sigma_3}\gamma^- \gamma^{\sigma_2} \left(\slashed{q} + \slashed{p} - \slashed{\ell}_2\right) \gamma^{\sigma_1} \left(\slashed{q} + \slashed{p}\right) \gamma^\nu\right].
\end{split}\label{eq:kernel2-W_gluon_photon_intermed2}
\end{equation}
%
Applying a similar algebraic manipulation to those presented in Eqs.~\ref{eq:lpq_4_mom_cons} through~\ref{eq:kernel1_z_deltaz_coordinate}, along with  the momentum fraction $y$ as $\ell^{-}_{2}=yq^{-}$, allows one to reduce the hadronic tensor $W^{\mu\nu}_{2,c}$ to
\begin{equation}
\begin{split}
W^{\mu\nu}_{2,c} & = \sum_f e^2 e^4_f g^2_{s} [C_F N_c] \int dx^- dy^- d^4z_2\,d^4z_3 \int \frac{dy d^2\ell_{2\perp}}{(2\pi)^3} \frac{d^2 p_{2\perp}}{(2\pi)^2}\delta\left(z^+_2-z^+_3\right) \left\langle P \left| \bar{\psi}_{_f}\left(z^+_3,y^-,\pmb{z}_{3\perp}\right) \frac{\gamma^{+}}{4} \psi_{_f}\left(z^+_2,x^-,\pmb{z}_{2\perp}\right)\right| P\right\rangle \\
& \times \frac{\theta(x^{-} - z^{-}_{2}) \theta(y^{-} - z^{-}_{3})}{ \left(2q^{-}\right)^2 \left[2(q^{-}-\ell^-_2)\right]^2} \left[\mathcal{G}^{(\ell_2)}_0\right]^{-2} e^{iq^{+}\left(y^{-}-x^{-}\right)} \left[ -1 + e^{i\mathcal{G}^{(\ell_2)}_{0}(x^{-}-z^{-}_2)} \right] \left[-1 + e^{-i\mathcal{G}^{(\ell_2)}_{0}(y^{-}-z^{-}_3)} \right]\\
& \times e^{i\left(z^-_2-z^-_3\right)\mathcal{H}^{\left(\ell_2,p_2\right)}_M} e^{-i\left(\pmb{p}_{2\perp}+\pmb{\ell}_{2\perp}\right)\cdot\left(\pmb{z}_{2\perp}-\pmb{z}_{3\perp}\right)} e^{-ik^-\left(z^+_3-z^+_2\right)} \langle P_{A-1} | \bar{\psi}_{_f}\left(z_2\right)\frac{\gamma^+}{4}\psi_{_f}\left(z_3\right)| P_{A-1} \rangle \frac{1}{2y} \frac{1}{2(1-y+\eta y)q^-} \\
& \times \mathrm{Tr}\left[\gamma^{-} \gamma^{\mu} (\slashed{q}+\slashed{p}') \gamma^{\sigma_4} \left(\slashed{q}+\slashed{p}'-\slashed{\ell}_2\right) \gamma^{\sigma_3}\gamma^- \gamma^{\sigma_2} \left(\slashed{q} + \slashed{p} - \slashed{\ell}_2\right) \gamma^{\sigma_1} \left(\slashed{q} + \slashed{p}\right) \gamma^\nu\right] d^{(\ell_2)}_{\sigma_{1} \sigma_{4}} d^{(p_2)}_{\sigma_{3} \sigma_{2}},
\end{split}\label{eq:kernel2-W_gluon_photon_intermed3}
\end{equation}
where $k^-=\eta\ell^-_2=\eta y q^-$ and 
\begin{eqnarray}
\mathcal{G}^{(\ell_2)}_0 &=& \frac{\pmb{\ell}^2_{2\perp} }{2y\left(1-y\right)q^-},\label{eq:kernel2_G0L2_final_aa}\\
\mathcal{H}^{(\ell_2,p_2)}_0 &=& \ell^+_2+p^+_2 =\frac{ \pmb{\ell}^{2}_{2\perp} }{2yq^{-}} + \frac{\pmb{p}^2_{2\perp} }{2q^{-}(1-y + \eta y)}.
\end{eqnarray}
In Eq.~\ref{eq:kernel2-W_gluon_photon_intermed3}, all phases that depend on the relative distances $y^{-}-x^{-}$ (i.e. $e^{iq^{+} (y^{-}-x^{-})}$ herein) are absorbed in the definition of the quark parton distribution function (PDF), while all phases that depend on $z^{-}_{3}-z^{-}_{2}$ (specifically $e^{-i(z^{-}_{3}-z^{-}_{2})\mathcal{H}^{(\ell_2,p_2)}_0}$ and $e^{i\pmb{k}_{\perp}\cdot (\pmb{z}_{3\perp}-\pmb{z}_{2\perp})}$) are part of the in-medium distribution function. Since the process (Fig.~\ref{fig:kernel2_ph_qgqm_gqqm_ph}) in the amplitude and the complex conjugate are identical, the associated $W^{\mu\nu}$ or the amplitude square should be a real number, therefore, the remaining phase factors $\left[ -1 + e^{i\mathcal{G}^{(\ell_2)}_{0}(x^{-}-z^{-}_{2})} \right] \left[ -1 + e^{-i\mathcal{G}^{(\ell_2)}_{0}(y^{-}-z^{-}_{3})} \right]$ must be real-valued. Thus, using the same arguments as in Eq.~\ref{eq:2_2cos_k1} yields
\begin{equation}
\begin{split}
\mathcal{R} = \left[ 2 -2 \cos\left\{\mathcal{G}^{(\ell_2)}_{0}\left(y^{-}-z^{-}_{3}\right)\right\}\right] = \left[ 2 -2 \cos\left\{\mathcal{G}^{(\ell_2)}_{0}\left(x^{-}-z^{-}_{2}\right)\right\}\right].
\label{eq:2_2cos_k2}
\end{split}    
\end{equation}
Of course, as in Eq.~\ref{eq:2_2cos_k1}, $\zeta^-=y^{-} - z^{-}_{3} = x^{-} - z^{-}_{2}$, given that $x^{-} - z^{-}_{2}$ represents the distance between the primary and secondary scattering vertex in the amplitude, while $y^{-} - z^{-}_{3}$ is that same distance in the complex conjugate.
%
%
%
%
Furthermore, using the same variables change in Eqs.~\ref{eq:kernel1_z_deltaz_coordinate} and \ref{eq:X_Delta_X_zeta}, 
as well as $\pmb{p}_{2\perp} + \pmb{\ell}_{2\perp} = \pmb{k}_{\perp}$, allows one to express the hadronic tensor as 
\begin{equation}
\begin{split}
W^{\mu\nu}_{2,c} & = \sum_f e^2 e^4_f g^2_{s} [C_F N_c] \int d (\Delta X^{-}) d \zeta^{-} d (\Delta z^{-})d^2 \Delta z_{\perp} \int \frac{dy}{2\pi}\frac{d^2 \ell_{2\perp}}{(2\pi)^2} \frac{d^2 k_{\perp}}{(2\pi)^2} e^{iq^{+}\Delta X^{-}} \left\langle P \left| \bar{\psi}_{_f}(\Delta X^-) \frac{\gamma^{+}}{4} \psi_{_f}(0)\right| P\right\rangle \\
& \times \frac{\theta(\zeta^{-}) }{ \left(2q^{-}\right)^2 \left[2(q^{-}-\ell^-_2)\right]^2} \left[\mathcal{G}^{(\ell_2)}_0\right]^{-2}   \left[ 2 -2 \cos\left\{\mathcal{G}^{(\ell_2)}_{0}\zeta^- \right\}\right] \frac{1}{2y} \frac{1}{2(1-y+\eta y)q^-}\\
& \times e^{-i\Delta z^-\mathcal{H}^{\left(\ell_2,p_2\right)}_0} e^{i\pmb{k}_{\perp}\cdot\Delta\pmb{z}_{\perp}}  \left\langle P_{A-1} \left| \bar{\psi}_{_f}\left(\zeta^-, 0\right)\frac{\gamma^+}{4}\psi_{_f}\left(\zeta^-, \Delta z^-, \Delta z_{\perp}\right)\right| P_{A-1} \right\rangle \\
& \times \mathrm{Tr}\left[\gamma^{-} \gamma^{\mu} (\slashed{q}+\slashed{p}') \gamma^{\sigma_4} \left(\slashed{q}+\slashed{p}'-\slashed{\ell}_2\right) \gamma^{\sigma_3}\gamma^- \gamma^{\sigma_2} \left(\slashed{q} + \slashed{p} - \slashed{\ell}_2\right) \gamma^{\sigma_1} \left(\slashed{q} + \slashed{p}\right) \gamma^\nu\right] d^{(\ell_2)}_{\sigma_{1} \sigma_{4}} d^{(p_2)}_{\sigma_{3} \sigma_{2}}.
\end{split}\label{eq:kernel2-W_gluon_photon_intermed44}
\end{equation}
The trace in the equation above (Eq.~\ref{eq:kernel2-W_gluon_photon_intermed44}) is now evaluated, giving
\begin{equation}
\begin{split}
& {\rm Tr} \left[ \gamma^- \gamma^\mu \left(\slashed{q}+\slashed{p}'\right)\gamma^{\sigma_4}\left(\slashed{q} +\slashed{p}'-\slashed{\ell}_2\right) \gamma^{\sigma_3}\gamma^-\gamma^{\sigma_2}\left(\slashed{q}+\slashed{p}-\slashed{\ell}_2\right)\gamma^{\sigma_1} \left(\slashed{q} + \slashed{p}\right) \gamma^\nu\right] d^{(\ell_2)}_{\sigma_4\sigma_1}d^{(p_2)}_{\sigma_3\sigma_2}\\
&= \left(q^-\right)^2 \left[-g^{\mu\nu}_{\perp \perp}\right] {\rm Tr}\left[\gamma^-\gamma^+\gamma^{\sigma_4}\left(\slashed{q}+\slashed{p}'-\slashed{\ell}_2\right)\gamma^{\sigma_3}\gamma^-\gamma^{\sigma_2}\left(\slashed{q}+\slashed{p}-\slashed{\ell}_2\right) \gamma^{\sigma_1}\gamma^+\right]d^{(\ell_2)}_{\sigma_4\sigma_1}d^{(p_2)}_{\sigma_3\sigma_2}.
\end{split}
\label{eq:tr_k2_init}
\end{equation}
The first non-vanishing term in $d^{(\ell_2)}_{\sigma_4\sigma_1}d^{(p_2)}_{\sigma_3\sigma_2}$ comes from $g_{\sigma_4\sigma_1}g_{\sigma_3\sigma_2}$, giving
\begin{equation}
\begin{split}
& \left(q^-\right)^2\left[-g^{\mu\nu}_{\perp\perp}\right] {\rm Tr} \left[\gamma^-\gamma^+\gamma^{\sigma_4}\left(\slashed{q}+\slashed{p}'-\slashed{\ell}_2\right)\gamma^{\sigma_3}\gamma^-\gamma^{\sigma_2}\left(\slashed{q}+\slashed{p}-\slashed{\ell}_2\right)\gamma^{\sigma_1}\gamma^+\right] g_{\sigma_4\sigma_1} g_{\sigma_3\sigma_2} \\
& = 32 [-g^{\mu\nu}_{\perp \perp}] (q^{-})^{2} \pmb{\ell}^{2}_{2\perp},
\end{split} \label{eq:kernel2_trace_parts}
\end{equation}
while the second stems from $\frac{n_{\sigma_{4}} \ell_{2\sigma_1}}{\ell^-_2}g_{\sigma_3 \sigma_2}$ yielding \footnote{Any term in $d^{(\ell_2)}_{\sigma_4\sigma_1}d^{(p_2)}_{\sigma_3\sigma_2}$ that is proportional to $n_{\sigma_3}\ell_{2\sigma_2}$ will contract with the $\gamma^{\sigma_3} \gamma^{-} \gamma^{\sigma_2}$ term in Eq.~\ref{eq:kernel2_trace_parts} to give $(n\cdot\gamma) \gamma^- (\ell_2\cdot\gamma)=n_+\left(\gamma^-\right)^2(\ell_2\cdot\gamma)\equiv0$.}
\begin{equation}
\begin{split}
& \left(q^-\right)^2\left[-g^{\mu\nu}_{\perp\perp}\right]{\rm Tr}\left[\gamma^-\gamma^+\gamma^{\sigma_4}\left(\slashed{q}+\slashed{p}'-\slashed{\ell}_2\right)\gamma^{\sigma_3}\gamma^-\gamma^{\sigma_2}\left(\slashed{q}+\slashed{p}-\slashed{\ell}_2\right)\gamma^{\sigma_1}\gamma^+\right]\frac{\left(n_{\sigma_4}\ell_{2\sigma_1}+n_{\sigma_1}\ell_{2\sigma_4}\right)}{n^+\ell^-_2}\left(-g_{\sigma_3\sigma_2}\right)\\
& = 2\left(q^-\right)^2\left[-g^{\mu\nu}_{\perp\perp}\right]{\rm Tr}\left[\gamma^-\gamma^+\gamma^{\sigma_4}\left(\slashed{q}+\slashed{p}'-\slashed{\ell}_2\right)\gamma^{\sigma_3}\gamma^-\gamma^{\sigma_2}\left(\slashed{q}+\slashed{p}-\slashed{\ell}_2\right)\gamma^{\sigma_1}\gamma^+\right]\frac{n_{\sigma_4}\ell_{2\sigma_1}}{\ell^-_2}\left(-g_{\sigma_3\sigma_2}\right)\\
& = 64 \left[-g^{\mu\nu}_{\perp \perp}\right] \left(q^-\right)^2\pmb{\ell}^2_{2\perp} \left[\frac{\left(1-y\right)^2}{y^2}+\frac{1-y}{y}\right].
\end{split}
\end{equation}
Combining terms together, the trace is expressed as
\begin{equation}
\begin{split}
& \mathrm{Tr}\left[\gamma^-\gamma^\mu\left(\slashed{q}+\slashed{p}'\right)\gamma^{\sigma_4}\left(\slashed{q}+\slashed{p}'- \slashed{\ell}_2\right)\gamma^{\sigma_3}\gamma^-\gamma^{\sigma_2}\left(\slashed{q}+\slashed{p}-\slashed{\ell}_2\right)\gamma^{\sigma_1}\left(\slashed{q}+\slashed{p}\right)\gamma^\nu\right]d^{(\ell_2)}_{\sigma_4\sigma_1}d^{(p_2)}_{\sigma_3\sigma_2}\\
& = 32 \left[-g^{\mu\nu}_{\perp\perp}\right]\left(q^{-}\right)^2\pmb{\ell}^2_{2\perp}\left[\frac{1+\left(1-y\right)^2}{y^2}\right].
\end{split}
\label{eq:tr_k2_final}
\end{equation}
The final expression for the hadronic tensor (see Fig.~\ref{fig:kernel2_ph_qgqm_gqqm_ph}) as
\begin{equation}
\begin{split}
W^{\mu\nu}_{2,c} & = \sum_f 2e^2 e^4_f g^2_s C_F N_c \left[-g^{\mu\nu}_{\perp\perp}\right]\int d(\Delta X^-)e^{iq^+(\Delta X^-)}\left\langle P \left| \bar{\psi}_{_f}(\Delta X^-)\frac{\gamma^+}{4}\psi_{_f}(0)\right| P\right\rangle\\
& \times \int d\left(\Delta z^-\right) d^2\left(\Delta z_{\perp}\right) \frac{dy}{2\pi} \frac{d^2\ell_{2\perp}}{(2\pi)^2}\frac{d^2 k_{\perp}}{(2\pi)^2} \left[\frac{1+\left(1-y\right)^2}{y}\right] e^{-i\Delta z^-\mathcal{H}^{(\ell_2,p_2)}_0}e^{i\pmb{k}_{\perp}\cdot\pmb{\Delta z_{\perp}}}\\
& \times \int d\zeta^- \frac{\theta(\zeta^-)}{(1-y+\eta y)q^-} \frac{1}{\pmb{\ell}^2_{2\perp}} \left[2-2\cos\left\{\mathcal{G}^{(\ell_2)}_0\zeta^-\right\}\right] \left\langle P_{A-1}\left|\bar{\psi}_{_f}\left(\zeta^-,0\right)\frac{\gamma^+}{4}\psi_{_f}(\zeta^-,\Delta z^-, \Delta\pmb{z}_{\perp})\right|P_{A-1}\right\rangle,
\end{split} 
\label{eq:kernel2-hadronic-tensor_final_dia1}
\end{equation}
where
\begin{equation}
\mathcal{H}^{(\ell_2,p_2)}_0=\ell^+_2 + p^+_2= \frac{\pmb{\ell}^2_{2\perp} }{2yq^-} + \frac{\left(\pmb{\ell}_{2\perp}-\pmb{k}_\perp\right)^2}{2q^-(1-y+y\eta)}.
\label{eq:kernel2_H0L2P2}
\end{equation}
The above expression of the hadronic tensor (Eq.~\ref{eq:kernel2-hadronic-tensor_final_dia1}) when compared to the diagram in kernel-1, differs through the appearance of a two-point fermionic correlator $\langle\bar{\psi}_{_f}\left(\zeta^-,0\right)\frac{\gamma^+}{4}\psi_{_f}(\zeta^-,\Delta z^-, \Delta\pmb{z}_{\perp})\rangle$, along with the factor of $(1-y+\eta y)q^-$ in the denominator. This indicates that the quark-to-gluon conversion processes are suppressed by the incoming energy of the quark, i.e., $(1-y+\eta y)q^-$. The five other diagrams contributing to kernel-2 are presented in Appendix~\ref{append:kernel2}.

\section{MEDIUM MODIFIED KV kernels for photon production}
In the preceding sections, we discussed in detail the steps involved in the derivation of the hadronic tensor for each kernel. In this section, we add contributions from all diagrams for each kernel and provide a full scattering kernel for photon emission from the jet parton for each category. 

\subsection{Full KV scattering kernel without collinear expansion}
In this section, the full algebraic form of the hadronic tensor is presented for each kernel. For kernel-1, a total of eight diagrams were identified, including the left-cut and right-cut diagrams. These are presented in Appendix~\ref{append:kernel1}. In order to add these diagrams, we institute $\Delta X^-=y^{-} -x^{-}$, $\Delta z^-=z^{-}_3 -z^{-}_2$, and $\zeta^- = y^{-}-z_{3}=x^{-}-z_{2}$. The exponentials that depend on $\Delta X^-$ are absorbed in the definition of the nucleon parton distribution function, whereas the exponentials that depend on the relative distance $\Delta z^- = z^{-}_{3} -z^{-}_2$ are absorbed in the definition of the gluon/quark distribution in the medium. Under these algebraic transformations, diagrams within each kernel can be summed. Including all diagrams for kernel type-1 (Fig.~\ref{fig:SESS_diagram_photon_quark}), the full hadronic tensor  is given as
\begin{eqnarray}
\begin{split}
W^{\mu\nu}_{1,\rm full}  =\sum_{f} 2 \left[ -g^{\mu\nu}_{\perp\perp}  \right] e^2_f  \int d (\Delta X^{-}) e^{i\Delta X^{-}\left(q^{+}-\frac{M^2}{2q^-}\right)} \left\langle P \left| \bar{\psi}_{_f}(\Delta X^-) \frac{\gamma^{+}}{4} \psi_{_f}(0)\right| P\right\rangle   \times \mathcal{K}^{\rm eff}_{1},
\end{split}\label{eq:Wmunu_final_type-I}   
\end{eqnarray}
where we define $\mathcal{K}^{\rm eff}_{1}$ as a effective medium-modified scattering kernel for type-1 process (Fig.~\ref{fig:SESS_diagram_photon_quark}) as
\begin{eqnarray}
 \mathcal{K}^{\rm eff}_{1} & =& e^{2}e^{2}_{f}  g^{2}_{s} \int d (\Delta z^{-})d^2 \Delta z_{\perp} \frac{dy}{2\pi}\frac{d^2 \ell_{2\perp}}{(2\pi)^2} \frac{d^2 k_{\perp}}{(2\pi)^2} e^{-i\Delta z^{-}\mathcal{H}^{(\ell_2,p_2)}_M} e^{i\pmb{k}_{\perp}\cdot \Delta\pmb{z}_{\perp} } \nonumber \\ 
  &  & \times \int d \zeta^{-} \theta(\zeta^-) \,\mathcal{S}^{\rm eff}_1 \langle P_{A-1} | {\rm Tr}[A^{+}(\zeta^-, \Delta z^-, \Delta z_{\perp})A^+(\zeta^-,0)] | P_{A-1} \rangle,
\end{eqnarray}

where $\mathcal{S}^{\rm eff}_1$ denotes the perturbative part in the integrand of the medium-modified kernel given as 
\begin{eqnarray}
 \mathcal{S}^{\rm eff}_{1} &=& \left[\frac{1+\left(1-y\right)^2}{y}\right] \left[\frac{\pmb{\ell}^2_{2\perp}+M^2 y^4 \kappa}{\left[\pmb{\ell}^2_{2\perp}+y^2 M^2\right]^2}\right]\left[1-\cos\left\{\mathcal{G}^{(\ell_2)}_{M}\zeta^- \right\}\right] \nonumber \\
 & & +  \left[\frac{\left(1+\eta y\right)^2 + \left(1-y+\eta y\right)^2}{y}\right]  \left[ \frac{ \left\{ (1+\eta y)\pmb{\ell}_{2\perp} -y \pmb{k}_\perp\right\}^2 + M^2 y^4 \kappa }{   J^2_1 } \right] \nonumber \\
 & & -  \left[\frac{1 + \left(1-y\right)^2 + \eta y\left(2-y\right)}{y}\right]  \left[ \frac{ \left(1+\eta y\right)\pmb{\ell}^2_{2\perp} -y \pmb{k}_\perp \cdot\pmb{\ell}_{2\perp} + M^2 y^4 \kappa }{  \left\{\pmb{\ell}^2_{2\perp} + M^2 y^2\right\} J_1 } \right]  \left[ 2 - 2 \cos\left\{ \mathcal{G}^{(\ell_2)}_{M}\zeta^- \right\}\right], 
\label{eq:s1_eff_full}
\end{eqnarray}
where
\begin{eqnarray}
J_1 &=& \left\{ (1+\eta y)\pmb{\ell}_{2\perp} -y \pmb{k}_\perp\right\}^2+y^2 M^2.
\end{eqnarray}
The quantity $\eta$ is defined in Eq.~\ref{eq:eta_definition} while $\kappa$ is in Eq.~\ref{eq:kappa}. The functions $\mathcal{G}^{(\ell_2)}_{M}$ and $\mathcal{H}^{(\ell_2,p_2)}_M$ in Eq.~\ref{eq:s1_eff_full} are provided in Eq.~\ref{eq:H_l2-p2-M}.
We notice similarities between our calculation and those presented in Ref.~\cite{Abir:2015hta}. The first line in Eq.~\ref{eq:s1_eff_full} represents the contribution from central-cut (Fig.~\ref{fig:kernel1_ph_qqgm_qqgm_ph_central}) and non-central-cut (Fig.~\ref{fig:kernel-1_ph_qqgm_qqgm_ph_left_right}) diagrams and contains identical formulas for the splitting function and the perturbative part of the scattering kernel when compared with the expression given in Eq. 26 of Ref.~\cite{Abir:2015hta}. The first line in Eq.~\ref{eq:s1_eff_full}, which contains a $\pmb{\ell}_{2\perp}$-dependent phase factor, is not present in Ref.~\cite{Abir:2015hta} as this term has been absorbed in the definition of the initial-state PDF used by Ref.~\cite{Abir:2015hta}. We decided to keep this $\pmb{\ell}_{2\perp}$-dependent phase factor within the scattering kernel herein as it contains $\zeta^-$ path length dependence between the first and second scattering. On the other hand, the modified splitting function and perturbative part in the second and third lines in Eq.~\ref{eq:s1_eff_full} are identical to Eqs. 30 and Eq. 32, respectively, in Ref.~\cite{Abir:2015hta}.

Next, kernel-2 diagrams are considered, of which six are central-cut diagrams. The hadronic tensor associated with each diagram is presented in Appendix~\ref{append:kernel2}. Adding all diagrams for kernel-2 depicted in Fig.~\ref{fig:SESS_diagram_photon_gluon}, yields the full hadronic tensor 
\begin{eqnarray}
\begin{split}
W^{\mu\nu}_{2,\rm full}  = \sum_f 2 \left[ -g^{\mu\nu}_{\perp\perp}  \right] e^2_f  \int d (\Delta X^{-}) e^{i\Delta X^{-}q^{+}} \left\langle P \left| \bar{\psi}_{_f}(\Delta X^-) \frac{\gamma^{+}}{4} \psi_{_f}(0)\right| P\right\rangle   \times \mathcal{K}^{\rm eff}_{2}
\end{split}\label{eq:Wmunu_final_type-II}   
\end{eqnarray}
where we define $\mathcal{K}^{\rm eff}_{2}$ as a effective medium-modified scattering kernel for type-2 processes via
\begin{eqnarray}
 \mathcal{K}^{\rm eff}_{2} & = & e^{2} e^{2}_{f}  g^{2}_{s} \left[ C_{F} N_{c}\right]\int d (\Delta z^{-})d^2 \Delta z_{\perp} \frac{dy}{2\pi}\frac{d^2 \ell_{2\perp}}{(2\pi)^2} \frac{d^2 k_{\perp}}{(2\pi)^2} e^{-i\Delta z^{-}\mathcal{H}^{(\ell_2,p_2)}_0} e^{i\pmb{k}_{\perp}\cdot \Delta\pmb{z}_{\perp} } \nonumber  \\ 
  &  &  \times  \int d \zeta^{-} \theta(\zeta^-)\, \mathcal{S}^{\rm eff}_2\, \left\langle P_{A-1} \left| \bar{\psi}_{_f}(\zeta^-,0) \frac{\gamma^+}{4}\psi_{_f}\left(\zeta^{-},\Delta z^{-}, \Delta \pmb{z}_{\perp}\right)\right| P_{A-1} \right\rangle.
\end{eqnarray}
$\mathcal{S}^{\rm eff}_2$ denotes the perturbative part of the medium modified kernel given by
\begin{eqnarray}
 \mathcal{S}^{\rm eff}_{2} &= &\left[\frac{1+ \left(1-y\right)^2}{y}\right]  \left[\frac{2-2\cos\left\{\mathcal{G}^{(\ell_2)}_{0}\zeta^- \right\} }{\pmb{\ell}^2_{2\perp}(1-y+\eta y)q^-} \right]  \nonumber \\
 & & + \left[\frac{1+ y^2\left( 1-\eta\right)^2 }{1-y\left(1+\eta\right) }\right]  \left[\frac{2-2\cos\left\{\mathcal{G}^{(p_2)}_{0}\zeta^- \right\} }{\left(\pmb{\ell}_{2\perp}-\pmb{k}_{\perp}\right)^2  yq^-} \right]  \nonumber \\
 & & -  \left[\frac{1- y\left(1 -2\eta\right) }{y}\right]  \left[\frac{ \pmb{\ell}^2_{2\perp} - \pmb{\ell}_{2\perp}\cdot\pmb{k}_{\perp} }{\left(\pmb{\ell}_{2\perp}-\pmb{k}_{\perp}\right)^2 \pmb{\ell}^2_{2\perp} \left(1-y+\eta y\right)q^-} \right] \left[ 2-2\cos\left\{\mathcal{G}^{(p_2)}_{0}\zeta^- \right\} - 2\cos\left\{\mathcal{G}^{(\ell_2)}_{0}\zeta^- \right\} + 2\cos\left\{ \Delta\mathcal{G}_0\zeta^- \right\}   \right], \nonumber \\
 && 
 \label{eq:s2_eff_full}
%
\end{eqnarray}
where $\Delta\mathcal{G}_0=\left(\mathcal{G}^{(p_2)}_{0}-\mathcal{G}^{(\ell_2)}_{0}\right)$, $\eta$ is defined as in Eq.~\ref{eq:eta_definition}, while $\mathcal{G}^{(\ell_2)}_{0}$  and $\mathcal{H}^{(\ell_2,p_2)}_0$ are provided in Eqs.~\ref{eq:kernel-2_G0L2} and ~\ref{eq:kernel2_H0L2P2}, respectively. Also, $\mathcal{G}^{(p_2)}_{0}$ is given as 
\begin{equation}
    \mathcal{G}^{(p_2)}_0 = p^{+}_{2} + \frac{\pmb{p}^{2}_{2\perp}}{2(q^{-}-p^{-}_{2})}  = \frac{\left(\pmb{\ell}_{2\perp}-\pmb{k}_{\perp}\right)^2 }{2y(1-y+\eta y)(1-\eta) q^-},
    \label{eq:kernel2_G0P2_S2eff}
\end{equation}
The first line in Eq.~\ref{eq:s2_eff_full} represents the contribution from the diagram in Fig.~\ref{fig:kernel2_ph_qgqm_gqqm_ph_both}(a), the second corresponds to diagram in Fig.~\ref{fig:kernel2_ph_qgqm_gqqm_ph_both}(b), while the third line corresponds to diagrams in Figs.~\ref{fig:Kernel2_g_qphqm_gqqm_gqq_both}(a,b). Each term in Eq.~\ref{eq:s2_eff_full} carries a suppression factor of $1/q^{-}$ compared to the scattering processes in kernel-1. These terms indicate that the medium-induced quark-to-gluon or quark-to-photon conversion is suppressed by a power of the incoming energy of the quark, i.e., $yq^-$ or $(1-y+\eta y)q^-$ in this case.

%
%

\subsection{Collinear expansion and jet transport coefficients at next-to-leading order (NLO) and next-to-leading twist (NLT)}
In the previous section, we presented a full scattering kernel for the hard quark traversing the nuclear medium without invoking any collinear expansion for the soft in-medium Glauber gluon/quark. In this section, we carry out the gradient expansion in $\pmb{k}_\perp$ and $k^-$ of the perturbative function $\mathcal{S}^{\rm eff}_i$ within the integrand of the scattering kernel. As in previous higher-twist calculations, such as \cite{Abir:2015hta,Sirimanna:2021sqx}, a Taylor expansion of $\mathcal{S}^{\rm eff}_i$ in $\pmb{k}_\perp$ and $k^-$, around $\pmb{k}_\perp = \pmb{0}_\perp$ and $k^-=0$, is performed, giving  
\begin{eqnarray}
    \mathcal{S}^{\rm eff}_{i}(\pmb{k}_{\perp},k^-) & = &\mathcal{S}^{\rm eff}_{i}(\pmb{k}_{\perp}=0,k^-=0) + \left. \frac{\partial \mathcal{S}^{\rm eff}_{i}}{ \partial k^{\rho}_{\perp}} \right|_{k=0}  k^{\rho}_{\perp}  +  \left. \frac{\partial^2 \mathcal{S}^{\rm eff}_{i}}{ \partial k^{\rho}_{\perp} \partial k^{\sigma}_{\perp}} \right| _{k=0} k^{\rho}_{\perp} k^{\sigma}_{\perp} + \cdots \nonumber \\
    & + & \left. \frac{\partial \mathcal{S}^{\rm eff}_{i}}{ \partial k^{-}} \right|_{k=0}  k^{-} + \left. \frac{\partial^2 \mathcal{S}^{\rm eff}_{i}}{\partial {k^{-}}^2} \right|_{k=0} \left(k^{-}\right)^2 + \cdots, %
\label{eq:H-taylor-expansion}
\end{eqnarray}
where $\vert_{k=0}$ is shorthand notation for all components of $k$ being evaluated to zero. In Eq.~\ref{eq:H-taylor-expansion}, the second term in the expansion will give a vanishing contribution to the scattering kernel (after integration over $k_\perp$ is performed), if the nuclear medium is assumed to be homogenous and isotropic. This homogeneity and isotropy assumption also ensures that the third term in Eq.~\ref{eq:H-taylor-expansion} is non-trivial solely when $\rho$ and $\sigma$ are identical.

Applying collinear expansion, the effective medium-modified scattering kernel ($\mathcal{K}^{\rm eff}_{1}$) for type-1 processes can be written as
\begin{eqnarray}
 \mathcal{K}^{\rm eff}_{1}  = e^{2} e^{2}_f  \int   \frac{dy}{2\pi}\frac{d^2 \ell_{2\perp}}{(2\pi)^2} d\zeta^{-} \left[ \mathcal{R}^{(1)}_0\hat{\mathcal{A}}_{0} + \left(\mathcal{R}^{(1)}_{T2}\hat{\mathcal{A}}_{T2} + \mathcal{R}^{(1)}_{T4}\hat{\mathcal{A}}_{T4} + \cdots\right) + \left(\mathcal{R}^{(1)}_{L1}\hat{\mathcal{A}}_{L1} + \mathcal{R}^{(1)}_{L2}\hat{\mathcal{A}}_{L2} + \cdots\right)  \right] 
\label{eq:kernel1-coll-expansion}
\end{eqnarray}
where $\mathcal{R}^{(1)}_{0}$ is the zeroth order term in the Taylor expansion of $\mathcal{S}^{\rm eff}_1$, $\mathcal{R}^{(1)}_{L,i}$ represents $i^{\rm th}$ order derivative of $\mathcal{S}^{\rm eff}_1$ along $k^-$ direction, and $\mathcal{R}^{(1)}_{T,i}$ denotes the $i^{\rm th}$ order derivative of $\mathcal{S}^{\rm eff}_1$ along $k_\perp$ direction. The operators $\hat{\mathcal{A}}_{0}$, $\hat{\mathcal{A}}_{T,i}$ and $\hat{\mathcal{A}}_{L,i}$ represent two-point gluonic jet-medium correlation functions (also called jet-medium transport coefficients), where the factors of $k_\perp$ and $k^-$ in the Taylor series expansion are converted into derivatives acting on $A^{+}$--field and are thereby absorbed in the definition of jet-medium transport coefficients. The operator $\hat{\mathcal{A}}_{T,2}$ represents the gluonic contributions to the jet-medium transport coefficient known as $\hat{q}$, characterizing the stochastic momentum broadening in the transverse direction. However, $\hat{\mathcal{A}}_{L,1}$ represents longitudinal momentum drag imparted by the nuclear medium on the jet parton; in other words, it affects the median expectation of the longitudinal momentum. Note that $\mathcal{R}^{(1)}_{i}$'s are independent of the momentum $k$ (and therefore independent of $\eta$), thus they only depend on the momentum fraction $y$, $\zeta^-$, $\pmb{\ell}^2_{2\perp}$, and quark mass $M$.
The functions $\mathcal{R}^{(1)}_{i}$ for kernel-1 are given as
\begin{eqnarray}
\mathcal{R}^{(1)}_{0} & = &  \mathcal{S}^{\rm eff}_{1}(k_{\perp}=0,k^-=0) =  \frac{1 }{\pmb{\ell}^2_{2\perp}} \left[ \frac{1+\left(1-y\right)^2}{y}\right] \left[ \frac{ 1+\chi y^2 \kappa }{ [1 +\chi]^2}\right]  \cos\left\{ \mathcal{G}^{(\ell_2)}_{M}\zeta^-\right\}, \\
%
%
\mathcal{R}^{(1)}_{T,2} &=& \left. \frac{\partial^2 \mathcal{S}^{\rm eff}_{1}}{ \partial k^2_{x}} \right| _{ k=0} + \left. \frac{\partial^2 \mathcal{S}^{\rm eff}_{1}}{ \partial k^2_{y}} \right| _{ k=0} \nonumber \\
&=&   \frac{4y^2}{\pmb{\ell}^4_{2\perp} [1+\chi]^4} \left[ \frac{1+\left(1-y\right)^2}{y}\right]   \left[ 9 + 12\chi y^2\kappa  + \chi^2    -2\cos\left\{ \mathcal{G}^{(\ell_2)}_{M}\zeta^-\right\} \left(3 + \chi\left\{1+4y^2\kappa \right\} + y^2 \chi^2\kappa \right) \right]
\end{eqnarray}

where $\chi=\frac{y^2M^2}{\ell^2_\perp}$ and $\kappa$ is defined in Eq.~\ref{eq:kappa}, while
\begin{eqnarray}
\mathcal{R}^{(1)}_{L,1} = \left. \frac{\partial \mathcal{S}^{\rm eff}_{1}}{ \partial k^{-}} \right| _{ k=0} & = &   \left[ \frac{2-y}{y}\right] \left[ \frac{1}{q^-\pmb{\ell}^2_{2\perp}}\right] \left[ \frac{-2 +2\cos{\left\{\mathcal{G}^{(\ell_2)}_{M}\zeta^-\right\}} }{1+\chi} + 2\left( \frac{1+\chi y^2\kappa}{\left(1+\chi\right)^2}\right) \right] \nonumber \\
& & + \left[  \frac{1+\left(1-y\right)^2}{y} \right]
\left[ \frac{-1 + \chi \left(1-2y^2\kappa\right)}{q^-\pmb{\ell}^2_{2\perp}\left(1+\chi\right)^3}\right] \left[ 2\cos\left\{\mathcal{G}^{(\ell_2)}_{M}\zeta^-\right\}\right]. 
\end{eqnarray}

The jet transport coefficients for kernel-1 are also the moments in $k$ momentum space of the in-medium gluon distribution, which, formally, are given by
\begin{eqnarray}
    \hat{ \mathcal{A}}_{0} & = &   g^{2}_{s}  \int d (\Delta z^{-})d^2 \Delta z_{\perp}  \frac{d^2 k_{\perp}}{(2\pi)^2}  e^{-i\Delta z^{-}\mathcal{H}^{(\ell_2,p_2)}_M} e^{i\pmb{k}_{\perp}\cdot \Delta\pmb{z}_{\perp} }  \nonumber  \\
& & \times  \theta(\zeta^-) \langle P_{A-1} | {\rm Tr}[A^{+}(\zeta^-, \Delta z^-, \Delta z_{\perp})A^+(\zeta^-, 0)] | P_{A-1} \rangle, \\
%
\hat{ \mathcal{A}}_{L,1} & = &   g^{2}_{s}  \int d (\Delta z^{-})d^2 \Delta z_{\perp}  \frac{d^2 k_{\perp}}{(2\pi)^2}  e^{-i\Delta z^{-}\mathcal{H}^{(\ell_2,p_2)}_M} e^{i\pmb{k}_{\perp}\cdot \Delta\pmb{z}_{\perp} }  \nonumber \\
& & \times  \theta(\zeta^-) \langle P_{A-1} | {\rm Tr}[i\partial^-  A^{+}(\zeta^-, \Delta z^-, \Delta z_{\perp})A^+(\zeta^-, 0)] | P_{A-1} \rangle, \\
%
\hat{ \mathcal{A}}_{T,2} & = &  g^{2}_{s}  \int d (\Delta z^{-})d^2 \Delta z_{\perp}  \frac{d^2 k_{\perp}}{(2\pi)^2}  e^{-i\Delta z^{-}\mathcal{H}^{(\ell_2,p_2)}_M} e^{i\pmb{k}_{\perp}\cdot \Delta\pmb{z}_{\perp} }   \nonumber \\
& & \times  \theta(\zeta^-) \langle P_{A-1} | {\rm Tr}[  \partial_{\perp} A^{+}(\zeta^-, \Delta z^-, \Delta z_{\perp}) \partial_{\perp}A^+(\zeta^-,0)]  | P_{A-1} \rangle.
\end{eqnarray}
In the above equations, the function $\mathcal{H}^{(\ell_2,p_2)}_M$ is defined as Eq.~\ref{eq:H_l2-p2-M}, while $\hat{\mathcal{A}}_{T,2}$ is similar to $\hat{q}$~\cite{Abir:2015hta}, which characterizes transverse momentum diffusion of the hard parton traversing the nuclear medium.
\footnote{The relative importance of the bosonic and fermionic contribution to $\hat{q}$ depends on the composition of the plasma. At very early times, the plasma in heavy-ion collisions is gluon-dominated as the gluonic PDF is much larger than quark PDFs. As the plasma evolves, quark population densities will increase to reach near thermal equilibrium in the QGP. So, at hydrodynamization time, both quarks and gluons contribute to $\hat{q}$, while at early times, the gluonic contribution to $\hat{q}$ is the only relevant one.}  
On the other hand, $\hat{\mathcal{A}}_{L,1}$ is the jet transport coefficient characterizing longitudinal momentum drag, which is similar to $\hat{e}~$\cite{Abir:2015hta}. Both $\hat{\mathcal{A}}_{L,1}$ and $\hat{\mathcal{A}}_{T,2}$ involve gluon correlators. Note that $\hat{\mathcal{A}}_{0}, \hat{\mathcal{A}}_{L,1}$, and $\hat{\mathcal{A}}_{T,2}$ depend explicitly on $\pmb{\ell}_{2\perp}$ via the function $\mathcal{H}^{(\ell_2,p_2)}_M$, thus these are transverse-momentum-dependent gluon parton distribution functions (TMD-gPDFs) \cite{Kumar:2025rsa}. We note in Eq.~\ref{eq:kernel1-coll-expansion}, the function $\hat{\mathcal{A}}_{0}$ gives rise to a gauge correction term to the nuclear PDF in the limit $\pmb{\ell}_{\perp}, \pmb{k}_{\perp} \rightarrow 0$~\cite{Wang:2001ifa}. 


Collinear expansion to kernel-2 is examined next. The effective medium-modified kernel for type-2 processes ($\mathcal{K}^{\rm eff}_{2}$), is
\begin{equation}
\begin{split}
 \mathcal{K}^{\rm eff}_{2} & = e^{2} e^{2}_{f}  \left[ C_F N_c\right] \int   \frac{dy}{2\pi}\frac{d^2 \ell_{2\perp}}{(2\pi)^2} d\zeta^{-} \left[ \mathcal{R}^{(2)}_0\hat{\mathcal{F}}_{0} + \left(\mathcal{R}^{(2)}_{T2}\hat{\mathcal{F}}_{T2} + \mathcal{R}^{(2)}_{T4}\hat{\mathcal{F}}_{T4} + \cdots\right) + \left(\mathcal{R}^{(2)}_{L1}\hat{\mathcal{F}}_{L1} + \mathcal{R}^{(2)}_{L2}\hat{\mathcal{F}}_{L2} + \cdots\right)  \right] 
  \end{split}
  \label{eq:kernel2-R0F0_RiFi}
\end{equation}
where $\mathcal{R}^{(2)}_{0}$ is the zeroth order term in the Taylor expansion of $\mathcal{S}^{\rm eff}_2$, $\mathcal{R}^{(2)}_{L,i}$ represents $i^{\rm th}$ order derivative of $\mathcal{S}^{\rm eff}_2$ along the $k^-$ direction, and $\mathcal{R}^{(2)}_{T,i}$ denotes the $i^{\rm th}$ order derivative of $\mathcal{S}^{\rm eff}_2$ along $k_\perp$ direction, as before. The operators $\hat{\mathcal{F}}_{0}$, $\hat{\mathcal{F}}_{T,i}$ and $\hat{\mathcal{F}}_{L,i}$ represent two-point fermionic jet-medium correlation functions (or transport coefficients), where the factors of $k_\perp$ and $k^-$ in the Taylor series expansion are converted into derivatives acting on Glauber fermionic $\psi$ fields, and are thereby absorbed in the definition of the jet-medium transport coefficients. 
%
The operator $\hat{\mathcal{F}}_{T,2}$ represents a subleading, transverse-momentum-dependent correction from the conversion process to the zeroth-order term in Eq.~\ref{eq:kernel2-R0F0_RiFi}.
%
%
As before, $\mathcal{R}^{(2)}_{i}$'s solely depend on the momentum fraction $y$, $\zeta^-$, and $\pmb{\ell}^2_{2\perp}$. The Taylor expansion above allows one to decouple $\pmb{\ell}_{\perp}$ and $\pmb{k}_{\perp}$ integrals.

The function $\mathcal{R}^{(2)}_{i}$ for kernel-2 are given as
\begin{eqnarray}
\mathcal{R}^{(2)}_{0} =\left. \mathcal{S}^{\rm eff}_{2}(k_{\perp},k^-)\right|_{ k=0} &=& \left[ \frac{1+\left(1-y\right)^2}{y}\right] \left[ \frac{2-2\cos\left\{\mathcal{G}^{(\ell_2)}_0\zeta^-\right\}}{\pmb{\ell}^2_{2\perp}(1-y)q^-}\right] + \left[ \frac{1+y^2}{1-y} \right] \left[ \frac{2-2\cos\left\{\mathcal{G}^{(\ell_2)}_0\zeta^-\right\}}{\pmb{\ell}^2_{2\perp}yq^-}\right] \nonumber \\
&-& \left[ \frac{1-y}{y}\right] \left[ \frac{4-4\cos\left\{\mathcal{G}^{(\ell_2)}_0\zeta^-\right\}}{\pmb{\ell}^2_{2\perp}(1-y)q^-}\right], \\
\mathcal{R}^{(2)}_{T,2} &=& \left. \frac{\partial^2 \mathcal{S}^{\rm eff}_{2}}{ \partial k_{x}^2} \right| _{ k=0} + \left. \frac{\partial^2 \mathcal{S}^{\rm eff}_{2}}{ \partial k_{y}^2} \right| _{ k=0}\nonumber\\
& = & \left[ \frac{1+y^2}{1-y}\right] \left[ \frac{1}{yq^-}\right] \left[ \frac{8 -8 \cos\left\{ \mathcal{G}^{(\ell_2)}_{0}\zeta^-\right\} }{\pmb{\ell}^4_{2\perp}} - \frac{8\beta \sin\left\{ \mathcal{G}^{(\ell_2)}_{0}\zeta^-\right\} }{\pmb{\ell}^2_{2\perp}} + 8\beta^2 \cos\left\{ \mathcal{G}^{(\ell_2)}_{0}\zeta^-\right\} \right] \\
& & + \frac{1}{yq^-} \left[ 8\beta^2 \left[1- \cos\left\{ \mathcal{G}^{(\ell_2)}_{0}\zeta^-\right\} \right] -\frac{16\beta}{\pmb{\ell}^2_{2\perp}}  \sin\left\{ \mathcal{G}^{(\ell_2)}_{0}\zeta^-\right\} \right] \\
\mathcal{R}^{(2)}_{L,1} = \left. \frac{\partial \mathcal{S}^{\rm eff}_{2}}{ \partial k^{-}} \right| _{ k=0} & =& \frac{\left[2-2\cos\left\{ \mathcal{G}^{(\ell_2)}_{0} \zeta^-\right\} \right]\left(2y-5\right) }{y\left(1-y\right)^2 \pmb{\ell}_{2\perp}^2 \left(q^{-}\right)^2 }       
- \frac{\sin\left\{\mathcal{G}^{(\ell_2)}_{0} \zeta^- \right\} \zeta^- \left(3+y\right) }{y\left(1-y\right)^3 \left(q^-\right)^3},
\end{eqnarray}
where
\begin{equation}
\beta = \frac{\zeta^-}{2y(1-y)q^-}. 
\label{eq:k_expansion_beta}
\end{equation}
The jet-medium transport coefficients for kernel-2 at NLO and NLT are in-medium two-point fermionic field distributions given by
\begin{eqnarray}
\hat{ \mathcal{F}}_{0} & = &   g^{2}_{s}  \int d (\Delta z^{-})d^2 \Delta z_{\perp}  \frac{d^2 k_{\perp}}{(2\pi)^2}  e^{-i\Delta z^{-}\mathcal{H}^{(\ell_2,p_2)}_0} e^{i\pmb{k}_{\perp}\cdot \Delta\pmb{z}_{\perp} }  \nonumber \\
& &\times \theta(\zeta^-) \left\langle P_{A-1}\left|\bar{\psi}_{_f}\left(\zeta^-,0\right)\frac{\gamma^+}{4}\psi_{_f}(\zeta^-, \Delta z^-, \Delta\pmb{z}_{\perp}) \right|P_{A-1}\right\rangle, \\
\hat{ \mathcal{F}}_{L,1} & = &  g^{2}_{s}  \int d (\Delta z^{-})d^2 \Delta z_{\perp}  \frac{d^2 k_{\perp}}{(2\pi)^2}  e^{-i\Delta z^{-}\mathcal{H}^{(\ell_2,p_2)}_0} e^{i\pmb{k}_{\perp}\cdot \Delta\pmb{z}_{\perp} }  \nonumber \\
& & \times \theta(\zeta^-) \left\langle P_{A-1}\left|i\partial^-\bar{\psi}_{_f}\left(\zeta^-,0\right)\frac{\gamma^+}{4}\psi_{_f}(\zeta^-, \Delta z^-, \Delta\pmb{z}_{\perp}) \right|P_{A-1}\right\rangle, \\
\hat{ \mathcal{F}}_{T,2} & =&   g^{2}_{s}  \int d (\Delta z^{-})d^2 \Delta z_{\perp}  \frac{d^2 k_{\perp}}{(2\pi)^2}  e^{-i\Delta z^{-}\mathcal{H}^{(\ell_2,p_2)}_0} e^{i\pmb{k}_{\perp}\cdot \Delta\pmb{z}_{\perp} }  \nonumber \\
& & \times \theta(\zeta^-) \left\langle P_{A-1}\left|\partial_{\perp}\bar{\psi}_{_f}\left(\zeta^-,0\right)\frac{\gamma^+}{4}\partial_{\perp}\psi_{_f}(\zeta^-, \Delta z^-, \Delta\pmb{z}_{\perp}) \right|P_{A-1}\right\rangle.
\end{eqnarray}
where $\mathcal{H}^{(\ell_2,p_2)}_{0}$ is given in Eq.~\ref{eq:kernel2_H0L2P2}. The $\hat{\mathcal{F}}_{0}, \hat{\mathcal{F}}_{L,1}$, and $\hat{\mathcal{F}}_{T,2}$ distribution functions depend explicitly on $\pmb{\ell}_{2\perp}$ via the function $\mathcal{H}^{(\ell_2,p_2)}_0$ as before; hence, these are transverse-momentum-dependent quark parton distribution functions (TMD-qPDFs). Note that the zeroth-order term $\mathcal{R}^{(1)}_{0}\hat{\mathcal{A}}_{0}$ in Eq.~\ref{eq:kernel1-coll-expansion} represents a gauge correction term that makes the nuclear PDF gauge invariant and therefore does not contribute to parton energy loss. Unlike this, the zeroth order term in Eq.~\ref{eq:kernel2-R0F0_RiFi} does not represent a gauge term and hence cannot be absorbed in the nuclear PDF. Indeed, this zeroth-order term is a genuine first non-vanishing leading term that contributes to the parton energy loss. 


Recent progress \cite{Kurkela:2018vqr,Kurkela:2018wud,Giacalone:2019ldn,Kamata:2020mka} shows that hydrodynamic attractors can provide a reasonable description of the transition of the energy-momentum tensor of the nuclear medium from highly off-equilibrium glasma-like dynamics at early times to a hydrodynamical evolution at later times. More studies are needed to understand the mechanisms responsible for the hydrodynamical behavior of quark flavors in the QGP, i.e., for flavor hydrodynamization. The photons from short-lived (highly-virtual) partons explored herein are uniquely positioned to discern the intricate dynamics involved in flavor hydrodynamization transitioning from glasma-like dynamics to hydrodynamics. In contrast, conversion processes giving photon production from on-shell (and nearly on-shell) partons, such as those studied in \cite{Ghiglieri:2013gia,Qin:2008rd}, probe not only the prehydrodynamical evolution, but also parts of the hydrodynamical evolution as well, owing to their longer lifetimes. 
If glasma-like dynamics are present at very early times, then operators $\hat{\mathcal{F}}_{0}$, $\hat{\mathcal{F}}_{T,i}$ and $\hat{\mathcal{F}}_{L,i}$ will play a key role in studying possible flavor hydrodynamization mechanisms: $\hat{\mathcal{A}}_{T,i}$ and $\hat{\mathcal{A}}_{L,i}$ are {\it not enough}. 

The first phenomenological study of photon's sensitivity to flavor hydrodynamization \cite{Gale:2021emg} used suppressed thermal photons rates in the pre-hydrodynamical stage using K\o MP\o ST.~\footnote{Prompt photons were included in Ref.~\cite{Gale:2021emg} that are only sensitive to nuclear-modified PDFs and not to the dynamical evolution of the nuclear medium.} The limited study in Ref.~\cite{Gale:2021emg} did not see large flavor hydordynamization effects. A more realistic calculation of photon production from early-time dynamics also included jet-medium photons from long-lived partons \cite{Yazdi:2022cuk}. However, no flavor hydrodynamization effects on jet-medium photons were shown in Ref.~\cite{Yazdi:2022cuk}. Combining jet-medium photons from highly virtual partons, discussed here, together with photons from long-lived partons, as well thermal photons explored in \cite{Yazdi:2022cuk}, all within a dynamical model that includes flavor hydrodynamization effects, promises to provide greater discriminating power to possible mechanisms involved in flavor hydrodynamization dynamics. 
The next section will quantify the importance of jet-medium photons from highly virtual partons. 
%



\subsection{Length dependence of energy loss}
A numerical evaluation of the zeroth-order, first-order and second-order coefficients $\mathcal{R}^{(i)}_{0}$,  $\mathcal{R}^{(i)}_{L,1}$, and $\mathcal{R}^{(i)}_{T,2}$ that arise in the Taylor expansion of the full scattering kernel is carried out for each kernel. The collinear expansion outlined in the preceding subsection allowed us to decouple  $k_\perp$ and $k^-$ dependence and absorb it in the definition of the jet transport coefficients.

\begin{figure}[!h]
    \begin{subfigure}[t]{0.49\textwidth}
\includegraphics[width=\textwidth]{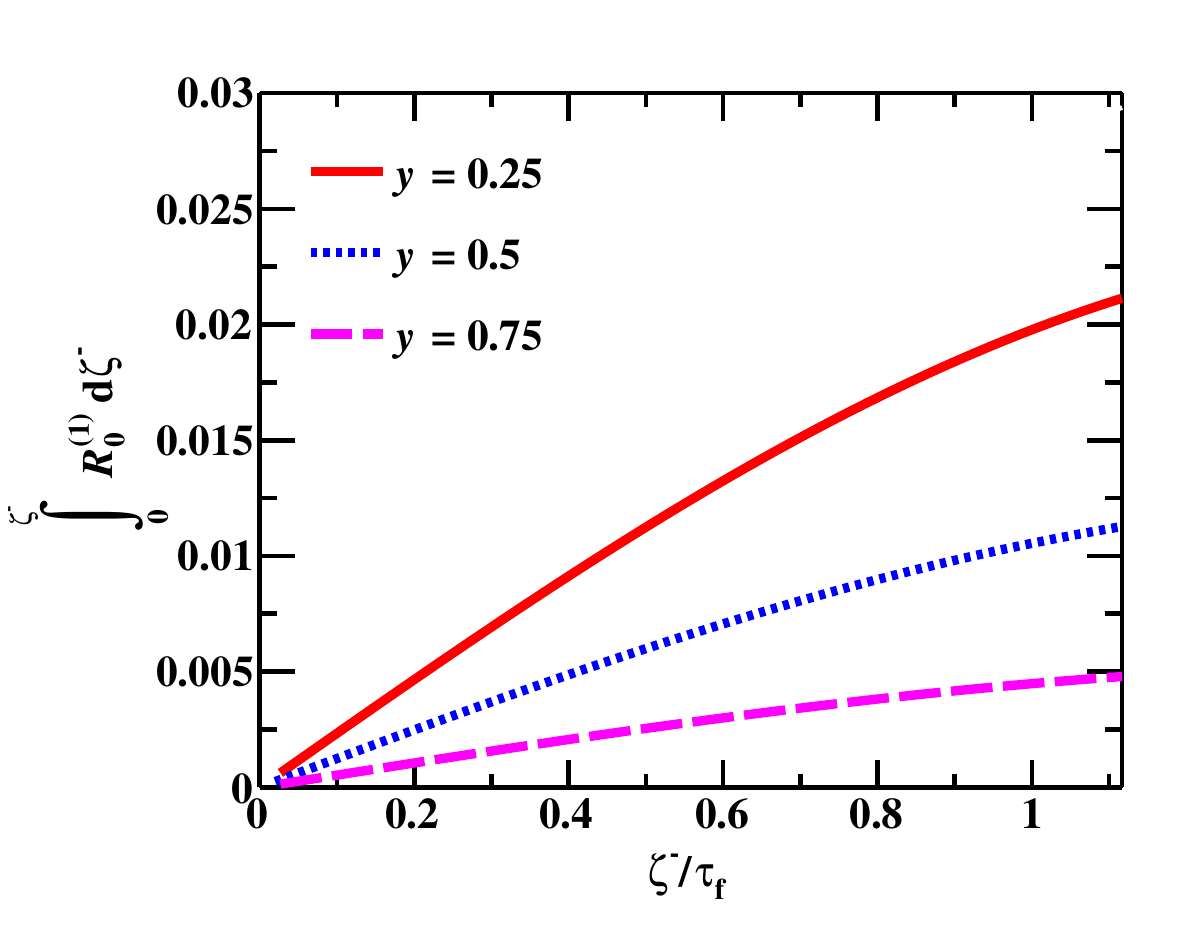}
        \caption{Zeroth-order coefficient ($\mathcal{R}^{(1)}_{0}$)  in Kernel-1.}
    \end{subfigure}%

    \begin{subfigure}[t]{0.49\textwidth}
\includegraphics[width=\textwidth]{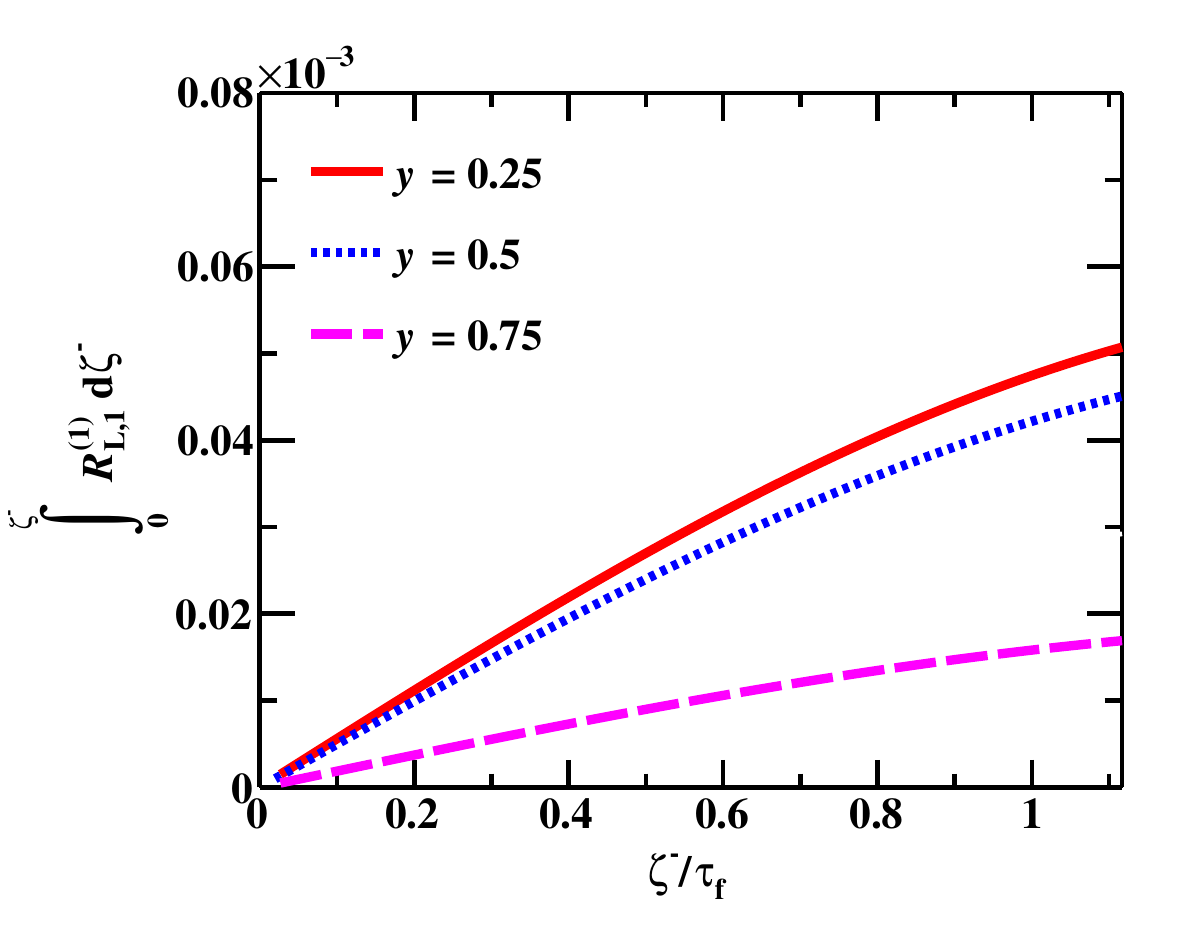}
        \caption{First-order coefficient ($\mathcal{R}^{(1)}_{L,1}$) in Kernel-1.}
    \end{subfigure}

    \begin{subfigure}[t]{0.49\textwidth}
        \centering        
\includegraphics[width=\textwidth]{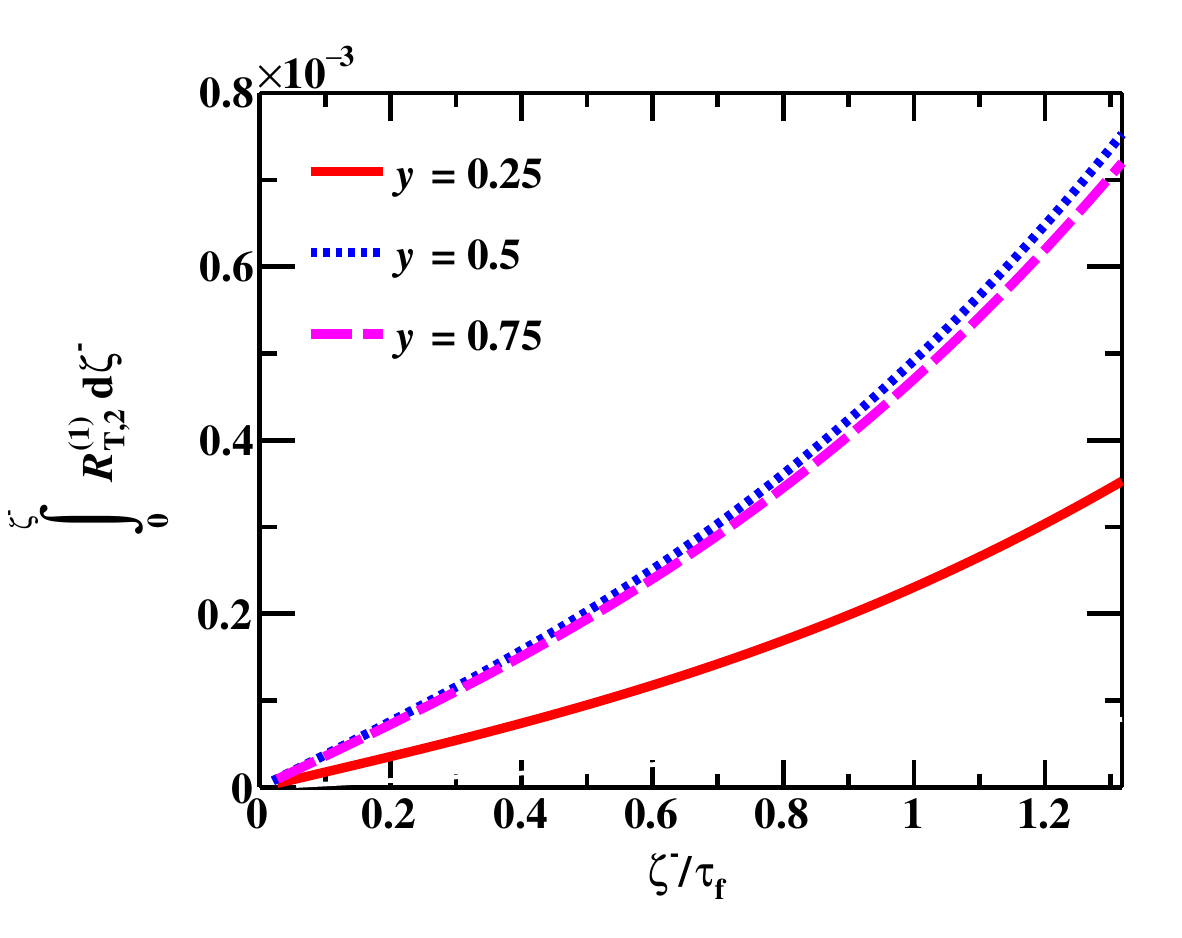}
        \caption{Second-order coefficient ($\mathcal{R}^{(1)}_{T,2}$) in Kernel-1.}
    \end{subfigure}
\caption{Path length dependence of zeroth-order ($\mathcal{R}^{(1)}_{0}$), first-order ($\mathcal{R}^{(1)}_{L,1}$) and second-order  ($\mathcal{R}^{(1)}_{T,2}$) perturbative coefficients in the gradient expansion of $\mathcal{S}^{\rm eff}_{1}$ as a function of $\zeta^-/\tau_f$, where $\tau_f$ is a formation time given as $\tau_f=2y(1-y)q^{-}/\ell^2_{2\perp}$. Here, $y$ is the momentum fraction carried away by the radiated photon. Other parameters are set to $q^{-}=100$ GeV, $\ell_x=10$ GeV, $\ell_y=0$ GeV, $M=0$ GeV.} 
\label{fig:kernel1_path_length_dependence}
\end{figure}
\begin{figure}[!h]
    \begin{subfigure}[t]{0.49\textwidth}
        \centering        \includegraphics[width=\textwidth]{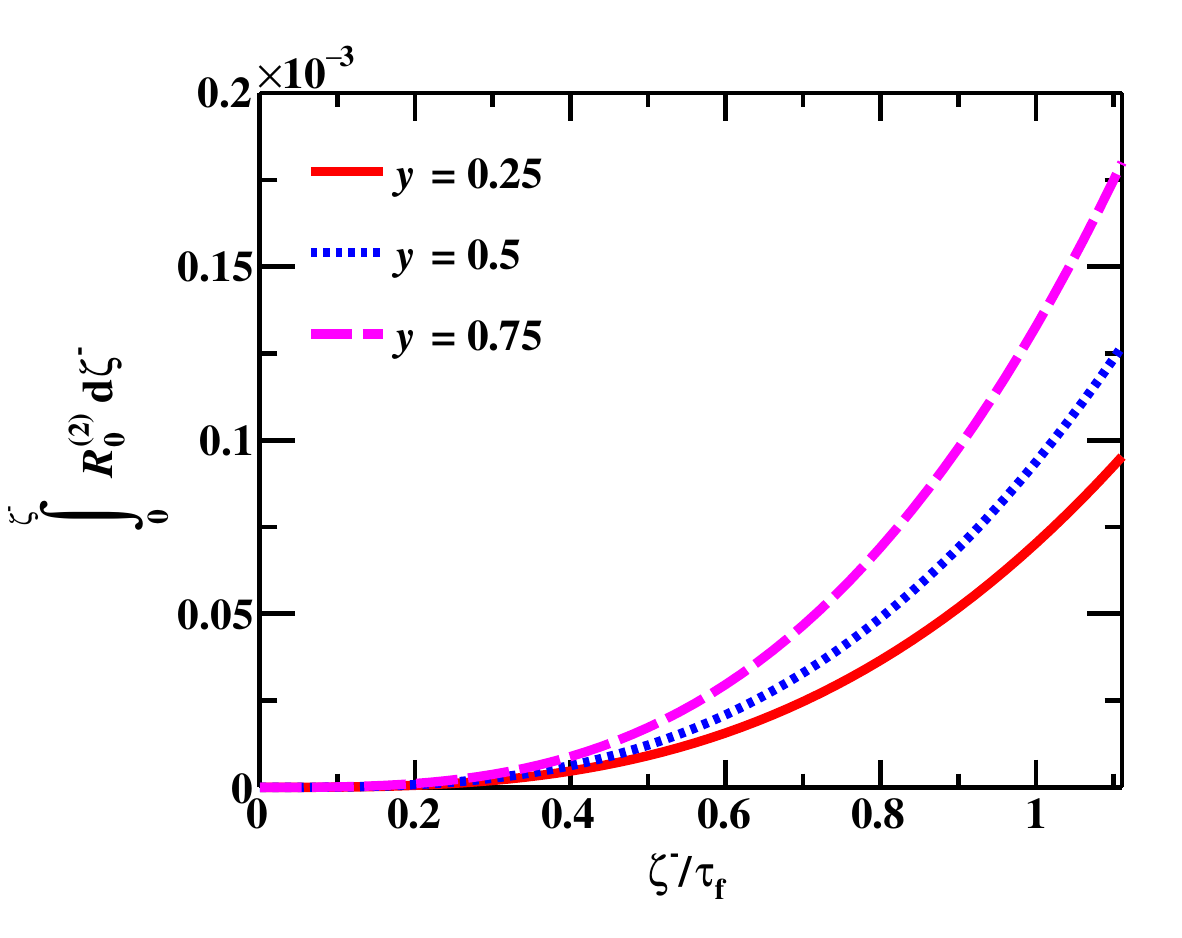}
        \caption{Zeroth-order coefficient in Kernel-2.}
    \end{subfigure}%
    
    \begin{subfigure}[t]{0.49\textwidth}
        \centering        \includegraphics[width=\textwidth]{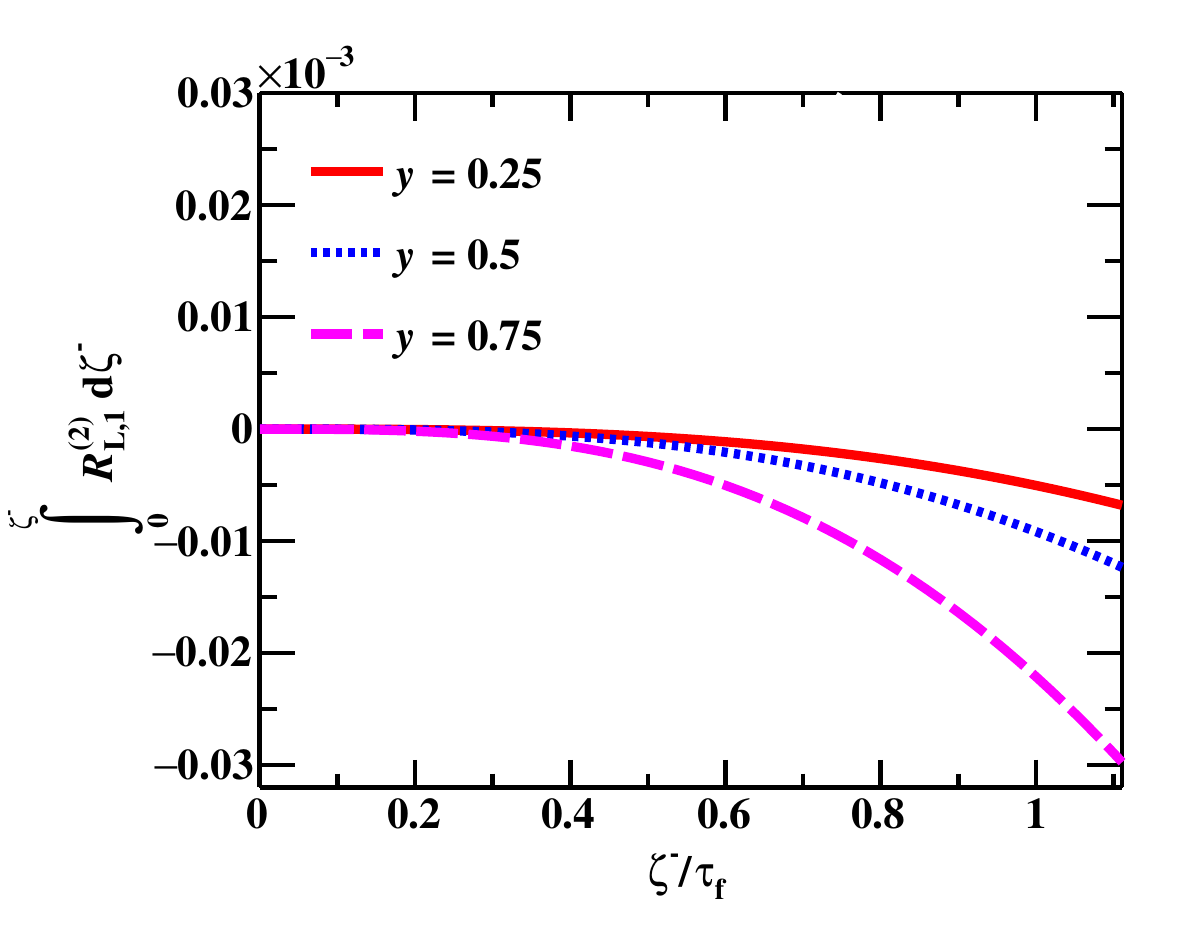}
        \caption{First-order coefficient in Kernel-2.}
    \end{subfigure}
   
    \begin{subfigure}[t]{0.49\textwidth}
        \centering        \includegraphics[width=\textwidth]{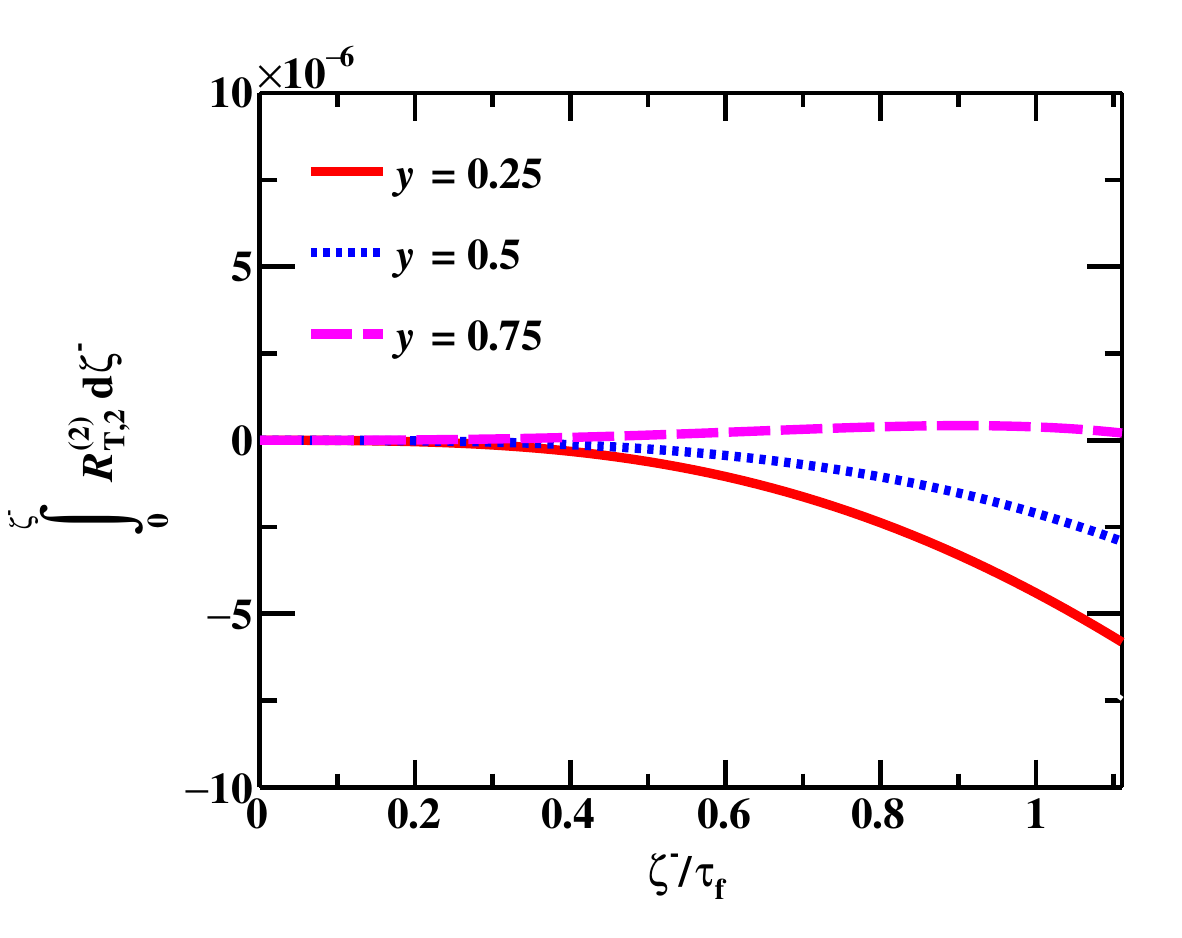}
        \caption{Second-order coefficient in Kernel-2.}
    \end{subfigure}
\caption{Path length dependence of zeroth-order ($\mathcal{R}^{(2)}_{0}$), first-order ($\mathcal{R}^{(2)}_{L,1}$) and second-order ($\mathcal{R}^{(2)}_{T,2}$) coefficients in the gradient expansion of $\mathcal{S}^{\rm eff}_{2}$  as a function of $\zeta^-/\tau_f$, where $\tau_f$ is a formation time given as $\tau_f=2y(1-y)q^{-}/\ell^2_{2\perp}$. Here, $y$ is the momentum fraction carried away by the radiated photon. Other parameters are set to $q^{-}=100$ GeV, $\ell_x=10$ GeV, $\ell_y=0$ GeV, $M=0$ GeV.} 
\label{fig:kernel2_path_length_dependence}
\end{figure}
First, we consider the perturbative functions $\mathcal{R}^{(1)}_{0}$, $\mathcal{R}^{(1)}_{L,1}$, and $\mathcal{R}^{(1)}_{T,2}$ in kernel-1. We note that the length integration variable $\zeta^-$ only appears in $\left[2-2\cos\left\{\mathcal{G}^{(\ell_2)}_{0}\zeta^-\right\}\right]$ term and the two-point correlator $\langle P_{A-1}|A^{+}(\zeta^-,\Delta z^-)A^{+}(\zeta^-,0)|P_{A-1} \rangle$. Under translational invariance around $\zeta^-$, the two-point correlator does not depend on the mean location $\zeta^-$, thus allowing to decouple the $d\zeta^-$ and $d\Delta z^-$ integrations. Using this result, Fig.~\ref{fig:kernel1_path_length_dependence} depicts the length-integrated $\mathcal{R}^{(1)}_{0}$, $\mathcal{R}^{(1)}_{L,1}$ and $\mathcal{R}^{(1)}_{T,2}$ for kernel-1, evaluated at three different momentum fractions: $y=0.25, 0.5$, and 0.75. Figure~\ref{fig:kernel1_path_length_dependence}(a) depicts the length dependence of the zeroth-order coefficient, $\mathcal{R}^{(1)}_{0}$. As mentioned in the previous section (Eq.~\ref{eq:kernel1-coll-expansion}),  $\mathcal{R}^{(1)}_0\hat{\mathcal{A}}_{0}$ term represents the gauge correction to the nuclear PDF and does not contribute to parton energy loss~\cite{Wang:2001ifa}. Figure~\ref{fig:kernel1_path_length_dependence}(b) and (c) depict the length dependence of the first-order and second-order perturbative coefficients in kernel-1, respectively, which are responsible for parton drag and diffusion as a function of path-length $(\zeta^-)$. Overall, the coefficient of longitudinal momentum drag ($\mathcal{R}^{(1)}_{L,1}$) is an order of magnitude smaller than the coefficient of transverse momentum broadening ($\mathcal{R}^{(1)}_{T,2}$), akin to Brownian diffusion. In these calculations, the quark mass is set to $M=0$.

Next, we consider the perturbative coefficients $\mathcal{R}^{(2)}_{0}$, $\mathcal{R}^{(2)}_{L,1}$, and $\mathcal{R}^{(2)}_{T,2}$ in kernel-2. This kernel represents contributions from fermion-to-boson conversion processes. 
To understand the physics implications of the relative magnitude of $\mathcal{R}^{(2)}_{0}$, $\mathcal{R}^{(2)}_{L,1}$, and $\mathcal{R}^{(2)}_{T,2}$ coefficients on the parton energy loss, it is important to have a lattice estimates of the transport coefficients $\hat{\mathcal{F}}_{0}$, $\hat{\mathcal{F}}_{L,1}$ and $\hat{\mathcal{F}}_{T,2}$. Although there exist lattice determination of the gluonic $\hat{q}$ \cite{Kumar:2020wvb}, the estimates of transport coefficients $\hat{\mathcal{F}}_{0}$, $\hat{\mathcal{F}}_{L,1}$ and $\hat{\mathcal{F}}_{T,2}$ are unknown. Therefore, we refrain from making concrete statements about the impact of these coefficients on the parton energy loss.
In Fig.~\ref{fig:kernel2_path_length_dependence}(a), zeroth-order perturbative function $\mathcal{R}^{(2)}_{0}$ is shown, whereas Figure~\ref{fig:kernel2_path_length_dependence}(b) and (c) represents the length dependence of first-order $\mathcal{R}^{(2)}_{L,1}$ and second-order $\mathcal{R}^{(2)}_{T,2}$, respectively. Note that in Eq.~\ref{eq:kernel2-R0F0_RiFi}, the term $\mathcal{R}^{(2)}_{0}\hat{ \mathcal{F}}_{0}$ represents the first non-vanishing contribution to parton energy loss for kernel-2. It is observed that both $\mathcal{R}^{(2)}_{L,1}$ and $\mathcal{R}^{(2)}_{T,2}$ are an order of magnitude smaller compared to the zeroth-order coefficient $\mathcal{R}^{(2)}_{0}$. Comparing $\mathcal{R}^{(2)}_{0}$ with coefficients in kernel-1, we observe that it is suppressed by a factor of 100 compared to $\mathcal{R}^{(1)}_{0}$ and is a factor of 4 smaller compared to $\mathcal{R}^{(1)}_{T,2}$ of kernel-1. Qualitatively, this signifies that the processes in kernel-2 are suppressed by an order of magnitude relative to $\mathcal{R}^{(1)}_{L,1}$ and $\mathcal{R}^{(1)}_{T,2}$.

\begin{figure}[!h]
    \begin{subfigure}[t]{0.49\textwidth}
        \centering        \includegraphics[width=\textwidth]{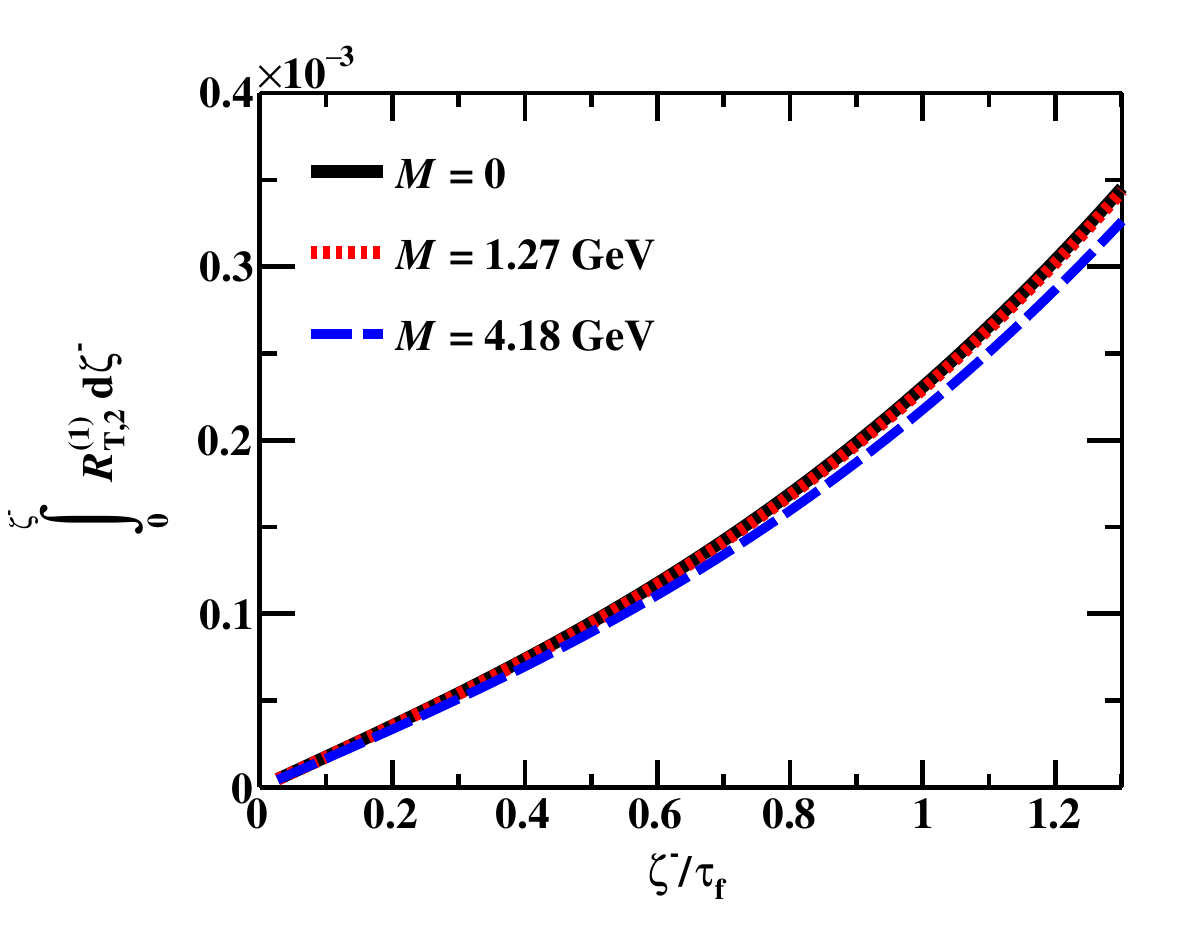}
        \caption{Length integrated  $\mathcal{R}^{(1)}_{T,2}$ for $y=0.25$.}
    \end{subfigure}%
    
    \begin{subfigure}[t]{0.49\textwidth}
        \centering        \includegraphics[width=\textwidth]{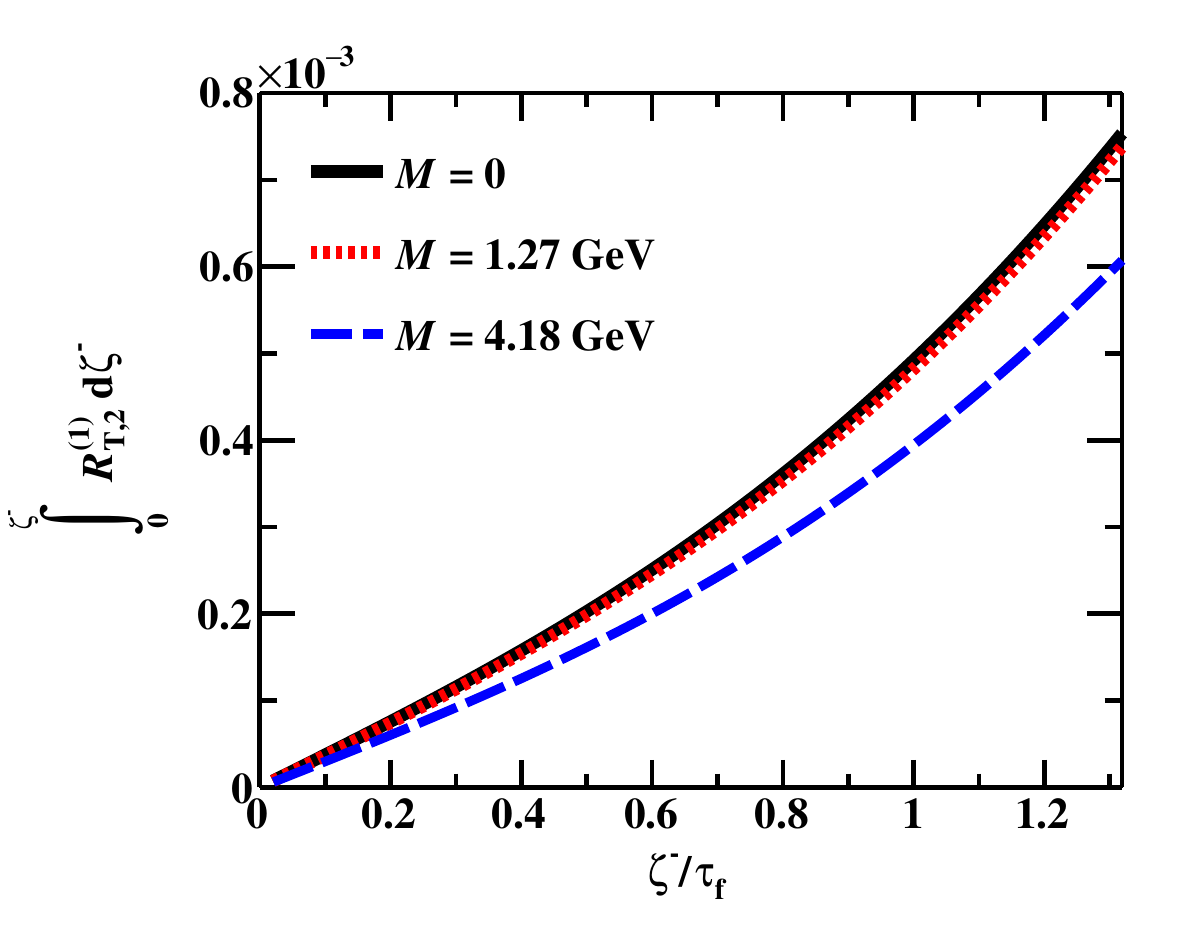}
        \caption{Length integrated $\mathcal{R}^{(1)}_{T,2}$ for $y=0.5$.}
    \end{subfigure}
      
    \begin{subfigure}[t]{0.49\textwidth}
        \centering        \includegraphics[width=\textwidth]{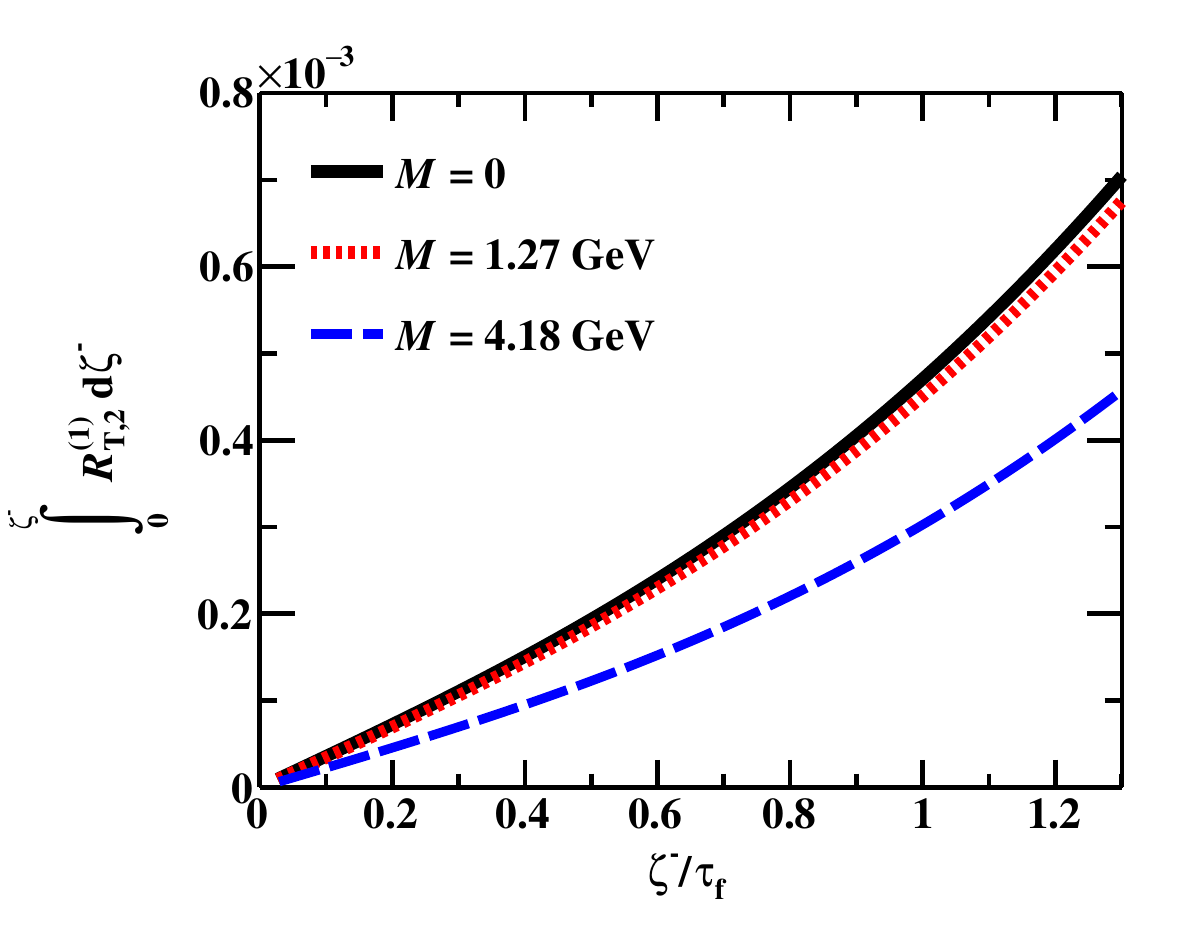}
        \caption{Length integrated $\mathcal{R}^{(1)}_{T,2}$ for $y=0.75$.}
    \end{subfigure} 
\caption{Path length dependence of  second order gradient term $\mathcal{R}^{(1)}_{T,2}$  (transverse direction) as a function of $\zeta^-/\tau_f$, where $\tau_f$ is a formation time given as $\tau_f=2y(1-y)q^{-}/(\ell^2_{2\perp}+y^2M^2)$. This is for kernel-1, and $y$ is the momentum fraction carried away by the radiated photon. Other parameters: $q^{-}=100$ GeV, $\ell_x=10$ GeV, and $\ell_y=0$ GeV.} 
\label{fig:Kernel1-Mass_dependence}
\end{figure}

\begin{figure}[!h]
    \begin{subfigure}[t]{0.49\textwidth}
        \centering        \includegraphics[width=\textwidth]{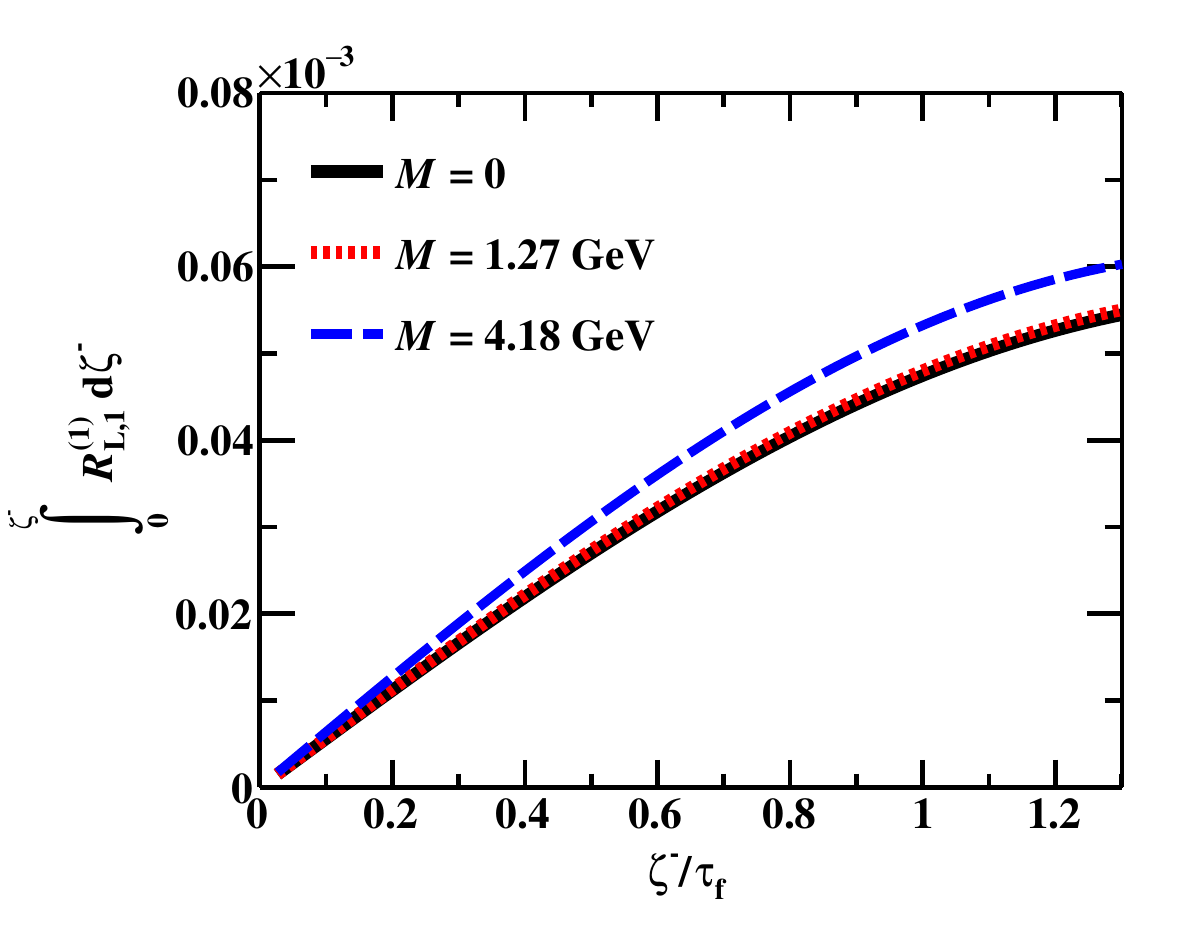}
        \caption{Length integrated  $\mathcal{R}^{(1)}_{L,1}$ for $y=0.25$.}
    \end{subfigure}%
    
    \begin{subfigure}[t]{0.49\textwidth}
        \centering        \includegraphics[width=\textwidth]{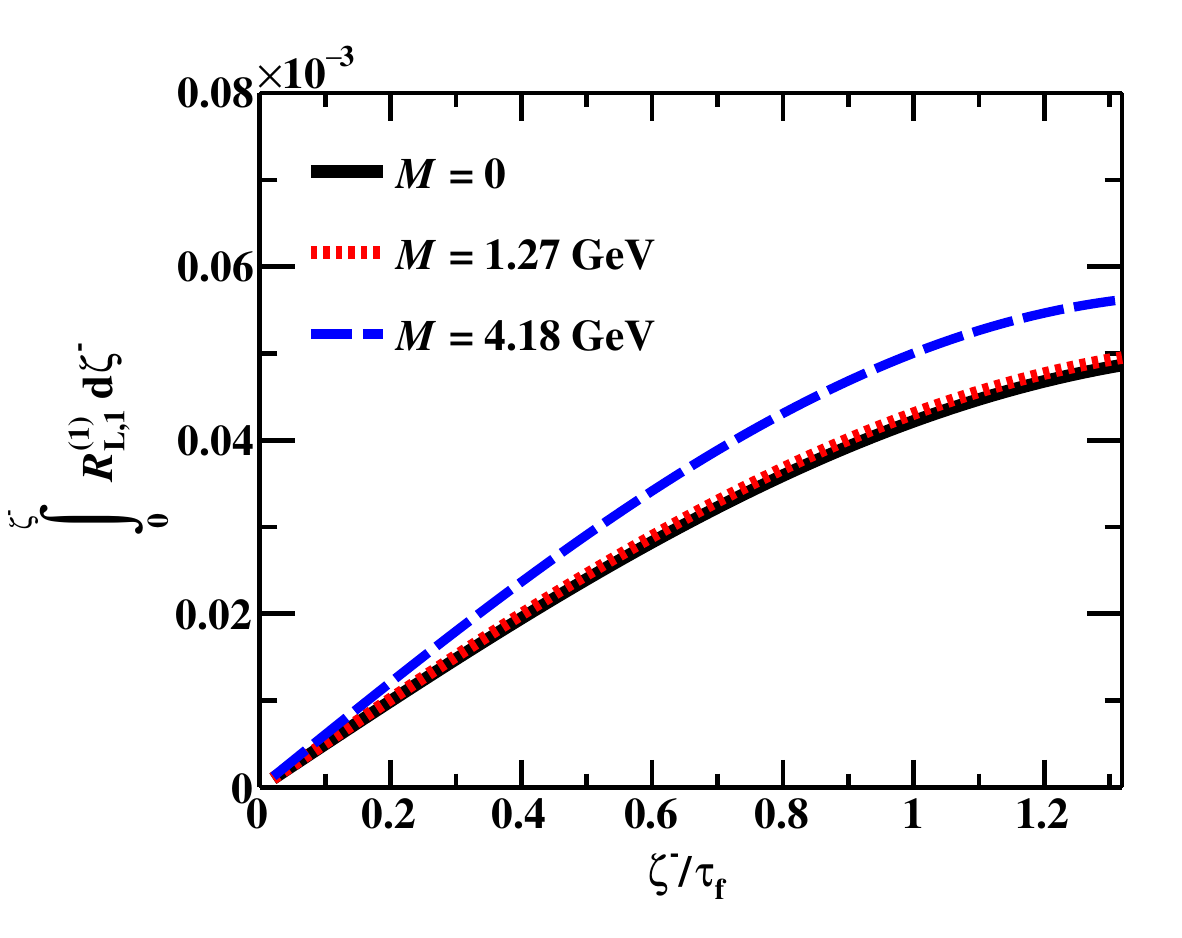}
        \caption{Length integrated $\mathcal{R}^{(1)}_{L,1}$ for $y=0.5$.}
    \end{subfigure}
     
    \begin{subfigure}[t]{0.49\textwidth}
        \centering        \includegraphics[width=\textwidth]{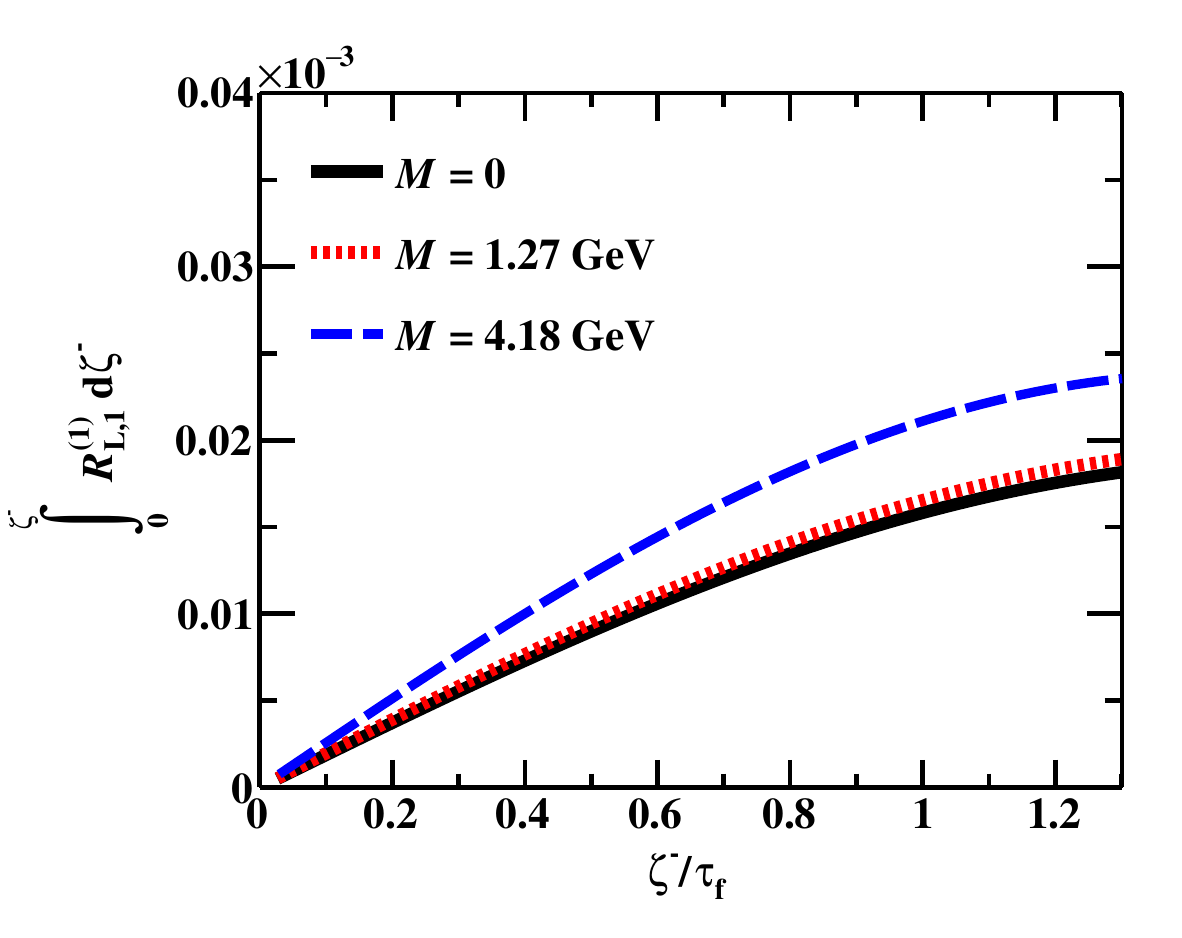}
        \caption{Length integrated $\mathcal{R}^{(1)}_{L,1}$ for $y=0.75$.}
    \end{subfigure} 
\caption{Path length dependence of  first order gradient term $\mathcal{R}^{(1)}_{L,1}$  (in $k^-$ direction) as a function of $\zeta^-/\tau_f$, where $\tau_f$ is a formation time given as $\tau_f=2y(1-y)q^{-}/(\ell^2_{2\perp}+y^2M^2)$. This is for kernel-1, and $y$ is the momentum fraction carried away by the radiated photon. Other parameters: $q^{-}=100$ GeV, $\ell_x=10$ GeV, and $\ell_y=0$ GeV.} 
\label{fig:Kernel1-Mass_dep_first_order_term}
\end{figure}

In Fig.~\ref{fig:Kernel1-Mass_dependence}, we present the quark mass dependence of the length-integrated second-order gradient $\mathcal{R}^{(1)}_{T,2}$ for kernel-1. Each panel represents a different momentum fraction, while containing three different quark masses. In the ${\rm \overline{MS}}$ scheme~\cite{PDG_qmass}, heavy-quark masses are set to: $M=1.27$ GeV (charm-quark) and $M=4.18$ GeV (bottom-quark). The results indicate no noticeable differences for the charm quark when compared to lighter quarks; however, a significant effect can be seen for the bottom quark for $y>0.25$.  In Fig.~\ref{fig:Kernel1-Mass_dep_first_order_term}, we present quark masses effects in the length-integrated first-order coefficient $\mathcal{R}^{(1)}_{L,1}$. Here as well, we observe a noticeable difference for the bottom quark compared to lighter quark flavors. Since the coefficient of longitudinal momentum drag $\left(\mathcal{R}^{(1)}_{L,1}\right)$ is smaller than the coefficient of transverse momentum broadening $\left(\mathcal{R}^{(1)}_{T,2}\right)$, Fig.~\ref{fig:Kernel1-Mass_dependence} shows that bottom quarks encounter smaller transverse momentum broadening compared to lighter flavors. For the case of kernel-2, the six possible central-cut diagrams do not contribute to heavy-quark energy loss, as typical temperatures reached in the QGP are insufficient to thermally pair-produce heavy quarks. 

Note that the calculations presented above are based on the collinear expansion approximation. A realistic numerical calculation of the full scattering kernel requires estimates of the non-perturbative correlators and will be carried out in the future. Given that the medium's contribution to all scattering kernels is encoded in the two-point correlation functions $\hat{\mathcal{A}}$ and $\hat{\mathcal{F}}$, our calculations are equally valid in cold nuclear matter or within hot QGP.

\section{Summary and Outlook}
In this paper, a first calculation of the Bremsstrahlung photon emission kernel is presented for a highly energetic and highly virtual quark traversing through a nuclear medium. At the perturbative scale $\mathcal{O}(\alpha_s \alpha_{em})$, two kernels have been identified. This classification is organized using the identity of particles in the final state. The first kernel represents the production of a real photon and a quark in a DIS between the highly off-shell virtual photon and the nucleus, whereas the second kernel consists of a real photon and a gluon in the final state with the fermion-to-boson conversion processes. 

The calculation is carried out within the framework of perturbative QCD and by computing the hadronic tensor ($W^{\mu\nu}$) for the case of deep-inelastic scattering between the virtual photon and the nucleus $A$. The calculation is performed in the Breit frame with a light-cone gauge $A^-=0$. The parton struck by the virtual photon coming off from the incoming electron is referred to as the primary hard parton, and the associated scattering is the primary hard scattering. It is assumed that the subsequent scatterings of this hard parton traversing the remainder of the nucleus are uncorrelated with the scattering from the first-struck nucleon and therefore, these can be factorized from the initial state nucleon parton distribution function. In the calculation of the hadronic tensor, the phase-space exponentials that contain relative distances ($\Delta x^{-}=y^{-}-x^-$) between the primary (hard) scattering in the amplitude ($x^-$) and its complex conjugate ($y^-$) are absorbed in the definition of the parton distribution function of the struck nucleon. Similarly, the phase-space exponentials that contain the relative distance ($\Delta z^{-}=z^{-}_3-z^-_2$) between the second scattering in the amplitude ($z^-_2$) and the complex conjugate ($z^-_3$) are absorbed in the definition of non-perturbative jet-medium transport coefficients. It is argued that the remainder of the phase-space exponentials should be a real number. This led us to define the path length integration variable $\zeta^-$ representing the relative distance ($x^{-}-z^{-}_{2}$ and $y^{-}-z^{-}_{3}$) between the first scattering and second scattering in the amplitude (and also its complex conjugate). In this calculation, the phase factors $\left[2-2\cos\left\{ \mathcal{G}^{(\ell_2)}_{0}\zeta^-\right\}\right]$ and $e^{-i\mathcal{H}^{(\ell_2,p_2)}_{0}\Delta z^-}$ contain explicit $\ell_{2\perp}$--dependence and have been kept in the definition of the scattering kernel. 

For all the calculations presented herein, it has been assumed that the second scattering occurs via an exchange of the Glauber gluon (or quark), which has a transverse momentum ($\pmb{k}_{\perp}\gg k^-, k^+$) larger than its (plus and minus) light-cone components. The hadronic tensor for kernel-2 involve an additional factor of $yq^-$ or $(1-y+\eta y)q^-$ in the denominator ~\cite{Ghiglieri:2015ala,Qin:2008rd} when compared to kernel-1 diagram, indicating that the fermion-to-boson conversion processes are suppressed by the hard quark energy scale. For each kernel, a full scattering kernel was presented first before a systematic (Taylor) expansion was employed. These kernels are planned to be implemented within a comprehensive Monte-Carlo simulation. Such a simulation will enable more precise constraints on parton energy-loss transport coefficients to be obtained. In addition to this, a Monte Carlo implementation of kernel-2 with the fermion-to-boson conversion processes would provide insight into the possible mechanism for dynamical quark flavor generation in the nuclear medium. In hot nuclear matter, photons are uniquely positioned to probe the attractor-like dynamics leading to hydrodynamical evolution at later times, for both the energy-momentum tensor of the QGP and its quark flavor content. The $\hat{\mathcal{A}}$ and $\hat{\mathcal{F}}$ operators in the scattering kernles for photon production should be used {\it together} to constrain the dynamics connecting early, glasma-like, evolution to later fluid evolution: i.e., the flavor-dependent hydrodynamization dynamics.   

Furthermore, the effects of quark masses have been studied for the first time at $\mathcal{O}(\alpha_s\alpha_{em})$. We have shown the sensitivity of kernel-1 to quark masses, and our findings indicate that the bottom quarks demonstrate a large effect on the second-order gradient terms $\left(\mathcal{R}^{(1)HQ}_{T,2}\right)$. Thus, the heavy-quark mass scale plays an important role in the parton energy loss at high virtuality.

One of the striking outcomes of this study has been the derivation of the nonperturbative (NP) function at NLO (and NLT), along with the appearance of the universal function $\mathcal{H}^{(\ell_2,p_2)}_{0}$ in phase space as $e^{-i\mathcal{H}^{(\ell_2,p_2)}_{0}\Delta z^-}$. We have shown that these NP correlators ($\hat{ \mathcal{A}}_{0}, \hat{ \mathcal{A}}_{T,2},\ldots,\hat{\mathcal{F}}_{0}, \hat{ \mathcal{F}}_{T,2},\ldots $ and so on) depend on the semi-hard scale $\ell_{2\perp} \gg \Lambda_{\rm QCD}$ momentum, i.e. on the transverse momentum generated in the radiative splitting. We also linked the NLO jet-medium transport coefficients to transverse-momentum-dependent PDFs (TMD-PDFs). In future, it would be interesting to study the transverse momentum dependence and the temperature dependence of these correlators using finite-temperature field theory and lattice gauge theory.

\section*{ACKNOWLEDGMENTS}
The authors would like to thank Abhijit Majumder and Chathuranga Sirimanna for invaluable discussions. This work was supported by the Canada Research Chair under Grant Number CRC-2022-00146 and the Natural Sciences and Engineering Research Council (NSERC) of Canada under Grant Number SAPIN-2023-00029. This work was also supported in part by the National Science Foundation (NSF) within the framework of the JETSCAPE collaboration, under Grant Number OAC-2004571 (CSSI:X-SCAPE).

\section*{DATA AVAILABILITY}
No data were created or analyzed in this study.

\begin{appendices}
\section{$\mbox{}$\!\!\!\!\!\!: Real photon emission from the quark: no in-medium scattering }
\label{append:wmunu_vacuum}

This section presents the hadronic tensors associated with a single photon emission from a quark in a DIS between the virtual photon and the nucleus. There are a total of four diagrams (Fig.~\ref{fig:W_0_vacuum}). The radiated photon ($\ell_2$) is traveling in the negative $z$ direction, i.e., along the direction of the final state quark, having momentum fraction $y=\ell^{-}_{2}/q^-$. We present the hadronic tensor in light-cone gauge $A^-=0$, with light-cone vector $n=[1,0,\pmb{0_{\perp}}]$.

The hadronic tensor for the diagram shown in Fig.~\ref{fig:W_0_vacuum}(a) is given as
\begin{eqnarray}
\begin{split}
W^{\mu\nu}_{0,a} & =\sum_f 2 \left[-g^{\mu\nu}_{\perp\perp}\right] e^2 e^4_f  \int d (\Delta X^{-}) e^{iq^{+}\Delta X^{-}} \left\langle AP \left| \bar{\psi}_{_f}(\Delta X^-) \frac{\gamma^{+}}{4} \psi_{_f}(0)\right| AP\right\rangle\\
  & \times  \int  \frac{dy}{2\pi}\frac{d^2 \ell_{2\perp}}{(2\pi)^2} e^{ -i\left\{ \frac{\pmb{\ell}^2_{2\perp} + yM^2}{2y(1-y)q^-}\right\} \Delta X^{-} } \left[\frac{1+ \left(1-y\right)^2}{y}\right]   \frac{[\pmb{\ell}^2_{2\perp} +M^2 y^4 \kappa ]}{ \left[  \pmb{\ell}^2_{2\perp}  + y^2  M^2 \right]^2} ,
\end{split}\label{eq:append_W_0_vacuum_a}
\end{eqnarray}
where 
\begin{eqnarray}
\kappa=\frac{1}{1+(1-y)^2}.
 \label{eq:kappa_def_append}
\end{eqnarray}
In Eq.~\ref{eq:append_W_0_vacuum_a} the quantity $\frac{1+(1-y)^2}{y}$ represents the splitting function characterizing the probability of a photon emission off from the quark carrying momentum fraction $y$ of the parent quark. 

The hadronic tensor for the diagram shown in Fig.~\ref{fig:W_0_vacuum}(b) has the following form:  
\begin{eqnarray}
\begin{split}
W^{\mu\nu}_{0,b} & =\sum_f 2 \left[-g^{\mu\nu}_{\perp\perp}\right] e^2 e^4_f  \int d (\Delta X^{-}) e^{iq^{+}\Delta X^{-}} \left\langle AP \left| \bar{\psi}_{_f}(\Delta X^-) \frac{\gamma^{+}}{4} \psi_{_f}(0)\right| AP\right\rangle\\
  & \times  \int  \frac{dy}{2\pi}\frac{d^2 \ell_{2\perp}}{(2\pi)^2} e^{ -i\left\{ \frac{\pmb{\ell}^2_{2\perp} + yM^2}{2y(1-y)q^-}\right\}\Delta X^{-} }    \frac{[\pmb{\ell}^2_{2\perp} +M^2(1-2y)^2 ]}{ y^3(1-y)^2Q^4}.
\end{split}\label{eq:append_W_0_vacuum_b}
\end{eqnarray}

The hadronic tensor for the diagrams shown in Figs.~\ref{fig:W_0_vacuum}(c) and (d) is identical and has the following form:
\begin{eqnarray}
\begin{split}
W^{\mu\nu}_{0,c(d)} & =-\sum_f 2 \left[-g^{\mu\nu}_{\perp\perp}\right] e^2 e^4_f  \int d (\Delta X^{-}) e^{iq^{+}\Delta X^{-}} \left\langle AP \left| \bar{\psi}_{_f}(\Delta X^-) \frac{\gamma^{+}}{4} \psi_{_f}(0)\right| AP\right\rangle\\
  & \times  \int  \frac{dy}{2\pi}\frac{d^2 \ell_{2\perp}}{(2\pi)^2} e^{ -i\left\{ \frac{\pmb{\ell}^2_{2\perp} + yM^2}{2y(1-y)q^-}\right\}\Delta X^{-} }    \left[ \frac{1}{y^2(1-y)}\right] \frac{[\pmb{\ell}^2_{2\perp} -M^2(1-2y) y^2 ]}{Q^2[\pmb{\ell}^2_{2\perp} +M^2y^2 ] } ,
\end{split}\label{eq:append_W_0_vacuum_c_d}
\end{eqnarray}

In leading-log approximation ($\ell_{2\perp}\rightarrow 0$), the dominant contribution comes from the final state radiation diagram [Fig.~\ref{fig:W_0_vacuum}(a)]. 

\section{$\mbox{}$\!\!\!\!\!\!: The kernel for single-scattering-induced emission: one photon and one quark in the final state}
\label{append:kernel1}

In this section, we summarize the calculation of all possible diagrams at next-leading-order (NLO) and next-leading-twist(NLT) contributing to  kernel-1 with a photon and a quark in the final state. We discuss singularity structure, contour integrations, and involved traces in the final calculation of the hadronic tensor.

\begin{figure}[!h]
    \centering
    \includegraphics[width=0.48\textwidth]{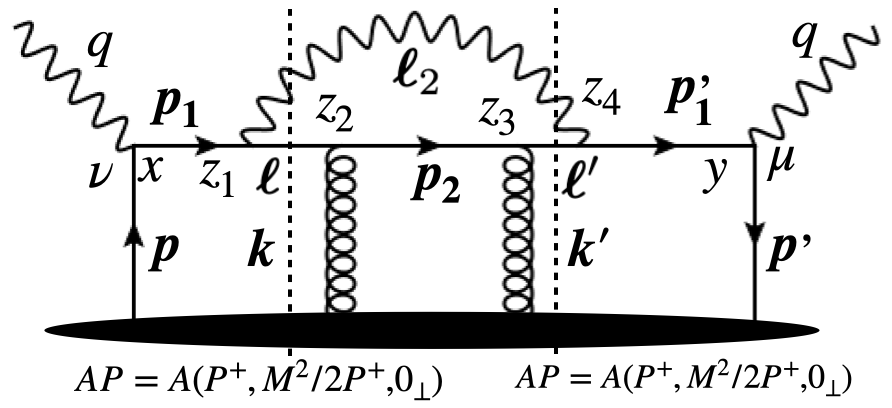}
    \caption{A forward scattering diagram contributing to kernel-1 at NLO. The left-cut line corresponds to an interference between the single emission no scattering process and the single emission double scattering process. The right-cut line generates a process that is the complex conjugate of the process generated by the left-cut. }
    \label{fig:kernel-1_ph_qqgm_qqgm_ph_left_right}
\end{figure}
The Fig.~\ref{fig:kernel-1_ph_qqgm_qqgm_ph_left_right} represents a forward scattering diagram contributing to the type-1 kernel at NLO and NLT. The left-cut gives rise to an interference between the single-photon emission with no scattering process and single photon pre-emission with double in-medium gluon scattering. The hadronic tensor for the left-cut diagram (Fig.~\ref{fig:kernel-1_ph_qqgm_qqgm_ph_left_right}) is given as 
\begin{equation}
\begin{split}
W^{\mu\nu}_{1,\ell} & = \sum_{f} e^2 e^4_f g^2_{s} \int d^4 x d^4 y d^4 z_{2} d^4 z_{3} \int \frac{d^4 \ell}{(2\pi)^4} \frac{d^4 p'}{(2\pi)^4}  \frac{d^4 \ell_{2}}{(2\pi)^4} \frac{d^4 p_{2}}{(2\pi)^4} e^{-ip'y} e^{i(-q+\ell+\ell_2)x}\left\langle P \left| \bar{\psi}_{_f}(y) \frac{\gamma^{+}}{4} \psi_{_f}(x)\right| P\right\rangle  \\
  & \times e^{i\left(q+p'-p_{2}-\ell_{2}\right) z_{3}} e^{i(p_{2} - \ell)z_{2}} \langle P_{A-1} | {\rm Tr}[A^{+}(z_3)A^+(z_2)]| P_{A-1} \rangle d^{(\ell_2)}_{\sigma_{1} \sigma_{4}} (2 \pi) \delta\left(\ell_{2}^{2}\right) (2 \pi)  \delta\left(\ell^{2}-M^{2}\right) \\
& \times \frac{  {\rm Tr} \left[\gamma^- \gamma^{\mu} \left(\slashed{q}+\slashed{p}'+M\right) \gamma^{\sigma_{4}}\left( \slashed{q} +\slashed{p}' - \slashed{\ell}_2 +  M\right)\gamma^{-}\left(\slashed{p}_2+M\right) \gamma^{-}\left(\slashed{\ell} +M\right)\gamma^{\sigma_1} \left(\slashed{\ell}_2 +\slashed{\ell}+M\right)\gamma^{\nu} \right]
}{  \left[\left(q+p'\right)^{2}-M^2-i\epsilon\right] \left[\left(q+p'-\ell_2\right)^2 - M^2 - i\epsilon\right]   \left[p_2^2 -M^2-i\epsilon\right] \left[\left(\ell_2+\ell\right)^2-M^2+i\epsilon\right]  }.
\end{split} \label{eq:kernel-1_ph_qqgm_qqgm_ph_left-right-cut-wi}
\end{equation}

The above expression admits singularities owing to the presence of two simple poles for momentum variable $p'^+$ and one simple pole for $p^+_2$. The contour integration for momentum $p^+_2$ gives
\begin{equation}
    C_{1} = \oint \frac{dp^+_2}{(2\pi)} \frac{e^{-ip^+_2(z^{-}_3-z^-_2)} }{[ p^2_2 -M^2 - i\epsilon]} =  \oint \frac{dp^+_2}{(2\pi)} \frac{e^{-ip^+_2(z^{-}_3-z^{-}_{2})} }{ 2p^-_2 \left[p^{+}_2 - \frac{\pmb{p}^2_{2\perp}+M^2}{2p^-_2} - i\epsilon\right]} = \frac{(-2\pi i)}{2\pi}\frac{\theta( z^{-}_3 - z^{-}_{2} )}{2p^-_2}  e^{-i\left( \frac{\pmb{p}^2_{2\perp}+M^2}{2p^-_2}\right)(z^{-}_3-z^{-}_{2})}.
\end{equation}
Similarly, the contour integration for $p'^+$ can be done as
\begin{equation}
\begin{split}
C_{2} & = \oint \frac{dp'^{+}}{(2\pi)} \frac{e^{-ip'^{+}(y^{-}-z^{-}_{3})}}{\left[\left(q+p'\right)^{2} - M^{2} - i \epsilon\right]\left[\left(q+p'-\ell_{2}\right)^{2} -M^2 - i\epsilon\right]} \\
      & = \oint \frac{dp'^{+}}{(2\pi)} \frac{e^{-ip'^{+}(y^{-} - z^{-}_{3}) }}{2q^{-}\left[ q^{+} + p'^{+} - \frac{M^2}{2q^{-}} - i \epsilon\right] 2(q^{-} -\ell^{-}_{2} )\left[ q^{+} + p'^{+} -\ell^{+}_{2} - \frac{\pmb{\ell}^{2}_{2\perp}+M^2}{2(q^{-}-\ell^{-}_{2})} - i \epsilon\right]}   \\
      & = \frac{(-2\pi i)}{2\pi} \frac{\theta(y^{-} - z^{-}_{3})}{4q^{-}(q^{-}-\ell^{-}_{2})}  e^{i\left(q^{+} - \frac{M^2}{2q^{-}})(y^{-}-z^{-}_{3}\right)}\left[ \frac{ -1 + e^{-i \mathcal{G}^{(\ell_2)}_M(y^{-}-z^{-}_{3})} }{ \mathcal{G}^{(\ell_2)}_M}  \right],
\end{split}
\label{eq:}
\end{equation}
where
\begin{equation}
    \mathcal{G}^{(\ell_2)}_M = \ell^{+}_{2} + \frac{\pmb{\ell}^{2}_{2\perp}+M^2}{2(q^{-}-\ell^{-}_{2})} - \frac{M^2}{2q^{-}} = \frac{\pmb{\ell}^2_{2\perp} + y^2M^2}{2y(1-y)q^-}.
    \label{eq:kernel1_GML2_appendix1}
\end{equation}
The trace in the numerator of the third line of Eq.~\ref{eq:kernel-1_ph_qqgm_qqgm_ph_left-right-cut-wi} yields 
\begin{equation}
 \begin{split}
 &   {\rm Tr} \left[\gamma^- \gamma^{\mu} \left(\slashed{q}+\slashed{p}'+M\right) \gamma^{\sigma_{4}}\left( \slashed{q} +\slashed{p}' - \slashed{\ell}_2 +  M\right)\gamma^{-}\left(\slashed{p}_2+M\right) \gamma^{-}\left(\slashed{\ell} +M\right)\gamma^{\sigma_1} \left(\slashed{\ell}_2 +\slashed{\ell}+M\right)\gamma^{\nu} \right] d^{(\ell_2)}_{\sigma_1 \sigma_4} \\
 & = 32 (q^-)^3 [-g^{\mu\nu}_{\perp\perp}] \left[ \frac{1-y+\eta y}{y}\right] \left[ \frac{1 + \left(1-y\right)^2}{y}\right]  \left[ \pmb{\ell}^2_{2\perp} + M^2y^4\kappa \right],
 \end{split}
\end{equation}
where  $\kappa$ is defined in Eq.~\ref{eq:kappa_def_append}.
%
%
The final expression of the hadronic tensor for the left-cut diagram (Fig.~\ref{fig:kernel-1_ph_qqgm_qqgm_ph_left_right}) is given by
\begin{equation}
\begin{split}
    W^{\mu\nu}_{1,\ell} & =\sum_f 2 [-g^{\mu\nu}_{\perp \perp }]\int d (\Delta X^{-})   e^{iq^{+}(\Delta X^{-} )} 
  \left\langle P \left| \bar{\psi}_{_f}( \Delta X^{-}) \frac{\gamma^{+}}{4} \psi_{_f}(0)\right| P\right\rangle  \\
  & \times e^2 e^4_f g^2_s \int  d \zeta^{-} d (\Delta z^{-}) d^2 \Delta z_{\perp} \frac{dy}{2\pi} \frac{d^2 \ell_{2\perp}}{(2\pi)^2} \frac{d^2 k_{\perp}}{(2\pi)^2} \left[ 1 - e^{-i\mathcal{G}^{(\ell_2)}_{M}(y^{-}-z^-_3)} \right]  e^{-i p^+_2 (\Delta z^{-})}     e^{i \pmb{k}_{\perp}\cdot \Delta\pmb{z}_{\perp}}  \\
  & \times \frac{\theta( z^{-}_3-z^{-}_2) \theta( y^{-}-z^{-}_3)   \left[  \pmb{\ell}^2_{2\perp} + \kappa y^4 M^2 \right]}{\left[  \pmb{\ell}^2_{2\perp} + M^2y^2 \right]^2 } \left\langle P_{A-1} \left| {\rm Tr}[A^+(\zeta^-,\Delta z^-, \Delta z_{\perp} ) A^+(\zeta^-,0)] \right| P_{A-1} \right\rangle  \\
  &
\times  \left[ \frac{ 1 + \left(1-y\right)^2 }{y}\right] e^{i(\ell^+_2 + \ell^+ )x^-} e^{-i\ell^+_2 z^-_3} e^{-i\ell^+ z^-_2} e^{-i[M^2/(2q^-)](y^{-}-z^{-}_{3})},
\end{split} \label{eq:kernel1_wmunu_final1st}
\end{equation}
where $\mathcal{G}^{(\ell_2)}_{M}$ is defined in Eq.~\ref{eq:kernel1_GML2_appendix1}, and 
%

\begin{equation}
    \ell^{+}_{2} = \frac{\pmb{\ell}^2_{2\perp}}{2yq^-}; \hspace{2em}  \ell^{+} = \frac{\pmb{\ell}^2_{2\perp}+M^2}{2(1-y)q^-};  \hspace{2em}   p^{+}_{2} = \frac{\left[\left(\pmb{\ell}_{2\perp}-\pmb{k}_{\perp}\right)^2 +M^2\right]}{2(1-y+\eta y)q^-}. 
\end{equation}
However, Eq.~\ref{eq:kernel1_wmunu_final1st} can be recast into the form 
\begin{equation}
\begin{split}
    W^{\mu\nu}_{1,\ell} & = \sum_f 2 [-g^{\mu\nu}_{\perp \perp }]\int d (\Delta X^{-})   e^{iq^{+}(\Delta X^{-} )} 
  \left\langle P \left| \bar{\psi}_{_f}( \Delta X^{-}) \frac{\gamma^{+}}{4} \psi_{_f}(0)\right| P\right\rangle  \\
  & \times e^2 e^4_f g^2_s \int  d \zeta^{-} d (\Delta z^{-}) d^2 \Delta z_{\perp} \frac{dy}{2\pi} \frac{d^2 \ell_{2\perp}}{(2\pi)^2} \frac{d^2 k_{\perp}}{(2\pi)^2} \left[ 1 - e^{-i\mathcal{G}^{(\ell_2)}_{M}(y^{-}-z^-_3)} \right]  e^{-i (\ell^{+}_{2} +p^+_2)(\Delta z^{-})}     e^{i \pmb{k}_{\perp}\cdot\Delta\pmb{z}_{\perp}}  \\
  & \times \frac{\theta( z^{-}_3-z^{-}_2) \theta( y^{-}-z^{-}_3)   \left[  \pmb{\ell}^2_{2\perp} + \kappa y^4 M^2 \right]}{\left[  \pmb{\ell}^2_{2\perp} + M^2y^2 \right]^2 } \langle P_{A-1} |{\rm Tr}[ A^+(\zeta^-, \Delta z^-, \Delta z_{\perp}) A^+(\zeta^-,0)] | P_{A-1} \rangle  \\
  &
\times  \left[ \frac{ 1 + \left(1-y\right)^2 }{y}\right] e^{i(\ell^+_2 + \ell^+ )(x^{-}-z^{-}_{2})} e^{-i[M^2/(2q^-)](y^{-}-z^{-}_{3})},
\end{split} 
\end{equation}
by using the change of variables $\Delta z^-=z^{-}_{3}-z^{-}_{2}$. 

The right-cut diagram shown in Fig.~\ref{fig:kernel-1_ph_qqgm_qqgm_ph_left_right} is a complex-conjugate of the left-cut diagram. The final expression of the hadronic tensor for the right-cut diagram reduces to the following form:
\begin{equation}
\begin{split}
    W^{\mu\nu}_{1,r} & =\sum_f 2 [-g^{\mu\nu}_{\perp \perp }]\int d (\Delta X^{-})   e^{iq^{+}(\Delta X^{-} )} 
  \left\langle P \left| \bar{\psi}_{_f}( \Delta X^{-}) \frac{\gamma^{+}}{4} \psi_{_f}(0)\right| P\right\rangle  \\
  & \times e^2 e^4_f g^2_s \int  d \zeta^{-} d (\Delta z^{-}) d^2 \Delta z_{\perp} \frac{dy}{2\pi} \frac{d^2 \ell_{2\perp}}{(2\pi)^2} \frac{d^2 k_{\perp}}{(2\pi)^2} \left[ 1 - e^{i\mathcal{G}^{(\ell_2)}_{M}(x^{-}-z^-_2)} \right]  e^{-i p^+_2 (\Delta z^{-})}     e^{i \pmb{k}_{\perp}\cdot \Delta\pmb{z}_{\perp}}  \\
  & \times \frac{\theta( -z^{-}_3+z^{-}_2) \theta( x^{-}-z^{-}_2)   \left[  \pmb{\ell}^2_{2\perp} + \kappa y^4 M^2 \right]}{\left[  \pmb{\ell}^2_{2\perp} + M^2y^2 \right]^2 } \langle P_{A-1} | {\rm Tr}[A^+(\zeta^-,\Delta z^-, \Delta z_{\perp}) A^+(\zeta^-,0)] | P_{A-1} \rangle  \\
  &
\times  \left[ \frac{ 1 + \left(1-y\right)^2 }{y}\right] e^{-i(\ell^+_2 + \ell'^+ )y^-} e^{i\ell'^+ z^-_3} e^{i\ell^+_2 z^-_2} e^{i[M^2/(2q^-)](x^{-}-z^{-}_{2})},
\end{split} 
\end{equation}
where $\mathcal{G}^{(\ell_2)}_{M}$ is defined in Eq.~\ref{eq:kernel1_GML2_appendix1}, $\kappa$ is in Eq.~\ref{eq:kappa_def_append} and
%
%
\begin{equation}
    \ell^{+}_{2} = \frac{\pmb{\ell}^2_{2\perp}}{2yq^-}; \hspace{2em}  \ell'^{+} = \frac{\pmb{\ell}^2_{2\perp}+M^2}{2\left(1-y\right)q^-};  \hspace{2em}   p^{+}_{2} = \frac{\left[\left(\pmb{\ell}_{2\perp}-\pmb{k}_{\perp}\right)^2 +M^2\right]}{2(1-y+\eta y)q^-}.
\end{equation}

The above expression can be recast, by instituting $\Delta z^-=z^{-}_{3}-z^{-}_{2}$, into the following form:
\begin{equation}
\begin{split}
    W^{\mu\nu}_{1,r} & =\sum_f 2 [-g^{\mu\nu}_{\perp \perp }]\int d (\Delta X^{-})   e^{iq^{+}(\Delta X^{-} )} 
  \left\langle P \left| \bar{\psi}_{_f}( \Delta X^{-}) \frac{\gamma^{+}}{4} \psi_{_f}(0)\right| P\right\rangle  \\
  & \times e^2 e^4_f g^2_s \int  d \zeta^{-} d (\Delta z^{-}) d^2 \Delta z_{\perp} \frac{dy}{2\pi} \frac{d^2 \ell_{2\perp}}{(2\pi)^2} \frac{d^2 k_{\perp}}{(2\pi)^2} \left[ 1 - e^{i\mathcal{G}^{(\ell_2)}_{M}(x^{-}-z^-_2)} \right]  e^{-i (\ell^+_2 + p^+_2) (\Delta z^{-})}     e^{i \pmb{k}_{\perp}\cdot\Delta\pmb{z}_{\perp}}  \\
  & \times \frac{\theta( -z^{-}_3+z^{-}_2) \theta( x^{-}-z^{-}_2)   \left[  \pmb{\ell}^2_{2\perp} + \kappa y^4 M^2 \right]}{\left[  \pmb{\ell}^2_{2\perp} + M^2y^2 \right]^2 } \langle P_{A-1} |{\rm Tr}[ A^+(\zeta^-, \Delta z^-, \Delta z_{\perp}) A^+(\zeta^-,0)] | P_{A-1} \rangle  \\
  &
\times  \left[ \frac{ 1 + \left(1-y\right)^2 }{y}\right] e^{-i(\ell^+_2 + \ell'^+ )(y^{-}-z^{-}_{3})}  e^{i[M^2/(2q^-)](x^{-}-z^{-}_{2})}.
\end{split} 
\end{equation}
\begin{figure}[!h]
    \centering
    \includegraphics[width=0.5\textwidth]{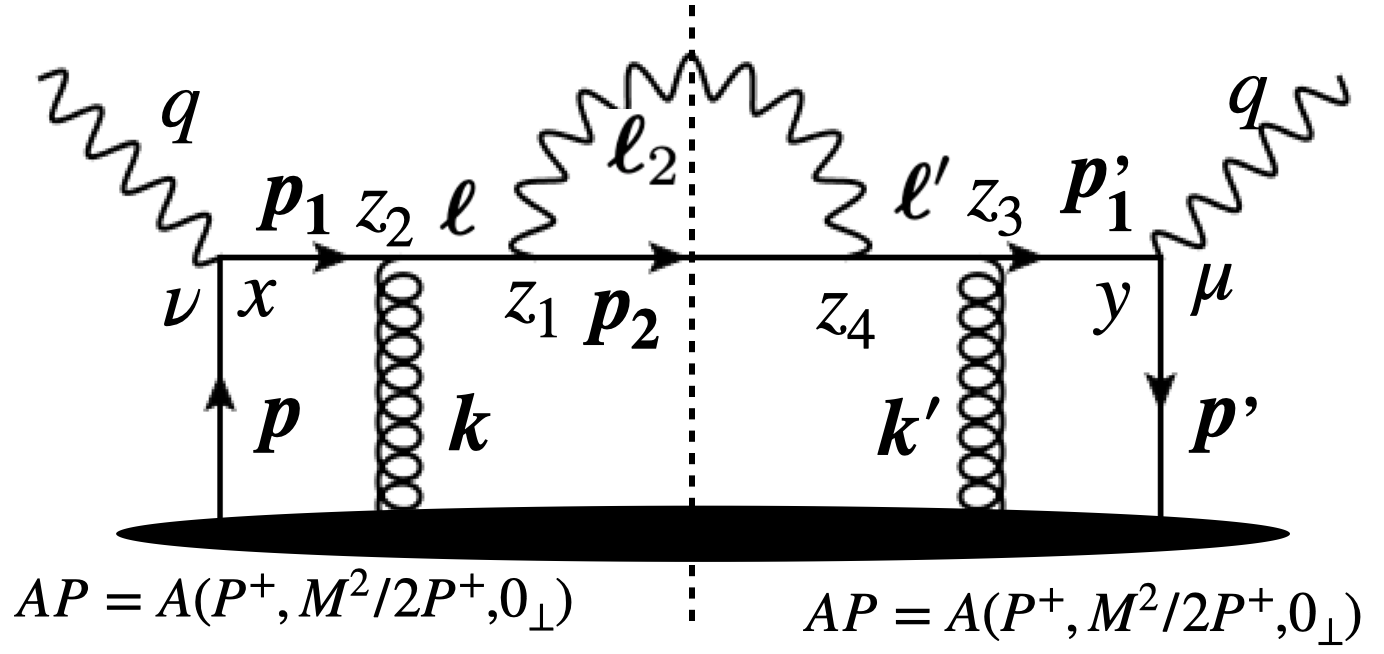}
    \caption{A forward scattering diagram contributing to kernel-1. }
    \label{fig:kernel-1_qqgm_ph_ph_qqgm}
\end{figure}

Next, the hadronic tensor for the central-cut diagram shown in Fig.~\ref{fig:kernel-1_qqgm_ph_ph_qqgm} can be written as
\begin{equation}
\begin{split}
W^{\mu\nu}_{1,c} & = \sum_f e^2 e^4_f g^2_{s} \int d^4 x d^4 y d^4 z_{2} d^4 z_{3} \int \frac{d^4 p}{(2\pi)^4} \frac{d^4 p'}{(2\pi)^4}  \frac{d^4 \ell_{2}}{(2\pi)^4} \frac{d^4 p_{2}}{(2\pi)^4} e^{-ip'y} e^{ipx}\left\langle P \left| \bar{\psi}_{_f}(y) \frac{\gamma^{+}}{4} \psi_{_f}(x)\right| P\right\rangle  \\
  & \times e^{i\left(q+p'-p_{2}-\ell_{2}\right) z_{3}} e^{i(\ell_{2}+p_{2} - q - p)z_{2}} \langle P_{A-1} | {\rm Tr}[A^{+}(z_3)A^+(z_2)] | P_{A-1} \rangle d^{(\ell_2)}_{\sigma_{1} \sigma_{4}} (2 \pi) \delta\left(\ell_{2}^{2}\right) (2 \pi)  \delta\left(p_{2}^{2}-M^{2}\right) \\
& \times \frac{  {\rm Tr} \left[\gamma^- \gamma^{\mu} \left(\slashed{q}+\slashed{p}'+M\right) \gamma^-\left( \slashed{\ell}_2 + \slashed{p}_2 + M\right)\gamma^{\sigma_{4}}\left(\slashed{p}_2+M\right) \gamma^{\sigma_1}\left(\slashed{\ell}_2 + \slashed{p}_2+M\right)\gamma^- \left(\slashed{q}+\slashed{p}+M\right)\gamma^{\nu} \right]
}{  \left[\left(q+p'\right)^{2}-M^2-i\epsilon\right] \left[\left(\ell_2+p_2\right)^2 - M^2 - i\epsilon\right]   \left[\left(\ell_2+p_2\right)^2 -M^2+i\epsilon\right] \left[\left(q+p\right)^2-M^2+i\epsilon\right]  }.
\end{split} \label{eq:kernel-1_qqgm_ph_ph_qqgm_wi-central}
\end{equation}
Equation~\ref{eq:kernel-1_qqgm_ph_ph_qqgm_wi-central} has singularity arising from the denominator of the quark propagator with momenta $p_{1}$ and $p'_{1}$. We identify one pole for each momentum variable $p^+$ and $p'^+$. 
 The contour integration for $p^+$ in the complex plane is given by
\begin{equation}
    C_{1} = \oint \frac{dp^+}{(2\pi)} \frac{e^{ip^+(x^{-}-z^-_2)} }{[ (q+p)^2 -M^2 + i\epsilon]} =  \oint \frac{dp^+}{(2\pi)} \frac{e^{ip^+(x^{-}-z^{-}_{2})} }{ 2q^{-}[ q^{+}+p^{+} - [M^2/(2q^-)]+ i\epsilon]} = \frac{(2\pi i)}{2\pi}\frac{\theta( x^{-} - z^{-}_{2} )}{2q^{-}}  e^{i\left(-q^{+} + \frac{M^2}{2q^-}\right)(x^{-}-z^{-}_{2})}. 
\end{equation}
Similarly, the contour integration for momentum $p'^+$ is carried out as 
\begin{equation}
    C_{2} = \oint \frac{dp'^+}{(2\pi)} \frac{e^{-ip'^+(y^{-}-z^{-}_3)} }{[ (q+p')^2 -M^2 -i\epsilon]} =  \oint \frac{dp'^+}{(2\pi)} \frac{e^{-ip'^+(y^{-}-z^{-}_3)} }{2q^- [ q^{+}+p'^{+} - [M^2/(2q^-)]-i\epsilon]} = \frac{(-2\pi i)}{2\pi} \frac{\theta( y^{-} - z^{-}_{3} )}{2q^-}  e^{i\left(q^+ - \frac{M^2}{2q^-}\right)(y^{-}-z^{-}_3)}.
\end{equation}
Including mass correction up to $\mathcal{O}(M^2)$, the trace yields
\begin{equation}
\begin{split}
& {\rm Tr} \left[\gamma^- \gamma^{\mu} \left(\slashed{q}+\slashed{p}'+M\right) \gamma^-\left( \slashed{\ell}_2 + \slashed{p}_2 + M\right) \gamma^{\sigma_{4}}  \left(\slashed{p}_2+M\right) \gamma^{\sigma_1} \left(\slashed{\ell}_2 + \slashed{p}_2+M\right) \gamma^-\left(\slashed{q}+\slashed{p}+M\right)\gamma^{\nu} \right] d^{(\ell_2)}_{\sigma_1 \sigma_4} \\
& = \frac{32 [-g^{\mu\nu}_{\perp\perp}](q^-)^3}{y\left(1-y+\eta y\right)} \left[ \frac{\left(1+\eta y\right)^2 + \left(1-y + \eta y\right)^2}{y}\right] \left[ \left\{\left(1+\eta y\right)\pmb{\ell}_{2\perp} - y\pmb{k}_{\perp}\right\}^2 + \kappa y^4 M^2 \right].
\end{split}
\end{equation}

The final expression of the hadronic tensor for the centralcut (Fig.~\ref{fig:kernel-1_qqgm_ph_ph_qqgm}) is
\begin{equation}
\begin{split}
    W^{\mu\nu}_{1,c} & =\sum_f 2 [-g^{\mu\nu}_{\perp \perp }]\int d (\Delta X^{-})   e^{iq^{+}(\Delta X^{-} )} 
    e^{-i[M^{2}/(2q^{-})](\Delta X^{-} )}
  \left\langle P \left| \bar{\psi}_{_f}( \Delta X^{-}) \frac{\gamma^{+}}{4} \psi_{_f}(0)\right| P\right\rangle  \\
  & \times e^2 e^4_f g^2_s \int  d \zeta^{-} d (\Delta z^{-}) d^2 \Delta z_{\perp} \frac{dy}{2\pi} \frac{d^2 \ell_{2\perp}}{(2\pi)^2} \frac{d^2 k_{\perp}}{(2\pi)^2}  e^{-i \mathcal{H}^{(\ell_2,p_2)}_{M} (\Delta z^{-})}     e^{i \pmb{k}_{\perp}\cdot\Delta\pmb{z}_{\perp}}   \\
  & \times \frac{\theta( x^{-}-z^{-}_2) \theta( y^{-}-z^{-}_3)\left[\left\{\left(1+\eta y\right)\pmb{\ell}_{2\perp} - y\pmb{k}_{\perp}\right\}^{2} + \kappa y^4 M^2 \right]}{ \left[\left(\pmb{\ell}_{2\perp} - y\pmb{k}_{\perp}\right)^2 +2y\eta  (\pmb{\ell}^2_{2\perp}- y\pmb{\ell}_{2\perp}\cdot \pmb{k}_{\perp})+ \eta^2 y^2 \pmb{\ell}^2_{2\perp} + y^2 M^2\right]^{2}}   
  \langle P_{A-1} | {\rm Tr}[A^+(\zeta^-, \Delta z^-, \Delta z_{\perp}) A^+(\zeta^-,0)] | P_{A-1} \rangle  \\
  & \times  \left[ \frac{ \left(1+\eta y\right)^2 + \left(1-y+\eta y\right)^2}{y}\right],
\end{split} 
\end{equation}
where
\begin{equation}
\mathcal{H}^{(\ell_2,p_2)}_{M} = \ell^{+}_{2} + p^{+}_{2} - \frac{M^2}{2q^-} =  \frac{ \pmb{\ell}^{2}_{2\perp} -yM^2 }{2yq^{-}}  + \frac{ (\pmb{\ell}_{2\perp} - \pmb{k}_{\perp})^2 + M^2  }{2q^{-}(1-y + \eta y)},
\label{eq:HL2P2M_append323}
\end{equation}
and $\kappa$ is defined in Eq.~\ref{eq:kappa_def_append}.
\begin{figure}[!h]
    \centering 
    \begin{subfigure}[t]{0.45\textwidth}
        \centering        \includegraphics[height=1.2in]{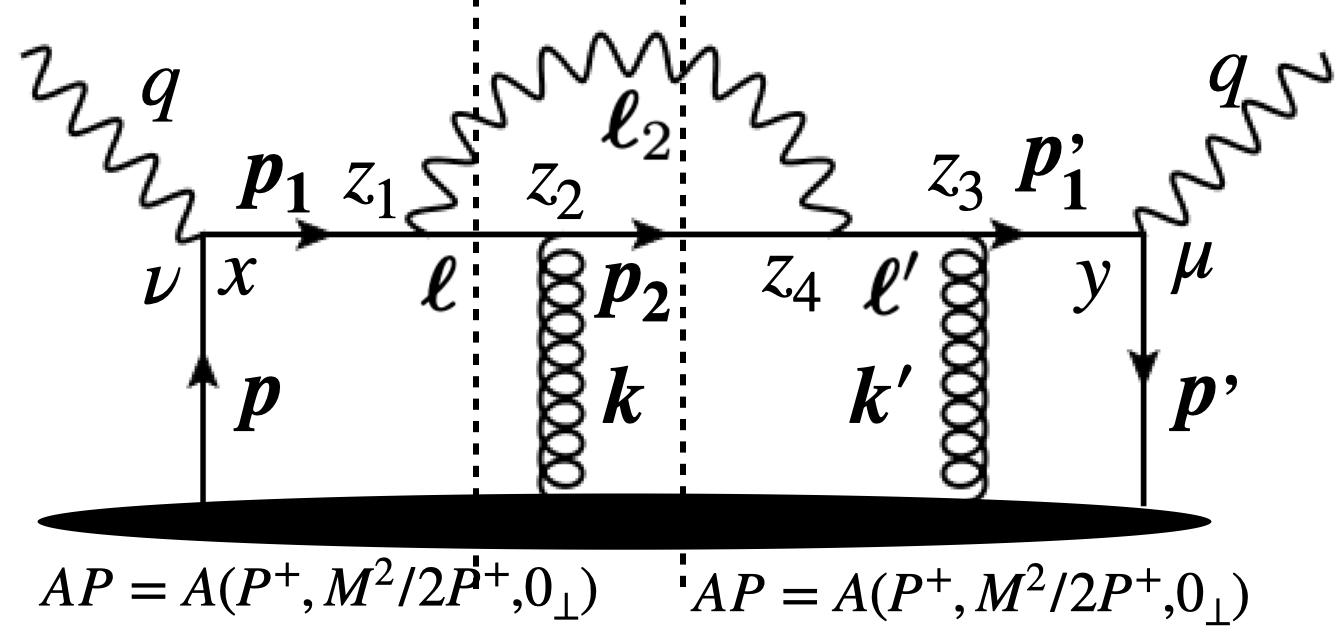}
        \caption{Interference diagram. There are two possible cuts leading to a photon and a quark as final state. }
    \end{subfigure}%
    \begin{subfigure}[t]{0.45\textwidth}
        \centering        \includegraphics[height=1.2in]{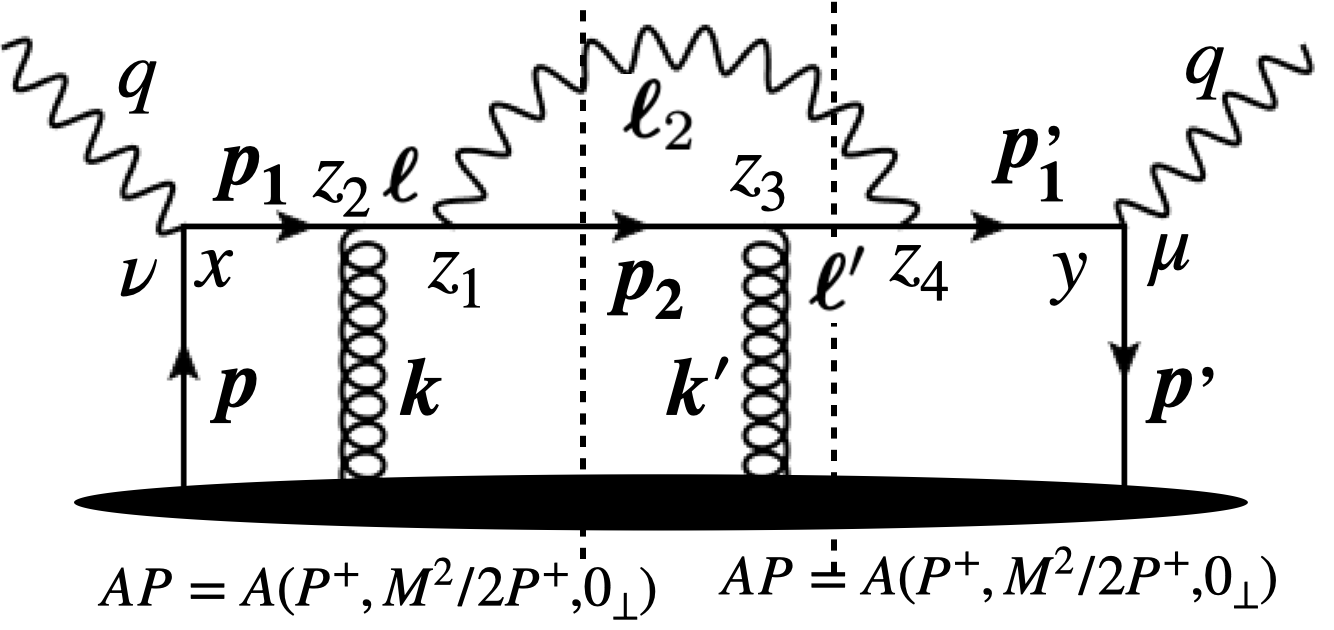}
        \caption{Complex conjugate of the diagram on the left panel. There are two possible cuts leading to a photon and a quark as final state.}
    \end{subfigure}
\caption{A forward scattering diagram contributing to kernel-1.}\label{fig:kernel-1_ph_qqgm_ph_qqgm_both}
\end{figure}

The hadronic tensor for the right-cut diagram shown in Fig.~\ref{fig:kernel-1_ph_qqgm_ph_qqgm_both}(a) is
\begin{equation}
\begin{split}
W^{\mu\nu}_{1,r} & =\sum_f e^2 e^4_f g^2_{s} \int d^4 x d^4 y d^4 z_{2} d^4 z_{3} \int \frac{d^4 p}{(2\pi)^4} \frac{d^4 p'}{(2\pi)^4}  \frac{d^4 \ell_{2}}{(2\pi)^4} \frac{d^4 p_{2}}{(2\pi)^4} e^{-ip'y} e^{ipx}\left\langle P \left| \bar{\psi}_{_f}(y) \frac{\gamma^{+}}{4} \psi_{_f}(x)\right| P\right\rangle  \\
  & \times e^{i\left(q+p'-p_{2}-\ell_{2}\right) z_{3}} e^{i(\ell_{2}+p_{2} - q - p)z_{2}} \langle P_{A-1} | {\rm Tr}[A^{+}(z_3)A^+(z_2)] | P_{A-1} \rangle d^{(\ell_2)}_{\sigma_{1} \sigma_{4}} (2 \pi) \delta\left(\ell_{2}^{2}\right) (2 \pi)  \delta\left(p_{2}^{2}-M^{2}\right) \\
& \times\frac{ {\rm Tr} \left[\gamma^- \gamma^{\mu} \left(\slashed{q}+\slashed{p}'+M\right) \gamma^-\left( \slashed{\ell}_2 + \slashed{p}_2 + M\right)\gamma^{\sigma_{4}}\left(\slashed{p}_2+M\right) \gamma^-\left(\slashed{q} + \slashed{p} - \slashed{\ell}_2+M\right)\gamma^{\sigma_1}  \left(\slashed{q}+\slashed{p}+M\right)\gamma^{\nu} \right] 
}{  \left[\left(q+p'\right)^{2}-M^2-i\epsilon\right] \left[\left(\ell_2+p_2\right)^2 - M^2 - i\epsilon\right]  \left[\left(q+p-\ell_2\right)^2 -M^2+i\epsilon\right] \left[\left(q+p\right)^2-M^2+i\epsilon\right]  }.
\end{split} \label{eq:kernel-1_ph_qqgm_ph_qqgm_both_wi_a_right}
\end{equation}
The above expression (Eq.~\ref{eq:kernel-1_ph_qqgm_ph_qqgm_both_wi_a_right}) has singularity arising from the denominator of the quark propagator with momenta $p_{1}$, $\ell$ and $p'_{1}$. It has two simple poles for the momentum variable $p^+$ and one simple pole for $p'^+$. 
The contour integration for momentum $p^+$ in the complex plane gives
\begin{equation}
\begin{split}
C_{1} & = \oint \frac{dp^{+}}{(2\pi)} \frac{e^{ip^{+}(x^{-}-z^{-}_{2})}}{\left[\left(q+p\right)^{2} - M^{2} + i \epsilon\right]\left[\left(q+p-\ell_{2}\right)^{2} -M^2 + i\epsilon\right]} \\
      & = \oint \frac{dp^{+}}{(2\pi)} \frac{e^{ip^{+}(x^{-} - z^{-}_{2}) }}{2q^{-}\left[ q^{+} + p^{+} - \frac{M^2}{2q^{-}} + i \epsilon\right] 2(q^{-} -\ell^{-}_{2} )\left[ q^{+} + p^{+} -\ell^{+}_{2} - \frac{\pmb{\ell}^{2}_{2\perp}+M^2}{2(q^{-}-\ell^{-}_{2})} + i \epsilon\right]}   \\
      & = \frac{(2\pi i)}{2\pi} \frac{\theta(x^{-} - z^{-}_{2})}{4q^{-}(q^{-}-\ell^{-}_{2})}  e^{i\left(-q^{+} + \frac{M^2}{2q^{-}}\right)(x^{-}-z^{-}_{2})}\left[ \frac{ -1 + e^{i \mathcal{G}^{(\ell_2)}_M(x^{-}-z^{-}_{2})} }{ \mathcal{G}^{(\ell_2)}_M}  \right],
\end{split}
\label{eq:}
\end{equation}
where $ \mathcal{G}^{(\ell_2)}_M $ is defined in Eq.~\ref{eq:kernel1_GML2_appendix1}.

Similarly, the contour integration for momentum $p'^+$ is carried out as 
\begin{equation}
    C_{2} = \oint \frac{dp'^+}{(2\pi)} \frac{e^{-ip'^+(y^{-}-z^{-}_3)} }{\left[\left(q+p'\right)^2 -M^2 -i\epsilon\right]} = \frac{(-2\pi i)}{2\pi} \frac{\theta( y^{-} - z^{-}_{3} )}{2q^-}  e^{i\left(q^+ - \frac{M^2}{2q^-}\right)(y^{-}-z^{-}_3)}.  
\end{equation}

Including mass correction up to $\mathcal{O}(M^2)$, the trace yields
\begin{equation}
\begin{split}
& {\rm Tr} \left[\gamma^- \gamma^{\mu} \left(\slashed{q}+\slashed{p}'+M\right) \gamma^-\left( \slashed{\ell}_2 + \slashed{p}_2 + M\right) \gamma^{\sigma_{4}}  \left(\slashed{p}_2+M\right) \gamma^- \left(\slashed{q}+ \slashed{p} - \slashed{\ell}_2+M\right) \gamma^{\sigma_1} \left(\slashed{q}+\slashed{p}+M\right)\gamma^{\nu} \right] d^{(\ell_2)}_{\sigma_1 \sigma_4} \\
& = \frac{32 [-g^{\mu\nu}_{\perp\perp}]\left(q^-\right)^3}{y} \left[ \frac{ 1  + \left(1-y\right)^2 + \eta y (2-y)}{y}\right] \left[ \left(1+\eta y\right)\pmb{\ell}^2_{2\perp} - y\pmb{k}_{\perp}\cdot \pmb{\ell}_{2\perp}  + \kappa y^4 M^2 \right],
\end{split}\label{eq:trace_right-cut_ph_qqgm_ph_qqgm}
\end{equation}
where $\kappa$ is defined in Eq.~\ref{eq:kappa_def_append}. The final expression of the hadronic tensor for the right-cut [Fig.~\ref{fig:kernel-1_ph_qqgm_ph_qqgm_both}(a)] is given as
\begin{equation}
\begin{split}
    W^{\mu\nu}_{1,r} & =\sum_f 2 [-g^{\mu\nu}_{\perp \perp }]\int d (\Delta X^{-})   e^{iq^{+}(\Delta X^{-} )} 
    e^{-i[M^{2}/(2q^{-})](\Delta X^{-} )}
  \left\langle P \left| \bar{\psi}_{_f}( \Delta X^{-}) \frac{\gamma^{+}}{4} \psi_{_f}(0)\right| P\right\rangle  \\
  & \times e^2 e^4_f g^2_s \int  d \zeta^{-} d (\Delta z^{-}) d^2 \Delta z_{\perp} \frac{dy}{2\pi} \frac{d^2 \ell_{2\perp}}{(2\pi)^2} \frac{d^2 k_{\perp}}{(2\pi)^2} \left[ -1 + e^{i\mathcal{G}^{(\ell_2)}_{M}(x^{-}-z^-_2)} \right] e^{-i \mathcal{H}^{(\ell_2,p_2)}_{M} (\Delta z^{-})}     e^{i \pmb{k}_{\perp}\cdot\Delta\pmb{z}_{\perp}}  \\
  & \times \frac{\theta( x^{-}-z^{-}_2) \theta( y^{-}-z^{-}_3)   \left[  (1+\eta y)\pmb{\ell}^2_{2\perp} - y\pmb{k}_{\perp}\pmb{\ell}_{2\perp}   + \kappa y^4 M^2 \right]}{\left[  \pmb{\ell}^2_{2\perp} + M^2y^2 \right] J_1} \langle P_{A-1} | {\rm Tr}[A^+(\zeta^-, \Delta z^-, \Delta z_{\perp}) A^+(\zeta^-,0)] | P_{A-1} \rangle  \\
  &
\times  \left[ \frac{ 1 + \left(1-y\right)^2 +\eta y(2-y)}{y}\right] ,
\end{split} 
\end{equation}
where $\mathcal{G}^{(\ell_2)}_M$ is defined in Eq.~\ref{eq:kernel1_GML2_appendix1}, $\mathcal{H}^{(\ell_2,p_2)}_{M}$ defined in Eq.~\ref{eq:HL2P2M_append323}, and  
\begin{equation}
    J_1 = \left[ \left(1+\eta y\right)\pmb{\ell}_{2\perp} - y\pmb{k}_{\perp} \right]^2  + y^2 M^2 .
    \label{eq:Kernel-1_J1}
\end{equation}

Next, we consider the left-cut diagram shown in Fig.~\ref{fig:kernel-1_ph_qqgm_ph_qqgm_both}(a). Its hadronic tensor can be written as
\begin{equation}
\begin{split}
W^{\mu\nu}_{1,\ell} & =\sum_f e^2 e^4_f g^2_{s} \int d^4 x d^4 y d^4 z_{2} d^4 z_{3} \int \frac{d^4 \ell}{(2\pi)^4} \frac{d^4 p'}{(2\pi)^4}  \frac{d^4 \ell_{2}}{(2\pi)^4} \frac{d^4 p_{2}}{(2\pi)^4} e^{-ip'y} e^{i(\ell_2 + \ell -q)x}\left\langle P \left| \bar{\psi}_{_f}(y) \frac{\gamma^{+}}{4} \psi_{_f}(x)\right| P\right\rangle  \\
  & \times e^{i\left(q+p'-p_{2}-\ell_{2}\right) z_{3}} e^{i( p_{2} - \ell )z_{2}} \langle P_{A-1} | {\rm Tr}[A^{+}(z_3)A^+(z_2)] | P_{A-1} \rangle d^{(\ell_2)}_{\sigma_{1} \sigma_{4}} (2 \pi) \delta\left(\ell_{2}^{2}\right) (2 \pi)  \delta\left(\ell^{2}-M^{2}\right) \\
& \times \frac{ {\rm Tr} \left[ \gamma^- \gamma^{\mu} \left(\slashed{q}+\slashed{p}'+M\right) \gamma^-\left( \slashed{\ell}_2 + \slashed{p}_2 + M\right)\gamma^{\sigma_{4}} \left(\slashed{p}_2+M\right)\gamma^-\left(\slashed{\ell} +M\right) \gamma^{\sigma_1}\left(\slashed{\ell}_2+\slashed{\ell}+M\right)\gamma^{\nu} \right] }{ \left[\left(q+p'\right)^{2}-M^2-i\epsilon\right] \left[\left(\ell_2+p_2\right)^2 - M^2 - i\epsilon\right] \left[p^2_2 -M^2-i\epsilon\right] \left[\left(\ell_2+\ell\right)^2-M^2+i\epsilon\right] } .
\end{split} \label{eq:kernel-1_ph_qqgm_ph_qqgm_both_wi_a_left}
\end{equation}
The above expression (Eq.~\ref{eq:kernel-1_ph_qqgm_ph_qqgm_both_wi_a_left}) has singularity arising from the denominator of the quark propagator with momenta $p'_{1}$, $\ell'$ and $p_{2}$. We identify two simple poles for the momentum variable $p^+_2$ and one simple pole for $p'^+$. We compute the integral in the complex plane of $p^+_2$ and $p'^{+}$.
The contour integration for momentum $p'^+$ is carried out as 
\begin{equation}
    C_{1} = \oint \frac{dp'^+}{(2\pi)} \frac{e^{-ip'^+(y^{-}-z^{-}_3)} }{[ (q+p')^2 -M^2 -i\epsilon]}  = \frac{(-2\pi i)}{2\pi} \frac{\theta( y^{-} - z^{-}_{3} )}{2q^-}  e^{i\left(q^+ - \frac{M^2}{2q^-}\right)(y^{-}-z^{-}_3)}  .
\end{equation}
Similarly, the contour integration for momentum $p^+_2$ is carried out as
\begin{equation}
\begin{split}
    C_{2} & = \oint \frac{dp^+_2}{(2\pi)} \frac{e^{-ip^+_2(z^{-}_3-z^{-}_2)} }{[ p^2_2 -M^2 -i\epsilon] [(\ell_2+p_2)^2 -M^2 - i\epsilon ] } \\
    & = \oint \frac{dp^+_2}{(2\pi)} \frac{e^{-ip^+_2(z^{-}_3-z^{-}_2)}}{[2p^-_2 p^+_2 - (\pmb{p}^2_{2\perp} + M^2) - i \epsilon ] [ 2(1+\eta y)q^-(\ell^+_2+p^+_2) - (\pmb{k}^2_\perp +M^2) -i\epsilon]}  \\ 
    & = \frac{(-2\pi i)}{2\pi} \frac{\theta( z^{-}_3 - z^{-}_{2} )}{4p^-_2 (1+\eta y)q^-}  e^{-i\left( \frac{\left(\pmb{\ell}_{2\perp}-\pmb{k}_{\perp}\right)^2+M^2}{2(1-y+\eta y)q^-}\right)(z^{-}_3-z^{-}_2)} \left[ \frac{1 - e^{i\mathcal{G}^{(\ell_2,p_2,k)}_{M}(z^-_3 - z^-_2)}}{\mathcal{G}^{(\ell_2,p_2,k)}_{M}}\right],
    \end{split}
\end{equation}
where
\begin{equation}
\mathcal{G}^{(\ell_2,p_2,k)}_{M} = \ell^+_2 + \frac{\pmb{p}^2_{2\perp}+M^2}{2p^-_2} - \frac{\pmb{k}^2_{\perp} + M^2}{2(1+\eta y)q^-} = \frac{J_{1}}{2(1+\eta y)y(1-y+\eta y)q^-},
\label{eq:GML2_J1}
\end{equation}
where $J_1$ is given in Eq.~\ref{eq:Kernel-1_J1}. The trace in the numerator of the third line of Eq.~\ref{eq:kernel-1_ph_qqgm_ph_qqgm_both_wi_a_left} is the same as the trace for the right-cut diagram given in Eq.~\ref{eq:trace_right-cut_ph_qqgm_ph_qqgm}. The final expression of the hadronic tensor for the left-cut diagram [Fig.~\ref{fig:kernel-1_ph_qqgm_ph_qqgm_both}(a)] is given by
\begin{equation}
\begin{split}
    W^{\mu\nu}_{1,\ell} & =\sum_f 2 [-g^{\mu\nu}_{\perp \perp }]\int d (\Delta X^{-})   e^{iq^{+}(\Delta X^{-} )} 
  \left\langle P \left| \bar{\psi}_{_f}( \Delta X^{-}) \frac{\gamma^{+}}{4} \psi_{_f}(0)\right| P\right\rangle  \\
  & \times e^2 e^4_f g^2_s \int  d \zeta^{-} d (\Delta z^{-}) d^2 \Delta z_{\perp} \frac{dy}{2\pi} \frac{d^2 \ell_{2\perp}}{(2\pi)^2} \frac{d^2 k_{\perp}}{(2\pi)^2} \left[ -1 + e^{i\mathcal{G}^{(\ell_2,p_2,k)}_{M}(z^{-}_3-z^-_2)} \right]  e^{-i p^+_2 (\Delta z^{-})}     e^{i \pmb{k}_{\perp}\cdot\Delta\pmb{z}_{\perp}}  \\
  & \times \frac{\theta( z^{-}_3-z^{-}_2) \theta( y^{-}-z^{-}_3)   \left[\left(1+\eta y\right)\pmb{\ell}^2_{2\perp} - y\pmb{k}_{\perp}\pmb{\ell}_{2\perp}   + \kappa y^4 M^2 \right]}{\left[  \pmb{\ell}^2_{2\perp} + M^2y^2 \right] J_1} \langle P_{A-1} | {\rm Tr}[A^+(\zeta^-, \Delta z^-, \Delta z_{\perp}) A^+(\zeta^-, 0)] | P_{A-1} \rangle  \\
  &
\times  \left[ \frac{ 1 + \left(1-y\right)^2 +\eta y(2-y)}{y}\right] e^{i(\ell^+_2 + \ell^+ )x^-} e^{-i\ell^+_2 z^-_3} e^{-i\ell^+ z^-_2} e^{-i[M^2/(2q^-)](y^{-}-z^{-}_{3})},
\end{split} \label{eq:kernel-1_ph_qqgm_ph_qqgm_both_left-cut-fin}
\end{equation}
where $\mathcal{G}^{(\ell_2,p_2,k)}_{M}$  is defined in Eq.~\ref{eq:GML2_J1}, $\kappa$ is in Eq.~\ref{eq:kappa_def_append}, and 
%
\begin{equation}
    \ell^{+}_{2} = \frac{\pmb{\ell}^2_{2\perp}}{2yq^-}; \hspace{2em}  \ell^{+} = \frac{\pmb{\ell}^2_{2\perp}+M^2}{2(1-y)q^-};  \hspace{2em}   p^{+}_{2} = \frac{\left[\left(\pmb{\ell}_{2\perp}-\pmb{k}_{\perp}\right)^2 +M^2\right]}{2(1-y+\eta y)q^-}.
\end{equation}
The above expression (Eq.~\ref{eq:kernel-1_ph_qqgm_ph_qqgm_both_left-cut-fin}) of the hadronic tensor can be recast by instituting $\Delta z^{-} = z^{-}_{3}-z^{-}_{2}$ into the following form:
\begin{equation}
\begin{split}
    W^{\mu\nu}_{1,\ell} & =\sum_f 2 [-g^{\mu\nu}_{\perp \perp }]\int d (\Delta X^{-})   e^{iq^{+}(\Delta X^{-} )} 
  \left\langle P \left| \bar{\psi}_{_f}( \Delta X^{-}) \frac{\gamma^{+}}{4} \psi_{_f}(0)\right| P\right\rangle  \\
  & \times e^2 e^4_f g^2_s \int  d \zeta^{-} d (\Delta z^{-}) d^2 \Delta z_{\perp} \frac{dy}{2\pi} \frac{d^2 \ell_{2\perp}}{(2\pi)^2} \frac{d^2 k_{\perp}}{(2\pi)^2} \left[ -1 + e^{i\mathcal{G}^{(\ell_2,p_2,k)}_{M}(z^{-}_3-z^-_2)} \right]  e^{-i (\ell^+_2 + p^+_2) \Delta z^{-}}     e^{i \pmb{k}_{\perp}\cdot \Delta\pmb{z}_{\perp}}  \\
  & \times \frac{\theta( z^{-}_3-z^{-}_2) \theta( y^{-}-z^{-}_3)   \left[  (1+\eta y)\pmb{\ell}^2_{2\perp} - y\pmb{k}_{\perp}\pmb{\ell}_{2\perp}   + \kappa y^4 M^2 \right]}{\left[  \pmb{\ell}^2_{2\perp} + M^2y^2 \right] J_1} \langle P_{A-1} | {\rm Tr}[A^+(\zeta^-, \Delta z^-, \Delta z_{\perp}) A^+(\zeta^-, 0)] | P_{A-1} \rangle  \\
  &
\times  \left[ \frac{ 1 + \left(1-y\right)^2 +\eta y(2-y)}{y}\right] e^{i(\ell^+_2 + \ell^+ )(x^{-}-z^-_{2})}  e^{-i[M^2/(2q^-)](y^{-}-z^{-}_{3})}.
\end{split} \label{eq:kernel-1_ph_qqgm_ph_qqgm_both_left-cut-final_exp}
\end{equation}

Next, we consider the diagram shown in the right panel of Fig.~\ref{fig:kernel-1_ph_qqgm_ph_qqgm_both}(b). The topology of the diagram is the same as the diagram on the left panel. Moreover, they are complex conjugates of each other. The hadronic tensor for the left-cut diagram shown in Fig.~\ref{fig:kernel-1_ph_qqgm_ph_qqgm_both}(b) can be written as
\begin{equation}
\begin{split}
    W^{\mu\nu}_{1,\ell} & = \sum_f 2 [-g^{\mu\nu}_{\perp \perp }]\int d (\Delta X^{-})   e^{iq^{+}(\Delta X^{-} )} 
    e^{-i[M^{2}/(2q^{-})](\Delta X^{-} )}
  \left\langle P \left| \bar{\psi}_{_f}( \Delta X^{-}) \frac{\gamma^{+}}{4} \psi_{_f}(0)\right| P\right\rangle  \\
  & \times e^2 e^4_f g^2_s \int  d \zeta^{-} d (\Delta z^{-}) d^2 \Delta z_{\perp} \frac{dy}{2\pi} \frac{d^2 \ell_{2\perp}}{(2\pi)^2} \frac{d^2 k_{\perp}}{(2\pi)^2} \left[ -1 + e^{-i\mathcal{G}^{(\ell_2)}_{M}(y^{-}-z^-_3)} \right] e^{-i \mathcal{H}^{(\ell_2,p_2)}_{M} (\Delta z^{-})}     e^{i \pmb{k}_{\perp}\cdot\Delta\pmb{z}_{\perp}}  \\
  & \times \frac{\theta( x^{-}-z^{-}_2) \theta( y^{-}-z^{-}_3)   \left[  \left(1+\eta y\right)\pmb{\ell}^2_{2\perp} - y\pmb{k}_{\perp}\pmb{\ell}_{2\perp}   + \kappa y^4 M^2 \right]}{\left[  \pmb{\ell}^2_{2\perp} + M^2y^2 \right] J_1} \langle P_{A-1} | {\rm Tr}[A^+(\zeta^-, \Delta z^-, \Delta z_{\perp}) A^+(\zeta^-,0)] | P_{A-1} \rangle  \\
  &
\times  \left[ \frac{ 1 + \left(1-y\right)^2 +\eta y(2-y)}{y}\right], 
\end{split} 
\end{equation}
where   $\mathcal{G}^{(\ell_2)}_M$ is defined in Eq.~\ref{eq:kernel1_GML2_appendix1}, $ \mathcal{H}^{(\ell_2,p_2)}_{M}$ is defined in Eq.~\ref{eq:HL2P2M_append323}, and $J_1$ is defined in Eq.~\ref{eq:Kernel-1_J1}.
Similarly, the final expression of the hadronic tensor for the right-cut diagram [Fig.~\ref{fig:kernel-1_ph_qqgm_ph_qqgm_both}(b)] is given as
\begin{equation}
\begin{split}
    W^{\mu\nu}_{1,r} & = \sum_f 2 [-g^{\mu\nu}_{\perp \perp }]\int d (\Delta X^{-})   e^{iq^{+}(\Delta X^{-} )} 
  \left\langle P \left| \bar{\psi}_{_f}( \Delta X^{-}) \frac{\gamma^{+}}{4} \psi_{_f}(0)\right| P\right\rangle  \\
  & \times e^2 e^4_f g^2_s \int  d \zeta^{-} d (\Delta z^{-}) d^2 \Delta z_{\perp} \frac{dy}{2\pi} \frac{d^2 \ell_{2\perp}}{(2\pi)^2} \frac{d^2 k_{\perp}}{(2\pi)^2} \left[ -1 + e^{i\mathcal{G}^{(\ell_2,p_2,k)}_{M}(z^{-}_3-z^-_2)} \right]  e^{-i p^+_2 (\Delta z^{-})}     e^{i \pmb{k}_{\perp}\cdot \Delta\pmb{z}_{\perp}}  \\
  & \times \frac{\theta( -z^{-}_3 + z^{-}_2) \theta( x^{-}-z^{-}_2)   \left[\left(1+\eta y\right)\pmb{\ell}^2_{2\perp} - y\pmb{k}_{\perp}\pmb{\ell}_{2\perp}   + \kappa y^4 M^2 \right]}{\left[  \pmb{\ell}^2_{2\perp} + M^2y^2 \right] J_1} \langle P_{A-1} |{\rm Tr}[ A^+(\zeta^-, \Delta z^-, \Delta z_{\perp}) A^+(\zeta^-, 0)] | P_{A-1} \rangle  \\
  &
\times  \left[ \frac{ 1 + \left(1-y\right)^2 +\eta y(2-y)}{y}\right] e^{-i(\ell^+_2 + \ell'^+ )y^-} e^{i\ell^+_2 z^-_2} e^{i\ell'^+ z^-_3} e^{i[M^2/(2q^-)](x^{-}-z^{-}_{2})},
\end{split} \label{eq:kernel-1_ph_qqgm_ph_qqgm_both_b_right-cut}
\end{equation}
where $\mathcal{G}^{(\ell_2,p_2,k)}_{M}$ is defined in Eq.~\ref{eq:GML2_J1}, and 
\begin{equation}
    \ell^{+}_{2} = \frac{\pmb{\ell}^2_{2\perp}}{2yq^-}; \hspace{2em}  \ell'^{+} = \frac{\pmb{\ell}^2_{2\perp}+M^2}{2(1-y)q^-};  \hspace{2em}   p^{+}_{2} = \frac{\left[\left(\pmb{\ell}_{2\perp}-\pmb{k}_{\perp}\right)^2 +M^2\right]}{2(1-y+\eta y)q^-}.
\end{equation}
The above expression (Eq.~\ref{eq:kernel-1_ph_qqgm_ph_qqgm_both_b_right-cut}) of the hadronic tensor can be recast by instituting $\Delta z^{-} = z^{-}_{3}-z^{-}_{2}$ into the following form:
\begin{equation}
\begin{split}
    W^{\mu\nu}_{1,r} & =\sum_f 2 [-g^{\mu\nu}_{\perp \perp }]\int d (\Delta X^{-})   e^{iq^{+}(\Delta X^{-} )} 
  \left\langle P \left| \bar{\psi}_{_f}( \Delta X^{-}) \frac{\gamma^{+}}{4} \psi_{_f}(0)\right| P\right\rangle  \\
  & \times e^2 e^4_f g^2_s \int  d \zeta^{-} d (\Delta z^{-}) d^2 \Delta z_{\perp} \frac{dy}{2\pi} \frac{d^2 \ell_{2\perp}}{(2\pi)^2} \frac{d^2 k_{\perp}}{(2\pi)^2} \left[ -1 + e^{i\mathcal{G}^{(\ell_2,p_2,k)}_{M}(z^{-}_3-z^-_2)} \right]  e^{-i (\ell^+_{2}+p^+_2) (\Delta z^{-})}     e^{i \pmb{k}_{\perp}\cdot \Delta\pmb{z}_{\perp}}  \\
  & \times \frac{\theta( -z^{-}_3 + z^{-}_2) \theta( x^{-}-z^{-}_2)   \left[  (1+\eta y)\pmb{\ell}^2_{2\perp} - y\pmb{k}_{\perp}\pmb{\ell}_{2\perp}   + \kappa y^4 M^2 \right]}{\left[  \pmb{\ell}^2_{2\perp} + M^2y^2 \right] J_1} \langle P_{A-1} |{\rm Tr}[A^+(\zeta^-, \Delta z^-, \Delta z_{\perp}) A^+(\zeta^-, 0)] | P_{A-1} \rangle  \\
  &
\times  \left[ \frac{ 1 + (1-y)^2 +\eta y(2-y)}{y}\right] e^{-i(\ell^+_2 + \ell'^+ )(y^{-}-z^{-}_{3})}  e^{i[M^2/(2q^-)](x^{-}-z^{-}_{2})}.
\end{split} \label{eq:kernel-1_ph_qqgm_ph_qqgm_both_b_right-cut_final}
\end{equation}
%

\section{$\mbox{}$\!\!\!\!\!\!: Single-emission single-scattering kernel: one photon and one gluon in the final state}
\label{append:kernel2}

%
This section summarizes the calculation of all possible diagrams contributing to kernel-2 in a DIS between the highly virtual photon and the nucleus. The final state consists of a photon and a gluon with an in-medium quark exchange from the medium. 

\begin{figure}[!h]
    \centering 
    \begin{subfigure}[t]{0.45\textwidth}
        \centering        \includegraphics[height=1.2in]{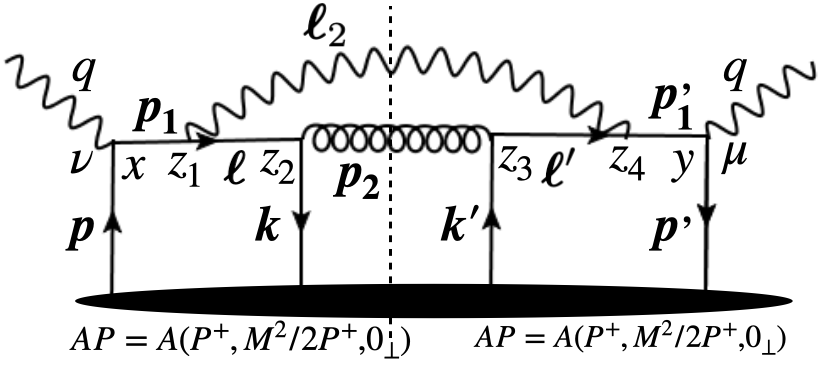}
        \caption{The final state contains a gluon generated from the conversion process and a bremsstrahlung photon. }
    \end{subfigure}%
    \begin{subfigure}[t]{0.45\textwidth}
        \centering        \includegraphics[height=1.2in]{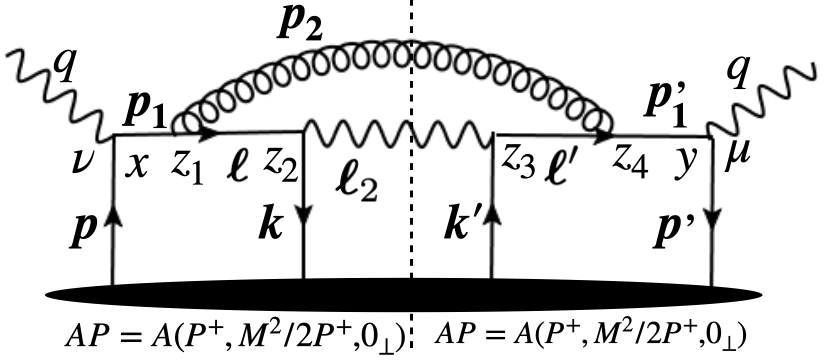}
        \caption{The final state contains a photon generated from the conversion process and a bremsstrahlung gluon.}
    \end{subfigure}
\caption{A forward scattering diagram contributing to kernel-2. }\label{fig:kernel2_ph_qgqm_gqqm_ph_both}
\end{figure}

The hadronic tensor for Fig.~\ref{fig:kernel2_ph_qgqm_gqqm_ph_both}(a) has the following form:

\begin{equation}
\begin{split}
W^{\mu\nu}_{2,c} & = \sum_f e^2 e^4_f g^2_{s} \int d^4 x d^4 y d^4 z_{2} d^4 z_{3} \int \frac{d^4 p}{(2\pi)^4} \frac{d^4 p'}{(2\pi)^4}  \frac{d^4 \ell_{2}}{(2\pi)^4} \frac{d^4 p_{2}}{(2\pi)^4} e^{-ip'y} e^{ipx}\left\langle P \left| \bar{\psi}_{_f}(y) \frac{\gamma^{+}}{4} \psi_{_f}(x)\right| P\right\rangle \delta^{bc} {\rm Tr}[t^{b}t^{c}] \\
  & \times e^{i\left(q+p'-p_{2}-\ell_{2}\right) z_{3}} e^{i(\ell_{2}+p_{2} - q - p)z_{2}}\left\langle P_{A-1} \left| \bar{\psi}_{_f}(z_2)\frac{\gamma^+}{4}\psi_{_f}(z_3)\right| P_{A-1} \right\rangle d^{(\ell_2)}_{\sigma_{1} \sigma_{4}} d^{(p_2)}_{\sigma_{3} \sigma_{2}} (2 \pi) \delta\left(\ell_{2}^{2}\right) (2 \pi)  \delta\left(p_{2}^{2}\right) \\
& \times \frac{ {\rm Tr} \left[\gamma^- \gamma^{\mu} \left(\slashed{q}+\slashed{p}'\right) \gamma^{\sigma_4}\left(\slashed{q}+\slashed{p}' - \slashed{\ell}_2\right)\gamma^{\sigma_{3}} \gamma^{-} \gamma^{\sigma_2} \left( \slashed{q} + \slashed{p} -\slashed{\ell}_2 \right) \gamma^{\sigma_1} \left( \slashed{q} + \slashed{p} \right) \gamma^{\nu} \right] 
}{  \left[\left(q+p'\right)^{2}-i\epsilon\right] \left[\left(q+p\right)^2  + i\epsilon\right]  \left[\left(q+p'-\ell_2\right)^2 - i\epsilon\right] \left[\left(q+p -\ell_2\right)^2+i\epsilon\right]  }.
\end{split} \label{eq:kernel2_ph_qgqm_gqqm_ph_both_wi_a}
\end{equation}

The above expression of the hadonic tensor has singularity when the denominator of the propagator for $p_1$, $\ell$, $\ell'$ and $p'_1$ becomes on-shell. It contains two simple poles for $p^+$ and $p'^+$. The contour integration for $p^+$ gives
\begin{equation}
\begin{split}
C_{1} & = \oint \frac{dp^{+}}{(2\pi)} \frac{e^{ip^{+}(x^{-}-z^{-}_{2})}}{\left[\left(q+p\right)^{2}  + i \epsilon\right]\left[\left(q+p-\ell_{2}\right)^{2}  + i\epsilon\right]} \\
      & = \oint \frac{dp^{+}}{(2\pi)} \frac{e^{ip^{+}(x^{-} - z^{-}_{2}) }}{2q^{-}[ q^{+} + p^{+}  + i \epsilon] 2(q^{-} -\ell^{-}_{2} )\left[ q^{+} + p^{+} -\ell^{+}_{2} - \frac{\pmb{\ell}^{2}_{2\perp}}{2(q^{-}-\ell^{-}_{2})} + i \epsilon\right]}   \\
      & = \frac{(2\pi i)}{2\pi} \frac{\theta(x^{-} - z^{-}_{2})}{4q^{-}(q^{-}-\ell^{-}_{2})}  e^{-iq^{+} (x^{-}-z^{-}_{2})}\left[ \frac{ -1 + e^{i \mathcal{G}^{(\ell_2)}_0(x^{-}-z^{-}_{2})} }{ \mathcal{G}^{(\ell_2)}_0}  \right],
\end{split}
\end{equation}
where
\begin{equation}
    \mathcal{G}^{(\ell_2)}_0 = \ell^{+}_{2} + \frac{\pmb{\ell}^{2}_{2\perp}}{2(q^{-}-\ell^{-}_{2})}  = \frac{\pmb{\ell}^2_{2\perp} }{2y(1-y) q^-}.
    \label{eq:GL20_kernel2_xy}
\end{equation}
Similarly, the contour integration for $p'^+$ gives
\begin{equation}
\begin{split}
C_{2} & = \oint \frac{dp'^{+}}{(2\pi)} \frac{e^{-ip'^{+}(y^{-}-z^{-}_{3})}}{\left[\left(q+p'\right)^{2}  - i \epsilon\right]\left[\left(q+p'-\ell_{2}\right)^{2}  - i\epsilon\right]} \\
      & = \oint \frac{dp'^{+}}{(2\pi)} \frac{e^{-ip'^{+}(y^{-} - z^{-}_{3}) }}{2q^{-}[ q^{+} + p'^{+} 
      - i \epsilon] 2(q^{-} -\ell^{-}_{2} )\left[ q^{+} + p'^{+} -\ell^{+}_{2} - \frac{\pmb{\ell}^{2}_{2\perp}}{2(q^{-}-\ell^{-}_{2})} - i \epsilon\right]}   \\
      & = \frac{(-2\pi i)}{2\pi} \frac{\theta(y^{-} - z^{-}_{3})}{4q^{-}(q^{-}-\ell^{-}_{2})}  e^{iq^{+}(y^{-}-z^{-}_{3})}\left[ \frac{ -1 + e^{-i \mathcal{G}^{(\ell_2)}_0(y^{-}-z^{-}_{3})} }{ \mathcal{G}^{(\ell_2)}_0}  \right] .
\end{split}
\end{equation}
The trace in the third line of Eq.~\ref{eq:kernel2_ph_qgqm_gqqm_ph_both_wi_a} simplifies to
\begin{equation}
\begin{split}    
&  {\rm Tr} \left[\gamma^- \gamma^{\mu} \left(\slashed{q}+\slashed{p}'\right) \gamma^{\sigma_4}\left(\slashed{q}+\slashed{p}' - \slashed{\ell}_2\right)\gamma^{\sigma_{3}} \gamma^{-} \gamma^{\sigma_2} \left( \slashed{q} + \slashed{p} -\slashed{\ell}_2 \right) \gamma^{\sigma_1} \left( \slashed{q} + \slashed{p} \right) \gamma^{\nu} \right] d^{(\ell_2)}_{\sigma_{1} \sigma_{4}} d^{(p_2)}_{\sigma_{3} \sigma_{2}} \\
& = 32 [-g^{\mu\nu}_{\perp\perp}](q^-)^2 \left[ \frac{1 + \left(1-y\right)^2}{y^2} \right]  \pmb{\ell}^2_{2\perp} .
\end{split}
\end{equation}
Finally, the hadronic tensor [Fig.~\ref{fig:kernel2_ph_qgqm_gqqm_ph_both}(a)] reduces to the following form:
\begin{equation}
\begin{split}
    W^{\mu\nu}_{2,c} & =\sum_f  2[-g^{\mu\nu}_{\perp \perp }]\int d (\Delta X^{-})   e^{iq^{+}(\Delta X^{-} )}     
  \left\langle P \left| \bar{\psi}_{_f}( \Delta X^{-}) \frac{\gamma^{+}}{4} \psi_{_f}(0)\right| P\right\rangle  \\
  & \times e^2 e^4_f g^2_s \int  d \zeta^{-} d (\Delta z^{-}) d^2 \Delta z_{\perp} \frac{dy}{2\pi} \frac{d^2 \ell_{2\perp}}{(2\pi)^2} \frac{d^2 k_{\perp}}{(2\pi)^2} \left[ 2 -2 \cos\left\{\mathcal{G}^{(\ell_2)}_{0}\zeta^-\right\} \right] e^{-i (\Delta z^{-})\mathcal{H}^{(\ell_2 , p_2)}_{0} }     e^{i \pmb{k}_{\perp} \cdot \Delta\pmb{z}_{\perp}}  \\
  & \times \frac{\theta( x^{-}-z^{-}_2) \theta( y^{-}-z^{-}_3)  }{  \pmb{\ell}^2_{2\perp}  } \frac{1}{(1-y+\eta y)q^-}\left\langle P_{A-1} \left| \bar{\psi}_{_f}(\zeta^-, 0) \frac{\gamma^+}{4} \psi_{_f}(\zeta^-, \Delta z^-, \Delta z_{\perp})\right| P_{A-1} \right\rangle  \\
  &
\times  \left[ \frac{ 1 + \left(1-y\right)^2 }{y}\right] [C_F N_c] ,
\end{split}     \label{eq:kernel2-wmunu-final-1st}
\end{equation}
where $ \mathcal{G}^{(\ell_2)}_0$ is given in Eq.~\ref{eq:GL20_kernel2_xy} and 
\begin{equation}    \mathcal{H}^{(\ell_2,p_2)}_{0} = \ell^+_2 + p^+_2 = \frac{\pmb{\ell}^2_{2\perp}}{2yq^-} + \frac{(\pmb{\ell}_{2\perp} - \pmb{k}_\perp)^2 }{2(1-y+\eta y)q^-} .
\label{eq:HL2P2M0_appendix}
\end{equation}

Next, we consider a forward scattering diagram [Fig.~\ref{fig:kernel2_ph_qgqm_gqqm_ph_both}(b)] where the final state contains a photon generated from the conversion process and a bremsstrahlung gluon. The associated hadronic tensor is given as
\begin{equation}
\begin{split}
W^{\mu\nu}_{2,c} & = \sum_f e^2 e^4_f g^2_{s} \int d^4 x d^4 y d^4 z_{2} d^4 z_{3} \int \frac{d^4 p}{(2\pi)^4} \frac{d^4 p'}{(2\pi)^4}  \frac{d^4 \ell_{2}}{(2\pi)^4} \frac{d^4 p_{2}}{(2\pi)^4} e^{-ip'y} e^{ipx}\left\langle P \left| \bar{\psi}_{_f}(y) \frac{\gamma^{+}}{4} \psi_{_f}(x)\right| P\right\rangle \delta^{ad} {\rm Tr}[t^{a}t^{d}] \\
  & \times e^{i\left(q+p'-p_{2}-\ell_{2}\right) z_{3}} e^{i(\ell_{2}+p_{2} - q - p)z_{2}} \left\langle P_{A-1}\left | \bar{\psi}_{_f}(z_2)\frac{\gamma^+}{4}\psi_{_f}(z_3)\right| P_{A-1} \right\rangle d^{(\ell_2)}_{\sigma_{3} \sigma_{2}} d^{(p_2)}_{\sigma_{4} \sigma_{1}} (2 \pi) \delta\left(\ell_{2}^{2}\right) (2 \pi)  \delta\left(p_{2}^{2}\right) \\
& \times \frac{ {\rm Tr} \left[\gamma^- \gamma^{\mu} \left(\slashed{q}+\slashed{p}'\right) \gamma^{\sigma_4}\left(\slashed{q}+\slashed{p}' - \slashed{p}_2\right)\gamma^{\sigma_{3}} \gamma^{-} \gamma^{\sigma_2} \left( \slashed{q} + \slashed{p} -\slashed{p}_2 \right) \gamma^{\sigma_1} \left( \slashed{q} + \slashed{p} \right) \gamma^{\nu} \right] 
}{  \left[\left(q+p'\right)^{2}-i\epsilon\right] \left[\left(q+p\right)^2  + i\epsilon\right]  \left[\left(q+p'-p_2\right)^2 - i\epsilon\right] \left[\left(q+p -p_2\right)^2+i\epsilon\right]  }.
\end{split} \label{eq:kernel2_ph_qgqm_gqqm_ph_both_wi_b}
\end{equation}
The above expression of the hadonic tensor has singularity when the denominator of the propagator for $p_1$, $\ell$, $\ell'$ and $p'_1$ becomes on-shell. It contains two simple poles for $p^+$ and $p'^+$. The contour integration for $p^+$ gives
\begin{equation}
\begin{split}
C_{1} & = \oint \frac{dp^{+}}{(2\pi)} \frac{e^{ip^{+}(x^{-}-z^{-}_{2})}}{\left[\left(q+p\right)^{2}  + i \epsilon\right]\left[\left(q+p-p_{2}\right)^{2}  + i\epsilon\right]} \\
      & = \oint \frac{dp^{+}}{(2\pi)} \frac{e^{ip^{+}(x^{-} - z^{-}_{2}) }}{2q^{-}[ q^{+} + p^{+}  + i \epsilon] 2(q^{-} -p^{-}_{2} )\left[q^{+} + p^{+} -p^{+}_{2} - \frac{\pmb{p}^{2}_{2\perp}}{2(q^{-}-p^{-}_{2})} + i \epsilon\right]}   \\
      & = \frac{(2\pi i)}{2\pi} \frac{\theta(x^{-} - z^{-}_{2})}{4q^{-}(q^{-}-p^{-}_{2})}  e^{-iq^{+} (x^{-}-z^{-}_{2})}\left[ \frac{ -1 + e^{i \mathcal{G}^{(p_2)}_0(x^{-}-z^{-}_{2})} }{ \mathcal{G}^{(p_2)}_0}  \right],
\end{split}
\end{equation}
where
\begin{equation}
    \mathcal{G}^{(p_2)}_0 = p^{+}_{2} + \frac{\pmb{p}^{2}_{2\perp}}{2(q^{-}-p^{-}_{2})}  = \frac{\pmb{p}^2_{2\perp} }{2y(1-y+\eta y)(1-\eta) q^-}.
\end{equation}
Similarly, the contour integration for $p'^+$ gives
\begin{equation}
\begin{split}
C_{2} & = \oint \frac{dp'^{+}}{(2\pi)} \frac{e^{-ip'^{+}(y^{-}-z^{-}_{3})}}{\left[\left(q+p'\right)^{2}  - i \epsilon\right]\left[\left(q+p'-p_{2}\right)^{2}  - i\epsilon\right]} \\
      & = \oint \frac{dp'^{+}}{(2\pi)} \frac{e^{-ip'^{+}(y^{-} - z^{-}_{3}) }}{2q^{-}\left[q^{+} + p'^{+}-i\epsilon\right] 2(q^{-} -p^{-}_{2} )\left[ q^{+} + p'^{+} -p^{+}_{2} - \frac{\pmb{p}^{2}_{2\perp}}{2(q^{-}-p^{-}_{2})} - i \epsilon\right]}   \\
      & = \frac{(-2\pi i)}{2\pi} \frac{\theta(y^{-} - z^{-}_{3})}{4q^{-}(q^{-}-p^{-}_{2})}  e^{iq^{+}(y^{-}-z^{-}_{3})}\left[ \frac{ -1 + e^{-i \mathcal{G}^{(p_2)}_0(y^{-}-z^{-}_{3})} }{ \mathcal{G}^{(p_2)}_0}  \right].
\end{split}
\end{equation}
The trace in the third line of Eq.~\ref{eq:kernel2_ph_qgqm_gqqm_ph_both_wi_b} simplifies to
\begin{equation}
\begin{split}    
&  {\rm Tr} \left[\gamma^- \gamma^{\mu} \left(\slashed{q}+\slashed{p}'\right) \gamma^{\sigma_4}\left(\slashed{q}+\slashed{p}' - \slashed{p}_2\right)\gamma^{\sigma_{3}} \gamma^{-} \gamma^{\sigma_2} \left( \slashed{q} + \slashed{p} -\slashed{p}_2 \right) \gamma^{\sigma_1} \left( \slashed{q} + \slashed{p} \right) \gamma^{\nu} \right]  d^{(p_2)}_{\sigma_{1} \sigma_{4}} d^{(\ell_2)}_{\sigma_{3} \sigma_{2}} \\
& = 32 [-g^{\mu\nu}_{\perp\perp}]\left(q^-\right)^2 \left[ \frac{1 + y^2 +\eta y^2 (\eta -2) }{\left(1-y+\eta y\right)^2} \right]  \pmb{p}^2_{2\perp}.
\end{split}
\end{equation}
Finally, the hadronic tensor [Fig.~\ref{fig:kernel2_ph_qgqm_gqqm_ph_both}(b)] reduces to the following form:
\begin{equation}
\begin{split}
    W^{\mu\nu}_{2,c} & =\sum_f  2[-g^{\mu\nu}_{\perp \perp }]\int d (\Delta X^{-})   e^{iq^{+}(\Delta X^{-} )}     
  \left\langle P \left| \bar{\psi}_{_f}( \Delta X^{-}) \frac{\gamma^{+}}{4} \psi_{_f}(0)\right| P\right\rangle  \\
  & \times e^2 e^4_f g^2_s \int  d \zeta^{-} d (\Delta z^{-}) d^2 \Delta z_{\perp} \frac{dy}{2\pi} \frac{d^2 \ell_{2\perp}}{(2\pi)^2} \frac{d^2 k_{\perp}}{(2\pi)^2} \left[ 2 -2 \cos\left\{\mathcal{G}^{(p_2)}_{0}\zeta^-\right\} \right] e^{-i (\Delta z^{-})\mathcal{H}^{(\ell_2 , p_2)}_{0} }     e^{i \pmb{k}_{\perp} \cdot \Delta\pmb{z}_{\perp}}  \\
  & \times \frac{\theta( x^{-}-z^{-}_2) \theta( y^{-}-z^{-}_3)  }{  \left(\pmb{\ell}_{2\perp} -\pmb{k}_{\perp} \right)^2} \frac{1}{yq^-}\left\langle P_{A-1} \left| \bar{\psi}_{_f}(\zeta^-,0) \frac{\gamma^+}{4} \psi_{_f}(\zeta^-, \Delta z^-, \Delta z_{\perp})\right| P_{A-1} \right\rangle  \\
  &
\times  \left[ \frac{ 1 + y^2 +\eta y^2\left(\eta-2\right) }{(1-y+\eta y)}\right] [C_F N_c], 
\end{split} 
\end{equation}
where $\mathcal{H}^{(\ell_2,p_2)}_{0}$ is defined in Eq.~\ref{eq:HL2P2M0_appendix}, and 
\begin{equation}
    \mathcal{G}^{(p_2)}_0 = p^{+}_{2} + \frac{\pmb{p}^{2}_{2\perp}}{2(q^{-}-p^{-}_{2})}  = \frac{(\pmb{\ell}_{2\perp}-\pmb{k}_{\perp})^2 }{2y(1-y+\eta y)(1-\eta) q^-},
    \label{eq:GP2M0_appendix}
\end{equation}
\begin{figure}[!h]
    \centering 
    \begin{subfigure}[t]{0.45\textwidth}
        \centering        \includegraphics[height=1.2in]{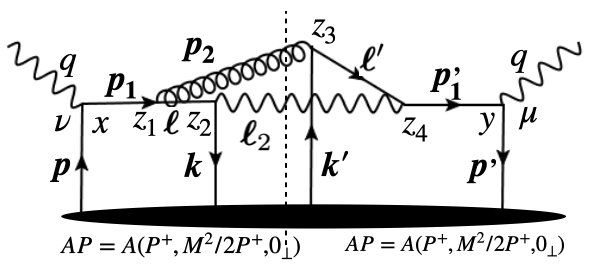}
        \caption{Interference diagram. }
    \end{subfigure}%
    \begin{subfigure}[t]{0.45\textwidth}
        \centering        \includegraphics[height=1.2in]{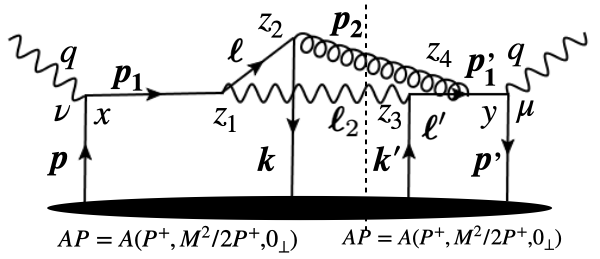}
        \caption{Complex conjugate of the diagram on the left panel.}
    \end{subfigure}
\caption{A forward scattering diagram contributing to kernel-2.}\label{fig:Kernel2_g_qphqm_gqqm_gqq_both}
\end{figure}

Next, we consider a forward scattering diagram as shown in Fig.~\ref{fig:Kernel2_g_qphqm_gqqm_gqq_both}. The hadronic tensor for Fig.~\ref{fig:Kernel2_g_qphqm_gqqm_gqq_both} (a) is
\begin{equation}
\begin{split}
W^{\mu\nu}_{2,c} & = \sum_f e^2 e^4_f g^2_{s} \int d^4 x d^4 y d^4 z_{2} d^4 z_{3} \int \frac{d^4 p}{(2\pi)^4} \frac{d^4 p'}{(2\pi)^4}  \frac{d^4 \ell_{2}}{(2\pi)^4} \frac{d^4 p_{2}}{(2\pi)^4} e^{-ip'y} e^{ipx}\left\langle P \left| \bar{\psi}_{_f}(y) \frac{\gamma^{+}}{4} \psi_{_f}(x)\right| P\right\rangle \delta^{ac} {\rm Tr}[t^{a}t^{c}] \\
  & \times e^{i\left(q+p'-p_{2}-\ell_{2}\right) z_{3}} e^{i(\ell_{2}+p_{2} - q - p)z_{2}} \left\langle P_{A-1} \left| \bar{\psi}_{_f}(z_2)\frac{\gamma^+}{4}\psi_{_f}(z_3)\right| P_{A-1} \right\rangle d^{(\ell_2)}_{\sigma_{2} \sigma_{4}} d^{(p_2)}_{\sigma_{3} \sigma_{1}} (2 \pi) \delta\left(\ell_{2}^{2}\right) (2 \pi)  \delta\left(p_{2}^{2}\right) \\
& \times \frac{ {\rm Tr} \left[\gamma^- \gamma^{\mu} \left(\slashed{q}+\slashed{p}'\right) \gamma^{\sigma_4}\left(\slashed{q}+\slashed{p}' - \slashed{\ell}_2\right)\gamma^{\sigma_{3}} \gamma^{-} \gamma^{\sigma_2} \left( \slashed{q} + \slashed{p} -\slashed{p}_2 \right) \gamma^{\sigma_1} \left( \slashed{q} + \slashed{p} \right) \gamma^{\nu} \right] 
}{  \left[\left(q+p'\right)^{2}-i\epsilon\right] \left[\left(q+p\right)^2  + i\epsilon\right]  \left[\left(q+p'-\ell_2\right)^2 - i\epsilon\right] \left[\left(q+p -p_2\right)^2+i\epsilon\right]  }.
\end{split} \label{eq:Kernel2_g_qphqm_gqqm_gqq_both_wi_a}
\end{equation}
The above expression has singularity when the denominator of the propagator for $p_1$, $\ell$, $\ell'$ and $p'_1$ becomes on-shell. It contains two simple poles for $p^+$ and $p'^+$. The contour integration for $p^+$ gives
\begin{equation}
\begin{split}
C_{1} & = \oint \frac{dp^{+}}{(2\pi)} \frac{e^{ip^{+}(x^{-}-z^{-}_{2})}}{\left[\left(q+p\right)^{2}  + i \epsilon\right]\left[\left(q+p-p_{2}\right)^{2}  + i\epsilon\right]} \\
      & = \oint \frac{dp^{+}}{(2\pi)} \frac{e^{ip^{+}(x^{-} - z^{-}_{2}) }}{2q^{-}\left[q^{+} + p^{+}  + i \epsilon\right] 2(q^{-} -p^{-}_{2} )\left[ q^{+} + p^{+} -p^{+}_{2} - \frac{\pmb{p}^{2}_{2\perp}}{2(q^{-}-p^{-}_{2})} + i \epsilon\right]}   \\
      & = \frac{(2\pi i)}{2\pi} \frac{\theta(x^{-} - z^{-}_{2})}{4q^{-}(q^{-}-p^{-}_{2})}  e^{-iq^{+} (x^{-}-z^{-}_{2})}\left[ \frac{ -1 + e^{i \mathcal{G}^{(p_2)}_0(x^{-}-z^{-}_{2})} }{ \mathcal{G}^{(p_2)}_0}  \right],
\end{split}
\end{equation}
where $\mathcal{G}^{(p_2)}_0 $ is defined in Eq.~\ref{eq:GP2M0_appendix}.

Similarly, the contour integration for $p'^+$ gives
\begin{equation}
\begin{split}
C_{2} & = \oint \frac{dp'^{+}}{(2\pi)} \frac{e^{-ip'^{+}(y^{-}-z^{-}_{3})}}{\left[\left(q+p'\right)^{2} - i\epsilon\right]\left[\left(q+p'-\ell_{2}\right)^{2} - i\epsilon\right]} \\
      & = \oint \frac{dp'^{+}}{(2\pi)} \frac{e^{-ip'^{+}(y^{-} - z^{-}_{3}) }}{2q^{-}[ q^{+} + p'^{+} 
      - i \epsilon] 2(q^{-} -\ell^{-}_{2} )\left[ q^{+} + p'^{+} -\ell^{+}_{2} - \frac{\pmb{\ell}^{2}_{2\perp}}{2\left(q^{-}-\ell^{-}_{2}\right)} - i\epsilon\right]}   \\
      & = \frac{(-2\pi i)}{2\pi} \frac{\theta(y^{-} - z^{-}_{3})}{4q^{-}(q^{-}-\ell^{-}_{2})} e^{iq^{+}(y^{-}-z^{-}_{3})}\left[ \frac{ -1 + e^{-i \mathcal{G}^{(\ell_2)}_0(y^{-}-z^{-}_{3})} }{ \mathcal{G}^{(\ell_2)}_0}  \right],
\end{split}
\end{equation}
where $\mathcal{G}^{(\ell_2)}_0$ is defined in Eq.~\ref{eq:GL20_kernel2_xy}.

The trace in the third line of Eq.~\ref{eq:Kernel2_g_qphqm_gqqm_gqq_both_wi_a} simplifies to
\begin{equation}
    \begin{split}
         & {\rm Tr} \left[\gamma^- \gamma^{\mu} \left(\slashed{q}+\slashed{p}'\right) \gamma^{\sigma_4}\left(\slashed{q}+\slashed{p}' - \slashed{\ell}_2\right)\gamma^{\sigma_{3}} \gamma^{-} \gamma^{\sigma_2} \left( \slashed{q} + \slashed{p} -\slashed{p}_2 \right) \gamma^{\sigma_1} \left( \slashed{q} + \slashed{p} \right) \gamma^{\nu} \right] d^{(\ell_2)}_{\sigma_{2} \sigma_{4}} d^{(p_2)}_{\sigma_{3} \sigma_{1}} \\
         & = \frac{32 \left(q^-\right)^2 \left[ -g^{\mu\nu}_{\perp\perp}\right] (1-y+2\eta y) }{(1-y+\eta y)y} \left[ -\pmb{\ell}^{2}_{2\perp} +  \pmb{\ell}_{2\perp}\cdot \pmb{k}_{\perp} \right].
    \end{split}
\end{equation}
The final expression of the hadronic tensor for Fig.~\ref{fig:Kernel2_g_qphqm_gqqm_gqq_both}(a) is given by
\begin{equation}
\begin{split}
    W^{\mu\nu}_{2,c} & = \sum_f 2[-g^{\mu\nu}_{\perp \perp }]\int d (\Delta X^{-})   e^{iq^{+}(\Delta X^{-} )}     
  \left\langle P \left| \bar{\psi}_{_f}( \Delta X^{-}) \frac{\gamma^{+}}{4} \psi_{_f}(0)\right| P\right\rangle  \\
  & \times e^2 e^4_f g^2_s \int  d \zeta^{-} d (\Delta z^{-}) d^2 \Delta z_{\perp} \frac{dy}{2\pi} \frac{d^2 \ell_{2\perp}}{(2\pi)^2} \frac{d^2 k_{\perp}}{(2\pi)^2} \left[ -1 + e^{i\mathcal{G}^{(p_2)}_{0}(x^{-} - z^{-}_{2})} \right] \left[ -1 + e^{-i\mathcal{G}^{(\ell_2)}_{0}(y^{-} - z^{-}_{3})} \right] e^{-i (\Delta z^{-})\mathcal{H}^{(\ell_2 , p_2)}_{0} }  \\
  & \times \frac{\theta( x^{-}-z^{-}_2) \theta( y^{-}-z^{-}_3)  }{  (1-y+\eta y)q^-  } \frac{\left[ -\pmb{\ell}^{2}_{2\perp} +  \pmb{\ell}_{2\perp}\cdot \pmb{k}_{\perp} \right]}{\left(\pmb{\ell}_{2\perp}-\pmb{k}_{\perp} \right)^2\pmb{\ell}^2_{2\perp}} e^{i \pmb{k}_{\perp} \cdot \Delta\pmb{z}_{\perp}} \left\langle P_{A-1} \left| \bar{\psi}_{_f}(\zeta^-,0) \frac{\gamma^+}{4} \psi_{_f}(\zeta^-, \Delta z^-, \Delta z_{\perp})\right| P_{A-1} \right\rangle  \\
  &
\times \left[ \frac{(1-y+2\eta y)}{y} \right] [C_F N_c], 
\end{split}     \label{Kernel2_g_qphqm_gqqm_gqq_both-wmunu-a_final}
\end{equation}
where $ \mathcal{G}^{(\ell_2)}_0$ is given in Eq.~\ref{eq:GL20_kernel2_xy}, $ \mathcal{G}^{(p_2)}_0$ is given in Eq.~\ref{eq:GP2M0_appendix}, and $\mathcal{H}^{(\ell_2,p_2)}_{0}$ is defined in Eq.~\ref{eq:HL2P2M0_appendix}.

The forward scattering diagram shown in Fig.~\ref{fig:Kernel2_g_qphqm_gqqm_gqq_both}(b) is a complex conjugate of the diagram in Fig.~\ref{fig:Kernel2_g_qphqm_gqqm_gqq_both} (a). The hadronic tensor of the diagram in Fig.~\ref{fig:Kernel2_g_qphqm_gqqm_gqq_both}(b) is given as
\begin{equation}
\begin{split}
W^{\mu\nu}_{2,c} & =\sum_f e^2 e^4_f g^2_{s} \int d^4 x d^4 y d^4 z_{2} d^4 z_{3} \int \frac{d^4 p}{(2\pi)^4} \frac{d^4 p'}{(2\pi)^4}  \frac{d^4 \ell_{2}}{(2\pi)^4} \frac{d^4 p_{2}}{(2\pi)^4} e^{-ip'y} e^{ipx}\left\langle P \left| \bar{\psi}_{_f}(y) \frac{\gamma^{+}}{4} \psi_{_f}(x)\right| P\right\rangle \delta^{ac} {\rm Tr}[t^{a}t^{c}] \\
  & \times e^{i\left(q+p'-p_{2}-\ell_{2}\right) z_{3}} e^{i(\ell_{2}+p_{2} - q - p)z_{2}} \left\langle P_{A-1} \left| \bar{\psi}_{_f}(z_2)\frac{\gamma^+}{4}\psi_{_f}(z_3)\right| P_{A-1} \right\rangle d^{(p_2)}_{\sigma_{2} \sigma_{4}} d^{(\ell_2)}_{\sigma_{3} \sigma_{1}} (2 \pi) \delta\left(\ell_{2}^{2}\right) (2 \pi)  \delta\left(p_{2}^{2}\right) \\
& \times \frac{ {\rm Tr} \left[\gamma^- \gamma^{\mu} \left(\slashed{q}+\slashed{p}'\right) \gamma^{\sigma_4}\left(\slashed{q}+\slashed{p}' - \slashed{p}_2\right)\gamma^{\sigma_{3}} \gamma^{-} \gamma^{\sigma_2} \left( \slashed{q} + \slashed{p} -\slashed{\ell}_2 \right) \gamma^{\sigma_1} \left( \slashed{q} + \slashed{p} \right) \gamma^{\nu} \right] 
}{  \left[\left(q+p'\right)^{2}-i\epsilon\right] \left[\left(q+p\right)^2  + i\epsilon\right]  \left[\left(q+p'-p_2\right)^2 - i\epsilon\right] \left[\left(q+p -\ell_2\right)^2+i\epsilon\right]  }.
\end{split} \label{eq:Kernel2_g_qphqm_gqqm_gqq_both_wi_b}
\end{equation}

The above expression of the hadonic tensor has singularity when the denominator of the propagator for $p_1$, $\ell$, $\ell'$ and $p'_1$ becomes on-shell. It contains two simple poles for $p^+$ and $p'^+$. The contour integration for $p^+$ gives
\begin{equation}
\begin{split}
C_{1} & = \oint \frac{dp^{+}}{(2\pi)} \frac{e^{ip^{+}(x^{-}-z^{-}_{2})}}{\left[\left(q+p\right)^{2} + i\epsilon\right]\left[\left(q+p-\ell_{2}\right)^{2} + i\epsilon\right]} \\
      & = \oint \frac{dp^{+}}{(2\pi)} \frac{e^{ip^{+}(x^{-} - z^{-}_{2}) }}{2q^{-}\left[q^{+} + p^{+}  + i \epsilon\right]2(q^{-} -\ell^{-}_{2})\left[ q^{+} + p^{+} -\ell^{+}_{2} - \frac{\pmb{\ell}^{2}_{2\perp}}{2(q^{-}-\ell^{-}_{2})} + i \epsilon\right]}   \\
      & = \frac{(2\pi i)}{2\pi} \frac{\theta(x^{-} - z^{-}_{2})}{4q^{-}(q^{-}-\ell^{-}_{2})}  e^{-iq^{+} (x^{-}-z^{-}_{2})}\left[ \frac{ -1 + e^{i \mathcal{G}^{(\ell_2)}_0(x^{-}-z^{-}_{2})} }{ \mathcal{G}^{(\ell_2)}_0}  \right],
\end{split}
\end{equation}
where $ \mathcal{G}^{(\ell_2)}_0$ is given in Eq.~\ref{eq:GL20_kernel2_xy}.
%
%

Similarly, the contour integration for $p'^+$ gives
\begin{equation}
\begin{split}
C_{2} & = \oint \frac{dp'^{+}}{(2\pi)} \frac{e^{-ip'^{+}(y^{-}-z^{-}_{3})}}{\left[\left(q+p'\right)^{2} - i\epsilon\right]\left[\left(q+p'-p_{2}\right)^{2}  - i\epsilon\right]} \\
      & = \oint \frac{dp'^{+}}{(2\pi)} \frac{e^{-ip'^{+}(y^{-} - z^{-}_{3}) }}{2q^{-}[ q^{+} + p'^{+} 
      - i \epsilon] 2(q^{-} -p^{-}_{2} )\left[ q^{+} + p'^{+} -p^{+}_{2} - \frac{\pmb{p}^{2}_{2\perp}}{2(q^{-}-p^{-}_{2})} - i \epsilon\right]}   \\
      & = \frac{(-2\pi i)}{2\pi} \frac{\theta(y^{-} - z^{-}_{3})}{4q^{-}(q^{-}-p^{-}_{2})}  e^{iq^{+}(y^{-}-z^{-}_{3})}\left[ \frac{ -1 + e^{-i \mathcal{G}^{(p_2)}_0(y^{-}-z^{-}_{3})} }{ \mathcal{G}^{(p_2)}_0}  \right],
\end{split}
\end{equation}
where $\mathcal{G}^{(p_2)}_0$ is given in Eq.~\ref{eq:GP2M0_appendix}.
%
%
The trace in third line of Eq.~\ref{eq:Kernel2_g_qphqm_gqqm_gqq_both_wi_b} simplifies to
\begin{equation}
    \begin{split}
         & {\rm Tr} \left[\gamma^- \gamma^{\mu} \left(\slashed{q}+\slashed{p}'\right) \gamma^{\sigma_4}\left(\slashed{q}+\slashed{p}' - \slashed{p}_2\right)\gamma^{\sigma_{3}} \gamma^{-} \gamma^{\sigma_2} \left( \slashed{q} + \slashed{p} -\slashed{\ell}_2 \right) \gamma^{\sigma_1} \left( \slashed{q} + \slashed{p} \right) \gamma^{\nu} \right] d^{(p_2)}_{\sigma_{2} \sigma_{4}} d^{(\ell_2)}_{\sigma_{3} \sigma_{1}} \\
         & = \frac{32 \left(q^-\right)^2 \left[ -g^{\mu\nu}_{\perp\perp}\right] (1-y+2\eta y) }{(1-y+\eta y)y} \left[ -\pmb{\ell}^{2}_{2\perp} +  \pmb{\ell}_{2\perp}\cdot \pmb{k}_{\perp} \right].
    \end{split}
\end{equation}
The final expression of the hadronic tensor for Fig.~\ref{fig:Kernel2_g_qphqm_gqqm_gqq_both}(b) is given by
\begin{equation}
\begin{split}
    W^{\mu\nu}_{2,c} & =\sum_f  2[-g^{\mu\nu}_{\perp \perp }]\int d (\Delta X^{-})   e^{iq^{+}(\Delta X^{-} )}     
  \left\langle P \left| \bar{\psi}_{_f}( \Delta X^{-}) \frac{\gamma^{+}}{4} \psi_{_f}(0)\right| P\right\rangle  \\
  & \times e^2 e^4_f g^2_s \int  d \zeta^{-} d (\Delta z^{-}) d^2 \Delta z_{\perp} \frac{dy}{2\pi} \frac{d^2 \ell_{2\perp}}{(2\pi)^2} \frac{d^2 k_{\perp}}{(\pi)^2} \left[ -1 + e^{i\mathcal{G}^{(\ell_2)}_{0}(x^{-} - z^{-}_{2})} \right] \left[ -1 + e^{-i\mathcal{G}^{(p_2)}_{0}(y^{-} - z^{-}_{3})} \right] e^{-i (\Delta z^{-})\mathcal{H}^{(\ell_2 , p_2)}_{0} }    \\
  & \times \frac{\theta( x^{-}-z^{-}_2) \theta( y^{-}-z^{-}_3)  }{  (1-y+\eta y)q^-  } \frac{\left[ -\pmb{\ell}^{2}_{2\perp} +  \pmb{\ell}_{2\perp}\cdot \pmb{k}_{\perp} \right]}{\left(\pmb{\ell}_{2\perp}-\pmb{k}_{\perp} \right)^2\pmb{\ell}^2_{2\perp} }    e^{i \pmb{k}_{\perp} \cdot \Delta\pmb{z}_{\perp}} \left\langle P_{A-1} \left| \bar{\psi}_{_f}(\zeta^-, 0) \frac{\gamma^+}{4} \psi_{_f}(\zeta^-, \Delta z^-, \Delta z_{\perp})\right| P_{A-1} \right\rangle  \\
  &
\times \left[ \frac{(1-y+2\eta y)}{y} \right] [C_F N_c], 
\end{split}     \label{eq:Kernel2_g_qphqm_gqqm_gqq_both-wmunu-b_final}
\end{equation}
where $\mathcal{G}^{(\ell_2)}_0$, $\mathcal{G}^{(p_2)}_0$, and $\mathcal{H}^{(\ell_2,p_2)}_{0}$ are given in Eq.~\ref{eq:GL20_kernel2_xy}, Eq.~\ref{eq:GP2M0_appendix}, Eq.~\ref{eq:HL2P2M0_appendix}, respectively.
\begin{figure}[!h]
    \centering 
    \begin{subfigure}[t]{0.45\textwidth}
        \centering        \includegraphics[height=1.2in]{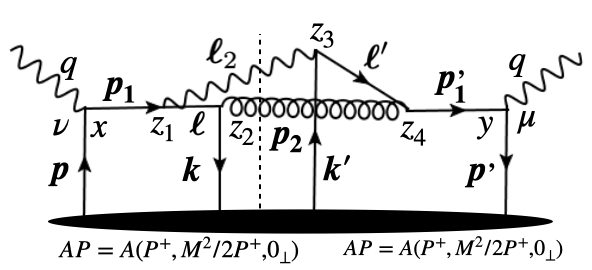}
        \caption{Interference diagram. }
    \end{subfigure}%
    \begin{subfigure}[t]{0.45\textwidth}
        \centering        \includegraphics[height=1.2in]{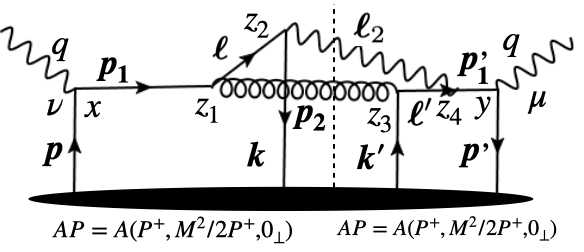}
        \caption{Complex conjugate of the diagram on the left panel.}
    \end{subfigure}
\caption{A forward scattering diagram contributing to kernel-2.}\label{fig:Kernel2_ph_qgqm_phqqm_gqq_both}
\end{figure}

The diagram presented in Fig.~\ref{fig:Kernel2_ph_qgqm_phqqm_gqq_both}(a) is identical to the diagram shown in Fig~\ref{fig:Kernel2_g_qphqm_gqqm_gqq_both}(b); therefore, the hadronic tensors are identical. Similarly, The diagram presented in Fig.~\ref{fig:Kernel2_ph_qgqm_phqqm_gqq_both}(b) is identical to the diagram shown in Fig~\ref{fig:Kernel2_g_qphqm_gqqm_gqq_both}(a); thus, the corresponding hadronic tensor is the same. We do not include the contributions to kernel-2 from  Fig.~\ref{fig:Kernel2_ph_qgqm_phqqm_gqq_both}(a) and Fig.~\ref{fig:Kernel2_ph_qgqm_phqqm_gqq_both}(b), because doing so leads to double-counting. 

\section{$\mbox{}$\!\!\!\!\!\!: Single-emission single-scattering kernel: Virtual photon corrections with a quark and anti-quark in final state}
\label{append:kernel3}

\begin{figure}[!h]
    \centering
    \includegraphics[width=0.45\textwidth]{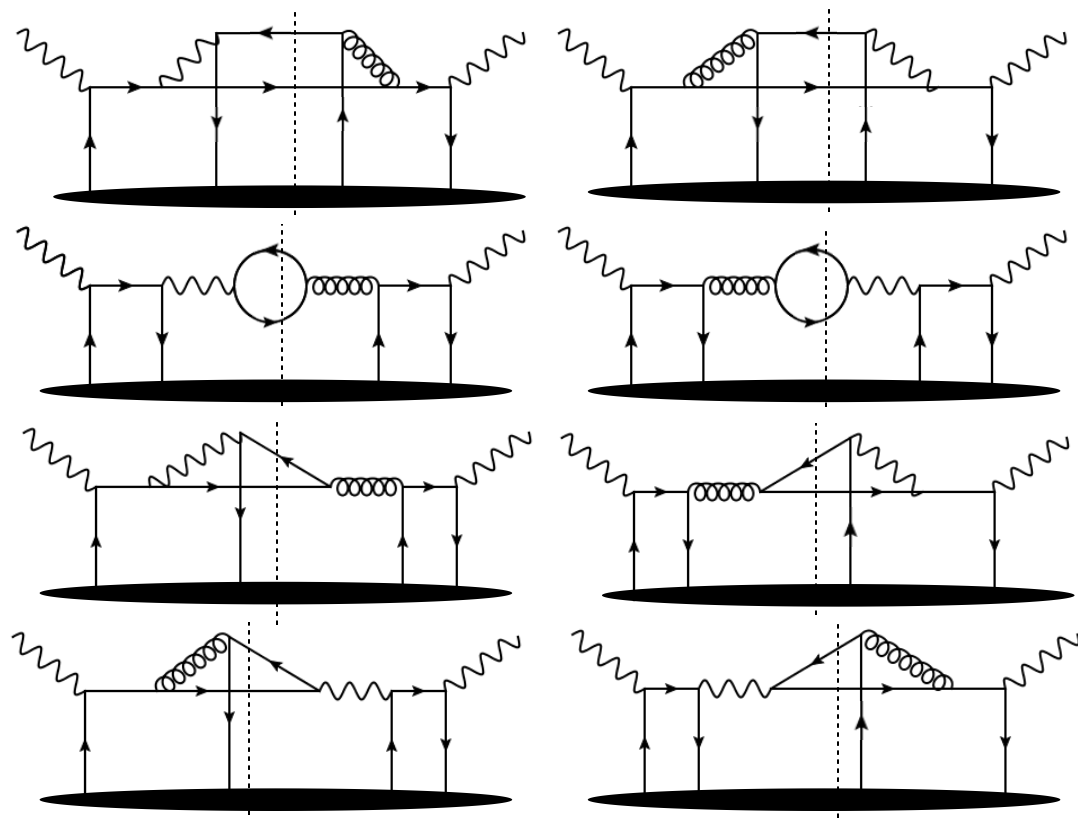}
    \caption{Diagrams for single emission and single scattering kernel (kernel-3) giving quark and antiquark final states through a virtual photon.}
    \label{fig:SESS_diagram_virtual_photon_corrections}
\end{figure}

This section summarizes the calculation of all possible diagrams at $\mathcal{O}(\alpha_s\alpha_{\rm em})$ involving a virtual photon and producing a quark-antiquark final state. The diagrams are presented in Fig.~\ref{fig:SESS_diagram_virtual_photon_corrections}, all of which contribute to kernel-3. Graphically, there are eight central-cut diagrams, each containing a photon propagator and a gluon propagator. Although these diagrams lead to correction $\mathcal{O}(\alpha_s\alpha_{em})$, they don't contribute to real photon emission scattering rates and therefore will only be presented for completeness.

In Fig.~\ref{fig:SESS_diagram_virtual_photon_corrections}, the diagrams in the first and second row have vanishing contributions to the hadronic tensor due to the appearance of the trace over a single Gell-Mann matrix. The matrices in SU(3) group are traceless.

\begin{figure}[!h]
    \centering 
    \begin{subfigure}[t]{0.45\textwidth}
        \centering        \includegraphics[height=1.2in]{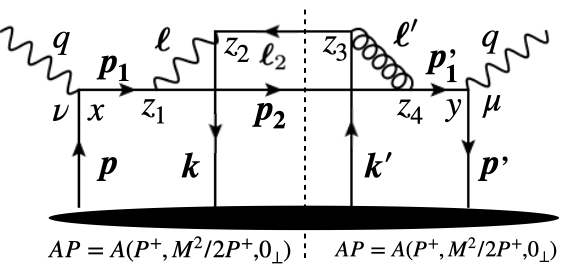}
        \caption{Interference diagram. }
    \end{subfigure}%
    \begin{subfigure}[t]{0.45\textwidth}
        \centering        \includegraphics[height=1.2in]{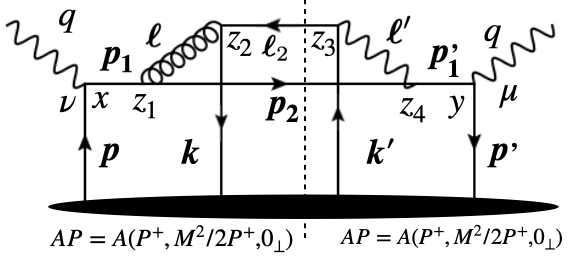}
        \caption{Complex conjugate of the diagram on the left panel.}
    \end{subfigure}
\caption{A forward scattering diagram contributing to kernel-3.}\label{fig:kernel-3_ph_phqqm_qqmg_g_both}
\end{figure}

The hadronic tensor for Fig.~\ref{fig:kernel-3_ph_phqqm_qqmg_g_both}(a) has the following form:
\begin{equation}
\begin{split}
W^{\mu\nu}_{3,c} & =\sum_f \sum_{f'} e^2 e^3_f e_{f'} g^2_{s} \int d^4 x d^4 y d^4 z_{2} d^4 z_{3} \int \frac{d^4 p}{(2\pi)^4} \frac{d^4 p'}{(2\pi)^4}  \frac{d^4 \ell_{2}}{(2\pi)^4} \frac{d^4 p_{2}}{(2\pi)^4} e^{-ip'y} e^{ipx}\left\langle P \left| \bar{\psi}_{_f}(y) \frac{\gamma^{+}}{4} \psi_{_f}(x)\right| P\right\rangle \delta^{cd} {\rm Tr}[t^{c}]{\rm Tr}[t^{d}] \\
  & \times e^{i\left(q+p'-p_{2}-\ell_{2}\right) z_{3}} e^{i(\ell_{2}+p_{2} - q - p)z_{2}} \left\langle P_{A-1} \left| \bar{\psi}_{_{f'}}(z_2)\frac{\gamma^+}{4}\psi_{_{f'}}(z_3)\right| P_{A-1} \right\rangle d^{(q+p-p_2)}_{\sigma_{1} \sigma_{2}} d^{(q+p'-p_2)}_{\sigma_{3} \sigma_{4}} (2 \pi) \delta\left(\ell_{2}^{2}\right) (2 \pi)  \delta\left(p_{2}^{2}-M^{2}\right) \\
& \times\frac{ {\rm Tr} \left[\gamma^- \gamma^{\mu} \left(\slashed{q}+\slashed{p}'+M\right) \gamma^{\sigma_4}\left(\slashed{p}_2+M\right)\gamma^{\sigma_{1}} \left( \slashed{q} + \slashed{p} + M\right) \gamma^{\nu}  \right] {\rm Tr} \left[ \gamma^- \gamma^{\sigma_2} \slashed{\ell}_2 \gamma^{\sigma_3} \right] 
}{  \left[\left(q+p'\right)^{2}-M^2-i\epsilon\right] \left[\left(q+p\right)^2 - M^2 + i\epsilon\right]  \left[\left(q+p'-p_2\right)^2 - i\epsilon\right] \left[\left(q+p -p_2\right)^2+i\epsilon\right]  }.
\end{split} \label{eq:kernel-3_ph_phqqm_qqmg_g_both_wi_a}
\end{equation}
The above expression of the hadonic tensor vanishes due to the color traces ${\rm Tr}[t^c]$ and ${\rm Tr}[t^d]$ being zero.

Note that the hadronic tensor for the central-cut diagram in Fig~\ref{fig:kernel-3_ph_phqqm_qqmg_g_both}(b) is identical to the diagram in Fig~\ref{fig:kernel-3_ph_phqqm_qqmg_g_both}(a) and yields a vanishing contribution due to the color trace being zero.

\begin{figure}[!h]
    \centering 
    \begin{subfigure}[t]{0.45\textwidth}
        \centering        \includegraphics[height=1.2in]{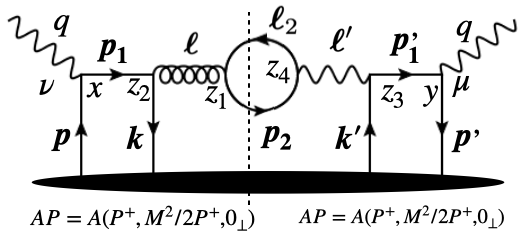}
        \caption{Interference diagram.  }
    \end{subfigure}%
    \begin{subfigure}[t]{0.45\textwidth}
        \centering        \includegraphics[height=1.2in]{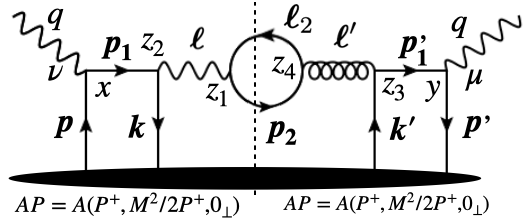}
        \caption{Complex conjugate of the diagram on the left panel. }
    \end{subfigure}
\caption{A forward scattering diagram contributing to kernel-3.}\label{fig:kernel3_qqmg_gqq_qqg_phqq}
\end{figure}
Now, we consider a central-cut diagram shown in Fig.~\ref{fig:kernel3_qqmg_gqq_qqg_phqq}(a). The hadronic tensor has the following form:
\begin{equation}
\begin{split}
W^{\mu\nu}_{3,c} & = \sum_f \sum_{f'} e^2 e^3_f e_{f'} g^2_{s} \int d^4 x d^4 y d^4 z_{2} d^4 z_{3} \int \frac{d^4 p}{(2\pi)^4} \frac{d^4 p'}{(2\pi)^4}  \frac{d^4 \ell_{2}}{(2\pi)^4} \frac{d^4 p_{2}}{(2\pi)^4} e^{-ip'y} e^{ipx}\left\langle P \left| \bar{\psi}_{_f}(y) \frac{\gamma^{+}}{4} \psi_{_f}(x)\right| P\right\rangle \delta^{ab} {\rm Tr}[t^a] {\rm Tr}[t^b] \\
  & \times e^{i\left(q+p'-p_{2}-\ell_{2}\right) z_{3}} e^{i(\ell_{2}+p_{2} - q - p)z_{2}} \left \langle P_{A-1} \left| \bar{\psi}_{_f}(z_2) \frac{\gamma^+}{4}\psi_{_f}(z_3)\right| P_{A-1} \right\rangle d^{(\ell_2+p_2)}_{\sigma_{1} \sigma_{2}} d^{(\ell_2+p_2)}_{\sigma_{3} \sigma_{4}} (2 \pi) \delta\left(\ell_{2}^{2}-M^2\right) (2 \pi)  \delta\left(p_{2}^{2}-M^{2}\right) \\
& \times \frac{  {\rm Tr} \left[\gamma^- \gamma^{\mu} \left(\slashed{q}+\slashed{p}'\right) \gamma^{\sigma_3}\gamma^-\gamma^{\sigma_2}\left( \slashed{q} + \slashed{p}\right) \gamma^{\nu} \right] {\rm Tr} \left[ \left(\slashed{\ell}_2+M\right) \gamma^{\sigma_{4}}\left(\slashed{p}_2+M\right) \gamma^{\sigma_1} \right]
}{  \left[\left(q+p'\right)^{2}-i\epsilon\right] \left[\left(\ell_2+p_2\right)^2  - i\epsilon\right]   \left[\left(\ell_2+p_2\right)^2 +i\epsilon\right] \left[\left(q+p\right)^2+i\epsilon\right]  }.
\end{split} \label{eq:kernel3_qqmg_gqq_qqg_phqq_wi-central}
\end{equation}
The expression in Eq.~\ref{eq:kernel3_qqmg_gqq_qqg_phqq_wi-central} vanishes since the color traces ${\rm Tr}[t^a]={\rm Tr} [t^b]=0$. The diagram shown in Fig.~\ref{fig:kernel3_qqmg_gqq_qqg_phqq}(b) is identical to  Fig.~\ref{fig:kernel3_qqmg_gqq_qqg_phqq}(a), and also gives zero contribution since the color traces vanish.

\begin{figure}[!h]
    \centering 
    \begin{subfigure}[t]{0.45\textwidth}
        \centering        \includegraphics[height=1.2in]{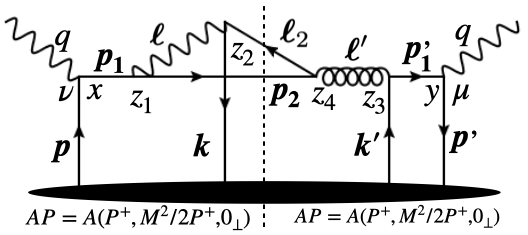}
        \caption{Interference diagram.  }
    \end{subfigure}%
    \begin{subfigure}[t]{0.45\textwidth}
        \centering        \includegraphics[height=1.2in]{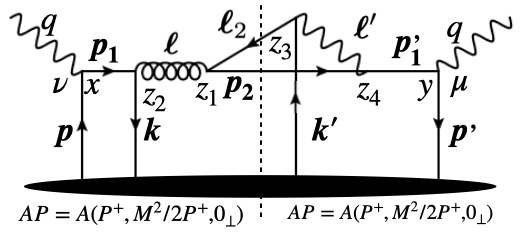}
        \caption{Complex conjugate of the diagram on the left panel. }
    \end{subfigure}
\caption{A forward scattering diagram contributing to kernel-3.}\label{fig:kernel3_ph_phqqm_qqg_gqqm}
\end{figure}

Next, we consider an interference diagram shown in Fig~\ref{fig:kernel3_ph_phqqm_qqg_gqqm}(a).

The hadronic tensor is given as
\begin{equation}
    \begin{split}
         W^{\mu \nu}_{3,c} & =\sum_f e^2 e^4_f g^{2}_s  \int d^{4} x  d^{4} y  
 \int  d^{4} z_{2}  d^{4} z_{3}    \int \frac{d^{4} p}{(2 \pi)^{4}}  \frac{d^{4} p'}{(2 \pi)^{4}}    \frac{d^{4} \ell_{2}}{(2 \pi)^{4}}  \frac{d^{4} p_{2}}{(2 \pi)^{4}} 
e^{-ip' y}  e^{ip x}  \langle P| \bar{\psi}_{_f}(y) \frac{\gamma^+}{4} \psi_{_f}(x)   |P  \rangle\\
&  \times   e^{iz_{2}(\ell_{2} +p_{2} - q -p )}  e^{iz_{3}( -p_{2} -\ell_{2}  +q + p' )}  \left \langle P_{A-1} \middle| \bar{\psi}_{_f}\left( z_{2} \right) 
\frac{\gamma^{+}}{4} 
\psi_{_f} \left( z_{3} \right) 
   \middle| P_{A-1} \right \rangle \\
& \times  \frac{\mathrm{Tr} \left[  \gamma^{-} \gamma^{\mu} (\slashed{q} + \slashed{p}' )   \gamma^{\sigma_{3}} 
\gamma^{-}
\gamma^{\sigma_{2}}  \slashed{\ell}_{2}  
\gamma^{\sigma_{4}} \slashed{p}_{2}  
\gamma^{\sigma_{1}}  (\slashed{q} + \slashed{p} )  \gamma^{\nu} \right]}{ \left[\left(q+p'\right)^{2}-i\epsilon\right] \left[\left(q+p\right)^{2}+ i\epsilon\right] } \\
&    \times \frac{d^{(\ell_2+p_2)} _{\sigma_{4}\sigma_{3}}}{\left[\left(\ell_{2}+p_{2}\right)^{2}-i\epsilon\right]}\frac{d^{(q+p-p_2)}_{\sigma_{1} \sigma_{2}}}{\left[\left(q+ p - p_{2}\right)^{2}+i \epsilon\right]}{\rm Tr}\left[t^c t^d\right] \delta^{cd} (2\pi) \delta(\ell^{ 2}_{2}) (2\pi) \delta(p^{ 2}_{2} ).
    \end{split}
\end{equation}
The above expression has singularity when the denominator of the parton propagator for $p_{1}$, $\ell$, and $p'_{1}$ becomes zero. We identify two poles for the momentum variable $p$ and one pole for $p'$. We compute the integral in the complex plane of $p^{+}$ and $p'^{+}$.

In this central-cut diagram, the momenta for the final state partons are $\ell^{-}_{2}=yq^{-}$ and $p^{-}_{2}=(1-y + \eta y)q^{-}$.
The contour integration for $p^{+}$ is given as
\begin{equation}
    \begin{split}
    C_{1} & = \oint \frac{dp^{+}}{2\pi} \frac{e^{ip^{+}( x^{-} - z^{-}_{2})}}{\left[\left(q+ p\right)^2  +i\epsilon\right]\left[\left(q+ p - p_{2}\right)^2 + i\epsilon\right] } \\
    & = \left( \frac{2\pi i}{2\pi}\right) \frac{\theta(x^{-} - z^{-}_{2} )}{4q^{-} (1-\eta)yq^{-}} e^{-i q^{+} ( x^{-} - z^{-}_{2})}  \left[ \frac{-1 + e^{i\mathcal{G}(x^{-} - z^{-}_{2})}}{\mathcal{G}^{(p_2)}_{0}}  \right],
    \end{split}
\end{equation}
where  $\mathcal{G}^{(p_2)}_{0} $ is given in Eq.~\ref{eq:GP2M0_appendix}. 

Similarly, the contour integration for $p'^{+}$ is 
\begin{equation}
    \begin{split}
    C_{2} & = \oint \frac{dp'^{+}}{2\pi} \frac{e^{ip'^{+}(-y^{-} + z^{-}_{3})}}{\left[\left(q+p'\right)^{2} - i\epsilon\right]} 
     = \left( \frac{-2\pi i}{2\pi}\right) \frac{\theta( y^{-} -z^{-}_{3} )}{2q^{-}}  e^{-iq^{+}(-y^{-} + z^{-}_{3})}.  
    \end{split}
\end{equation}
Simplifying the trace yields the following expression:
\begin{equation}
    \begin{split}
       & \mathrm{Tr} \left[  \gamma^{-} \gamma^{\mu}  (\slashed{q} + \slashed{p}' )   \gamma^{\sigma_{3}} 
\gamma^{-}
\gamma^{\sigma_{2}}  \slashed{\ell}_{2}  
\gamma^{\sigma_{4}} \slashed{p}_{2}  
\gamma^{\sigma_{1}}  (\slashed{q} + \slashed{p} )  \gamma^{\nu} \right] 
         \times  \left[ -g _{\sigma_{4} \sigma_{3}} + \frac{n_{\sigma_{4}} \ell'_{\sigma_{3}} + n_{\sigma_{3}} \ell'_{\sigma_{4}}}{ \ell'^{-}  } \right] \\
         & \times \left[ -g _{\sigma_{1} \sigma_{2}} + \frac{n_{\sigma_{1}} \ell_{\sigma_{2}} + n_{\sigma_{2}} \ell_{\sigma_{1}}}{ \ell^{-}  } \right] \\
       & = 32 (q^{-})^{2} [-g^{\mu\nu}_{\perp\perp}] \left[ \frac{1-y}{y}\right] \left[ \frac{J_2}{y(1-\eta)(1+\eta y)}  \right],
    \end{split}
\end{equation}
where
\begin{equation}
    \begin{split}
    J_2  = \pmb{\ell}^2_{2\perp}\{ -1 + y - \eta y (1 - y +\eta y)\} +y\pmb{k}^2_\perp \{-1 +y-\eta y\} + \pmb{k}_\perp \cdot \pmb{\ell}_{2\perp}\{1-y^2 + 2\eta y + \eta^2 y^2 \}  .
     \end{split}
     \label{eq:kernel3-J3_diagram5-6}
\end{equation}
The final expression for the hadronic tensor [Fig~\ref{fig:kernel3_ph_phqqm_qqg_gqqm}(a)] is
\begin{equation}
       \begin{split}
    W^{\mu\nu}_{3,c} & =\sum_f 2 \left[ -g^{\mu\nu}_{\perp \perp} \right] \int d (\Delta X^{-})   e^{iq^{+}(\Delta X^{-} )}    
  \left\langle P \left| \bar{\psi}_{_f}( \Delta X^{-}) \frac{\gamma^{+}}{4} \psi_{_f}(0)\right| P\right\rangle  \\
  & \times e^2 e^4_{f} g^2_s \int  d \zeta^{-} d (\Delta z^{-})d^2 \Delta z_{\perp} \frac{dy}{2\pi} \frac{d^2 \ell_{2\perp}}{(2\pi)^2} \frac{d^2 k_{\perp}}{(2\pi)^2} \left[ -1 + e^{i\mathcal{G}^{(p_2)}_{0}(x^{-}-z^-_2)} \right] e^{-i \mathcal{H}^{(\ell_2,p_2)}_{0} (\Delta z^{-})}     e^{i \pmb{k}_{\perp}\cdot \Delta\pmb{z}_{\perp}}   \\
  & \times \frac{\theta( x^{-} -z^{-}_2) \theta( y^{-} -z^{-}_{3})}{ yq^{-} }  \frac{J_2}{ \left[ \pmb{\ell}_{2\perp} - \pmb{k}_\perp\right]^2 \left[ \left(1+\eta y\right)\pmb{\ell}_{2\perp} - y\pmb{k}_{\perp} \right]^2 } \left\langle P_{A-1} \left| \bar{\psi}_{_f}(\zeta^{-},0) \frac{\gamma^+}{4}  \psi_{_f}(\zeta^{-},\Delta z^{-}, \Delta \pmb{z}_{\perp}) \right| P_{A-1} \right\rangle  \\
  &
\times  \left[ \frac{1-y+\eta y}{(1+\eta y)(1-\eta)}\right] \left[ C_{F}N_{c} \right],
    \end{split}\label{eq:kernel3_ph_phqqm_qqg_gqqm_wf_final_a} 
\end{equation}
where $\mathcal{G}^{(p_2)}_0$ is given in Eq.~\ref{eq:GP2M0_appendix} and $\mathcal{H}^{(\ell_2,p_2)}_{0}$ is given in Eq.~\ref{eq:HL2P2M0_appendix}.

Note that the diagram in Fig~\ref{fig:kernel3_ph_phqqm_qqg_gqqm}(b) and Fig~\ref{fig:kernel3_ph_phqqm_qqg_gqqm}(a) are complex-conjugates of each other. They differ only in contour integration over variables $p^{+}$ and $p'^{+}$.
 The calculation of the hadronic tensor for Fig~\ref{fig:kernel3_ph_phqqm_qqg_gqqm}(b) involves the contour integration for $p^{+}$ and is given as
\begin{equation}
    \begin{split}
    C_{1} & = \oint \frac{dp^{+}}{2\pi} \frac{e^{ip^{+}( x^{-} - z^{-}_{2})}}{\left[\left(q+p\right)^{2}+i\epsilon\right]} 
     = \left( \frac{2\pi i}{2\pi}\right) \frac{\theta(x^{-} - z^{-}_{2} )}{2q^{-}}  e^{-i q^{+} ( x^{-} - z^{-}_{2}) }  ,
    \end{split}
\end{equation}

and the contour integration for $p'^{+}$ is given as
\begin{equation}
    \begin{split}
    C_{2} & = \oint \frac{dp'^{+}}{2\pi} \frac{e^{ip'^{+}(-y^{-} + z^{-}_{3})}}{\left[\left(q+ p'\right)^2  -i\epsilon\right]\left[\left(q+ p' - p_{2}\right)^2-i\epsilon\right] } \\
    & = \left( \frac{-2\pi i}{2\pi}\right) \frac{\theta( y^{-} -z^{-}_{3} )}{4q^{-} (q^{-}-p^{-}_2)} e^{iq^{+}(y^{-} - z^{-}_{3})}  \left[ \frac{-1 + e^{-i\mathcal{G}^{(p_2)}_0(y^{-} - z^{-}_{3})}}{\mathcal{G}^{(p_2)}_0}  \right],
    \end{split}
\end{equation}
where $\mathcal{G}^{(p_2)}_0$ is given in Eq.~\ref{eq:GP2M0_appendix}.
%
%
The final expression for the hadronic tensor [Fig~\ref{fig:kernel3_ph_phqqm_qqg_gqqm}(b)] yields
\begin{equation}
       \begin{split}
    W^{\mu\nu}_{3,c} & = \sum_f 2 \left[ -g^{\mu\nu}_{\perp \perp} \right] \int d (\Delta X^{-})   e^{iq^{+}(\Delta X^{-} )}    
  \left\langle P \left| \bar{\psi}_{_f}( \Delta X^{-}) \frac{\gamma^{+}}{4} \psi_{_f}(0)\right| P\right\rangle  \\
  & \times e^2 e^4_{f} g^2_s \int  d \zeta^{-} d (\Delta z^{-})d^2 \Delta z_{\perp} \frac{dy}{2\pi} \frac{d^2 \ell_{2\perp}}{(2\pi)^2} \frac{d^2 k_{\perp}}{(2\pi)^2} \left[ -1 + e^{-i\mathcal{G}^{(p_2)}_{0}(y^{-}-z^-_3)} \right] e^{-i \mathcal{H}^{(\ell_2,p_2)}_{0} (\Delta z^{-})}     e^{i \pmb{k}_{\perp}\cdot \Delta\pmb{z}_{\perp}}   \\
  & \times \frac{\theta( x^{-} -z^{-}_2) \theta( y^{-} -z^{-}_{3})}{ yq^{-} }  \frac{J_2}{ \left[ \pmb{\ell}_{2\perp} - \pmb{k}_\perp\right]^2 \left[ \left(1+\eta y\right)\pmb{\ell}_{2\perp} - y\pmb{k}_{\perp} \right]^2 } \left\langle P_{A-1} \left| \bar{\psi}_{_f}(\zeta^{-},0) \frac{\gamma^+}{4}  \psi_{_f}(\zeta^{-},\Delta z^{-}, \Delta \pmb{z}_{\perp}) \right| P_{A-1} \right\rangle  \\
  &
\times  \left[ \frac{1-y+\eta y}{(1+\eta y)(1-\eta)}\right] \left[ C_{F}N_{c} \right],
    \end{split} \label{eq:kernel3_ph_phqqm_qqg_gqqm_wf_final_b}
\end{equation}
where $\mathcal{G}^{(p_2)}_0$ is given in Eq.~\ref{eq:GP2M0_appendix} and $\mathcal{H}^{(\ell_2,p_2)}_{0}$ is given in Eq.~\ref{eq:HL2P2M0_appendix}.

\begin{figure}[!h]
    \centering 
    \begin{subfigure}[t]{0.45\textwidth}
        \centering        \includegraphics[height=1.2in]{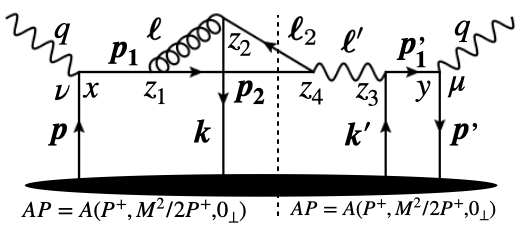}
        \caption{Interference diagram.  }
    \end{subfigure}%
    \begin{subfigure}[t]{0.45\textwidth}
        \centering        \includegraphics[height=1.2in]{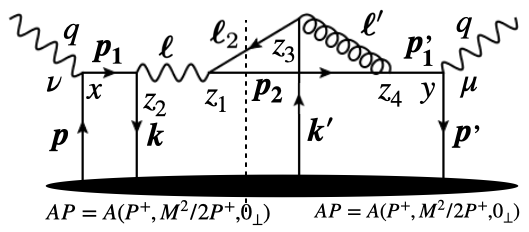}
        \caption{Complex conjugate of the diagram on the left panel. }
    \end{subfigure}
\caption{A forward scattering diagram contributing to kernel-3.} \label{fig:kernel3_g_gqqm_qqph_phqqm}
\end{figure}

Next, we consider the interference diagram with virtual photon as shown in Fig.~\ref{fig:kernel3_g_gqqm_qqph_phqqm}(a). 
The diagram is identical to Fig.~\ref{fig:kernel3_ph_phqqm_qqg_gqqm}(a), except that the photon propagator and the gluon propagator are interchanged between the amplitude side and the complex conjugate side, therefore, the hadronic tensor is the same as that given in Eq.~\ref{eq:kernel3_ph_phqqm_qqg_gqqm_wf_final_a}.

Similarly, the interference diagram shown in 
Fig.~\ref{fig:kernel3_g_gqqm_qqph_phqqm}(b) is identical to Fig.~\ref{fig:kernel3_ph_phqqm_qqg_gqqm}(b), except that the photon propagator and the gluon propagator are interchanged between the amplitude side and the complex conjugate side; therefore, the hadronic tensor is the same as given in Eq.~\ref{eq:kernel3_ph_phqqm_qqg_gqqm_wf_final_b}.

\section{$\mbox{}$\!\!\!\!\!\!: Single-emission single-scattering kernel:  Virtual  photon corrections with two quarks in the final state}
\label{append:kernel4}
The possible diagrams at $\mathcal{O}(\alpha_s\alpha_{\rm em})$ involving a virtual photon with two quarks in the final state are given in Fig.~\ref{fig:SESS_diagram_virtual_photon_qq_final_state} below. There are a total of four possible diagrams leading to photon correction at $\mathcal{O}(\alpha_{em}\alpha_s)$.
The in-medium quark exchanged with the medium is considered to be a Glauber quark. Each forward scattering diagram consists of a photon propagator and a gluon propagator leading to correction $\mathcal{O}(\alpha_{em}\alpha_s)$.
\begin{figure}[!h]
    \centering    \includegraphics[width=0.45\textwidth]{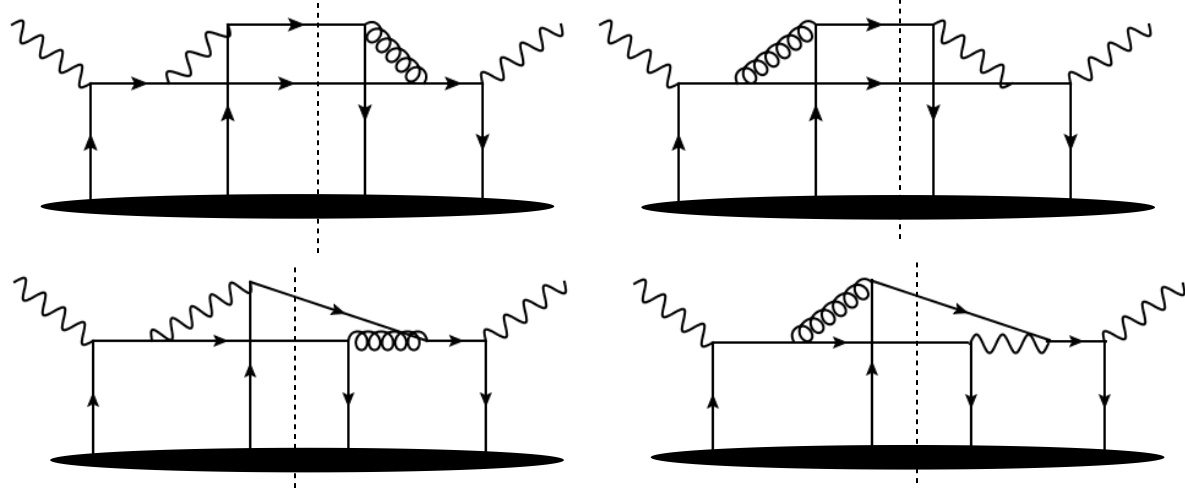}
    \caption{All diagrams with quark-quark final states contributing to kernel-4. }    
    \label{fig:SESS_diagram_virtual_photon_qq_final_state}
\end{figure}

\begin{figure}[h!]
    \begin{subfigure}[t]{0.45\textwidth}
        \centering        \includegraphics[height=1.2in]{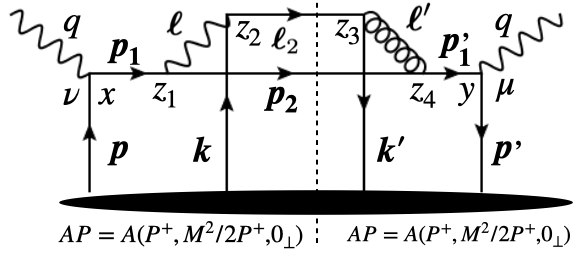}
        \caption{Interference diagram.  }
    \end{subfigure}%
    \begin{subfigure}[t]{0.45\textwidth}
        \centering        \includegraphics[height=1.2in]{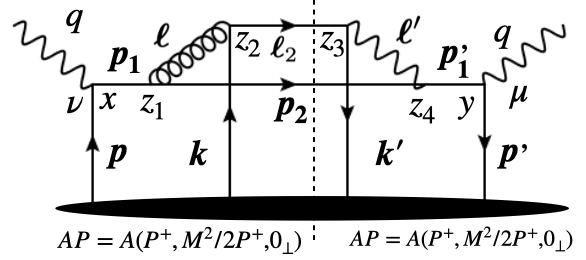}
        \caption{Complex conjugate of the diagram on the left panel. }
    \end{subfigure}
\caption{A forward scattering diagram contributing to kernel-4.} \label{fig:kernel4_ph_phqqm_qgqm_g_both}
\end{figure}

The hadronic tensor associated with the forward scattering diagram [Fig.~\ref{fig:kernel4_ph_phqqm_qgqm_g_both}(a)] is given as 
\begin{equation}
\begin{split}
W^{\mu\nu}_{4,c} & = \sum_f\sum_{f'} e^2 e^3_f e_{f'} g^2_{s} \int d^4 x d^4 y d^4 z_{2} d^4 z_{3} \int \frac{d^4 p}{(2\pi)^4} \frac{d^4 p'}{(2\pi)^4}  \frac{d^4 \ell_{2}}{(2\pi)^4} \frac{d^4 p_{2}}{(2\pi)^4} e^{-ip'y} e^{ipx}\left\langle P \left| \bar{\psi}_{_f}(y) \frac{\gamma^{+}}{4} \psi_{_f}(x)\right| P\right\rangle \delta^{cd} {\rm Tr}[t^{c}]{\rm Tr}[t^{d}] \\
  & \times e^{i\left(q+p'-p_{2}-\ell_{2}\right) z_{3}} e^{i(\ell_{2}+p_{2} - q - p)z_{2}} \left\langle P_{A-1} \left| \bar{\psi}_{f'}(z_3)\frac{\gamma^+}{4}\psi_{f'}(z_2) \right| P_{A-1} \right\rangle d^{(q+p-p_2)}_{\sigma_{1} \sigma_{2}} d^{(q+p'-p_2)}_{\sigma_{3} \sigma_{4}} (2 \pi) \delta\left(\ell_{2}^{2}\right) (2 \pi)  \delta\left(p_{2}^{2}-M^{2}\right) \\
& \times\frac{ {\rm Tr} \left[\gamma^- \gamma^{\mu} \left(\slashed{q}+\slashed{p}'+M\right) \gamma^{\sigma_4}\left(\slashed{p}_2+M\right)\gamma^{\sigma_{1}} \left( \slashed{q} + \slashed{p} + M\right) \gamma^{\nu}  \right] {\rm Tr} \left[ \gamma^- \gamma^{\sigma_3} \slashed{\ell}_2 \gamma^{\sigma_2} \right] 
}{  \left[\left(q+p'\right)^{2}-M^2-i\epsilon\right] \left[\left(q+p\right)^2 - M^2 + i\epsilon\right]  \left[\left(q+p'-p_2\right)^2 - i\epsilon\right] \left[\left(q+p -p_2\right)^2+i\epsilon\right]  }.
\end{split} \label{eq:kernel-4_ph_phqqm_qqmg_g_both_wi_a}
\end{equation}

The above expression of the hadronic tensor vanishes since the color traces ${\rm Tr}[t^c]$ and ${\rm Tr}[t^d]$ become zero. Note that the diagram shown in Fig.~\ref{fig:kernel4_ph_phqqm_qgqm_g_both}(a) is identical to Fig.~\ref{fig:kernel4_ph_phqqm_qgqm_g_both}(b) and has vanishing contribution since the color traces yield zero.  

\begin{figure}[!h]
    \centering 
    \begin{subfigure}[t]{0.45\textwidth}
        \centering        \includegraphics[height=1.2in]{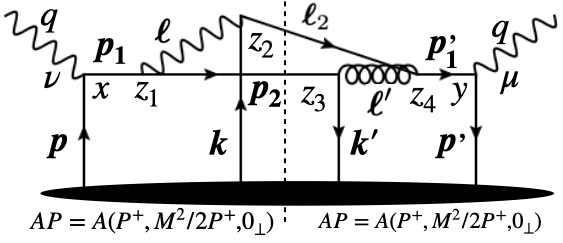}
        \caption{Interference diagram.  }
    \end{subfigure}%
    \begin{subfigure}[t]{0.45\textwidth}
        \centering        \includegraphics[height=1.2in]{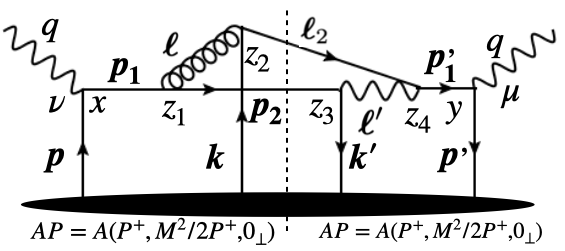}
        \caption{Complex conjugate of the diagram on the left panel. }
    \end{subfigure}
\caption{A forward scattering diagram contributing to kernel-4.} \label{kernel4_ph_phqqm_qgqm_qgq_both}
\end{figure}

Next, we consider a forward scattering diagram shown in Fig.~\ref{kernel4_ph_phqqm_qgqm_qgq_both}.

The hadronic tensor for Fig.~\ref{kernel4_ph_phqqm_qgqm_qgq_both}(a) is given as
\begin{equation}
\begin{split}
W^{\mu\nu}_{4,c} & =\sum_f e^2 e^4_f g^2_{s} \int d^4 x d^4 y d^4 z_{2} d^4 z_{3} \int \frac{d^4 p}{(2\pi)^4} \frac{d^4 p'}{(2\pi)^4}  \frac{d^4 \ell_{2}}{(2\pi)^4} \frac{d^4 p_{2}}{(2\pi)^4} e^{-ip'y} e^{ipx}\left\langle P \left| \bar{\psi}_{_f}(y) \frac{\gamma^{+}}{4} \psi_{_f}(x)\right| P\right\rangle \delta^{cd} {\rm Tr}[t^{c}t^{d}] \\
  & \times e^{i\left(q+p'-p_{2}-\ell_{2}\right) z_{3}} e^{i(\ell_{2}+p_{2} - q - p)z_{2}} \left\langle P_{A-1} \left| \bar{\psi}_{_f}(z_3)\frac{\gamma^+}{4}\psi_{_f}(z_2)\right| P_{A-1} \right\rangle d^{(q+p-p_2)}_{\sigma_{1} \sigma_{2}} d^{(q+p'-\ell_2)}_{\sigma_{3} \sigma_{4}} (2 \pi) \delta\left(\ell_{2}^{2}\right) (2 \pi)  \delta\left(p_{2}^{2}\right) \\
& \times\frac{ {\rm Tr} \left[\gamma^- \gamma^{\mu} \left(\slashed{q}+\slashed{p}'\right) \gamma^{\sigma_4}\slashed{\ell}_2\gamma^{\sigma_{2}} \gamma^-\gamma^{\sigma_{3}}\slashed{p}_2\gamma^{\sigma_{1}}(\slashed{q}+\slashed{p})\gamma^{\nu}  \right]  
}{  \left[\left(q+p'\right)^{2}-i\epsilon\right] \left[\left(q+p\right)^2  + i\epsilon\right]  \left[\left(q+p'-\ell_2\right)^2 - i\epsilon\right] \left[\left(q+p -p_2\right)^2+i\epsilon\right]  }.
\end{split} \label{eq:kernel4_ph_phqqm_qgqm_qgq_both_wi_a}
\end{equation}

The above expression of the hadronic tensor has singularity when the denominator of the propagator for $p_1$, $\ell$, $\ell'$, and $p'_1$ becomes on-shell. It contains two simple poles for $p^+$ and $p'^+$. The contour integration for $p^+$ gives
\begin{equation}
\begin{split}
C_{1} & = \oint \frac{dp^{+}}{(2\pi)} \frac{e^{ip^{+}(x^{-}-z^{-}_{2})}}{\left[\left(q+p\right)^{2}  + i\epsilon\right]\left[\left(q+p-p_{2}\right)^{2}+i\epsilon\right]} \\
      & = \oint \frac{dp^{+}}{(2\pi)} \frac{e^{ip^{+}(x^{-} - z^{-}_{2}) }}{2q^{-}[ q^{+} + p^{+}  + i \epsilon] 2(q^{-} -p^{-}_{2} )\left[ q^{+} + p^{+} -p^{+}_{2} - \frac{\pmb{p}^{2}_{2\perp}}{2(q^{-}-p^{-}_{2})} + i \epsilon\right]}   \\
      & = \frac{(2\pi i)}{2\pi} \frac{\theta(x^{-} - z^{-}_{2})}{4q^{-}(q^{-}-p^{-}_{2})}  e^{-iq^{+} (x^{-}-z^{-}_{2})}\left[ \frac{ -1 + e^{i \mathcal{G}^{(p_2)}_0(x^{-}-z^{-}_{2})} }{ \mathcal{G}^{(p_2)}_0}  \right],
\end{split}
\end{equation}
where $ \mathcal{G}^{(p_2)}_0$ is defined in Eq.~\ref{eq:GP2M0_appendix}.

Similarly, the contour integration for $p'^+$ gives
\begin{equation}
\begin{split}
C_{2} & = \oint \frac{dp'^{+}}{(2\pi)} \frac{e^{-ip'^{+}(y^{-}-z^{-}_{3})}}{\left[\left(q+p'\right)^{2} - i\epsilon\right]\left[\left(q+p'-\ell_{2}\right)^{2} - i\epsilon\right]} \\
      & = \oint \frac{dp'^{+}}{(2\pi)} \frac{e^{-ip'^{+}(y^{-} - z^{-}_{3}) }}{2q^{-}[ q^{+} + p'^{+} 
      - i \epsilon] 2(q^{-} -\ell^{-}_{2} )\left[ q^{+} + p'^{+} -\ell^{+}_{2} - \frac{\pmb{\ell}^{2}_{2\perp}}{2(q^{-}-\ell^{-}_{2})} - i \epsilon\right]}   \\
      & = \frac{(-2\pi i)}{2\pi} \frac{\theta(y^{-} - z^{-}_{3})}{4q^{-}(q^{-}-\ell^{-}_{2})}  e^{iq^{+}(y^{-}-z^{-}_{3})}\left[ \frac{ -1 + e^{-i \mathcal{G}^{(\ell_2)}_0(y^{-}-z^{-}_{3})} }{ \mathcal{G}^{(\ell_2)}_0}  \right],
\end{split}
\end{equation}
where $\mathcal{G}^{(\ell_2)}_0$ is defined in Eq.~\ref{eq:GL20_kernel2_xy}.

The trace in the third line of Eq.~\ref{eq:kernel4_ph_phqqm_qgqm_qgq_both_wi_a} simplifies as
\begin{equation}
    \begin{split}
         & {\rm Tr} \left[\gamma^- \gamma^{\mu} \left(\slashed{q}+\slashed{p}'\right) \gamma^{\sigma_4}\slashed{\ell}_2\gamma^{\sigma_{2}} \gamma^-\gamma^{\sigma_{3}}\slashed{p}_2\gamma^{\sigma_{1}}(\slashed{q}+\slashed{p})\gamma^{\nu}  \right] d^{(q+p-p_2)}_{\sigma_{2} \sigma_{1}} d^{(q+p'-\ell_2)}_{\sigma_{4} \sigma_{3}} \\
         & = \frac{32 (q^-)^2 \left[ -g^{\mu\nu}_{\perp\perp}\right]  }{(1-\eta )y(1-y)q^-} \left[ -\pmb{\ell}^{2}_{2\perp} +  \pmb{\ell}_{2\perp}\cdot \pmb{k}_{\perp} \right].
    \end{split}
\end{equation}
The final expression of the hadronic tensor for Fig.~\ref{kernel4_ph_phqqm_qgqm_qgq_both}(a) is given by
\begin{equation}
\begin{split}
    W^{\mu\nu}_{4,c} & =\sum_f  2[-g^{\mu\nu}_{\perp \perp }]\int d (\Delta X^{-})   e^{iq^{+}(\Delta X^{-} )}     
  \left\langle P \left| \bar{\psi}_{_f}( \Delta X^{-}) \frac{\gamma^{+}}{4} \psi_{_f}(0)\right| P\right\rangle  \\
  & \times e^2 e^4_f g^2_s \int  d \zeta^{-} d (\Delta z^{-}) d^2 \Delta z_{\perp} \frac{dy}{2\pi} \frac{d^2 \ell_{2\perp}}{(2\pi)^2} \frac{d^2 k_{\perp}}{(2\pi)^2} \left[ -1 + e^{i\mathcal{G}^{(p_2)}_{0}(x^{-} - z^{-}_{2})} \right] \left[ -1 + e^{-i\mathcal{G}^{(\ell_2)}_{0}(y^{-} - z^{-}_{3})} \right]   \\
  & \times \frac{\theta( x^{-}-z^{-}_2) \theta( y^{-}-z^{-}_3)  }{  (1-\eta )(1-y)yq^-  } \frac{\left[ -\pmb{\ell}^{2}_{2\perp} +  \pmb{\ell}_{2\perp}\cdot \pmb{k}_{\perp} \right]}{(\pmb{\ell}_{2\perp}-\pmb{k}_{\perp} )^2\pmb{\ell}^2_{2\perp} } e^{-i (\Delta z^{-})\mathcal{H}^{(\ell_2 , p_2)}_{0} }     e^{i \pmb{k}_{\perp} \cdot \Delta\pmb{z}_{\perp}} \left\langle P_{A-1} \left| \bar{\psi}_{_f}(\zeta^-, \Delta z^-, \Delta z_{\perp}) \frac{\gamma^+}{4} \psi_{_f}(\zeta^-, 0)\right| P_{A-1} \right\rangle  \\
  & \times  [C_F N_c] , 
\end{split}     
\end{equation}
where $\mathcal{G}^{(\ell_2)}_0$ is defined in Eq.~\ref{eq:GL20_kernel2_xy},  $ \mathcal{G}^{(p_2)}_0$ is given in Eq.~\ref{eq:GP2M0_appendix}, and $\mathcal{H}^{(\ell_2,p_2)}_{0}$ is given in Eq.~\ref{eq:HL2P2M0_appendix}.

Note that the diagram shown in Fig.~\ref{kernel4_ph_phqqm_qgqm_qgq_both}(a) is identical to Fig.~\ref{kernel4_ph_phqqm_qgqm_qgq_both}(b) except the photon propagator and gluon propagator are interchanged between the amplitude side and complex-conjugate side. Hence, the corresponding hadronic tensors are identical. 

\end{appendices}


\bibliography{references}

\begin{thebibliography}{51}%
\makeatletter
\providecommand \@ifxundefined [1]{%
 \@ifx{#1\undefined}
}%
\providecommand \@ifnum [1]{%
 \ifnum #1\expandafter \@firstoftwo
 \else \expandafter \@secondoftwo
 \fi
}%
\providecommand \@ifx [1]{%
 \ifx #1\expandafter \@firstoftwo
 \else \expandafter \@secondoftwo
 \fi
}%
\providecommand \natexlab [1]{#1}%
\providecommand \enquote  [1]{``#1''}%
\providecommand \bibnamefont  [1]{#1}%
\providecommand \bibfnamefont [1]{#1}%
\providecommand \citenamefont [1]{#1}%
\providecommand \href@noop [0]{\@secondoftwo}%
\providecommand \href [0]{\begingroup \@sanitize@url \@href}%
\providecommand \@href[1]{\@@startlink{#1}\@@href}%
\providecommand \@@href[1]{\endgroup#1\@@endlink}%
\providecommand \@sanitize@url [0]{\catcode `\\12\catcode `\$12\catcode
  `\&12\catcode `\#12\catcode `\^12\catcode `\_12\catcode `\%12\relax}%
\providecommand \@@startlink[1]{}%
\providecommand \@@endlink[0]{}%
\providecommand \url  [0]{\begingroup\@sanitize@url \@url }%
\providecommand \@url [1]{\endgroup\@href {#1}{\urlprefix }}%
\providecommand \urlprefix  [0]{URL }%
\providecommand \Eprint [0]{\href }%
\providecommand \doibase [0]{http://dx.doi.org/}%
\providecommand \selectlanguage [0]{\@gobble}%
\providecommand \bibinfo  [0]{\@secondoftwo}%
\providecommand \bibfield  [0]{\@secondoftwo}%
\providecommand \translation [1]{[#1]}%
\providecommand \BibitemOpen [0]{}%
\providecommand \bibitemStop [0]{}%
\providecommand \bibitemNoStop [0]{.\EOS\space}%
\providecommand \EOS [0]{\spacefactor3000\relax}%
\providecommand \BibitemShut  [1]{\csname bibitem#1\endcsname}%
\let\auto@bib@innerbib\@empty
\bibitem [{\citenamefont {Adcox}\ \emph {et~al.}(2002)\citenamefont {Adcox}
  \emph {et~al.}}]{PHENIX:2001hpc}%
  \BibitemOpen
  \bibfield  {author} {\bibinfo {author} {\bibfnamefont {K.}~\bibnamefont
  {Adcox}} \emph {et~al.} (\bibinfo {collaboration} {PHENIX}),\ }\bibfield
  {title} {\enquote {\bibinfo {title} {{Suppression of hadrons with large
  transverse momentum in central Au+Au collisions at $\sqrt{s_{NN}}$ =
  130-GeV}},}\ }\href {\doibase 10.1103/PhysRevLett.88.022301} {\bibfield
  {journal} {\bibinfo  {journal} {Phys. Rev. Lett.}\ }\textbf {\bibinfo
  {volume} {88}},\ \bibinfo {pages} {022301} (\bibinfo {year} {2002})},\
  \Eprint {http://arxiv.org/abs/nucl-ex/0109003} {arXiv:nucl-ex/0109003}
  \BibitemShut {NoStop}%
\bibitem [{\citenamefont {Adler}\ \emph {et~al.}(2004)\citenamefont {Adler}
  \emph {et~al.}}]{PHENIX:2003djd}%
  \BibitemOpen
  \bibfield  {author} {\bibinfo {author} {\bibfnamefont {S.~S.}\ \bibnamefont
  {Adler}} \emph {et~al.} (\bibinfo {collaboration} {PHENIX}),\ }\bibfield
  {title} {\enquote {\bibinfo {title} {{High $p_{T}$ charged hadron suppression
  in Au + Au collisions at $\sqrt{s}_{NN} = 200$ GeV}},}\ }\href {\doibase
  10.1103/PhysRevC.69.034910} {\bibfield  {journal} {\bibinfo  {journal} {Phys.
  Rev. C}\ }\textbf {\bibinfo {volume} {69}},\ \bibinfo {pages} {034910}
  (\bibinfo {year} {2004})},\ \Eprint {http://arxiv.org/abs/nucl-ex/0308006}
  {arXiv:nucl-ex/0308006} \BibitemShut {NoStop}%
\bibitem [{\citenamefont {Adler}\ \emph {et~al.}(2003)\citenamefont {Adler}
  \emph {et~al.}}]{PHENIX:2003qdj}%
  \BibitemOpen
  \bibfield  {author} {\bibinfo {author} {\bibfnamefont {S.~S.}\ \bibnamefont
  {Adler}} \emph {et~al.} (\bibinfo {collaboration} {PHENIX}),\ }\bibfield
  {title} {\enquote {\bibinfo {title} {{Suppressed $\pi^0$ production at large
  transverse momentum in central Au+ Au collisions at $\sqrt{S_{NN}}$ = 200
  GeV}},}\ }\href {\doibase 10.1103/PhysRevLett.91.072301} {\bibfield
  {journal} {\bibinfo  {journal} {Phys. Rev. Lett.}\ }\textbf {\bibinfo
  {volume} {91}},\ \bibinfo {pages} {072301} (\bibinfo {year} {2003})},\
  \Eprint {http://arxiv.org/abs/nucl-ex/0304022} {arXiv:nucl-ex/0304022}
  \BibitemShut {NoStop}%
\bibitem [{\citenamefont {Adler}\ \emph {et~al.}(2002)\citenamefont {Adler}
  \emph {et~al.}}]{STAR:2002ggv}%
  \BibitemOpen
  \bibfield  {author} {\bibinfo {author} {\bibfnamefont {C.}~\bibnamefont
  {Adler}} \emph {et~al.} (\bibinfo {collaboration} {STAR}),\ }\bibfield
  {title} {\enquote {\bibinfo {title} {{Centrality dependence of high $p_{T}$
  hadron suppression in Au+Au collisions at $\sqrt{s}_{NN}$ = 130-GeV}},}\
  }\href {\doibase 10.1103/PhysRevLett.89.202301} {\bibfield  {journal}
  {\bibinfo  {journal} {Phys. Rev. Lett.}\ }\textbf {\bibinfo {volume} {89}},\
  \bibinfo {pages} {202301} (\bibinfo {year} {2002})},\ \Eprint
  {http://arxiv.org/abs/nucl-ex/0206011} {arXiv:nucl-ex/0206011} \BibitemShut
  {NoStop}%
\bibitem [{\citenamefont {Adams}\ \emph {et~al.}(2003)\citenamefont {Adams}
  \emph {et~al.}}]{STAR:2003fka}%
  \BibitemOpen
  \bibfield  {author} {\bibinfo {author} {\bibfnamefont {J.}~\bibnamefont
  {Adams}} \emph {et~al.} (\bibinfo {collaboration} {STAR}),\ }\bibfield
  {title} {\enquote {\bibinfo {title} {{Transverse momentum and collision
  energy dependence of high p(T) hadron suppression in Au+Au collisions at
  ultrarelativistic energies}},}\ }\href {\doibase
  10.1103/PhysRevLett.91.172302} {\bibfield  {journal} {\bibinfo  {journal}
  {Phys. Rev. Lett.}\ }\textbf {\bibinfo {volume} {91}},\ \bibinfo {pages}
  {172302} (\bibinfo {year} {2003})},\ \Eprint
  {http://arxiv.org/abs/nucl-ex/0305015} {arXiv:nucl-ex/0305015} \BibitemShut
  {NoStop}%
\bibitem [{\citenamefont {Aaboud}\ \emph
  {et~al.}(2019{\natexlab{a}})\citenamefont {Aaboud} \emph
  {et~al.}}]{ATLAS:2018gwx}%
  \BibitemOpen
  \bibfield  {author} {\bibinfo {author} {\bibfnamefont {Morad}\ \bibnamefont
  {Aaboud}} \emph {et~al.} (\bibinfo {collaboration} {ATLAS}),\ }\bibfield
  {title} {\enquote {\bibinfo {title} {{Measurement of the nuclear modification
  factor for inclusive jets in Pb+Pb collisions at $\sqrt{s_\mathrm{NN}}=5.02$
  TeV with the ATLAS detector}},}\ }\href {\doibase
  10.1016/j.physletb.2018.10.076} {\bibfield  {journal} {\bibinfo  {journal}
  {Phys. Lett. B}\ }\textbf {\bibinfo {volume} {790}},\ \bibinfo {pages}
  {108--128} (\bibinfo {year} {2019}{\natexlab{a}})},\ \Eprint
  {http://arxiv.org/abs/1805.05635} {arXiv:1805.05635 [nucl-ex]} \BibitemShut
  {NoStop}%
\bibitem [{\citenamefont {Acharya}\ \emph
  {et~al.}(2020{\natexlab{a}})\citenamefont {Acharya} \emph
  {et~al.}}]{ALICE:2019qyj}%
  \BibitemOpen
  \bibfield  {author} {\bibinfo {author} {\bibfnamefont {Shreyasi}\
  \bibnamefont {Acharya}} \emph {et~al.} (\bibinfo {collaboration} {ALICE}),\
  }\bibfield  {title} {\enquote {\bibinfo {title} {{Measurements of inclusive
  jet spectra in pp and central Pb-Pb collisions at $\sqrt{s_{\rm{NN}}}$ = 5.02
  TeV}},}\ }\href {\doibase 10.1103/PhysRevC.101.034911} {\bibfield  {journal}
  {\bibinfo  {journal} {Phys. Rev. C}\ }\textbf {\bibinfo {volume} {101}},\
  \bibinfo {pages} {034911} (\bibinfo {year} {2020}{\natexlab{a}})},\ \Eprint
  {http://arxiv.org/abs/1909.09718} {arXiv:1909.09718 [nucl-ex]} \BibitemShut
  {NoStop}%
\bibitem [{\citenamefont {Khachatryan}\ \emph {et~al.}(2017)\citenamefont
  {Khachatryan} \emph {et~al.}}]{CMS:2016uxf}%
  \BibitemOpen
  \bibfield  {author} {\bibinfo {author} {\bibfnamefont {Vardan}\ \bibnamefont
  {Khachatryan}} \emph {et~al.} (\bibinfo {collaboration} {CMS}),\ }\bibfield
  {title} {\enquote {\bibinfo {title} {{Measurement of inclusive jet cross
  sections in $pp$ and PbPb collisions at $\sqrt{s_{NN}}=$ 2.76 TeV}},}\ }\href
  {\doibase 10.1103/PhysRevC.96.015202} {\bibfield  {journal} {\bibinfo
  {journal} {Phys. Rev. C}\ }\textbf {\bibinfo {volume} {96}},\ \bibinfo
  {pages} {015202} (\bibinfo {year} {2017})},\ \Eprint
  {http://arxiv.org/abs/1609.05383} {arXiv:1609.05383 [nucl-ex]} \BibitemShut
  {NoStop}%
\bibitem [{\citenamefont {Sirunyan}\ \emph {et~al.}(2021)\citenamefont
  {Sirunyan} \emph {et~al.}}]{CMS:2021vui}%
  \BibitemOpen
  \bibfield  {author} {\bibinfo {author} {\bibfnamefont {Albert~M}\
  \bibnamefont {Sirunyan}} \emph {et~al.} (\bibinfo {collaboration} {CMS}),\
  }\bibfield  {title} {\enquote {\bibinfo {title} {{First measurement of large
  area jet transverse momentum spectra in heavy-ion collisions}},}\ }\href
  {\doibase 10.1007/JHEP05(2021)284} {\bibfield  {journal} {\bibinfo  {journal}
  {JHEP}\ }\textbf {\bibinfo {volume} {05}},\ \bibinfo {pages} {284} (\bibinfo
  {year} {2021})},\ \Eprint {http://arxiv.org/abs/2102.13080} {arXiv:2102.13080
  [hep-ex]} \BibitemShut {NoStop}%
\bibitem [{\citenamefont {Aaboud}\ \emph
  {et~al.}(2019{\natexlab{b}})\citenamefont {Aaboud} \emph
  {et~al.}}]{ATLAS:2018dgb}%
  \BibitemOpen
  \bibfield  {author} {\bibinfo {author} {\bibfnamefont {Morad}\ \bibnamefont
  {Aaboud}} \emph {et~al.} (\bibinfo {collaboration} {ATLAS}),\ }\bibfield
  {title} {\enquote {\bibinfo {title} {{Measurement of photon\textendash{}jet
  transverse momentum correlations in 5.02 TeV Pb + Pb and $pp$ collisions with
  ATLAS}},}\ }\href {\doibase 10.1016/j.physletb.2018.12.023} {\bibfield
  {journal} {\bibinfo  {journal} {Phys. Lett. B}\ }\textbf {\bibinfo {volume}
  {789}},\ \bibinfo {pages} {167--190} (\bibinfo {year}
  {2019}{\natexlab{b}})},\ \Eprint {http://arxiv.org/abs/1809.07280}
  {arXiv:1809.07280 [nucl-ex]} \BibitemShut {NoStop}%
\bibitem [{\citenamefont {Aad}\ \emph {et~al.}(2023)\citenamefont {Aad} \emph
  {et~al.}}]{ATLAS:2023iad}%
  \BibitemOpen
  \bibfield  {author} {\bibinfo {author} {\bibfnamefont {Georges}\ \bibnamefont
  {Aad}} \emph {et~al.} (\bibinfo {collaboration} {ATLAS}),\ }\bibfield
  {title} {\enquote {\bibinfo {title} {{Comparison of inclusive and
  photon-tagged jet suppression in 5.02 TeV Pb+Pb collisions with ATLAS}},}\
  }\href {\doibase 10.1016/j.physletb.2023.138154} {\bibfield  {journal}
  {\bibinfo  {journal} {Phys. Lett. B}\ }\textbf {\bibinfo {volume} {846}},\
  \bibinfo {pages} {138154} (\bibinfo {year} {2023})},\ \Eprint
  {http://arxiv.org/abs/2303.10090} {arXiv:2303.10090 [nucl-ex]} \BibitemShut
  {NoStop}%
\bibitem [{\citenamefont {Sirunyan}\ \emph
  {et~al.}(2018{\natexlab{a}})\citenamefont {Sirunyan} \emph
  {et~al.}}]{CMS:2017ehl}%
  \BibitemOpen
  \bibfield  {author} {\bibinfo {author} {\bibfnamefont {Albert~M}\
  \bibnamefont {Sirunyan}} \emph {et~al.} (\bibinfo {collaboration} {CMS}),\
  }\bibfield  {title} {\enquote {\bibinfo {title} {{Study of jet quenching with
  isolated-photon+jet correlations in PbPb and pp collisions at
  $\sqrt{s_{_{\mathrm{NN}}}} =$ 5.02 TeV}},}\ }\href {\doibase
  10.1016/j.physletb.2018.07.061} {\bibfield  {journal} {\bibinfo  {journal}
  {Phys. Lett. B}\ }\textbf {\bibinfo {volume} {785}},\ \bibinfo {pages}
  {14--39} (\bibinfo {year} {2018}{\natexlab{a}})},\ \Eprint
  {http://arxiv.org/abs/1711.09738} {arXiv:1711.09738 [nucl-ex]} \BibitemShut
  {NoStop}%
\bibitem [{\citenamefont {Adamczyk}\ \emph {et~al.}(2016)\citenamefont
  {Adamczyk} \emph {et~al.}}]{STAR:2016jdz}%
  \BibitemOpen
  \bibfield  {author} {\bibinfo {author} {\bibfnamefont {L.}~\bibnamefont
  {Adamczyk}} \emph {et~al.} (\bibinfo {collaboration} {STAR}),\ }\bibfield
  {title} {\enquote {\bibinfo {title} {{Jet-like Correlations with
  Direct-Photon and Neutral-Pion Triggers at $\sqrt{s_{_{NN}}} = 200$ GeV}},}\
  }\href {\doibase 10.1016/j.physletb.2016.07.046} {\bibfield  {journal}
  {\bibinfo  {journal} {Phys. Lett. B}\ }\textbf {\bibinfo {volume} {760}},\
  \bibinfo {pages} {689--696} (\bibinfo {year} {2016})},\ \Eprint
  {http://arxiv.org/abs/1604.01117} {arXiv:1604.01117 [nucl-ex]} \BibitemShut
  {NoStop}%
\bibitem [{\citenamefont {Acharya}\ \emph
  {et~al.}(2020{\natexlab{b}})\citenamefont {Acharya} \emph
  {et~al.}}]{PHENIX:2020alr}%
  \BibitemOpen
  \bibfield  {author} {\bibinfo {author} {\bibfnamefont {U.}~\bibnamefont
  {Acharya}} \emph {et~al.} (\bibinfo {collaboration} {PHENIX}),\ }\bibfield
  {title} {\enquote {\bibinfo {title} {{Measurement of jet-medium interactions
  via direct photon-hadron correlations in Au$+$Au and $d$ $+$Au collisions at
  $\sqrt{s_{_{NN}}}=200$ GeV}},}\ }\href {\doibase 10.1103/PhysRevC.102.054910}
  {\bibfield  {journal} {\bibinfo  {journal} {Phys. Rev. C}\ }\textbf {\bibinfo
  {volume} {102}},\ \bibinfo {pages} {054910} (\bibinfo {year}
  {2020}{\natexlab{b}})},\ \Eprint {http://arxiv.org/abs/2005.14270}
  {arXiv:2005.14270 [hep-ex]} \BibitemShut {NoStop}%
\bibitem [{\citenamefont {Sirunyan}\ \emph
  {et~al.}(2018{\natexlab{b}})\citenamefont {Sirunyan} \emph
  {et~al.}}]{CMS:2017xgk}%
  \BibitemOpen
  \bibfield  {author} {\bibinfo {author} {\bibfnamefont {A.~M.}\ \bibnamefont
  {Sirunyan}} \emph {et~al.} (\bibinfo {collaboration} {CMS}),\ }\bibfield
  {title} {\enquote {\bibinfo {title} {{Azimuthal anisotropy of charged
  particles with transverse momentum up to 100 GeV/ c in PbPb collisions at
  $\sqrt {s}_{{NN}}$=5.02 TeV}},}\ }\href {\doibase
  10.1016/j.physletb.2017.11.041} {\bibfield  {journal} {\bibinfo  {journal}
  {Phys. Lett. B}\ }\textbf {\bibinfo {volume} {776}},\ \bibinfo {pages}
  {195--216} (\bibinfo {year} {2018}{\natexlab{b}})},\ \Eprint
  {http://arxiv.org/abs/1702.00630} {arXiv:1702.00630 [hep-ex]} \BibitemShut
  {NoStop}%
\bibitem [{\citenamefont {Chatrchyan}\ \emph {et~al.}(2012)\citenamefont
  {Chatrchyan} \emph {et~al.}}]{CMS:2012tqw}%
  \BibitemOpen
  \bibfield  {author} {\bibinfo {author} {\bibfnamefont {Serguei}\ \bibnamefont
  {Chatrchyan}} \emph {et~al.} (\bibinfo {collaboration} {CMS}),\ }\bibfield
  {title} {\enquote {\bibinfo {title} {{Azimuthal Anisotropy of Charged
  Particles at High Transverse Momenta in PbPb Collisions at $\sqrt{s_{NN}}=
  2.76$ TeV}},}\ }\href {\doibase 10.1103/PhysRevLett.109.022301} {\bibfield
  {journal} {\bibinfo  {journal} {Phys. Rev. Lett.}\ }\textbf {\bibinfo
  {volume} {109}},\ \bibinfo {pages} {022301} (\bibinfo {year} {2012})},\
  \Eprint {http://arxiv.org/abs/1204.1850} {arXiv:1204.1850 [nucl-ex]}
  \BibitemShut {NoStop}%
\bibitem [{\citenamefont {Aaboud}\ \emph {et~al.}(2018)\citenamefont {Aaboud}
  \emph {et~al.}}]{ATLAS:2018ezv}%
  \BibitemOpen
  \bibfield  {author} {\bibinfo {author} {\bibfnamefont {Morad}\ \bibnamefont
  {Aaboud}} \emph {et~al.} (\bibinfo {collaboration} {ATLAS}),\ }\bibfield
  {title} {\enquote {\bibinfo {title} {{Measurement of the azimuthal anisotropy
  of charged particles produced in $\sqrt{s_{_\text {NN}}}$ = 5.02 TeV Pb+Pb
  collisions with the ATLAS detector}},}\ }\href {\doibase
  10.1140/epjc/s10052-018-6468-7} {\bibfield  {journal} {\bibinfo  {journal}
  {Eur. Phys. J. C}\ }\textbf {\bibinfo {volume} {78}},\ \bibinfo {pages} {997}
  (\bibinfo {year} {2018})},\ \Eprint {http://arxiv.org/abs/1808.03951}
  {arXiv:1808.03951 [nucl-ex]} \BibitemShut {NoStop}%
\bibitem [{\citenamefont {Putschke}\ \emph {et~al.}(2019)\citenamefont
  {Putschke} \emph {et~al.}}]{Putschke:2019yrg}%
  \BibitemOpen
  \bibfield  {author} {\bibinfo {author} {\bibfnamefont {J.~H.}\ \bibnamefont
  {Putschke}} \emph {et~al.},\ }\bibfield  {title} {\enquote {\bibinfo {title}
  {{The JETSCAPE framework}},}\ }\href@noop {} {\  (\bibinfo {year} {2019})},\
  \Eprint {http://arxiv.org/abs/1903.07706} {arXiv:1903.07706 [nucl-th]}
  \BibitemShut {NoStop}%
\bibitem [{\citenamefont {Kumar}\ \emph {et~al.}(2020)\citenamefont {Kumar}
  \emph {et~al.}}]{JETSCAPE:2019udz}%
  \BibitemOpen
  \bibfield  {author} {\bibinfo {author} {\bibfnamefont {A.}~\bibnamefont
  {Kumar}} \emph {et~al.} (\bibinfo {collaboration} {JETSCAPE}),\ }\bibfield
  {title} {\enquote {\bibinfo {title} {{JETSCAPE framework: $p+p$ results}},}\
  }\href {\doibase 10.1103/PhysRevC.102.054906} {\bibfield  {journal} {\bibinfo
   {journal} {Phys. Rev. C}\ }\textbf {\bibinfo {volume} {102}},\ \bibinfo
  {pages} {054906} (\bibinfo {year} {2020})},\ \Eprint
  {http://arxiv.org/abs/1910.05481} {arXiv:1910.05481 [nucl-th]} \BibitemShut
  {NoStop}%
\bibitem [{\citenamefont {Kumar}\ \emph {et~al.}(2023)\citenamefont {Kumar}
  \emph {et~al.}}]{JETSCAPE:2022jer}%
  \BibitemOpen
  \bibfield  {author} {\bibinfo {author} {\bibfnamefont {A.}~\bibnamefont
  {Kumar}} \emph {et~al.} (\bibinfo {collaboration} {JETSCAPE}),\ }\bibfield
  {title} {\enquote {\bibinfo {title} {{Inclusive jet and hadron suppression in
  a multistage approach}},}\ }\href {\doibase 10.1103/PhysRevC.107.034911}
  {\bibfield  {journal} {\bibinfo  {journal} {Phys. Rev. C}\ }\textbf {\bibinfo
  {volume} {107}},\ \bibinfo {pages} {034911} (\bibinfo {year} {2023})},\
  \Eprint {http://arxiv.org/abs/2204.01163} {arXiv:2204.01163 [hep-ph]}
  \BibitemShut {NoStop}%
\bibitem [{\citenamefont {Park}(2019)}]{Park:2019sdn}%
  \BibitemOpen
  \bibfield  {author} {\bibinfo {author} {\bibfnamefont {Chanwook}\
  \bibnamefont {Park}} (\bibinfo {collaboration} {JETSCAPE}),\ }\bibfield
  {title} {\enquote {\bibinfo {title} {{Multi-stage jet evolution through QGP
  using the JETSCAPE framework: inclusive jets, correlations and leading
  hadrons}},}\ }\href {\doibase 10.22323/1.345.0072} {\bibfield  {journal}
  {\bibinfo  {journal} {PoS}\ }\textbf {\bibinfo {volume} {HardProbes2018}},\
  \bibinfo {pages} {072} (\bibinfo {year} {2019})},\ \Eprint
  {http://arxiv.org/abs/1902.05934} {arXiv:1902.05934 [nucl-th]} \BibitemShut
  {NoStop}%
\bibitem [{\citenamefont {Braaten}\ and\ \citenamefont
  {Pisarski}(1992)}]{Braaten:1991gm}%
  \BibitemOpen
  \bibfield  {author} {\bibinfo {author} {\bibfnamefont {Eric}\ \bibnamefont
  {Braaten}}\ and\ \bibinfo {author} {\bibfnamefont {Robert~D.}\ \bibnamefont
  {Pisarski}},\ }\bibfield  {title} {\enquote {\bibinfo {title} {{Simple
  effective Lagrangian for hard thermal loops}},}\ }\href {\doibase
  10.1103/PhysRevD.45.R1827} {\bibfield  {journal} {\bibinfo  {journal} {Phys.
  Rev. D}\ }\textbf {\bibinfo {volume} {45}},\ \bibinfo {pages} {R1827}
  (\bibinfo {year} {1992})}\BibitemShut {NoStop}%
\bibitem [{\citenamefont {Arnold}\ \emph
  {et~al.}(2003{\natexlab{a}})\citenamefont {Arnold}, \citenamefont {Moore},\
  and\ \citenamefont {Yaffe}}]{Arnold:2002zm}%
  \BibitemOpen
  \bibfield  {author} {\bibinfo {author} {\bibfnamefont {Peter~Brockway}\
  \bibnamefont {Arnold}}, \bibinfo {author} {\bibfnamefont {Guy~D.}\
  \bibnamefont {Moore}}, \ and\ \bibinfo {author} {\bibfnamefont {Laurence~G.}\
  \bibnamefont {Yaffe}},\ }\bibfield  {title} {\enquote {\bibinfo {title}
  {{Effective kinetic theory for high temperature gauge theories}},}\ }\href
  {\doibase 10.1088/1126-6708/2003/01/030} {\bibfield  {journal} {\bibinfo
  {journal} {JHEP}\ }\textbf {\bibinfo {volume} {01}},\ \bibinfo {pages} {030}
  (\bibinfo {year} {2003}{\natexlab{a}})},\ \Eprint
  {http://arxiv.org/abs/hep-ph/0209353} {arXiv:hep-ph/0209353} \BibitemShut
  {NoStop}%
\bibitem [{\citenamefont {Arnold}\ \emph {et~al.}(2001)\citenamefont {Arnold},
  \citenamefont {Moore},\ and\ \citenamefont {Yaffe}}]{Arnold:2001ms}%
  \BibitemOpen
  \bibfield  {author} {\bibinfo {author} {\bibfnamefont {Peter~Brockway}\
  \bibnamefont {Arnold}}, \bibinfo {author} {\bibfnamefont {Guy~D.}\
  \bibnamefont {Moore}}, \ and\ \bibinfo {author} {\bibfnamefont {Laurence~G.}\
  \bibnamefont {Yaffe}},\ }\bibfield  {title} {\enquote {\bibinfo {title}
  {{Photon emission from quark gluon plasma: Complete leading order
  results}},}\ }\href {\doibase 10.1088/1126-6708/2001/12/009} {\bibfield
  {journal} {\bibinfo  {journal} {JHEP}\ }\textbf {\bibinfo {volume} {12}},\
  \bibinfo {pages} {009} (\bibinfo {year} {2001})},\ \Eprint
  {http://arxiv.org/abs/hep-ph/0111107} {arXiv:hep-ph/0111107} \BibitemShut
  {NoStop}%
\bibitem [{\citenamefont {Ghiglieri}\ \emph {et~al.}(2013)\citenamefont
  {Ghiglieri}, \citenamefont {Hong}, \citenamefont {Kurkela}, \citenamefont
  {Lu}, \citenamefont {Moore},\ and\ \citenamefont
  {Teaney}}]{Ghiglieri:2013gia}%
  \BibitemOpen
  \bibfield  {author} {\bibinfo {author} {\bibfnamefont {Jacopo}\ \bibnamefont
  {Ghiglieri}}, \bibinfo {author} {\bibfnamefont {Juhee}\ \bibnamefont {Hong}},
  \bibinfo {author} {\bibfnamefont {Aleksi}\ \bibnamefont {Kurkela}}, \bibinfo
  {author} {\bibfnamefont {Egang}\ \bibnamefont {Lu}}, \bibinfo {author}
  {\bibfnamefont {Guy~D.}\ \bibnamefont {Moore}}, \ and\ \bibinfo {author}
  {\bibfnamefont {Derek}\ \bibnamefont {Teaney}},\ }\bibfield  {title}
  {\enquote {\bibinfo {title} {{Next-to-leading order thermal photon production
  in a weakly coupled quark-gluon plasma}},}\ }\href {\doibase
  10.1007/JHEP05(2013)010} {\bibfield  {journal} {\bibinfo  {journal} {JHEP}\
  }\textbf {\bibinfo {volume} {05}},\ \bibinfo {pages} {010} (\bibinfo {year}
  {2013})},\ \Eprint {http://arxiv.org/abs/1302.5970} {arXiv:1302.5970
  [hep-ph]} \BibitemShut {NoStop}%
\bibitem [{\citenamefont {Caron-Huot}(2009)}]{Caron-Huot:2008zna}%
  \BibitemOpen
  \bibfield  {author} {\bibinfo {author} {\bibfnamefont {Simon}\ \bibnamefont
  {Caron-Huot}},\ }\bibfield  {title} {\enquote {\bibinfo {title} {{O(g) plasma
  effects in jet quenching}},}\ }\href {\doibase 10.1103/PhysRevD.79.065039}
  {\bibfield  {journal} {\bibinfo  {journal} {Phys. Rev. D}\ }\textbf {\bibinfo
  {volume} {79}},\ \bibinfo {pages} {065039} (\bibinfo {year} {2009})},\
  \Eprint {http://arxiv.org/abs/0811.1603} {arXiv:0811.1603 [hep-ph]}
  \BibitemShut {NoStop}%
\bibitem [{\citenamefont {Jackson}\ and\ \citenamefont
  {Laine}(2019)}]{Jackson:2019yao}%
  \BibitemOpen
  \bibfield  {author} {\bibinfo {author} {\bibfnamefont {G.}~\bibnamefont
  {Jackson}}\ and\ \bibinfo {author} {\bibfnamefont {M.}~\bibnamefont
  {Laine}},\ }\bibfield  {title} {\enquote {\bibinfo {title} {{Testing thermal
  photon and dilepton rates}},}\ }\href {\doibase 10.1007/JHEP11(2019)144}
  {\bibfield  {journal} {\bibinfo  {journal} {JHEP}\ }\textbf {\bibinfo
  {volume} {11}},\ \bibinfo {pages} {144} (\bibinfo {year} {2019})},\ \Eprint
  {http://arxiv.org/abs/1910.09567} {arXiv:1910.09567 [hep-ph]} \BibitemShut
  {NoStop}%
\bibitem [{\citenamefont {Ali}\ \emph {et~al.}(2024)\citenamefont {Ali},
  \citenamefont {Bala}, \citenamefont {Francis}, \citenamefont {Jackson},
  \citenamefont {Kaczmarek}, \citenamefont {Turnwald}, \citenamefont {Ueding},\
  and\ \citenamefont {Wink}}]{Ali:2024xae}%
  \BibitemOpen
  \bibfield  {author} {\bibinfo {author} {\bibfnamefont {Sajid}\ \bibnamefont
  {Ali}}, \bibinfo {author} {\bibfnamefont {Dibyendu}\ \bibnamefont {Bala}},
  \bibinfo {author} {\bibfnamefont {Anthony}\ \bibnamefont {Francis}}, \bibinfo
  {author} {\bibfnamefont {Greg}\ \bibnamefont {Jackson}}, \bibinfo {author}
  {\bibfnamefont {Olaf}\ \bibnamefont {Kaczmarek}}, \bibinfo {author}
  {\bibfnamefont {Jonas}\ \bibnamefont {Turnwald}}, \bibinfo {author}
  {\bibfnamefont {Tristan}\ \bibnamefont {Ueding}}, \ and\ \bibinfo {author}
  {\bibfnamefont {Nicolas}\ \bibnamefont {Wink}} (\bibinfo {collaboration}
  {HotQCD}),\ }\bibfield  {title} {\enquote {\bibinfo {title} {{Lattice QCD
  estimates of thermal photon production from the QGP}},}\ }\href {\doibase
  10.1103/PhysRevD.110.054518} {\bibfield  {journal} {\bibinfo  {journal}
  {Phys. Rev. D}\ }\textbf {\bibinfo {volume} {110}},\ \bibinfo {pages}
  {054518} (\bibinfo {year} {2024})},\ \Eprint
  {http://arxiv.org/abs/2403.11647} {arXiv:2403.11647 [hep-lat]} \BibitemShut
  {NoStop}%
\bibitem [{\citenamefont {Schenke}\ \emph {et~al.}(2009)\citenamefont
  {Schenke}, \citenamefont {Gale},\ and\ \citenamefont
  {Jeon}}]{Schenke:2009gb}%
  \BibitemOpen
  \bibfield  {author} {\bibinfo {author} {\bibfnamefont {Bjoern}\ \bibnamefont
  {Schenke}}, \bibinfo {author} {\bibfnamefont {Charles}\ \bibnamefont {Gale}},
  \ and\ \bibinfo {author} {\bibfnamefont {Sangyong}\ \bibnamefont {Jeon}},\
  }\bibfield  {title} {\enquote {\bibinfo {title} {{MARTINI: An Event generator
  for relativistic heavy-ion collisions}},}\ }\href {\doibase
  10.1103/PhysRevC.80.054913} {\bibfield  {journal} {\bibinfo  {journal} {Phys.
  Rev. C}\ }\textbf {\bibinfo {volume} {80}},\ \bibinfo {pages} {054913}
  (\bibinfo {year} {2009})},\ \Eprint {http://arxiv.org/abs/0909.2037}
  {arXiv:0909.2037 [hep-ph]} \BibitemShut {NoStop}%
\bibitem [{\citenamefont {Yazdi}\ \emph {et~al.}(2023)\citenamefont {Yazdi},
  \citenamefont {Shi}, \citenamefont {Gale},\ and\ \citenamefont
  {Jeon}}]{Yazdi:2022cuk}%
  \BibitemOpen
  \bibfield  {author} {\bibinfo {author} {\bibfnamefont {Rouzbeh~Modarresi}\
  \bibnamefont {Yazdi}}, \bibinfo {author} {\bibfnamefont {Shuzhe}\
  \bibnamefont {Shi}}, \bibinfo {author} {\bibfnamefont {Charles}\ \bibnamefont
  {Gale}}, \ and\ \bibinfo {author} {\bibfnamefont {Sangyong}\ \bibnamefont
  {Jeon}},\ }\bibfield  {title} {\enquote {\bibinfo {title} {{Jet-medium
  Photons as a Probe of Parton Dynamics}},}\ }\href {\doibase
  10.5506/APhysPolBSupp.16.1-A129} {\bibfield  {journal} {\bibinfo  {journal}
  {Acta Phys. Polon. Supp.}\ }\textbf {\bibinfo {volume} {16}},\ \bibinfo
  {pages} {1--A129} (\bibinfo {year} {2023})},\ \Eprint
  {http://arxiv.org/abs/2207.12513} {arXiv:2207.12513 [hep-ph]} \BibitemShut
  {NoStop}%
\bibitem [{\citenamefont {Gale}\ \emph {et~al.}(2022)\citenamefont {Gale},
  \citenamefont {Paquet}, \citenamefont {Schenke},\ and\ \citenamefont
  {Shen}}]{Gale:2021emg}%
  \BibitemOpen
  \bibfield  {author} {\bibinfo {author} {\bibfnamefont {Charles}\ \bibnamefont
  {Gale}}, \bibinfo {author} {\bibfnamefont {Jean-Fran\c{c}ois}\ \bibnamefont
  {Paquet}}, \bibinfo {author} {\bibfnamefont {Bj\"orn}\ \bibnamefont
  {Schenke}}, \ and\ \bibinfo {author} {\bibfnamefont {Chun}\ \bibnamefont
  {Shen}},\ }\bibfield  {title} {\enquote {\bibinfo {title} {{Multimessenger
  heavy-ion collision physics}},}\ }\href {\doibase
  10.1103/PhysRevC.105.014909} {\bibfield  {journal} {\bibinfo  {journal}
  {Phys. Rev. C}\ }\textbf {\bibinfo {volume} {105}},\ \bibinfo {pages}
  {014909} (\bibinfo {year} {2022})},\ \Eprint
  {http://arxiv.org/abs/2106.11216} {arXiv:2106.11216 [nucl-th]} \BibitemShut
  {NoStop}%
\bibitem [{\citenamefont {Paquet}\ \emph {et~al.}(2016)\citenamefont {Paquet},
  \citenamefont {Shen}, \citenamefont {Denicol}, \citenamefont {Luzum},
  \citenamefont {Schenke}, \citenamefont {Jeon},\ and\ \citenamefont
  {Gale}}]{Paquet:2015lta}%
  \BibitemOpen
  \bibfield  {author} {\bibinfo {author} {\bibfnamefont {Jean-Fran\c{c}ois}\
  \bibnamefont {Paquet}}, \bibinfo {author} {\bibfnamefont {Chun}\ \bibnamefont
  {Shen}}, \bibinfo {author} {\bibfnamefont {Gabriel~S.}\ \bibnamefont
  {Denicol}}, \bibinfo {author} {\bibfnamefont {Matthew}\ \bibnamefont
  {Luzum}}, \bibinfo {author} {\bibfnamefont {Bj\"orn}\ \bibnamefont
  {Schenke}}, \bibinfo {author} {\bibfnamefont {Sangyong}\ \bibnamefont
  {Jeon}}, \ and\ \bibinfo {author} {\bibfnamefont {Charles}\ \bibnamefont
  {Gale}},\ }\bibfield  {title} {\enquote {\bibinfo {title} {{Production of
  photons in relativistic heavy-ion collisions}},}\ }\href {\doibase
  10.1103/PhysRevC.93.044906} {\bibfield  {journal} {\bibinfo  {journal} {Phys.
  Rev. C}\ }\textbf {\bibinfo {volume} {93}},\ \bibinfo {pages} {044906}
  (\bibinfo {year} {2016})},\ \Eprint {http://arxiv.org/abs/1509.06738}
  {arXiv:1509.06738 [hep-ph]} \BibitemShut {NoStop}%
\bibitem [{\citenamefont {G\"otz}\ \emph {et~al.}(2022)\citenamefont {G\"otz},
  \citenamefont {Sch\"afer}, \citenamefont {Garcia-Montero}, \citenamefont
  {Paquet}, \citenamefont {Elfner},\ and\ \citenamefont {Gale}}]{Gotz:2021dco}%
  \BibitemOpen
  \bibfield  {author} {\bibinfo {author} {\bibfnamefont {Niklas}\ \bibnamefont
  {G\"otz}}, \bibinfo {author} {\bibfnamefont {Anna}\ \bibnamefont
  {Sch\"afer}}, \bibinfo {author} {\bibfnamefont {Oscar}\ \bibnamefont
  {Garcia-Montero}}, \bibinfo {author} {\bibfnamefont {Jean-Fran\c{c}ois}\
  \bibnamefont {Paquet}}, \bibinfo {author} {\bibfnamefont {Hannah}\
  \bibnamefont {Elfner}}, \ and\ \bibinfo {author} {\bibfnamefont {Charles}\
  \bibnamefont {Gale}},\ }\bibfield  {title} {\enquote {\bibinfo {title}
  {{Out-of-equilibrium photon production in the late stages of relativistic
  heavy-ion collisions}},}\ }\href {\doibase 10.1103/PhysRevC.105.044910}
  {\bibfield  {journal} {\bibinfo  {journal} {Phys. Rev. C}\ }\textbf {\bibinfo
  {volume} {105}},\ \bibinfo {pages} {044910} (\bibinfo {year} {2022})},\
  \bibinfo {note} {[Erratum: Phys.Rev.C 109, 049901 (2024)]},\ \Eprint
  {http://arxiv.org/abs/2111.13603} {arXiv:2111.13603 [hep-ph]} \BibitemShut
  {NoStop}%
\bibitem [{\citenamefont {Arnold}\ \emph {et~al.}(2000)\citenamefont {Arnold},
  \citenamefont {Moore},\ and\ \citenamefont {Yaffe}}]{Arnold:2000dr}%
  \BibitemOpen
  \bibfield  {author} {\bibinfo {author} {\bibfnamefont {Peter~Brockway}\
  \bibnamefont {Arnold}}, \bibinfo {author} {\bibfnamefont {Guy~D.}\
  \bibnamefont {Moore}}, \ and\ \bibinfo {author} {\bibfnamefont {Laurence~G.}\
  \bibnamefont {Yaffe}},\ }\bibfield  {title} {\enquote {\bibinfo {title}
  {{Transport coefficients in high temperature gauge theories. 1. Leading log
  results}},}\ }\href {\doibase 10.1088/1126-6708/2000/11/001} {\bibfield
  {journal} {\bibinfo  {journal} {JHEP}\ }\textbf {\bibinfo {volume} {11}},\
  \bibinfo {pages} {001} (\bibinfo {year} {2000})},\ \Eprint
  {http://arxiv.org/abs/hep-ph/0010177} {arXiv:hep-ph/0010177} \BibitemShut
  {NoStop}%
\bibitem [{\citenamefont {Arnold}\ \emph
  {et~al.}(2003{\natexlab{b}})\citenamefont {Arnold}, \citenamefont {Moore},\
  and\ \citenamefont {Yaffe}}]{Arnold:2003zc}%
  \BibitemOpen
  \bibfield  {author} {\bibinfo {author} {\bibfnamefont {Peter~Brockway}\
  \bibnamefont {Arnold}}, \bibinfo {author} {\bibfnamefont {Guy~D}\
  \bibnamefont {Moore}}, \ and\ \bibinfo {author} {\bibfnamefont {Laurence~G.}\
  \bibnamefont {Yaffe}},\ }\bibfield  {title} {\enquote {\bibinfo {title}
  {{Transport coefficients in high temperature gauge theories. 2. Beyond
  leading log}},}\ }\href {\doibase 10.1088/1126-6708/2003/05/051} {\bibfield
  {journal} {\bibinfo  {journal} {JHEP}\ }\textbf {\bibinfo {volume} {05}},\
  \bibinfo {pages} {051} (\bibinfo {year} {2003}{\natexlab{b}})},\ \Eprint
  {http://arxiv.org/abs/hep-ph/0302165} {arXiv:hep-ph/0302165} \BibitemShut
  {NoStop}%
\bibitem [{\citenamefont {Abir}\ and\ \citenamefont
  {Majumder}(2016)}]{Abir:2015hta}%
  \BibitemOpen
  \bibfield  {author} {\bibinfo {author} {\bibfnamefont {Raktim}\ \bibnamefont
  {Abir}}\ and\ \bibinfo {author} {\bibfnamefont {Abhijit}\ \bibnamefont
  {Majumder}},\ }\bibfield  {title} {\enquote {\bibinfo {title} {{Drag-induced
  radiative energy loss from semihard heavy quarks}},}\ }\href {\doibase
  10.1103/PhysRevC.94.054902} {\bibfield  {journal} {\bibinfo  {journal} {Phys.
  Rev. C}\ }\textbf {\bibinfo {volume} {94}},\ \bibinfo {pages} {054902}
  (\bibinfo {year} {2016})},\ \Eprint {http://arxiv.org/abs/1506.08648}
  {arXiv:1506.08648 [nucl-th]} \BibitemShut {NoStop}%
\bibitem [{\citenamefont {Ghiglieri}\ \emph {et~al.}(2016)\citenamefont
  {Ghiglieri}, \citenamefont {Moore},\ and\ \citenamefont
  {Teaney}}]{Ghiglieri:2015ala}%
  \BibitemOpen
  \bibfield  {author} {\bibinfo {author} {\bibfnamefont {Jacopo}\ \bibnamefont
  {Ghiglieri}}, \bibinfo {author} {\bibfnamefont {Guy~D.}\ \bibnamefont
  {Moore}}, \ and\ \bibinfo {author} {\bibfnamefont {Derek}\ \bibnamefont
  {Teaney}},\ }\bibfield  {title} {\enquote {\bibinfo {title} {{Jet-Medium
  Interactions at NLO in a Weakly-Coupled Quark-Gluon Plasma}},}\ }\href
  {\doibase 10.1007/JHEP03(2016)095} {\bibfield  {journal} {\bibinfo  {journal}
  {JHEP}\ }\textbf {\bibinfo {volume} {03}},\ \bibinfo {pages} {095} (\bibinfo
  {year} {2016})},\ \Eprint {http://arxiv.org/abs/1509.07773} {arXiv:1509.07773
  [hep-ph]} \BibitemShut {NoStop}%
\bibitem [{\citenamefont {Qin}\ \emph {et~al.}(2009)\citenamefont {Qin},
  \citenamefont {Ruppert}, \citenamefont {Gale}, \citenamefont {Jeon},\ and\
  \citenamefont {Moore}}]{Qin:2008rd}%
  \BibitemOpen
  \bibfield  {author} {\bibinfo {author} {\bibfnamefont {G.~Y.}\ \bibnamefont
  {Qin}}, \bibinfo {author} {\bibfnamefont {J.}~\bibnamefont {Ruppert}},
  \bibinfo {author} {\bibfnamefont {Charles}\ \bibnamefont {Gale}}, \bibinfo
  {author} {\bibfnamefont {S.}~\bibnamefont {Jeon}}, \ and\ \bibinfo {author}
  {\bibfnamefont {G.~D.}\ \bibnamefont {Moore}},\ }\bibfield  {title} {\enquote
  {\bibinfo {title} {{Radiative and Collisional Energy Loss, and Photon-Tagged
  Jets at RHIC}},}\ }\href {\doibase 10.1140/epjc/s10052-008-0841-x} {\bibfield
   {journal} {\bibinfo  {journal} {Eur. Phys. J. C}\ }\textbf {\bibinfo
  {volume} {61}},\ \bibinfo {pages} {819--823} (\bibinfo {year} {2009})},\
  \Eprint {http://arxiv.org/abs/0809.2030} {arXiv:0809.2030 [hep-ph]}
  \BibitemShut {NoStop}%
\bibitem [{\citenamefont {Sirimanna}\ \emph {et~al.}(2022)\citenamefont
  {Sirimanna}, \citenamefont {Cao},\ and\ \citenamefont
  {Majumder}}]{Sirimanna:2021sqx}%
  \BibitemOpen
  \bibfield  {author} {\bibinfo {author} {\bibfnamefont {Chathuranga}\
  \bibnamefont {Sirimanna}}, \bibinfo {author} {\bibfnamefont {Shanshan}\
  \bibnamefont {Cao}}, \ and\ \bibinfo {author} {\bibfnamefont {Abhijit}\
  \bibnamefont {Majumder}},\ }\bibfield  {title} {\enquote {\bibinfo {title}
  {{Final-state gluon emission in deep-inelastic scattering at next-to-leading
  twist}},}\ }\href {\doibase 10.1103/PhysRevC.105.024908} {\bibfield
  {journal} {\bibinfo  {journal} {Phys. Rev. C}\ }\textbf {\bibinfo {volume}
  {105}},\ \bibinfo {pages} {024908} (\bibinfo {year} {2022})},\ \Eprint
  {http://arxiv.org/abs/2108.05329} {arXiv:2108.05329 [hep-ph]} \BibitemShut
  {NoStop}%
\bibitem [{\citenamefont {Schafer}\ \emph {et~al.}(2007)\citenamefont
  {Schafer}, \citenamefont {Wang},\ and\ \citenamefont
  {Zhang}}]{Schafer:2007xh}%
  \BibitemOpen
  \bibfield  {author} {\bibinfo {author} {\bibfnamefont {Andreas}\ \bibnamefont
  {Schafer}}, \bibinfo {author} {\bibfnamefont {Xin-Nian}\ \bibnamefont
  {Wang}}, \ and\ \bibinfo {author} {\bibfnamefont {Ben-Wei}\ \bibnamefont
  {Zhang}},\ }\bibfield  {title} {\enquote {\bibinfo {title} {{Multiple Parton
  Scattering in Nuclei: Quark-quark Scattering}},}\ }\href {\doibase
  10.1016/j.nuclphysa.2007.06.009} {\bibfield  {journal} {\bibinfo  {journal}
  {Nucl. Phys. A}\ }\textbf {\bibinfo {volume} {793}},\ \bibinfo {pages}
  {128--170} (\bibinfo {year} {2007})},\ \Eprint
  {http://arxiv.org/abs/0704.0106} {arXiv:0704.0106 [hep-ph]} \BibitemShut
  {NoStop}%
\bibitem [{\citenamefont {Collins}(2011)}]{Collins:2011zzd}%
  \BibitemOpen
  \bibfield  {author} {\bibinfo {author} {\bibfnamefont {John}\ \bibnamefont
  {Collins}},\ }\href {\doibase 10.1017/9781009401845} {\emph {\bibinfo {title}
  {{Foundations of Perturbative QCD}}}},\ Vol.~\bibinfo {volume} {32}\
  (\bibinfo  {publisher} {Cambridge University Press},\ \bibinfo {year}
  {2011})\BibitemShut {NoStop}%
\bibitem [{\citenamefont {Cutkosky}(1960)}]{Cutkosky:1960sp}%
  \BibitemOpen
  \bibfield  {author} {\bibinfo {author} {\bibfnamefont {R.~E.}\ \bibnamefont
  {Cutkosky}},\ }\bibfield  {title} {\enquote {\bibinfo {title} {{Singularities
  and discontinuities of Feynman amplitudes}},}\ }\href {\doibase
  10.1063/1.1703676} {\bibfield  {journal} {\bibinfo  {journal} {J. Math.
  Phys.}\ }\textbf {\bibinfo {volume} {1}},\ \bibinfo {pages} {429--433}
  (\bibinfo {year} {1960})}\BibitemShut {NoStop}%
\bibitem [{\citenamefont {Peskin}\ and\ \citenamefont
  {Schroeder}(1995)}]{Peskin:1995ev}%
  \BibitemOpen
  \bibfield  {author} {\bibinfo {author} {\bibfnamefont {Michael~E.}\
  \bibnamefont {Peskin}}\ and\ \bibinfo {author} {\bibfnamefont {Daniel~V.}\
  \bibnamefont {Schroeder}},\ }\href {\doibase 10.1201/9780429503559} {\emph
  {\bibinfo {title} {{An Introduction to quantum field theory}}}}\ (\bibinfo
  {publisher} {Addison-Wesley},\ \bibinfo {address} {Reading, USA},\ \bibinfo
  {year} {1995})\BibitemShut {NoStop}%
\bibitem [{\citenamefont {Kumar}\ \emph {et~al.}(2025)\citenamefont {Kumar},
  \citenamefont {Majumder}, \citenamefont {Sirimanna},\ and\ \citenamefont
  {Tachibana}}]{Kumar:2025rsa}%
  \BibitemOpen
  \bibfield  {author} {\bibinfo {author} {\bibfnamefont {Amit}\ \bibnamefont
  {Kumar}}, \bibinfo {author} {\bibfnamefont {Abhijit}\ \bibnamefont
  {Majumder}}, \bibinfo {author} {\bibfnamefont {Chathuranga}\ \bibnamefont
  {Sirimanna}}, \ and\ \bibinfo {author} {\bibfnamefont {Yasuki}\ \bibnamefont
  {Tachibana}},\ }\bibfield  {title} {\enquote {\bibinfo {title} {{Modified
  Coherence and the Transverse Extent of Jets}},}\ }\href@noop {} {\  (\bibinfo
  {year} {2025})},\ \Eprint {http://arxiv.org/abs/2501.07823} {arXiv:2501.07823
  [hep-ph]} \BibitemShut {NoStop}%
\bibitem [{\citenamefont {Wang}\ and\ \citenamefont
  {Guo}(2001)}]{Wang:2001ifa}%
  \BibitemOpen
  \bibfield  {author} {\bibinfo {author} {\bibfnamefont {Xin-Nian}\
  \bibnamefont {Wang}}\ and\ \bibinfo {author} {\bibfnamefont {Xiao-feng}\
  \bibnamefont {Guo}},\ }\bibfield  {title} {\enquote {\bibinfo {title}
  {{Multiple parton scattering in nuclei: Parton energy loss}},}\ }\href
  {\doibase 10.1016/S0375-9474(01)01130-7} {\bibfield  {journal} {\bibinfo
  {journal} {Nucl. Phys. A}\ }\textbf {\bibinfo {volume} {696}},\ \bibinfo
  {pages} {788--832} (\bibinfo {year} {2001})},\ \Eprint
  {http://arxiv.org/abs/hep-ph/0102230} {arXiv:hep-ph/0102230} \BibitemShut
  {NoStop}%
\bibitem [{\citenamefont {Kurkela}\ \emph
  {et~al.}(2019{\natexlab{a}})\citenamefont {Kurkela}, \citenamefont
  {Mazeliauskas}, \citenamefont {Paquet}, \citenamefont {Schlichting},\ and\
  \citenamefont {Teaney}}]{Kurkela:2018vqr}%
  \BibitemOpen
  \bibfield  {author} {\bibinfo {author} {\bibfnamefont {Aleksi}\ \bibnamefont
  {Kurkela}}, \bibinfo {author} {\bibfnamefont {Aleksas}\ \bibnamefont
  {Mazeliauskas}}, \bibinfo {author} {\bibfnamefont {Jean-Fran{\c{c}}ois}\
  \bibnamefont {Paquet}}, \bibinfo {author} {\bibfnamefont {S{\"o}ren}\
  \bibnamefont {Schlichting}}, \ and\ \bibinfo {author} {\bibfnamefont {Derek}\
  \bibnamefont {Teaney}},\ }\bibfield  {title} {\enquote {\bibinfo {title}
  {{Effective kinetic description of event-by-event pre-equilibrium dynamics in
  high-energy heavy-ion collisions}},}\ }\href {\doibase
  10.1103/PhysRevC.99.034910} {\bibfield  {journal} {\bibinfo  {journal} {Phys.
  Rev. C}\ }\textbf {\bibinfo {volume} {99}},\ \bibinfo {pages} {034910}
  (\bibinfo {year} {2019}{\natexlab{a}})},\ \Eprint
  {http://arxiv.org/abs/1805.00961} {arXiv:1805.00961 [hep-ph]} \BibitemShut
  {NoStop}%
\bibitem [{\citenamefont {Kurkela}\ \emph
  {et~al.}(2019{\natexlab{b}})\citenamefont {Kurkela}, \citenamefont
  {Mazeliauskas}, \citenamefont {Paquet}, \citenamefont {Schlichting},\ and\
  \citenamefont {Teaney}}]{Kurkela:2018wud}%
  \BibitemOpen
  \bibfield  {author} {\bibinfo {author} {\bibfnamefont {Aleksi}\ \bibnamefont
  {Kurkela}}, \bibinfo {author} {\bibfnamefont {Aleksas}\ \bibnamefont
  {Mazeliauskas}}, \bibinfo {author} {\bibfnamefont {Jean-Fran{\c{c}}ois}\
  \bibnamefont {Paquet}}, \bibinfo {author} {\bibfnamefont {S{\"o}ren}\
  \bibnamefont {Schlichting}}, \ and\ \bibinfo {author} {\bibfnamefont {Derek}\
  \bibnamefont {Teaney}},\ }\bibfield  {title} {\enquote {\bibinfo {title}
  {{Matching the Nonequilibrium Initial Stage of Heavy Ion Collisions to
  Hydrodynamics with QCD Kinetic Theory}},}\ }\href {\doibase
  10.1103/PhysRevLett.122.122302} {\bibfield  {journal} {\bibinfo  {journal}
  {Phys. Rev. Lett.}\ }\textbf {\bibinfo {volume} {122}},\ \bibinfo {pages}
  {122302} (\bibinfo {year} {2019}{\natexlab{b}})},\ \Eprint
  {http://arxiv.org/abs/1805.01604} {arXiv:1805.01604 [hep-ph]} \BibitemShut
  {NoStop}%
\bibitem [{\citenamefont {Giacalone}\ \emph {et~al.}(2019)\citenamefont
  {Giacalone}, \citenamefont {Mazeliauskas},\ and\ \citenamefont
  {Schlichting}}]{Giacalone:2019ldn}%
  \BibitemOpen
  \bibfield  {author} {\bibinfo {author} {\bibfnamefont {Giuliano}\
  \bibnamefont {Giacalone}}, \bibinfo {author} {\bibfnamefont {Aleksas}\
  \bibnamefont {Mazeliauskas}}, \ and\ \bibinfo {author} {\bibfnamefont
  {S{\"o}ren}\ \bibnamefont {Schlichting}},\ }\bibfield  {title} {\enquote
  {\bibinfo {title} {{Hydrodynamic attractors, initial state energy and
  particle production in relativistic nuclear collisions}},}\ }\href {\doibase
  10.1103/PhysRevLett.123.262301} {\bibfield  {journal} {\bibinfo  {journal}
  {Phys. Rev. Lett.}\ }\textbf {\bibinfo {volume} {123}},\ \bibinfo {pages}
  {262301} (\bibinfo {year} {2019})},\ \Eprint
  {http://arxiv.org/abs/1908.02866} {arXiv:1908.02866 [hep-ph]} \BibitemShut
  {NoStop}%
\bibitem [{\citenamefont {Kamata}\ \emph {et~al.}(2020)\citenamefont {Kamata},
  \citenamefont {Martinez}, \citenamefont {Plaschke}, \citenamefont
  {Ochsenfeld},\ and\ \citenamefont {Schlichting}}]{Kamata:2020mka}%
  \BibitemOpen
  \bibfield  {author} {\bibinfo {author} {\bibfnamefont {Syo}\ \bibnamefont
  {Kamata}}, \bibinfo {author} {\bibfnamefont {Mauricio}\ \bibnamefont
  {Martinez}}, \bibinfo {author} {\bibfnamefont {Philip}\ \bibnamefont
  {Plaschke}}, \bibinfo {author} {\bibfnamefont {Stephan}\ \bibnamefont
  {Ochsenfeld}}, \ and\ \bibinfo {author} {\bibfnamefont {S{\"o}ren}\
  \bibnamefont {Schlichting}},\ }\bibfield  {title} {\enquote {\bibinfo {title}
  {{Hydrodynamization and nonequilibrium Green{\textquoteright}s functions in
  kinetic theory}},}\ }\href {\doibase 10.1103/PhysRevD.102.056003} {\bibfield
  {journal} {\bibinfo  {journal} {Phys. Rev. D}\ }\textbf {\bibinfo {volume}
  {102}},\ \bibinfo {pages} {056003} (\bibinfo {year} {2020})},\ \Eprint
  {http://arxiv.org/abs/2004.06751} {arXiv:2004.06751 [hep-ph]} \BibitemShut
  {NoStop}%
\bibitem [{\citenamefont {Kumar}\ \emph {et~al.}(2022)\citenamefont {Kumar},
  \citenamefont {Majumder},\ and\ \citenamefont {Weber}}]{Kumar:2020wvb}%
  \BibitemOpen
  \bibfield  {author} {\bibinfo {author} {\bibfnamefont {Amit}\ \bibnamefont
  {Kumar}}, \bibinfo {author} {\bibfnamefont {Abhijit}\ \bibnamefont
  {Majumder}}, \ and\ \bibinfo {author} {\bibfnamefont {Johannes~Heinrich}\
  \bibnamefont {Weber}},\ }\bibfield  {title} {\enquote {\bibinfo {title} {{Jet
  transport coefficient q\textasciicircum{} in lattice QCD}},}\ }\href
  {\doibase 10.1103/PhysRevD.106.034505} {\bibfield  {journal} {\bibinfo
  {journal} {Phys. Rev. D}\ }\textbf {\bibinfo {volume} {106}},\ \bibinfo
  {pages} {034505} (\bibinfo {year} {2022})},\ \Eprint
  {http://arxiv.org/abs/2010.14463} {arXiv:2010.14463 [hep-lat]} \BibitemShut
  {NoStop}%
\bibitem [{PDG()}]{PDG_qmass}%
  \BibitemOpen
  \href@noop {} {\bibinfo  {journal}
  {https://pdg.lbl.gov/2022/tables/rpp2022-sum-quarks.pdf}\ }\BibitemShut
  {NoStop}%
\end{thebibliography}%
\end{document}